\documentclass[twocolumn,secnumarabic,amssymb, aps,prx,showpacs]{revtex4-2}

\newcommand{\subsubsubsection}[1]{\paragraph{#1}}
\newcommand{\revisecolor}[2]{\textcolor{#1}{#2}}
\usepackage{float}
\usepackage{comment}
\usepackage{amsmath,amsthm,amssymb}
\usepackage{listings} 
\usepackage{graphicx}
\usepackage{color}
\usepackage[abs]{overpic}
\usepackage[labelfont=bf]{caption} 
\usepackage{subcaption} 
\usepackage{xparse}

\ExplSyntaxOn
\NewDocumentCommand \emoji { m } { \emoji_bridge:n { #1 } }
\cs_new:Npn \emoji_bridge:n #1
  {
    \str_case:nnF { #1 }
      {
        {hollow-red-circle}{\coloremojicode{2B55}} 
        {anchor}{\coloremojicode{2693}}            
        {tent}{\coloremojicode{26FA}}              
        {high-voltage}{\coloremojicode{26A1}}      
        {baseball}{\coloremojicode{26BE}}          
        {part-alternation-mark}{\coloremojicode{303D}} 
      }
      { \coloremoji{#1} } 
  }
\ExplSyntaxOff

\usepackage{bxcoloremoji}
\usepackage{multirow}

\usepackage{newtxtext} 
\usepackage{newtxmath} 
\usepackage{tcolorbox}
\usepackage[utf8]{inputenc}
\usepackage{CJKutf8}

\newenvironment{JPlisting}{%
  \par\noindent
  \begin{tcolorbox}[colback=white,boxrule=0.5pt,sharp corners,
                    left=6pt,right=6pt,top=6pt,bottom=6pt]
  \ttfamily\small\setlength{\parindent}{0pt}%
  \raggedright\obeylines\obeyspaces
}{%
  \end{tcolorbox}%
}

\usepackage[colorlinks=true, allcolors=blue]{hyperref}

\makeatletter
\let\oldappendix\appendix

\renewcommand{\appendix}{%
  \oldappendix
  
  \gdef\p@subsection{\thesection.}%
  \gdef\p@subsubsection{\thesection.\thesubsection.}%
}
\makeatother

\begin{document}
\begin{CJK*}{UTF8}{min}
\title{
Sub-exponential Growth Dynamics in Complex Systems: A Piecewise Power-Law Model for the Diffusion of New Words and Names
}
\author{Hayafumi Watanabe$^{1,2}$}\email[E-mail: ]{hayafumi.watanabe@gmail.com}

\affiliation{Department of Economics, Seijo University, Setagaya-ku, Tokyo 157-8511, Japan}
\affiliation{The Institute of Statistical Mathematics, Tachikawa-shi, Tokyo 190-8562,Japan}

\begin{abstract}
The diffusion of ideas and language represents a typical example of collective dynamics in complex systems, conventionally described by S-shaped models such as the logistic function and its generalizations. However, the role of sub-exponential growth---a slower-than-exponential pattern known in epidemiology---has been largely overlooked in social diffusion processes. Here, we present a piecewise power-law model to characterize complex growth curves with a few parameters. We systematically analyzed a large-scale dataset of approximately one billion Japanese blog articles linked to Wikipedia vocabulary, and observed consistent patterns in web search trend data (English, Spanish, and Japanese). Our analysis of 2,963 items, selected for reliable estimation (e.g., sufficient duration/peak, monotonic growth), reveals that 1,625 (55\%) diffusion patterns without abrupt level shifts were adequately described by one or two segments. For single-segment curves, we found that (i) the mode of the shape parameter $\alpha$ was near 0.5, indicating prevalent sub-exponential growth; (ii) the peak diffusion scale is primarily determined by the growth rate $R$, with minor contributions from $\alpha$ or the duration $T$; and (iii) $\alpha$ showed a tendency to vary with the nature of the topic, being smaller for niche/local topics and larger for widely shared ones. Furthermore, a micro-behavioral model of outward (stranger) vs. inward (community) contact suggests that $\alpha$ can be interpreted as an index of the preference for outward-oriented communication. These findings suggest that sub-exponential growth is a common pattern of social diffusion, and our model provides a practical framework for consistently describing, comparing, and interpreting complex and diverse growth curves.
\end{abstract}
\maketitle
\section{Introduction}
Diffusion in complex systems is a central theme in statistical physics. Human society represents a typical complex system and network; therefore, analyzing information diffusion in this context is a significant challenge relevant to both fundamental physics and practical applications. As a result, researchers in diverse fields—ranging from physics and epidemiology to social and information sciences—have worked to describe and understand these phenomena. \par
Spreading phenomena—such as ideas, products, language dynamics, and infectious diseases—represent one of the most typical forms of diffusion in society.
A basic way to quantify these phenomena is to analyze growth curves that track the level of diffusion over time. In particular, S-shaped growth models, typified by the logistic function, are widely recognized as a common pattern because they appear across many systems, from social to biological \cite{fokas2007growth,mendez2015stochastic,ryan1943diffusion,denison2003log,marchetti1994millenarian}. For example, in sociology and management science, Rogers's theory of the diffusion of innovations \cite{Rogers2003Diffusion} and the Bass model that formalized it \cite{Bass1969} have served as classic, basic frameworks and are still often used today \cite{balash2020comparative}. \par
Recent advances in information technology and large-scale data analysis have opened new directions in this classic field. Particularly in the fields of complex systems physics and information science, many studies now investigate how factors such as social network structure and geographic or social constraints shape diffusion dynamics \cite{balash2020comparative, burridge2021inferring, tarrade2024position, lorenz2019accelerating, stewart2018making}. 
\revisecolor{black}{In the latest developments, research has expanded the scope beyond pairwise interactions, examining how higher-order interactions, such as peer pressure and group-level influences, are related to diffusion processes \cite{battiston2025higher}. These developments are closely related to the field of social physics, which aims to understand social phenomena using methods from statistical physics and network science. For comprehensive overviews of this approach, see Ref.~\cite{JUSUP20221}.} \par
At the same time, there is active work to extend the traditional S-shaped models to capture real-world complexity more precisely. Approaches include models that incorporate asymmetric curves \cite{ghanbarnejad2014extracting}, capture bursty or deadline-driven dynamics \cite{alfi2007conference}, utilize generalized logistic equations \cite{watanabe2023minor, vasconcelos2021power}, apply epidemic models to social phenomena \cite{sidorov2023extension, jiang2021neologisms}, and employ machine-learning-based classifications \cite{lehmann2012dynamical}. 
Despite these advances, sub-exponential growth has attracted limited interest in complex systems research, presumably due to the scarcity of empirically observed cases. A rare exception is the field of epidemiology, where sub-exponential growth is occasionally identified during the early stages of outbreaks. In this domain, sub-exponential growth—often analyzed using the General Growth Model—is often theoretically attributed to factors such as spatial effects and heterogeneity in individual susceptibility. 
However, in the realm of social processes—especially regarding the diffusion of innovations and culture—this pattern has been largely overlooked.
Reports of such growth in social contexts are rare, with examples mostly limited to specific case studies, such as shifts in the pronunciation of several words in Philadelphia \cite{labov2013one} and archaeological cases \cite{crema2024modelling}. \par
However, recent studies using large-scale language data suggest that sub-exponential growth may in fact be one of the main diffusion patterns in society \cite{watanabe2023minor}. This points to a need to revisit the conventional view of its importance. Yet its generality and properties remain largely unexplored. In particular, key questions are still open: (i) how the shape of the growth curve relates to the scale and duration of diffusion; (ii) what social meaning the curve's shape reflects; and (iii) how macro patterns arise from individual-level behavior. 
In addition, no established mathematical model existed to systematically describe the diverse growth curves observed in online language data, including sub-exponential growth. \par
To address these gaps, we present a simple, general representation: a piecewise power-law model (a piecewise generalized growth model). Using this model and a systematic analysis of large-scale online language data, we show that sub-exponential growth is one of the common patterns of social diffusion. We also investigate how the curve shape (the power exponent) relates to quantitative features such as growth scale and duration, and to qualitative aspects such as a word's topic appeal, through parameter analysis. 
Further, by connecting to a micro-level, infection-style model with the notion of ``inwardness'', we propose one mechanism by which individual interactions can generate macro-level sub-exponential curves. \par
Diffusion analysis utilizing large-scale language data finds broad application across diverse domains, from computational social science to marketing, spanning both academic disciplines and practical fields. Our findings offer a unified way to handle growth curves across these areas and provide a step toward a physics-based understanding of the common dynamics of social diffusion. \par
\begin{figure*}[t]
    \centering
    \begin{overpic}[width=5.7cm]{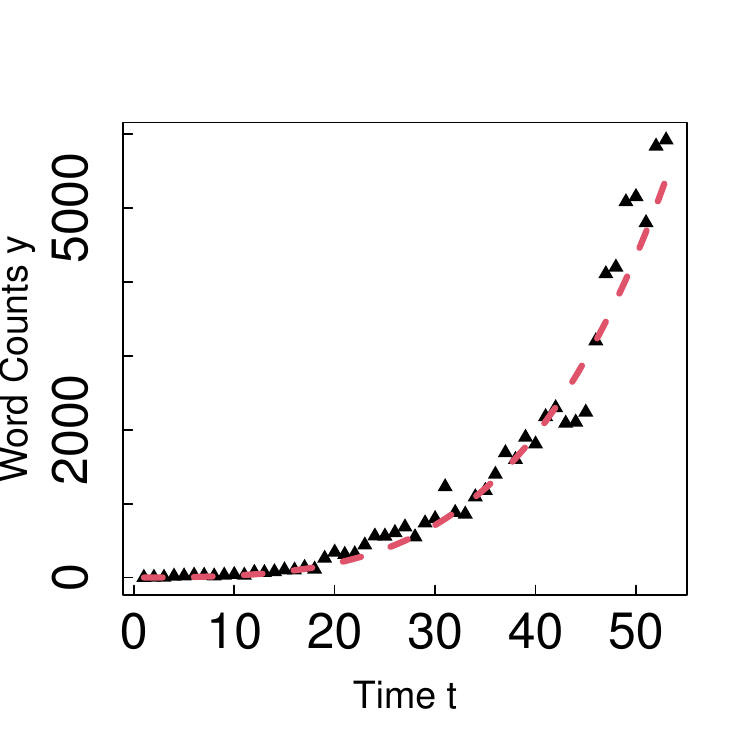}
        \put(30,120){\color{black}\Large\bfseries (a)}
    \end{overpic}
     \begin{overpic}[width=5.7cm]{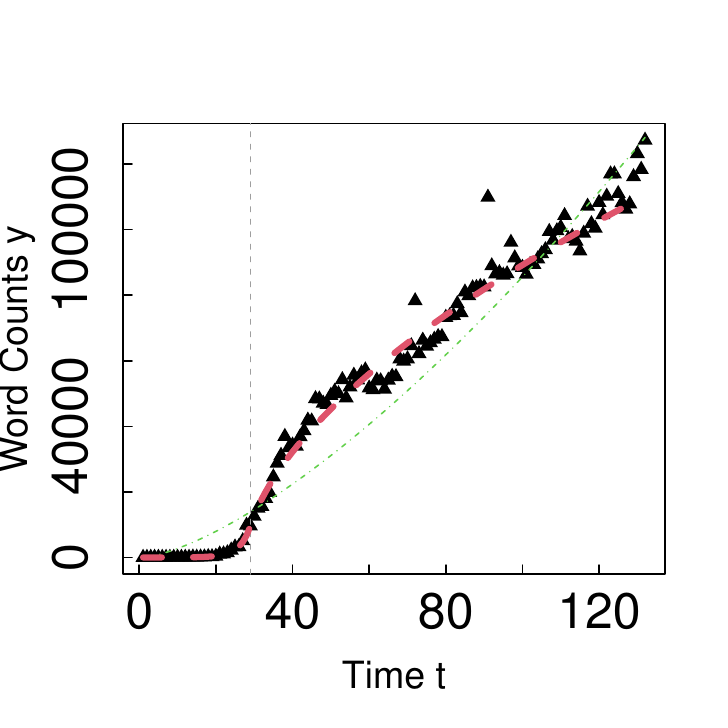}
        \put(30,120){\color{black}\Large\bfseries (b)}
    \end{overpic} 
     \begin{overpic}[width=5.7cm]{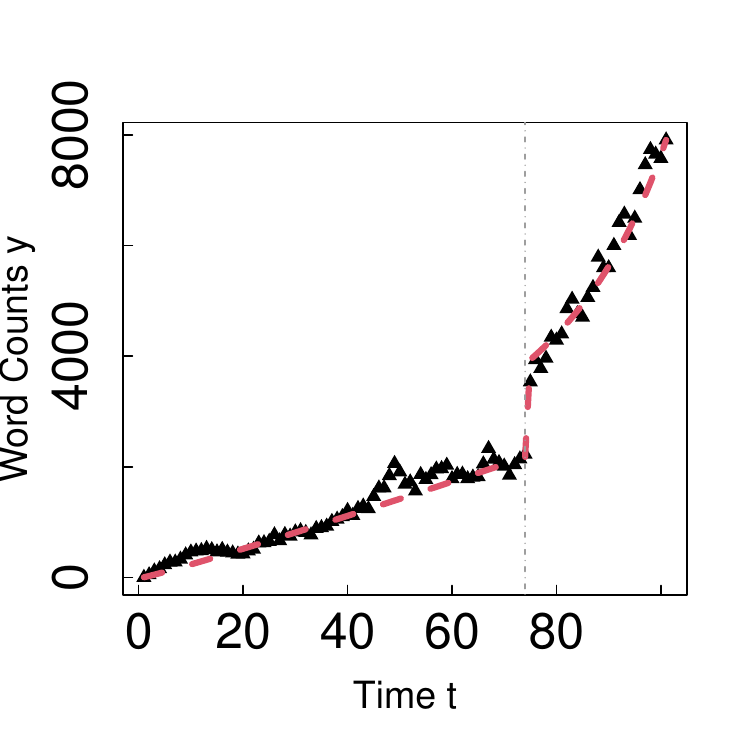}
        \put(30,120){\color{black}\Large\bfseries (c)}
    \end{overpic}
    \begin{overpic}[width=5.7cm]{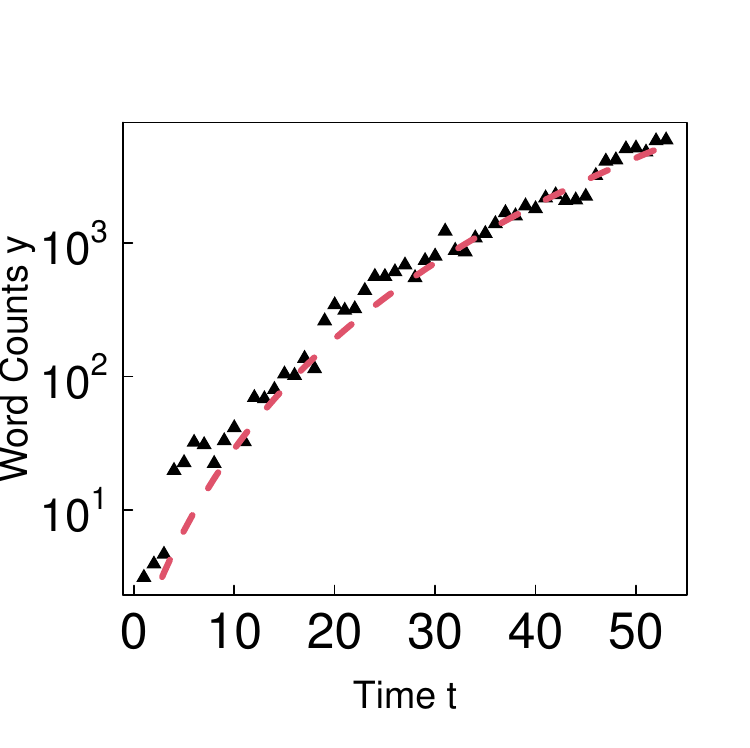}
        \put(30,120){\color{black}\Large\bfseries (d)}
    \end{overpic}
    \begin{overpic}[width=5.7cm]{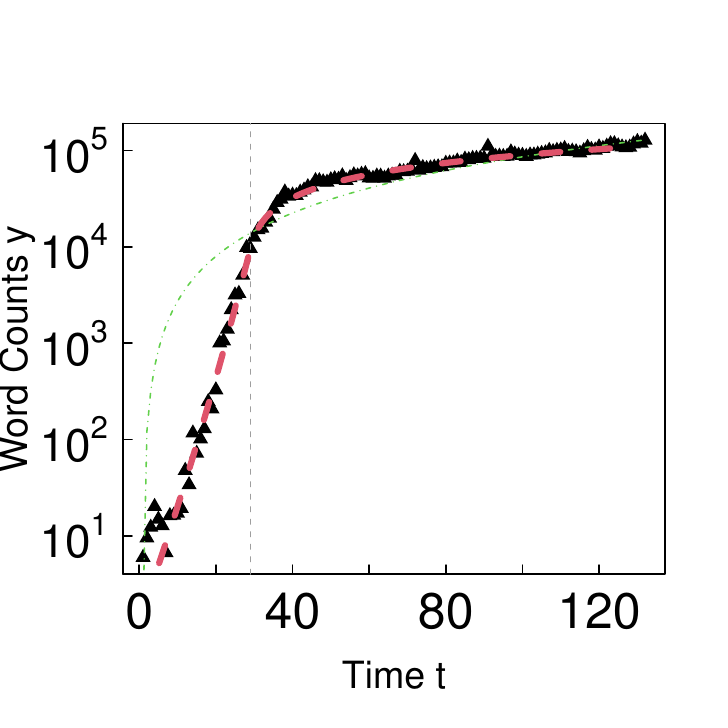}
        \put(30,120){\color{black}\Large\bfseries (e)}
    \end{overpic}
    \begin{overpic}[width=5.7cm]{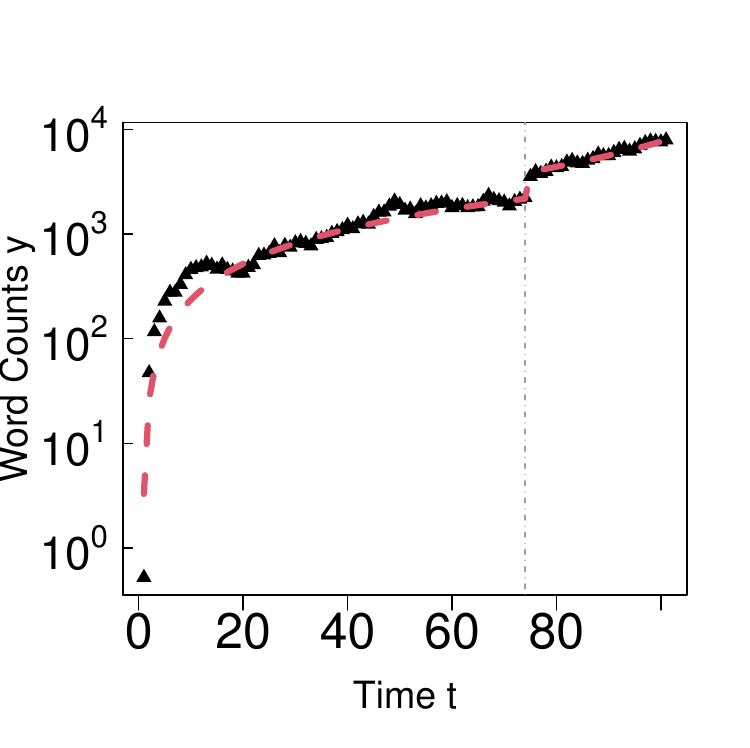}
        \put(30,120){\color{black}\Large\bfseries (f)}
    \end{overpic}
   \caption{Examples of keyword time series (normalized by the total number of articles; one step = 30 days).  Black triangles denote empirical data; the red dashed line is the piecewise power-law model; the green dotted line is the single power-law model (Eq.~\ref{eq_base0}).  Keywords are English translations; the original Japanese keywords are given in Appendix~\ref{app_sec_fig_word}.  (a) \textbf{``Low-cost SIM''}: $\alpha_i^{(1)}=0.77,\ R_i^{(1)}=0.22,\ T_i^{(1)}=53$. Adequately captured by the single power-law model (Section~\ref{sec_base}). (b) \textbf{``Smartphone''}: $\alpha_i^{(1)}=1.03$, $R_i^{(1)}=0.30$, $\alpha_i^{(2)}=-0.72$, $R_i^{(2)}=4411$, $T_i=131$. The single model (green dotted) is insufficient, but the continuous piecewise power-law model (Section~\ref{sec_kubun}) fits well. The changepoint is $t=29$ (late August 2009). (c) ``\emoji{hollow-red-circle}'' \textbf{(red circle emoji)}: $\alpha_i^{(1)}=0.066$, $R_i^{(1)}=0.59$, $\alpha_i^{(2)}=1.77$, $R_i^{(2)}=6.28\times 10^{-4}$. A typical case with a discontinuous jump at $t=74$ (September 2017) (see Section~\ref{sec_jump} for the model with jumps). The vertical gray line marks the jump time. 
   A potential contributing factor is improvements in emoji input tied to smartphone OS updates. (d)-(f) are the corresponding semi-log plots.
   }
    \label{fig_all_time_series}
\end{figure*}

\begin{figure*}[tp]
    \centering
    \begin{overpic}[width=5.0cm]{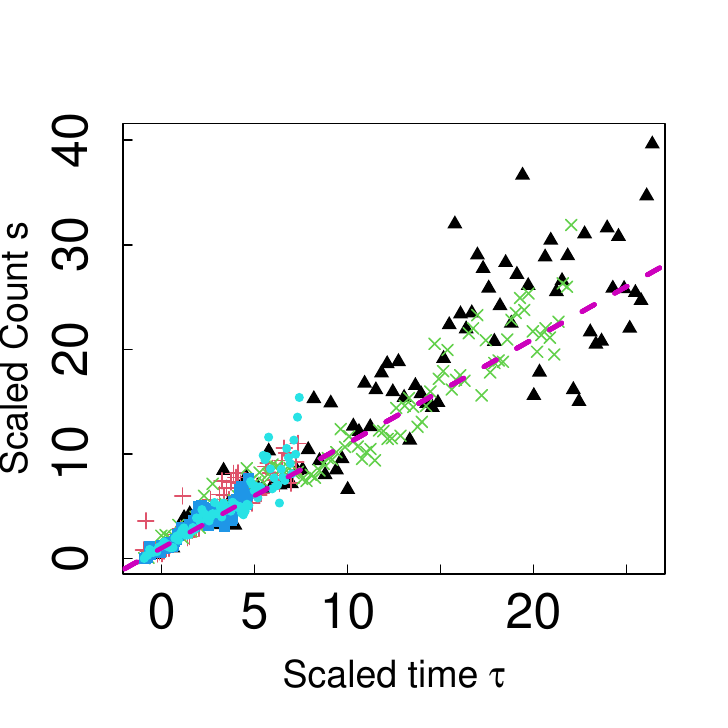}
        \put(30,103){\color{black}\Large\bfseries (a)}
    \end{overpic}
    \begin{overpic}[width=5.0cm]{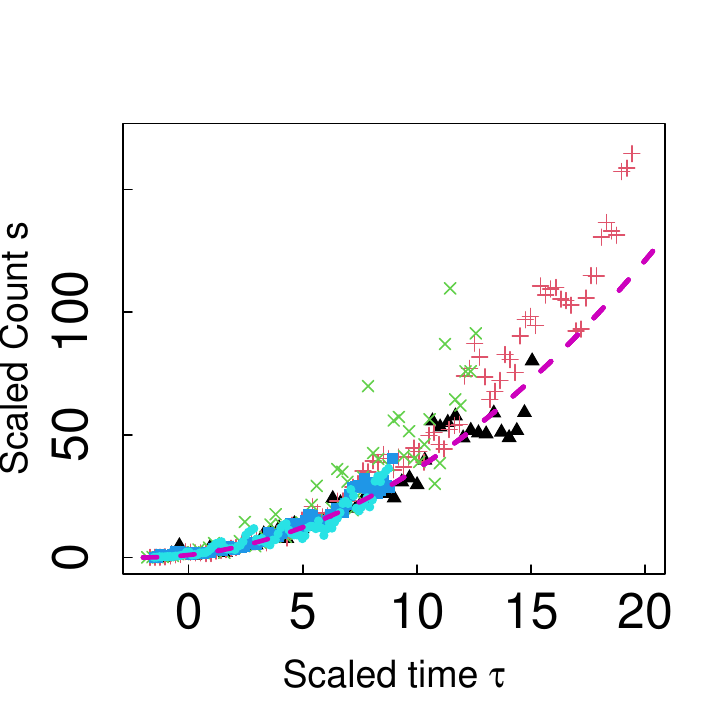}
        \put(30,103){\color{black}\Large\bfseries (b)}
    \end{overpic}
    \begin{overpic}[width=5.0cm]{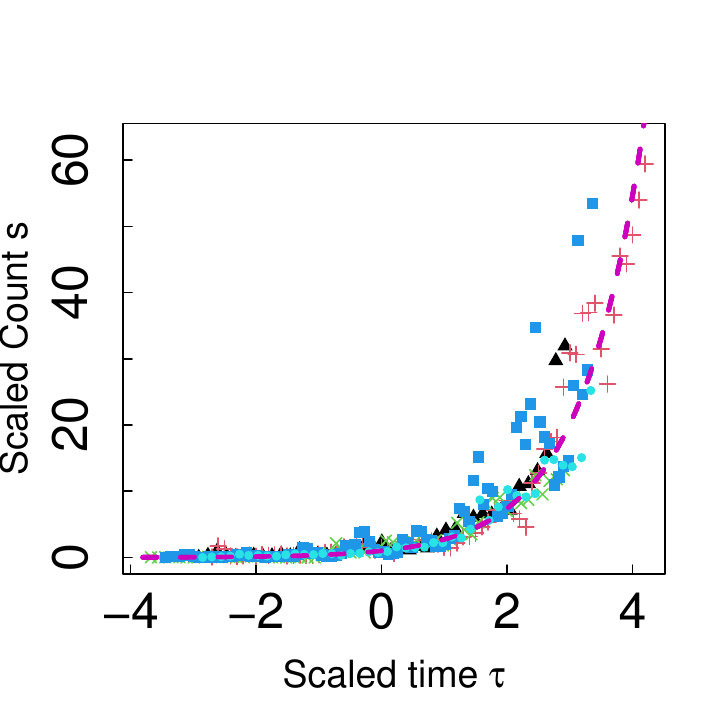}
        \put(30,103){\color{black}\Large\bfseries (c)}
    \end{overpic}
    \begin{overpic}[width=5.0cm]{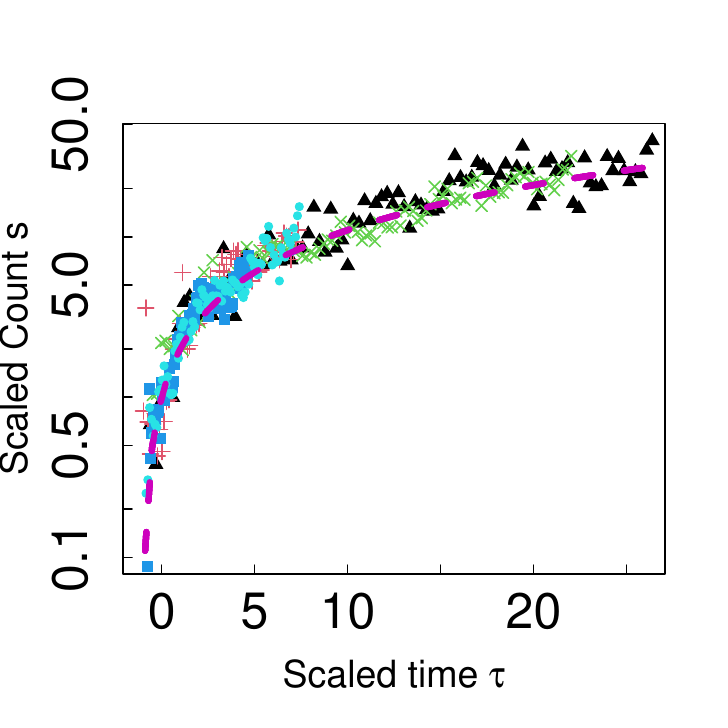}
        \put(30,103){\color{black}\Large\bfseries (d)}
    \end{overpic}
    \begin{overpic}[width=5.0cm]{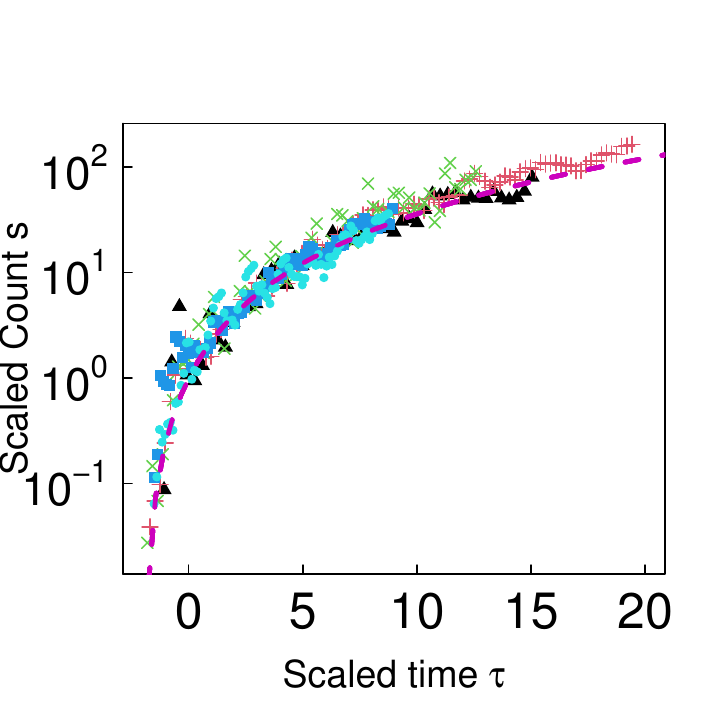}
        \put(30,103){\color{black}\Large\bfseries (e)}
    \end{overpic}
    \begin{overpic}[width=5.0cm]{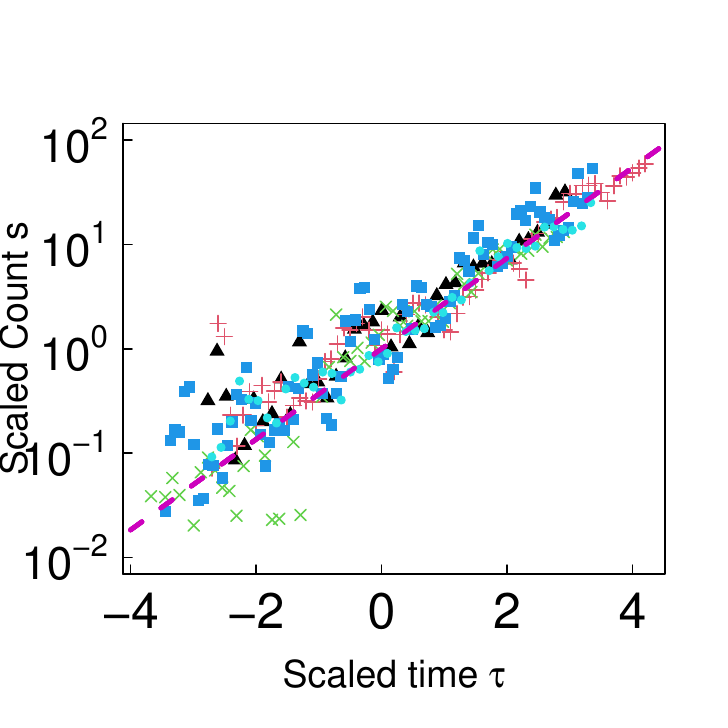}
        \put(30,103){\color{black}\Large\bfseries (f)}
    \end{overpic}
 \begin{overpic}[height=3.8cm,width=4.5cm]{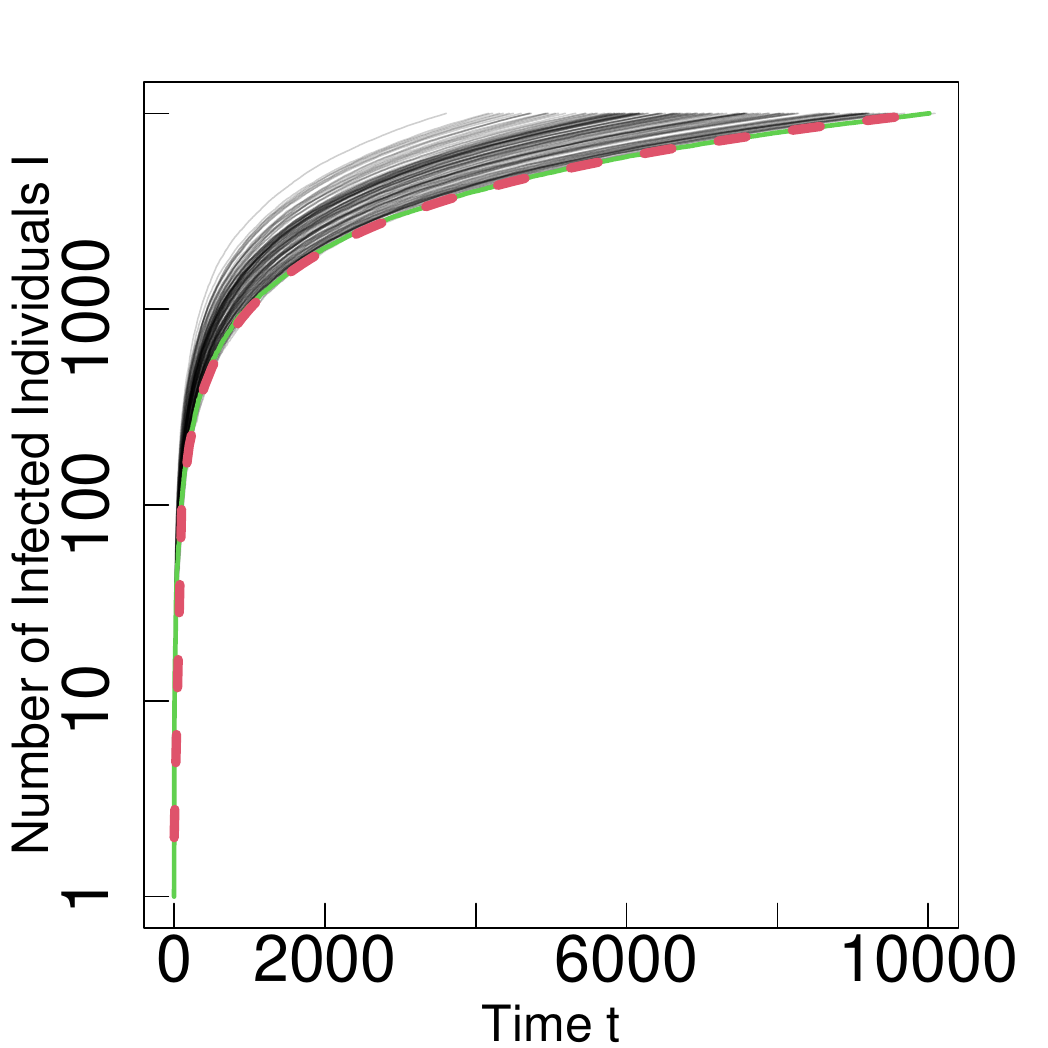}
        \put(20,91){\color{black}\Large\bfseries (g)}
    \end{overpic}
    \begin{overpic}[height=3.8cm,width=4.5cm] {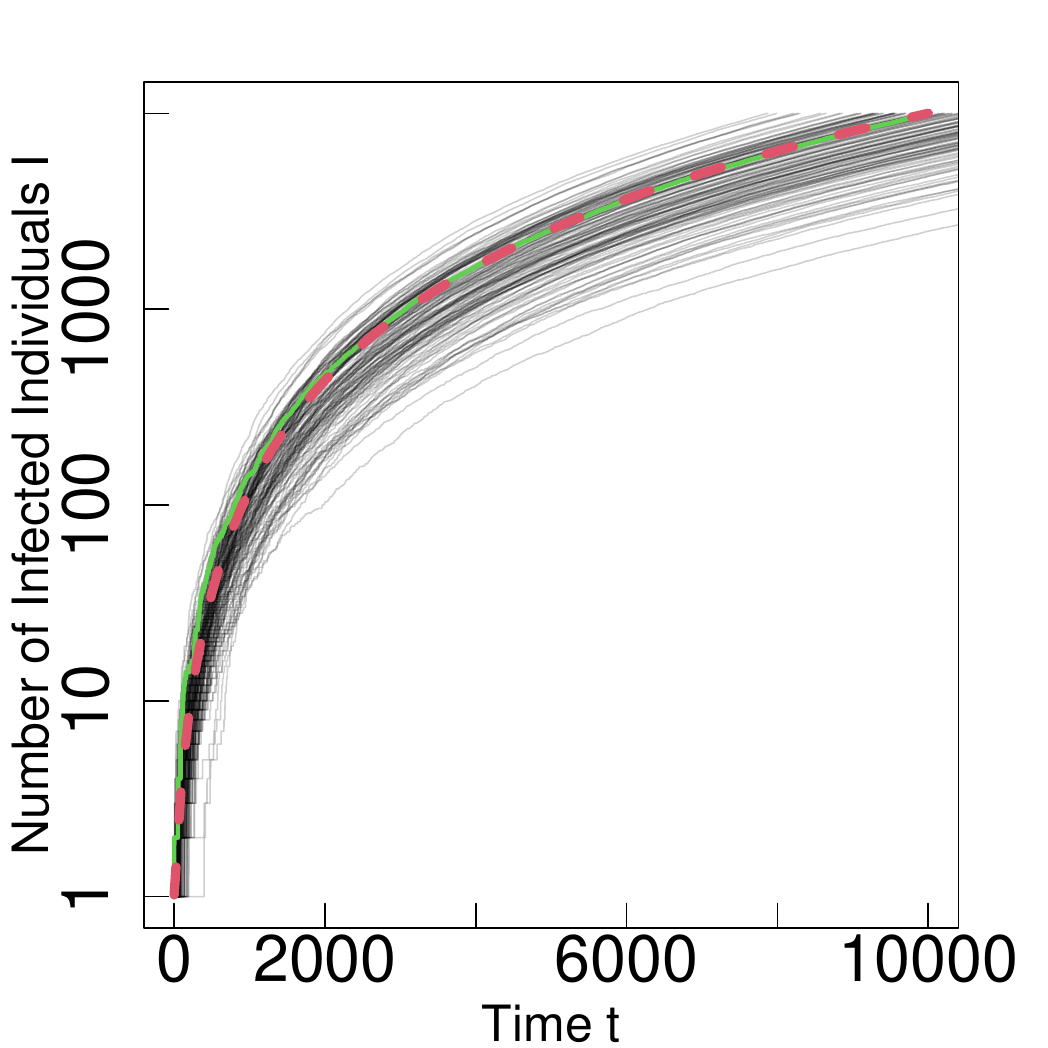}
        \put(20,88){\color{black}\Large\bfseries (h)}
    \end{overpic}
    \begin{overpic}[height=3.8cm,width=4.5cm] {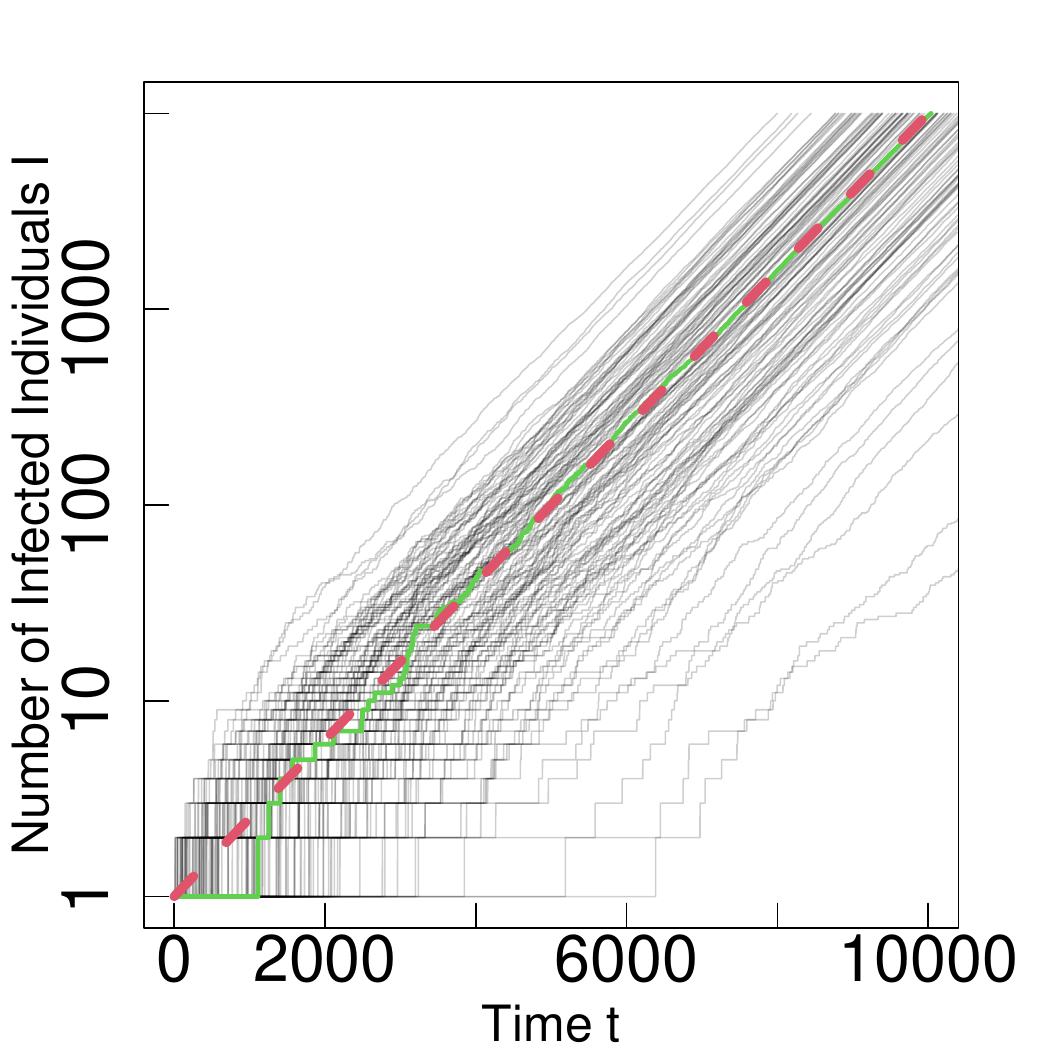}
        \put(20,88){\color{black}\Large\bfseries (i)}
    \end{overpic}
    \begin{overpic}[width=4.6cm,heigth=3.4cm]{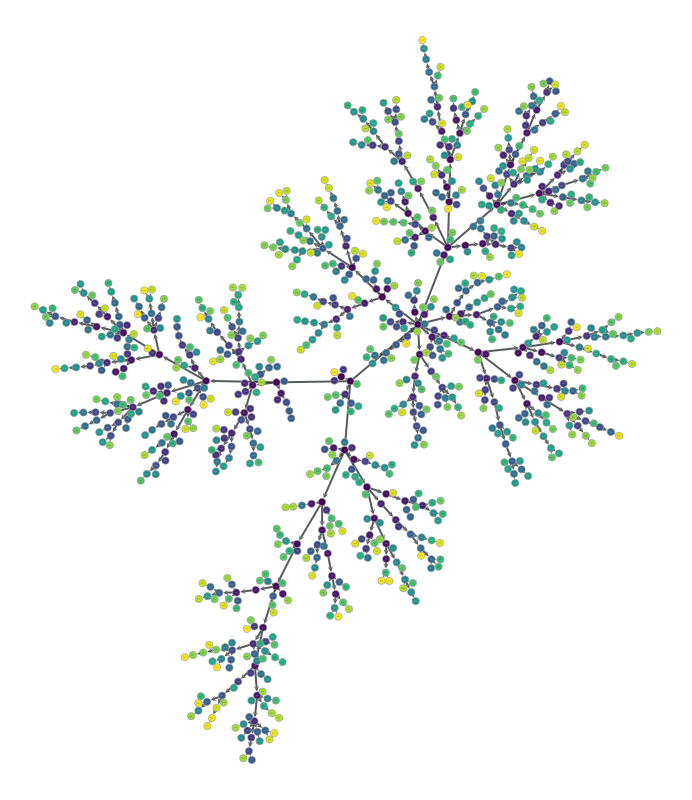}
        \put(30,120){\color{black}\Large\bfseries (j)}
    \end{overpic}
    \begin{overpic}[width=4.6cm,heigth=3.6cm]{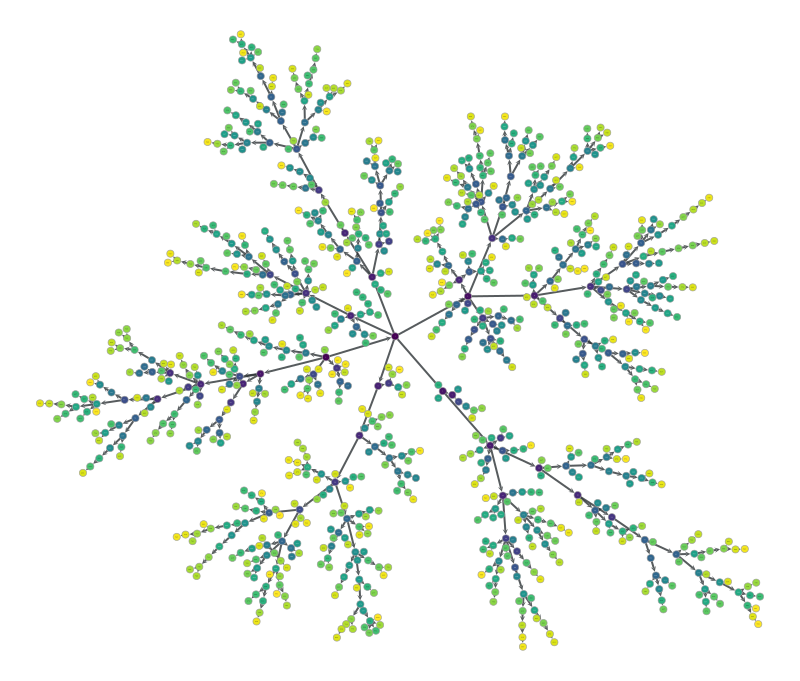}
        \put(30,120){\color{black}\Large\bfseries (k)}
    \end{overpic}
    \begin{overpic}[width=4.6cm,heigth=3.6cm]{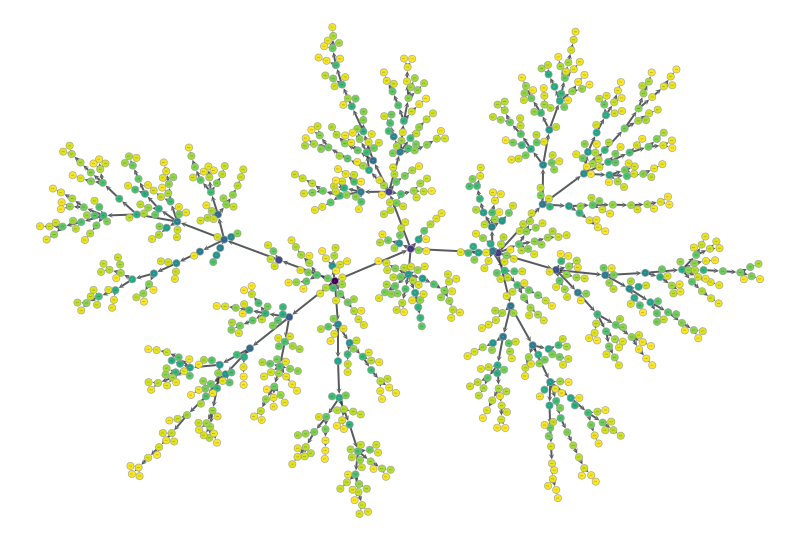}
        \put(30,120){\color{black}\Large\bfseries (l)}
    \end{overpic}
\caption{Time series captured by a single power-law model: scaled empirical data and corresponding simulations.
\textbf{(a)--(c) Scaled count series.} Points show empirical data $s_i(t)$ (Section~\ref{sec_base}); the red solid line is the scaled single power-law model (Eq.~\ref{eq_scale}); many words collapse onto a common curve.
Left: near-linear ($\alpha_i\approx 0$); middle: typical ($\alpha_i\approx 0.5$); right: exponential-like ($\alpha_i\approx 1$).
In each panel, five items are shown in the order black triangle, red cross, green cross, blue square, and light-blue circle, labeled as word ($\alpha_i,\,R_i$; brief note). Keywords are English translations; the original Japanese keywords are given in Appendix~\ref{app_sec_fig_word}. 
\textbf{(a)$\alpha_i\approx 0$:} ``Erika Ikuta'' ($0.00,0.59$; Japanese idol name), ``NicoNico Seiga'' ($0.09,0.12$; illustration sharing service), ``Chuo Ward, Sagamihara City'' ($-0.02,0.23$; new place name), ``Labor pain taxi'' ($-0.08,0.072$; maternity taxi service), ``beLEGEND'' ($0.01,0.097$; protein supplement brand).
\textbf{(b) $\alpha_i\approx 0.5$:} ``Tablet device'' ($0.47,0.34$), ``Crowdfunding'' ($0.53,0.22$), ``BABYMETAL'' ($0.55,0.23$; metal idol group), ``Rescue cat cafe'' ($0.50,0.14$), ``\emoji{anchor}'' ($0.45,0.12$; anchor emoji).
\textbf{(c)$\alpha_i\approx 1$:} ``Shale gas'' ($1.03,0.15$), ``Acai bowl'' ($0.98,0.10$), ``Fumika Baba'' ($0.91,0.11$; actress), ``\emoji{tent}'' ($1.01,0.076$;  outdoors-related emoji), ``Net-juu'' ($0.93,0.15$; slang: fulfilled online life). 
\textbf{(d)--(f) Corresponding log plots.}
\textbf{(g)--(i) Simulations of the infection model} (Section~\ref{sec_infection}). Black thin solid line: 128 sample paths ($Q=1$); red dotted line: theoretical approximation (Eq.~\ref{eq_infect_ans}); Green thick solid line: the simulation path closest to the theoretical prediction.  (g) $\gamma_i=1, J_i=1$; (h) $\gamma_i=0.5, J_i=0.020$; (i) $\gamma_i=0, J_i=9.2\times 10^{-4}$.
\textbf{(j)--(l) Corresponding diffusion-path networks} (directed edges from a recruiter to their recruits; first 1000 nodes shown; internal links via ``exchanges'' are excluded). Colors indicate infection time: older nodes are blue and newer nodes are yellow, varying linearly with time $t$.}
    \label{fig_scale_time_series}
\end{figure*}

\begin{figure}[t]
    \centering
    \begin{overpic}[width=5.7cm]{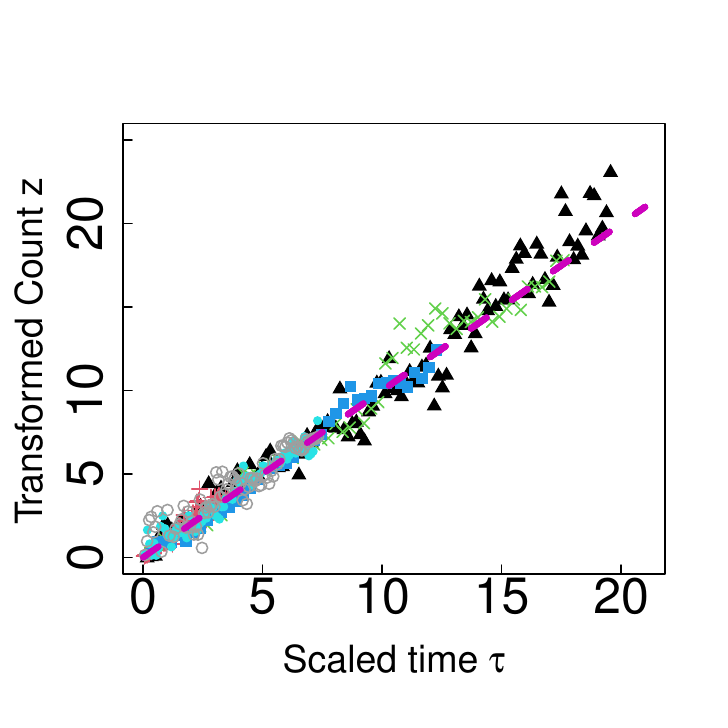}
    \end{overpic}
\caption{Linearized word counts $z_i(t)$ (Eq.~\ref{eq_boxcox}). Points show empirical data; the pink dashed line is $z=\tau$. 
Items are noted as word ($\alpha_i, R_i$; brief gloss). 
Keywords are English translations; the original Japanese keywords are given in Appendix~\ref{app_sec_fig_word}. 
Black triangle: ``Minami Ward, Sagamihara City'' ($0.17,0.39$; new place name). 
Red cross: ``SoundCloud'' ($0.180,0.15$; music sharing site). 
Green cross: ``Instagrammer'' ($0.44,2.38$; person popular on Instagram). 
Blue square: ``\textit{Komyushō}'' ($0.64,0.42$; net slang: poor at communication). 
Light-blue circle: ``MicroUSB'' ($0.80,0.080$; electronic interface). 
Gray hollow circle: ``Microplastics'' ($1.1,0.0043$; small plastic debris).
    }
    \label{fig_boxcox_time_series}
\end{figure}

\begin{figure*}[t]
     \begin{overpic}[width=5.7cm]{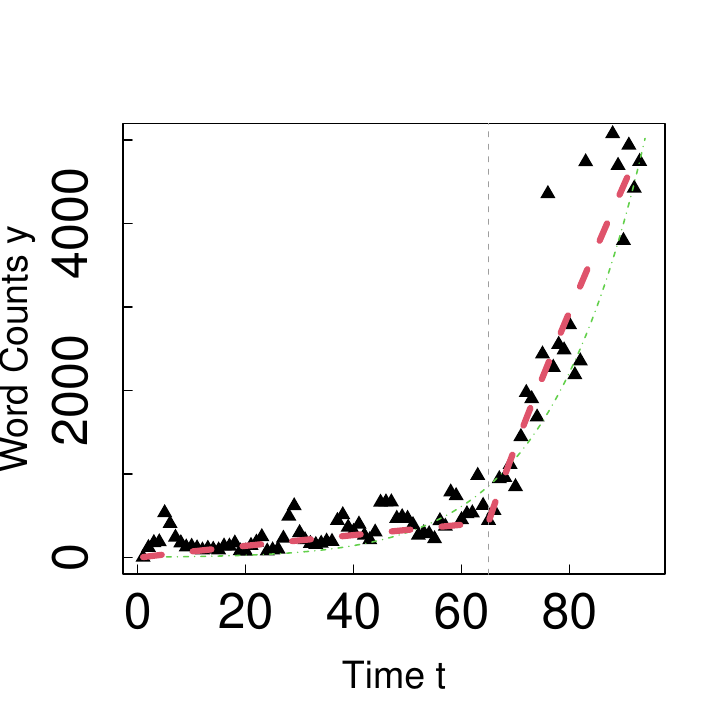}
        \put(30,120){\color{black}\Large\bfseries (a)}
    \end{overpic}
    \begin{overpic}[width=5.7cm]{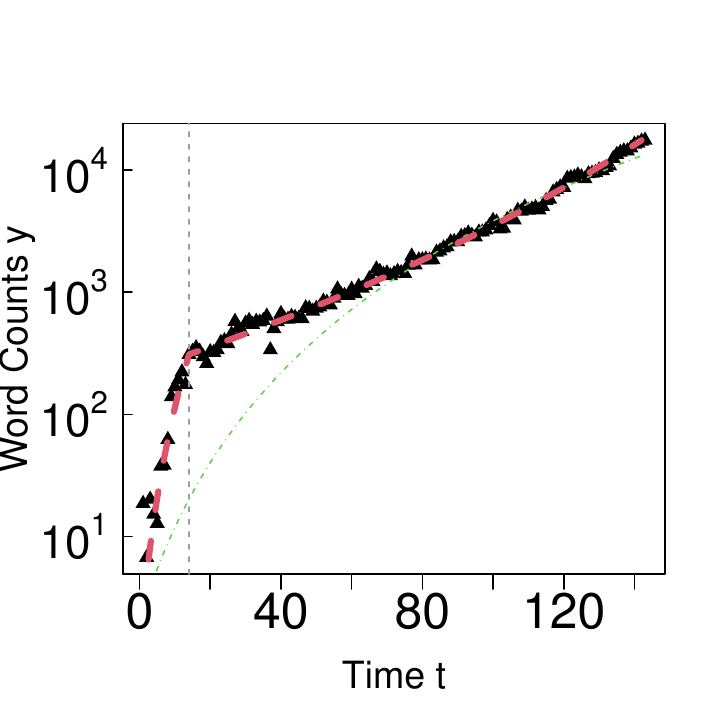}
        \put(50,120){\color{black}\Large\bfseries (b)}
    \end{overpic}
   \begin{overpic}[width=5.7cm]{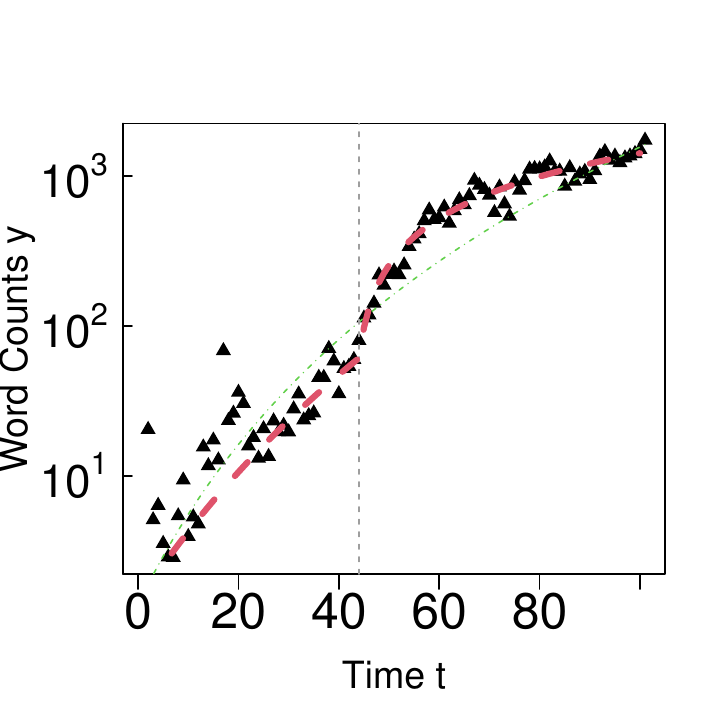}
        \put(30,120){\color{black}\Large\bfseries (c)}
    \end{overpic}
    \caption{Examples of growth curves with two segments ($N=2$). Parenthetical tuples list $(\alpha_i^{(1)}, R_i^{(1)};\ \alpha_i^{(2)}, R_i^{(2)};\ \text{brief gloss})$. 
    The black triangles denote the data, the red dash-dotted line is the $N=2$ piecewise power-law model, and the green dash-dotted line is the single power-law model ($N=1$).
     Keywords are English translations; the original Japanese keywords are given in Appendix~\ref{app_sec_fig_word}. 
\textbf{(a) ``Kenshi Yonezu''} ($-0.077,0.18; -0.12,6.40$; singer). Changepoint $t=65$ (November 2016). An example that is nearly linear, with the slope changing at the boundary; the slope change is plausibly related to increased exposure following a record label transfer.
\textbf{(b) ``\textit{Arafifu}''} ($0.77,0.36; 1.15,0.018$; slang: around age 50). Changepoint $t=14$ (April 2009). An example transitioning from sub-exponential to exponential growth; the shift likely reflects broader recognition after winning the 2008 ``Buzzword of the Year'' award.
\textbf{(c) ``Facebook Messenger''} ($0.83,0.067; -0.20,1.0$; messaging app). Changepoint $t=44$ (March 2015). The change is likely associated with major feature updates, such as adding video and enabling use without a Facebook account.}
    \label{fig_n_count2}
\end{figure*}

\begin{figure*}[hpt]
    \centering
    \begin{overpic}[width=5.7cm]{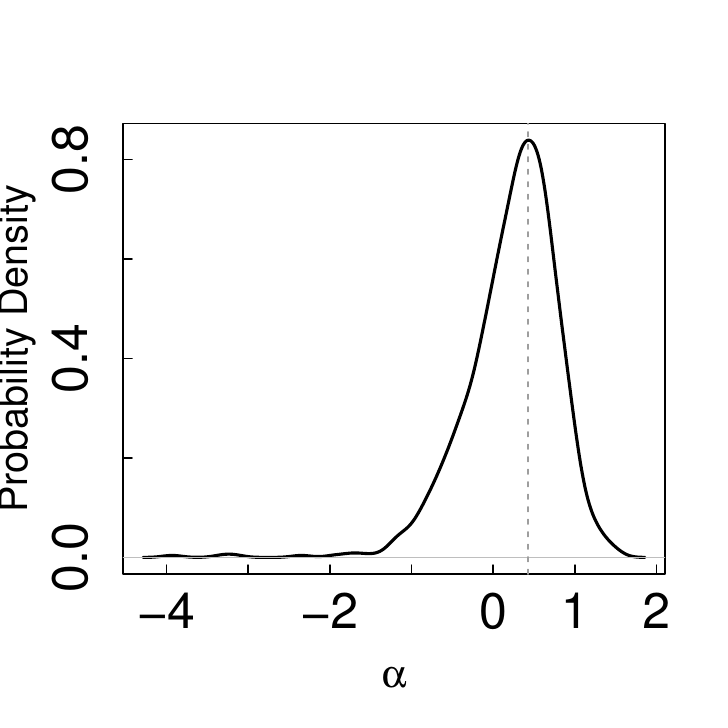}
        \put(30,120){\color{black}\Large\bfseries (a)} 
    \end{overpic}
    \begin{overpic}[width=5.7cm]{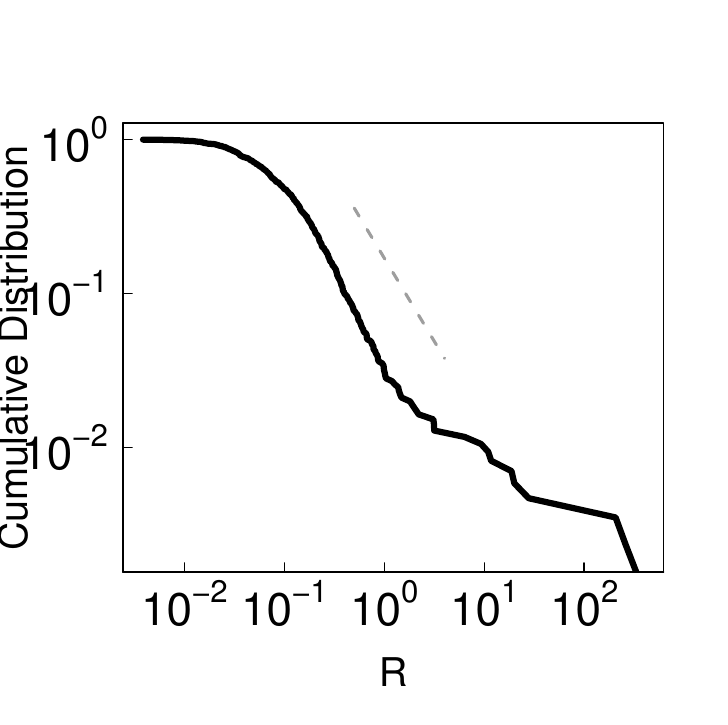}
        \put(30,110){\color{black}\Large\bfseries (b)}
    \end{overpic}
     \begin{overpic}[width=5.7cm]{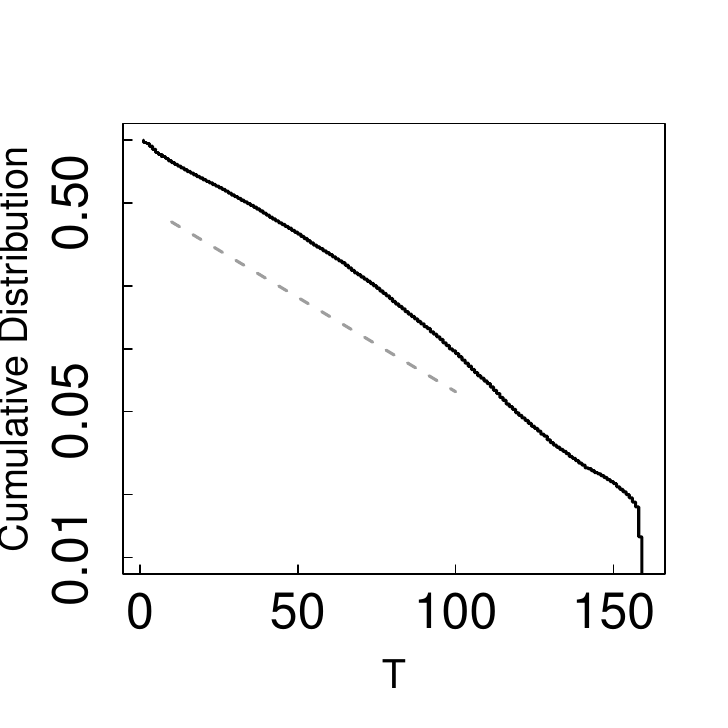}
        \put(60,120){\color{black}\Large\bfseries (c)}
    \end{overpic}
%
    \begin{overpic}[width=5.7cm]{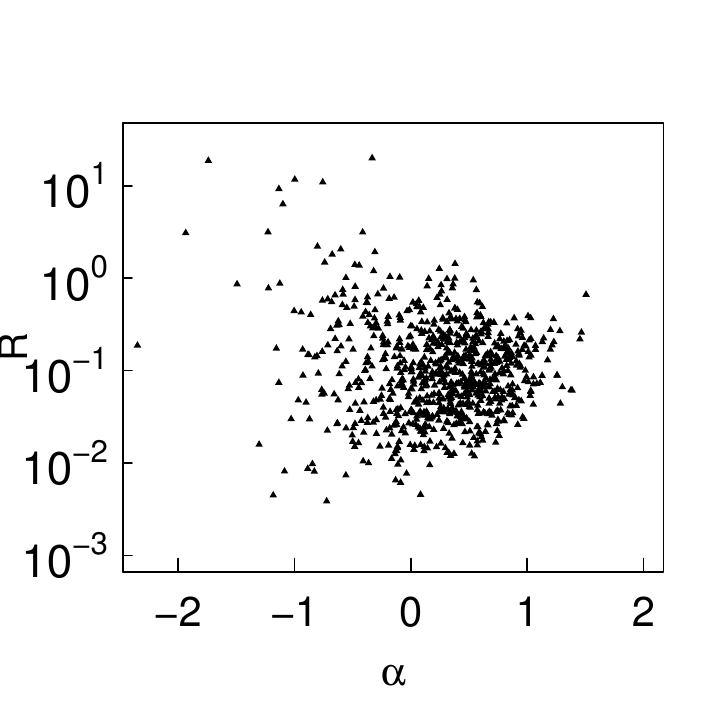}
        \put(30,120){\color{black}\Large\bfseries (d)}
    \end{overpic}
    \begin{overpic}[width=5.7cm]{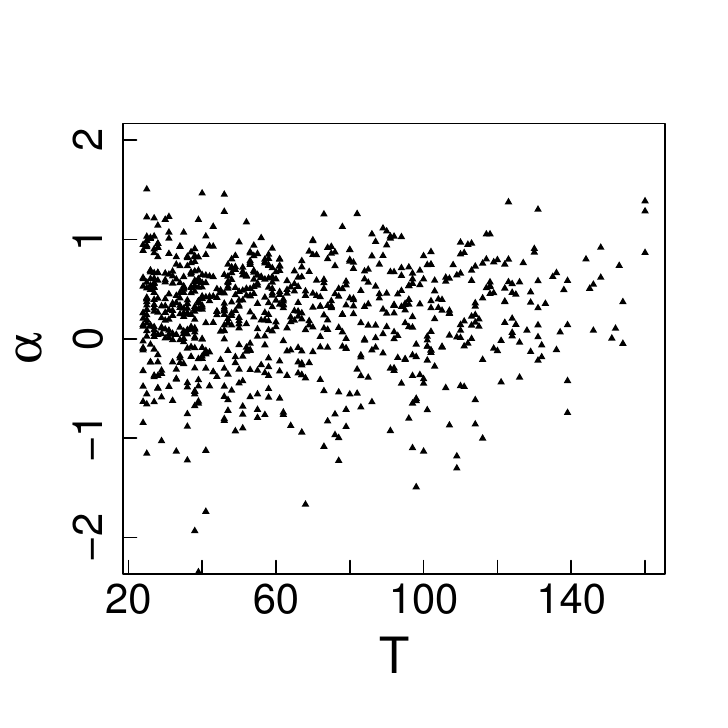}
        \put(30,120){\color{black}\Large\bfseries (e)}
    \end{overpic}
    \begin{overpic}[width=5.7cm]{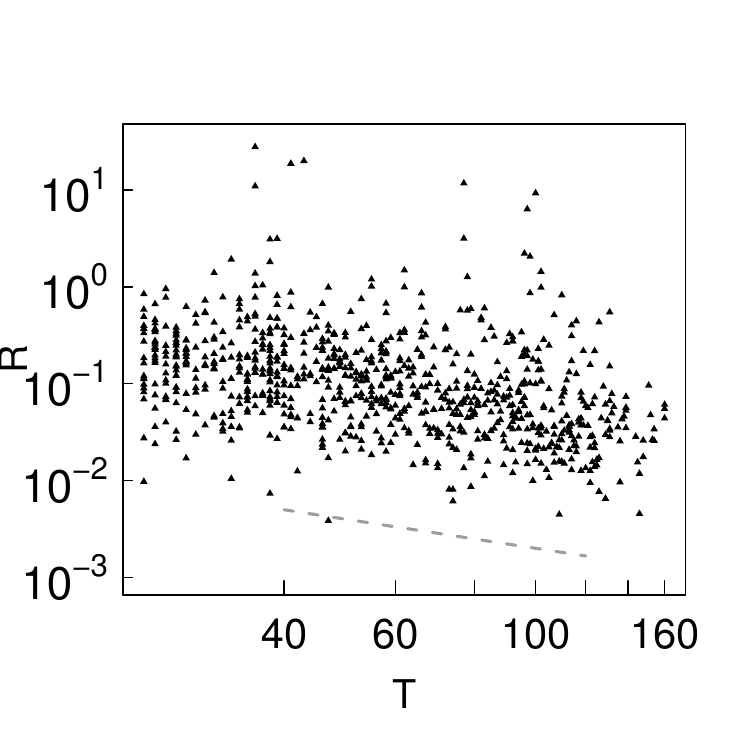}
        \put(30,120){\color{black}\Large\bfseries (f)}
    \end{overpic}
    \begin{overpic}[width=5.7cm]{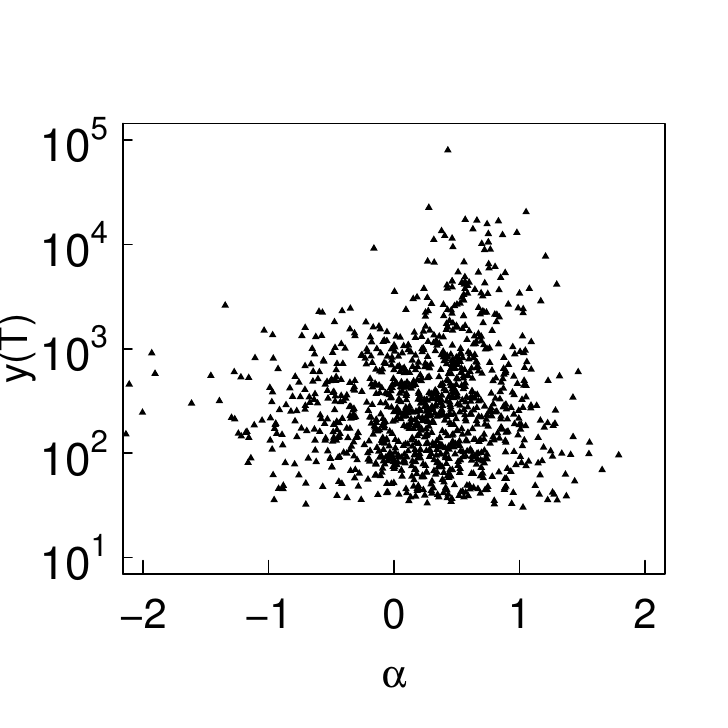}
        \put(30,120){\color{black}\Large\bfseries (g)} 
    \end{overpic}
    \begin{overpic}[width=5.7cm]{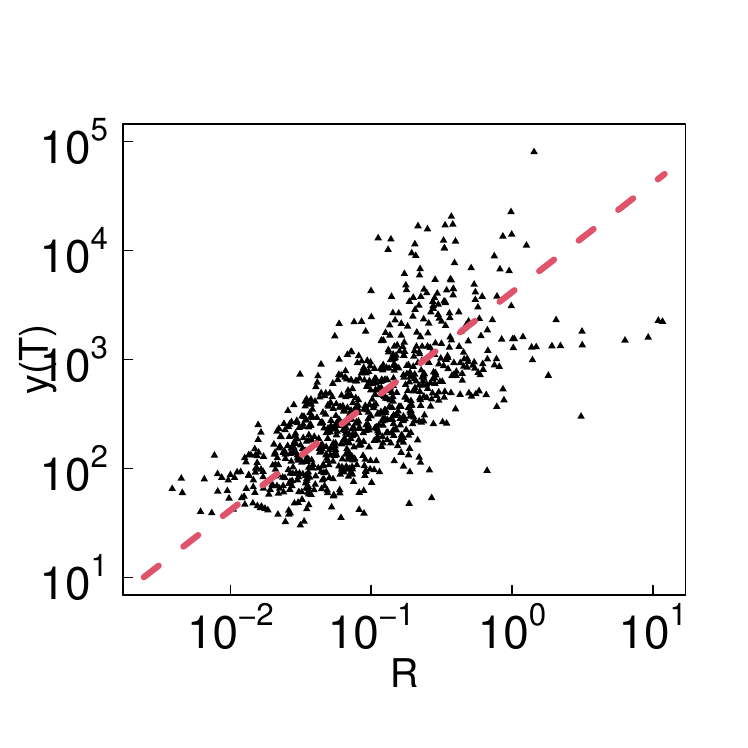}
        \put(30,120){\color{black}\Large\bfseries (h)}
    \end{overpic}
    \begin{overpic}[width=5.7cm]{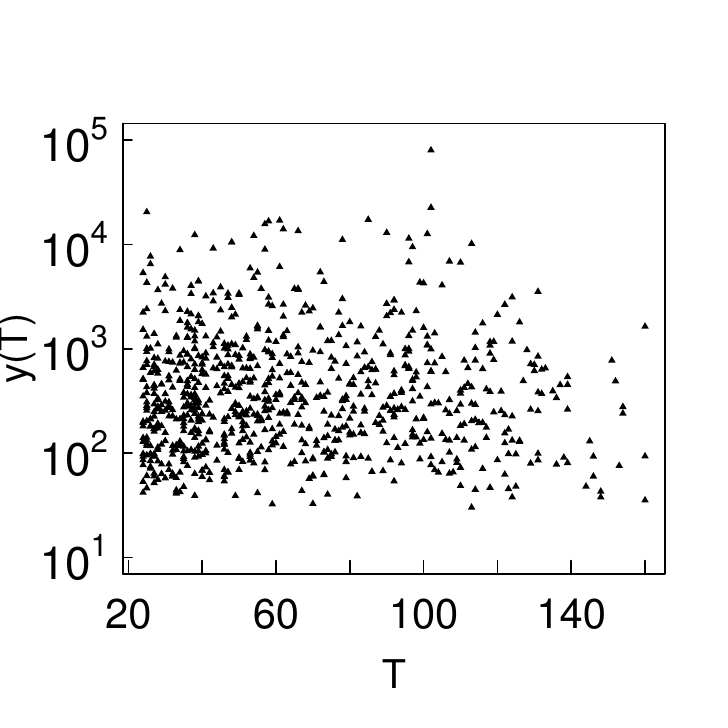}
        \put(30,120){\color{black}\Large\bfseries (i)}
    \end{overpic}
    \caption{Statistics of parameters for single-segment words ($N=1$; Section~\ref{sec_stat_main}). 
\textbf{(a) Probability density of $\alpha_i$.} The vertical dashed line marks the mode at $0.43$. 
\textbf{(b) Cumulative distribution of $R_i$.} The dashed guide follows $\propto R_i^{-1.1}$; the cumulative distribution of $R_i$ is close to a power law with exponent $1$ (Zipf's law). 
\textbf{(c) Cumulative distribution of $T_i$}, which is close to exponential; the dashed guide follows $\propto \exp(-x/30)$. 
\textbf{(d) Correlation between $\alpha_i$ and $R_i$.} No clear correlation is observed $(\tau=-0.017,\ p=0.47)$ (Kendall's $\tau$ and $p$-value for the null of zero correlation; same notation below). 
\textbf{(e) Correlation between $\alpha_i$ and $T_i$.} No clear correlation is observed $(\tau=-0.00,\ p=0.91)$. 
\textbf{(f) Correlation between $R_i$ and $T_i$.} A weak negative correlation is detected, approximately consistent with $R_i \propto 1/T_i$ $(\tau=-0.11,\ p<10^{-16})$, indicating that faster growth tends to be sustained for shorter durations. 
\textbf{(g) Correlation between $\alpha_i$ and $y_i(T)$.} No correlation is detected $(\tau=0.034,\ p=0.13)$. 
\textbf{(h) Correlation between $R_i$ and $y_i(T)$.} A strong positive, near-proportional relationship is detected $(\tau=0.55,\ p=2.2\times 10^{-16})$, showing that the growth rate $R_i$ is closely related to the peak value. 
\textbf{(i) Correlation between $T_i$ and $y_i(T)$.} No correlation is detected $(\tau=0.017,\ p=0.47)$. Further discussion of the lack of correlation is provided in Section~\ref{sec_stat_main}.}
\label{fig_para}
\end{figure*}
\begin{figure}[t]
    \centering
    \begin{overpic}[width=5.7cm]{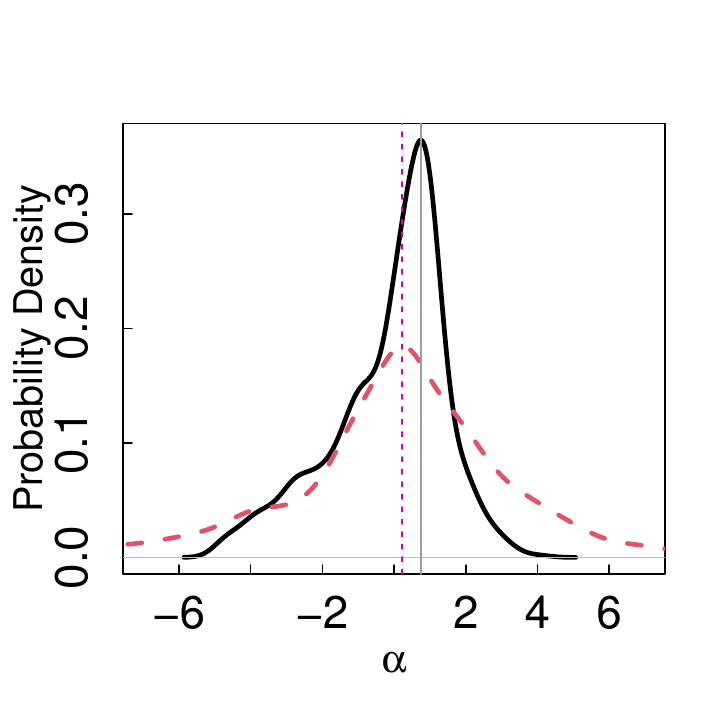}
    \end{overpic}
\caption{Distribution of $\alpha_i$ in the piecewise power-law model with two segments ($N=2$). 
The black solid curve shows the distribution of the first-segment exponents $\alpha_i^{(1)}$, whose mode is indicated by the gray vertical solid line at $\alpha_i^{(1)}=0.75$. 
The red dashed curve shows the distribution of the second-segment exponents $\alpha_i^{(2)}$, with the pink vertical dashed line at $\alpha_i^{(2)}=0.23$. 
The distributions of $\alpha_i$ differ between the first and second segments.}
    \label{fig_para_alpha_n2}
\end{figure}

\section{Overview of the Keyword Time Series}
In this study, we analyze the diffusion of newly introduced lexical items in Japanese blogs. The set of words we analyze consists of 20,764 items drawn from Wikipedia entries that had zero appearances in our blog corpus during 2006. For each item $i$, we define $y_i(t)$ as the number of blog posts containing $i$ in bin $t$, normalized by the total number of posts in the same 30-day bin, and construct 30-day-binned time series. For analysis, we further extract the growth interval—from its beginning to its end (see Fig.~\ref{app_fig_comp_google_blog_same} for examples of the extracted intervals). Details of the dataset and the preprocessing procedures are provided in Appendix~\ref{app_sec_data_base} and \ref{app_sec_cut}.
Fig.~\ref{fig_all_time_series} illustrates representative growth curves:
(a) “\textit{kakuyasu keitai}” (budget mobile phones), exhibiting smooth long-term growth;
(b) “\textit{sumaho}” (smartphone), showing continuous growth with a mid-course change in behavior; and
(c) \emoji{hollow-red-circle} (the red hollow circle emoji), exhibiting an abrupt jump.
Case (c) is plausibly attributable to improvements in smartphone predictive text that made this emoji easier to input.
Sections~\ref{sec_base}, \ref{sec_kubun}, and \ref{sec_jump} describe how these patterns (a),(b),(c) are captured by our models. Cross-lingual comparative analyses using additional datasets are summarized in the Appendix, Sections \ref{app_sec_blog_google_time} and \ref{app_sec_google_stat}. \par

\section{Core Components: A Power-Law Growth Model for Continuous Change}
\label{sec_base}
If there are no sudden external changes, empirical time series of word usage are expected to exhibit continuous, smooth growth. In this section, we introduce a power-law growth model as a foundational description of such growth segments. This model serves as a building block for more complex time series with discontinuities, providing the basis for the piecewise power-law model in section~\ref{sec_kubun} and for modeling discontinuous changes in section~\ref{sec_jump}.

For an individual lexical item $i$, the temporal evolution of its usage frequency $y_i(t)$ over a continuous, smooth growth interval is described by the following power-law growth equation:
\begin{equation}
\frac{dy_i(t)}{dt}
= R_i\,Y \left( \frac{y_i(t)}{Y} \right)^{\alpha_i}.
\label{eq_base0}
\end{equation}
This model is equivalent to the \emph{generalized growth model} (GGM) for $Y=1$ used in infectious-disease epidemiology to describe
sub-exponential growth in the early phase of an outbreak \cite{chowell2016mathematical}. \par
Here,
\begin{itemize}
\item \textbf{$Y$} is an observation-scale parameter, which could be interpreted as corresponding to the size of the data-collection platform (e.g., total posts or total users). It is intended to compensate for scale differences in term frequencies across platforms of different sizes. For the Japanese blog data, the empirically optimal value was estimated as $Y=41.3$ (see Appendix, section~\ref{app_sec_r}).
\item \textbf{$R_i$} is a term-specific rate parameter that characterizes the diffusion speed of item $i$. Its inverse, $1/R_i$, provides a characteristic time scale of the diffusion process. As clarified by the variable transformation introduced later (Eq.~\ref{eq_boxcox}), $R_i$ becomes the slope (linear growth rate) in the transformed space.
\item \textbf{$\alpha_i$} is a shape parameter of the growth curve that controls the nonlinearity of the diffusion dynamics. Its relationship to human behavior in real social settings is discussed in sections~\ref{sec_meaning}, \ref{sec_llm} and \ref{sec_infection}.
\begin{itemize}
    \item If $0 \le \alpha_i < 1$, the model describes sub-exponential growth (with $\alpha_i=0$ corresponding to linear growth).
    \item If $\alpha_i=1$, the equation reduces to exponential growth.
    \item If $\alpha_i>1$, growth accelerates further and, in theory, diverges to infinity in finite time. This regime can account for phenomena such as sharp increases toward a specific date (the ``deadline effect'' \cite{alfi2007conference}).
\end{itemize}
\end{itemize}

This study builds upon an earlier extended logistic model given by Eq.~\ref{app_eq_old_model} for diverse keyword time series growth patterns \cite{watanabe2023minor}. The proposed power-law model is positioned as a mathematical form that approximates the parameter region most typically observed in empirical data under that framework (see \revisecolor{black}{section~\ref{sec_motivation_piecewise}} and Appendix \ref{app_sec_relationOldmodel}). While this simplification captures the essential behavior, it has limited expressive power; this limitation is addressed by the piecewise power-law model introduced in section~\ref{sec_kubun}. \par
We assess the validity of the growth model using several keyword time series; the results are shown in Fig.~\ref{fig_scale_time_series} and Fig.~\ref{fig_boxcox_time_series} and are detailed below.
To extract the model's essential behavior, we perform a nondimensionalization. Define the normalized frequency
$s_i(t)=y_i(t)/Y$ and the rescaled time $\tau_i = R_i \cdot (t - t_i^{(0)})$, where the reference time $t_i^{(0)}$ is chosen so that
$s_i(t_i^{(0)})=1$. Under this change of variables, the solution of Eq.~\ref{eq_base0} for $\alpha_i\neq 1$ collapses to a one-parameter family that depends only on $\alpha_i$ (see Appendix, section~\ref{app_sec_henkan}):
\begin{equation}
s_i(\tau_i)=\left((1-\alpha_i)\,\tau_i+1\right)^{\frac{1}{1-\alpha_i}}.
\label{eq_scale}
\end{equation}
In the limit $\alpha_i \to 1$, Eq.~\ref{eq_scale} reduces to the exponential form $s_i(\tau_i)=\exp(\tau_i)$.
This collapse implies that the fundamental growth curve is determined solely by the shape parameter $\alpha_i$ and is
independent of the observation scale $Y$ and the rate parameter $R_i$. Consistent with this prediction,
Fig.~\ref{fig_scale_time_series} shows that the nondimensionalized data closely follow the curve given by Eq.~\ref{eq_scale} for $\alpha_i=0$ (linear; (a),(d)), $\alpha_i=0.5$ (a typical sub-exponential growth; (b),(e)), and $\alpha_i=1.0$ (exponential; (c),(f), supporting the
adequacy of the model in Eq.~\ref{eq_base0}.

To further test the model across a range of $\alpha_i$ values, we apply a linearizing transformation of the growth curve.
Specifically, with
$z_i = \bigl(s_i^{\,1-\alpha_i}-1\bigr)/(1-\alpha_i)$, Eq.~\ref{eq_scale} simplifies to a linear relation
\begin{equation}
z_i(\tau_i)=\tau_i,
\label{eq_boxcox}
\end{equation}
as detailed in Appendix, section~\ref{app_sec_henkan}. In the limit $\alpha_i\to 1$, this transformation becomes the
logarithmic function $z_i=\log s_i$. Notably, this is the Box-Cox transformation \cite{box1964analysis}. The transformation
enables a unified linear comparison across diverse growth shapes with different $\alpha_i$. As shown in Fig.~\ref{fig_boxcox_time_series}, the transformed data $z_i$ fall on the predicted line $z_i=\tau_i$, indicating that the
model holds over a wide range of $\alpha_i$. Moreover, from the definition of $\tau_i$ we obtain
$z_i(t)=R_i \cdot t+\mathrm{const.}$, confirming that $R_i$ corresponds to the linear growth rate (slope) with respect to time in the transformed space. \par

\section{Describing Complex Growth Dynamics}
\subsection{Piecewise Power-Law Model}
\label{sec_kubun}
\revisecolor{black}{
To accommodate complex time series that are not well captured by a single power-law model, we propose a ``piecewise power-law model'' in which the series is partitioned into $N$ segments, and an independent power-law is fitted to each segment. Concretely, the time series is divided into $N$ intervals, and each segment $k=1,2,\dots,N$ is described by a power-law with its own parameters $(\alpha_i^{(k)}, R_i^{(k)})$. The general form of the model and the parameter-estimation procedure are detailed in Appendix~\ref{app_sec_kubun} and \ref{app_sec_est_no_jumps}. \revisecolor{black}{Furthermore, the R script provided in the Supplemental Material \cite{supp_material} includes code demonstrating this parameter-estimation procedure.
}}
\par
\revisecolor{black}{
Fig.~\ref{fig_all_time_series}(b) shows the time series for the keyword ``\textit{sumaho}'' (smartphone). The single power-law fit (green) fails to adequately describe the data, whereas the piecewise model with $N=2$ segments (red)---split at $t=29$ (August 2009) with parameters $\alpha_i^{(1)}=1.03$, $R_i^{(1)}=0.30$, and $\alpha_i^{(2)}=-0.72$, $R_i^{(2)}=4411$---reproduces the observed trajectory well. Furthermore, Fig.~\ref{fig_n_count2} illustrates other typical growth patterns well described by this piecewise approach: (i) piecewise linear growth with slope changes (e.g., ``Yonezu Kenshi'', singer); (ii) transitions from sub-exponential to super-exponential growth (e.g., ``\textit{arafifu}'', slang); and (iii) shock-induced rapid increases followed by deceleration after exogenous events (e.g., ``Facebook Messenger'').  As a point of comparison, the growth patterns shown in Fig.~\ref{fig_n_count2} cannot be captured by our previously proposed ``extended logistic model'' \cite{watanabe2023minor}. The reasons for this limitation are discussed in Section~\ref{sec_motivation_piecewise}.
}
\revisecolor{black}{
\subsubsection{Motivation for the Piecewise Power-Law Formulation}
\label{sec_motivation_piecewise}
The introduction of the piecewise power-law model is motivated by two perspectives: data-driven modeling insights and the intrinsic phenomenological properties of social diffusion.
}
\par
\revisecolor{black}{
\paragraph{Data-driven viewpoint}
In our previous work, we proposed an ``extended logistic model'' based on the standard logistic equation to capture diverse growth patterns \cite{watanabe2023minor}:
\begin{equation}
\frac{dy_i(t)}{dt}=\rho_i \cdot y_i(t)\left(1+\frac{y_i(t)}{V_i}\right)^{\zeta_i}
\end{equation}
Note that this model reduces to the standard logistic equation when $\zeta_i=1$ and $V_i<0$, where $-V_i$ represents the carrying capacity.
Empirical analysis using this model revealed redundancies. First, for approximately 90\% of the observed words, $V_i > 0$; in this parameter regime, the extended logistic model behaves approximately as a piecewise power-law model. Second, many of these cases have $V_i$ close to zero, in which the extended logistic model effectively reduces to a single power-law growth model ($N=1$).  In fact, for many words, the predictive accuracy of the single power-law model ($N=1$) surpassed that of the extended logistic model. 
} 
\par
\revisecolor{black}{
Furthermore, fitting the extended logistic model near $V_i \sim 0$ causes instability in parameter estimation. The piecewise power-law model avoids this problem by performing sequential estimation starting from $N=1$. Another advantage is that the complexity of the model is measured by the intuitive number of segments $N$, rather than a difficult-to-interpret metric (such as regularization strength). Based on these findings, we concluded that it is more appropriate to take the power-law growth model as the basic form and extend it piecewise when necessary rather than extending the logistic function (see Appendix~\ref{app_sec_relationOldmodel} for a detailed comparison).
} 
\par
\revisecolor{black}{
Conversely, as detailed in Appendix~\ref{app_sec_relationOldmodel}, the extended logistic model can only approximately describe special cases of the piecewise power-law model: either $N=1$ (when $V_i \sim 0$), or $N=2$ with $\alpha_i^{(1)}=1$ (where the period $0 < y_i(t) \ll V_i$ corresponds to the segment with $\alpha_i^{(1)}=1$). Consequently, outside these parameter regimes, there are empirical word trajectories that the extended logistic model fails to describe, which can only be captured by the piecewise power-law model. For example, as seen in Fig.~\ref{fig_n_count2}, the extended logistic model cannot describe patterns such as (i) piecewise linear growth with slope changes (e.g., ``Yonezu Kenshi'', singer); or (ii) transitions from sub-exponential ($\alpha_i^{(1)} < 1$) to super-exponential ($\alpha_i^{(2)} > 1$) growth (e.g., ``\textit{arafifu}'', slang).
} \par
\revisecolor{black}{
\paragraph{Phenomenological viewpoint}
Beyond statistical fitting, the piecewise formulation is necessary to flexibly capture two types of structural changes observed in real-world diffusion phenomena.
} \par
\revisecolor{black}{
The first is \emph{exogenous change}. External events---such as an artist switching record labels or a product feature rollout---cause discontinuous and persistent shifts in the underlying properties of the diffusion process (e.g., Fig.~\ref{fig_all_time_series}(b)). Because the timing and frequency of such external shocks are unpredictable, they cannot, in principle, be described by a single continuous growth equation. A piecewise description ($N \ge 2$) is therefore necessary for systematically handling these diverse time series.
} \par
\revisecolor{black}{
The second is \emph{endogenous regime shift}. As previously seen in the ``\textit{sumaho} (smartphone)'' example (Fig.~\ref{fig_all_time_series}(b)), the diffusion process can qualitatively transition from an initial exponential-like growth phase to a subsequent linear-like establishment phase. Although the segment boundaries approximate such transitions in a stepwise manner, they provide a means to describe qualitatively different growth regimes in a practical and interpretable way.
} \par
\revisecolor{black}{
More generally, one could formulate a continuous-parameter model where $\alpha_i(t)$ and $R_i(t)$ act as continuous functions of time; our piecewise model corresponds to the special case where these parameters are piecewise constant. From this viewpoint, rather than introducing complex continuous functions, the piecewise model serves as a practical, step-function approximation of complex temporal dynamics, ensuring both flexibility and interpretability. Note that, if needed, sparse modeling may provide an intermediate approach between fully continuous and piecewise formulations.
}\par
\revisecolor{black}{
As a possible analogy, this type of piecewise description could be compared to geometric optics, where the path of a light ray passing through different media (e.g., air, water, glass) is determined piecewise by the refractive index of each medium.
}
\par
\subsection{Jump Effects}
\label{sec_jump}
We observe nonnegligible discontinuous increases even within an overall monotonic increase—see the vertical gray line in Fig.~\ref{fig_all_time_series}(c). We define a jump effect as a discontinuous increase followed by a sustained rise in the usage level. Transient surges that quickly revert to the prior baseline (``spikes'')—for example, short-lived news-driven bursts—are excluded. \par
Such jump effects often occur when a term's status shifts abruptly—for example, after an earthquake, a major system update, or the inauguration of a national leader (president or prime minister). In Fig.~\ref{fig_all_time_series}(c), the keyword is the emoji ``\emoji{hollow-red-circle}''. The observed discontinuity is plausibly attributable to a smartphone OS update that improved predictive text, making this emoji easier to input. \par
The red curve shows a theoretical fit from a piecewise power-law model augmented with a jump component, which reproduces the empirical trajectory well. Details of the jump-detection procedure and how the jump is incorporated into the model (as a discontinuity at a segmentation point) are provided in the Appendix, Sections~\ref{app_sec_detect_jump} and \ref{app_sec_kubun}, respectively.

\section{Statistical Properties of Growth Curves: Parameter Statistics}
\label{sec_parameters}
Using the proposed piecewise power-law model, we analyze the statistical properties of empirical growth curves. \par
\subsection{Data Selection}
To ensure reliable parameter estimation and robust statistical analysis, we restricted the sample to 2,963 lexical items that satisfy all three of the following criteria:
\begin{enumerate}
\item growth interval length of at least 24 time points (i.e., $\sim$2 years with 30-day bins);
\item peak usage frequency of at least 30 articles;
\item Spearman's rank correlation between time and usage frequency of at least $0.6$. 
\end{enumerate}
Criteria 1 (12,727 items) and 2 (17,613 items) were imposed to secure estimation accuracy: when the number of observations is small or usage is low, random noise and exogenous shocks have relatively large effects, making parameter estimates unstable. Criterion 3 (7,917 items) selects series that approximately satisfy the model's assumption of monotonic increase. This condition excludes typical time series patterns such as: (a) sequences that grow, diffuse, then decline and later regrow; and (b) sequences whose underlying smooth growth trend is obscured by excessive exogenous noise. \par

\subsection{Distribution of the Number of Segments ($N$)}
\label{sec_stat_n}
As shown in Table~\ref{app_tab_n_table} in Appendix~\ref{app_sec_n_table}, the distribution of the number of segments, $N$, for the 1,701 items where no jump was detected is highest at $N=1$ (852 items), followed by $N=2$ (773 items). Cases with $N\ge 3$ are rare, comprising only 76 items. These results suggest that most series can be adequately described with a small number of segments ($N \le 2$). This conclusion is based on a composite decision procedure that combines multiple error measures, as outlined in the following paragraphs.

We determine $N$ using a stepwise procedure based on several groups of error criteria. Specifically, we first compare the fit errors for $N=1$ and $N=2$. If any single group of criteria satisfies its preset conditions, we adopt the smaller value of $N$. If these conditions are not satisfied, we compare $N=2$ and $N=3$ and iterate this process. This approach is adopted to improve robustness against diverse error structures in the data. Detailed definitions of the criteria are provided in Appendix~\ref{app_sec_n_estimation}.

We deliberately do not use standard information criteria such as AIC for model selection. Such criteria typically require an explicit, or at least approximate, specification of the noise structure. In practice, our data contain complex, heterogeneous fluctuations---for instance, transient responses to news events---that are difficult to model concisely. We therefore employ a staged evaluation that systematically combines multiple error measures.

\subsection{Statistical Properties of Growth Curves: Shape Distribution, Peak Relationships, and Correlation Structure}
\label{sec_stat_main}
Our analysis yields three main findings about the growth curves of word usage.

\noindent\textbf{(1) Shape parameter $\alpha_i$.}
For single-segment fits ($N=1$), the distribution of $\alpha_i$ has a mode at $\alpha_i \approx 0.43$, between linear ($\alpha_i=0$) and exponential ($\alpha_i=1$) growth (Fig.~\ref{fig_para}a). This tendency is broadly consistent with prior results obtained using an extended logistic model \cite{watanabe2023minor}. Moreover, more than 95\% of items satisfy $\alpha_i<1$, indicating that sub-exponential growth is pervasive in word-usage diffusion.

A notable contrast with epidemic dynamics is that whereas sub-exponential growth  is typically confined to the early phase of an outbreak, many lexical items exhibit sub-exponential behavior that persists throughout the entire diffusion period.

For two-segment fits ($N=2$; Fig.~\ref{fig_para_alpha_n2}), the mode value of $\alpha$ in the first segment is $\approx 0.8$ (black solid line), while in the second segment it is $\approx 0.2$ (red dashed line). Thus, the early phase tends to be closer to exponential, whereas the later phase tends to be closer to linear. The second-segment distribution also extends to $\alpha<0$, which we interpret as capturing sublinear shapes—deceleration following earlier acceleration—often associated with responses to exogenous events near the segment boundary (see examples in Fig.~\ref{fig_n_count2}(c)).

\noindent\textbf{(2) Relationships between the peak value and model parameters ($R_i, \alpha_i, T_i$).}
For items well described by a single-segment power-law model ($N=1$), the peak value $y_i(T_i)$ shows a strong positive correlation with the growth rate $R_i$ (Fig.~\ref{fig_para}h), whereas no clear correlation is observed with the shape parameter $\alpha_i$ (Fig.~\ref{fig_para}g) or the growth duration $T_i$ (Fig.~\ref{fig_para}i). These observations suggest that the eventual diffusion scale is tied more closely to the rate of growth $R_i$ than to the curve shape $\alpha_i$ or duration $T_i$.

As shown in Fig.~\ref{fig_para}i, $T_i$ and $y_i(T_i)$ appear uncorrelated overall. In contrast, Eqs.~\ref{eq_scale} and \ref{eq_boxcox} suggest that $y_i(T_i)$ should increase monotonically with $T_i$. This apparent discrepancy can be explained by two factors. First, as discussed in item (3) below, $R_i$ and $T_i$ are negatively correlated; thus, even when $T_i$ is large, a smaller $R_i$ can offset the positive effect of $T_i$. Second, the variability of $R_i$ substantially exceeds that of $T_i$: as noted in (3), $R_i$ is heavy-tailed (approximately power-law), whereas $T_i$ is closer to exponential and hence less variable. Consequently, the contribution of $T_i$ is often masked by that of $R_i$ (cf.\ the approximate relation $\log y_i(T_i) \propto \tfrac{1}{1-\alpha_i}\,(\log R_i + \log T_i)$). Empirically, Kendall's partial $\tau$ between $T_i$ and $y_i(T_i)$ controlling for $R_i$ is $0.22$ (two-sided $p<10^{-22}$), indicating a positive association; conversely, the partial correlation between $R_i$ and $y_i(T_i)$ controlling for $T_i$ is $0.58$ ($p<10^{-139}$).

\noindent\textbf{(3) Parameter distributions and correlations.}
For items well described by a single-segment power-law model ($N=1$), several characteristic features emerge. The growth rate $R_i$ is well approximated by a power-law distribution with tail exponent $\approx 1.1$, whereas the growth duration $T_i$ is well approximated by an exponential distribution (Fig.~\ref{fig_para}b,c). The former is broadly consistent with Zipf's law (a power law with exponent near 1) commonly observed for word frequencies, and the latter is consistent with approximately memoryless termination over time.

Pairwise correlations between $(\alpha_i,R_i)$ and $(\alpha_i,T_i)$ are generally weak, with no pronounced relationships detected (Fig.~\ref{fig_para}d,e). In particular, the near absence of correlation between $R_i$ (rate) and $\alpha_i$ (shape) can be attributed to the role of the scale parameter $Y$ in our model: $Y$ suppresses spurious scale-induced correlations and helps isolate intrinsic parameter characteristics. Operationally, we choose $Y$ to minimize the correlation between $R_i$ and $\alpha_i$; when $Y$ is fixed at 1, a clear correlation reappears (Fig.~\ref{app_fig_alpha_r}; see Section~\ref{app_sec_r}). By contrast, for $R_i$ and $T_i$ (Fig.~\ref{fig_para}f), we observe a negative relationship close to $R_i \propto 1/T_i$, indicating that faster growth tends to be sustained for shorter durations.


\begin{table*}[!p]
\centering
\begin{tabular}{p{0.31\textwidth}p{0.31\textwidth}p{0.31\textwidth}}
\hline
\textbf{$\alpha \approx 0$} & \textbf{$\alpha \approx 0.5$} & \textbf{$\alpha \approx 1.0$} \\
\textbf{(linear growth, local/niche)} & \textbf{(sub‑exponential, diverse/mixed)} & \textbf{(exponential growth, global/shared)} \\
\hline
\begin{minipage}[t]{\linewidth}
\textbf{[1. Proper names (persons, organizations, places)]} \\
\textcolor{red}{{\Large $\blacktriangle$}} Erika Ikuta (-0.00; AKB48-related group member), \textcolor{red}{{\Large $\blacktriangle$}} Mai Shinuchi (0.02; AKB48-related group member), \textcolor{red}{{\Large $\blacktriangle$}} Himeka Nakamoto (0.08; AKB48-related group member), \textcolor{red}{{\Large $\blacktriangle$}} Ranze Terada (0.09; AKB48-related group member), \textcolor{magenta}{{\huge $\bullet$}} Kita Ward, Okayama City (0.02; place name), \textcolor{magenta}{{\huge $\bullet$}} Midori Ward, Sagamihara City (-0.10; place name), \textcolor{magenta}{{\huge $\bullet$}} Chuo Ward, Sagamihara City (-0.01; place name), \textcolor{magenta}{{\huge $\bullet$}} Itsukushima Shrine (0.08; a Shinto shrine on Itsukushima Island), Saori Hayami (0.08; voice actress) \\
\vspace{0.5\baselineskip}

\textbf{[2. Culture / Media / Subculture/ Slang]} \\
\textcolor{orange}{$\star$} AppBank (-0.08; Japanese app review and media site), \textcolor{orange}{$\star$} appbank (-0.08; Japanese app review and media site, all lower case letters),  \textcolor{orange}{$\star$} NewsPicks (-0.09; Japanese business news platform), \textcolor{orange}{$\star$} Japan Business Press (-0.06; Japanese online business magazine), NicoNico Seiga (0.09; Image sharing service),  left on read (0.07; message seen but not replied to), Kusozako namekuji (-0.06; internet slang: Fucking useless weakling), Hack and Slash (-0.08; video game genre focusing on combat), S.H.Figuarts (-0.09; Bandai's action figure line), Bundodo (0.03; playing with action figures and making sound effects) \\
\vspace{0.5\baselineskip}

\textbf{[3. ICT / Technology]} \\
Pinterest (-0.06; visual discovery engine) \\
\vspace{0.5\baselineskip}

\textbf{[4. Society / Lifestyle]} \\
Seven \& I Holdings' private brand (0.01; private label products), Ideathon (-0.10; idea generation workshop), galaxxxy (0.02; Japanese fashion brand), Labor pain taxi (-0.08; taxi service for pregnant women going into labor), beLEGEND (0.01; Japanese sports nutrition brand) \\
\vspace{0.5\baselineskip}

\textbf{[5. International / Public / Affairs]} \\
(no terms in this regime) \\
\end{minipage}
&
\begin{minipage}[t]{\linewidth}
\textbf{[1. Proper names (persons, organizations, places)]} \\
\textcolor{red}{{\Large $\blacktriangle$}} Asuka Saito (0.46; AKB48-related group member), Shouta Aoi (0.45; Japanese singer and voice actor), Kardashian (0.48; family name of American reality television personalities), Sora Tokui (0.41; Japanese voice actress and singer), Kiryu (0.52; Japanese visual kei band), \textcolor{blue}{$\blacklozenge$}  BABYMETAL (0.55; Japanese girl metal band), Wednesday Campanella (0.47; Japanese girl music group) \\
\vspace{0.5\baselineskip}

\textbf{[2. Culture / Media / Subculture/ Slang]} \\
\textcolor{blue}{$\blacklozenge$} Instagrammer (0.44; person popular on Instagram), Netizen (0.55; active internet user), \textcolor{blue}{$\blacklozenge$}  Crowdfunding (0.53; funding a project by raising small amounts of money from many people),  \textcolor{blue}{$\blacklozenge$} Shiotaio (0.41; Slang: Giving the cold shoulder), Nuitadori (0.51; taking photos with stuffed animals) \\
\vspace{0.5\baselineskip}

\textbf{[3. ICT / Technology]} \\
Tablet device (0.47; portable computer with touchscreen), Smartphone app (0.42; application software for mobile devices), Twitter account (0.56; user profile on Twitter), Push notification (0.48; message sent by an app to a device), TripAdvisor (0.45; travel website for reviews and bookings), Facebook account (0.46; user profile on Facebook), Coconala (0.54; Japanese online marketplace for skills/services), Jimdo (0.57; website builder platform) \\
\vspace{0.5\baselineskip}

\textbf{[4. Society / Lifestyle]} \\
Rakuten Bank (0.60; Japanese online bank), Tsukada Nojo (0.53; Japanese izakaya restaurant chain), "\emoji{high-voltage}" (0.56; emoji lightning bolt: symbol for electricity or quickness), "\emoji{baseball}" (0.53; emoji baseball: symbol for the sport of baseball), "\emoji{anchor}" (0.45; emoji anchor: symbol for stability or nautical themes), Community Revitalization Cooperator (0.46; program for urban residents to support rural areas), \textcolor{blue}{$\blacklozenge$} Rescue cat cafe (0.50; cafe where rescued cats can be adopted), Dry aging (0.46; meat preservation technique), Celecoxx (0.47; anti-inflammatory drug), Takecab (0.49; drug for acid-related disorders) \\
\vspace{0.5\baselineskip}

\textbf{[5. International / Public / Affairs]} \\
(no terms in this regime) \\
\end{minipage}
&
\begin{minipage}[t]{\linewidth}
\textbf{[1. Proper names (persons, organizations, places)]} \\
Fumika Baba (0.91; actress/model), Chinatsu Akasaki (0.93; voice actress), Tsubasa Sakiyama (1.03; actor/singer), Hikaru Yu (0.91; Takarazuka Revue star), Izakaya Hanako (0.94; Japanese pub chain), Chita Musume (0.92; local idol group/mascot), Mitsuru Kurayama (0.93; historian/commentator), SHU-I (0.94; South Korean boy band), \textcolor{magenta}{{\huge $\bullet$}} TIAT (0.92; Tokyo International Air Terminal Corporation), Eric Chu (1.05; Taiwanese politician) \\
\vspace{0.5\baselineskip}

\textbf{[2. Culture / Media / Subculture / Slang]} \\
XFLAG (0.90; mixi's gaming brand), Okanehira (0.98; a famous Japanese sword), Hunger Games (0.95; novel/film series), \textcolor{blue}{$\blacklozenge$} \textit{Tsuratan} (1.01; slang for “it's tough/sad”), Bonbonribbon (0.93; Sanrio character), "\emoji{part-alternation-mark}" (0.95; Part alternation mark; Unicode U+303D: often for traditional Japanese poetry), \textit{Net-juu} (0.93; slang for someone who enjoys online life), \textcolor{blue}{$\blacklozenge$} Instagrammable (1.03; visually appealing for Instagram), \textcolor{blue}{$\blacklozenge$} Online Salon (0.94; paid online community) \\
\vspace{0.5\baselineskip}

\textbf{[3. ICT / Technology]} \\
\textcolor{blue}{$\blacklozenge$}  Airbnb (0.99; online lodging marketplace), Twitter Client (1.07; app for Twitter access), Udemy (0.92; online learning platform), Mobatwi (0.97; a former mobile Twitter client), \textcolor{cyan}{$\blacksquare$} Ethereum (0.92; cryptocurrency/blockchain platform), \textcolor{cyan}{$\blacksquare$} VR Goggles (1.03; virtual reality headset) \\
\vspace{0.5\baselineskip}

\textbf{[4. Society / Lifestyle]} \\
\textcolor{blue}{$\blacklozenge$}  Acai Bowl (0.98; a fruit bowl with acai berries), \textcolor{blue}{$\blacklozenge$} Raycop (1.02; brand of futon cleaner), \textcolor{blue}{$\blacklozenge$} "\emoji{tent}" (1.01; emoji Tent: camping equipment symbol), Certified Public Psychologist (0.99; national qualification) \\
\vspace{0.5\baselineskip}

\textbf{[5. International / Public / Affairs]} \\
\textcolor{cyan}{$\blacksquare$} Shale Gas (1.03; natural gas from shale formations), \textcolor{cyan}{$\blacksquare$} Microplastic (1.05; tiny plastic debris), \textcolor{cyan}{$\blacksquare$} Shale Gas Revolution (0.96; major energy shift) \\
\end{minipage}
\\
\hline
\end{tabular}
\caption{
\revisecolor{black}{Examples of keywords classified by the growth-shape parameter $\alpha_i$, reorganized into five semantic categories: (1) proper names (persons, organizations, places), (2) culture/media/subculture, (3) ICT/technology, (4) society/lifestyle, and (5) international/public/affairs. Columns correspond to $\alpha_i \approx 0$ (linear growth, local/niche), $\alpha_i \approx 0.5$ (sub‑exponential, diverse/mixed), and $\alpha_i \approx 1.0$ (exponential growth, global/shared).
The left column ($\alpha_i \approx 0$) is dominated by niche and local terms. Specifically, these include AKB48 (a Japanese pop idol group) -related person names (\textcolor{red}{{\Large $\blacktriangle$}}), place names (\textcolor{magenta}{{\huge $\bullet$}}), and game/anime/subculture terms. In addition, some news sites and media platforms (\textcolor{orange}{$\star$}) also stand out. In contrast, the right column ($\alpha_i \approx 1.0$) contains globally shared, high‑profile terms, such as global news terms (\textcolor{cyan}{$\blacksquare$}) and domestic buzzwords (\textcolor{blue}{$\blacklozenge$}). The middle column ($\alpha_i \approx 0.5$) shows a diverse mixture of topics.
A version listing the original Japanese words is provided in Appendix Table~\ref{tab:examples_by_alpha_japanese}.
}}
\label{tab:examples_by_alpha}
\end{table*}

\begin{table*}[t]
\centering
\begin{tabular}{p{0.48\textwidth}p{0.48\textwidth}}
\hline
\textbf{Small $\alpha_i$} & \textbf{Large $\alpha_i$} \\
\textbf{(linear growth, local/niche)} & \textbf{(exponential growth, global/shared)} \\
\hline
\begin{minipage}[t]{\linewidth}
\textbf{[AKB48-centered idol culture terms]} \\
AKB48 trainee (-0.33, 0.0010); Birthday festival (-0.25, 0.0014); Photo session (-0.21, 0.015); Training (-0.24, 0.012); Dancer (-0.24, 0.025); News anchor (-0.27, 0.019); Takagi:surname (-0.27, 0.0081); Matsumura:surname (-0.23, 0.042); 
Hori:surname (-0.22, 0.023); Ishida:surname (-0.22, 0.016); Hiro:name (-0.22, 0.048);
Aya:given name (-0.23, 0.022); 
Grand prize (-0.21, 0.031); New program/show (-0.20, 0.045); Under (-0.23, 0.035); Friendship (-0.24, 0.034) \\
\vspace{0.5\baselineskip}

\textbf{[Place-related terms]} \\
Miyagi:prefecture name (-0.24, 0.040); Kanagawa:prefecture name (-0.23, 0.0091); In front of station (-0.21, 0.011); Regular holiday (-0.26, 0.023); Telephone (-0.37, 0.00088); Piano (-0.21, 0.0066);\\
\vspace{0.5\baselineskip}

\textbf{[Game, anime, subculture–related terms]} \\
Form/Mode (-0.25, 0.012); Creator (-0.24, 0.031); x:a symbol used for multiplication or combinations (-0.24, 0.018); Advent/Descent (-0.24, 0.011); Equipment/Gear (-0.23, 0.015); Multi (-0.23, 0.026); Second term/period (-0.21, 0.024); Playable (-0.22, 0.017)
\vspace{0.5\baselineskip}

\textbf{[Other]} \\
Yahoo! News (-0.21, 0.020); Calorie (-0.23, 0.044); Writing/Posting (-0.23, 0.041); SM (-0.22, 0.019);  Comedy (-0.21, 0.030); Introduction/Beginner's course (-0.21, 0.041); Jean/Genre (-0.21, 0.032); Number 3/Third (-0.23, 0.049)\\
\end{minipage}
&
\begin{minipage}[t]{\linewidth}
\textbf{[Global -related terms]} \\
Worldwide/Global (0.27, 0.0019); France (0.22, 0.048);  Europe (0.23, 0.042);  \\
\vspace{0.5\baselineskip}

\textbf{[Economic and Political terms]} \\ Profit/Benefit (0.23, 0.024); Borrower/Tenant (0.22, 0.018);  Transaction/Deal (0.24, 0.024); Regulation/Control (0.23, 0.010);  Establish/Set up (0.24, 0.016);  Establishment/Enactment (0.28, 0.0051);  Political party (0.27, 0.018); Independence (0.23, 0.034); Axis/Pivot (0.21, 0.017);  \\
\vspace{0.5\baselineskip}

\textbf{[News media -related terms]} \\
Press (0.22, 0.033); OA (On Air) (0.22, 0.0093); On air (0.20, 0.023); ITmedia:news site (0.20, 0.039); Janetter:Twitter client (0.22, 0.014);  \\
\vspace{0.5\baselineskip}

\textbf{[Other]} \\
 Older sister (0.21, 0.022); Sound (0.20, 0.040); Stay/Night (0.21, 0.014); Official (0.20, 0.030); Premium/Exclusive (0.23, 0.023);  House/Eaves (0.23, 0.011); Great East Japan Earthquake (0.25, 0.021); Honda:company name (0.24, 0.024); \\
\end{minipage}
\\
\hline
\end{tabular}
\caption{
\revisecolor{black}{
Co-occurring words associated with the growth-shape parameter $\alpha_i$, organized by semantic categories that differ between small and large $\alpha_i$ regimes. Co-occurring words are those appearing within 40 characters of a focal new word with a given $\alpha_i$. The left column lists words that tend to co-occur with terms having small $\alpha_i$ (negative correlation), grouped into AKB48 (a Japanese pop idol group)-related terms, place-related terms, game/anime/subculture-related terms, and others. The right column lists words that tend to co-occur with terms having large $\alpha_i$ (positive correlation), grouped into global/public affairs terms, news media/journalism terms, and others. Entries are listed as: Co-occurring word (Kendall's $\tau$, $p$-value). The table shows that small $\alpha$ co-occurrences are dominated by niche, local terms such as AKB48-related terms, place names, and anime/game-related terms, whereas large $\alpha$ co-occurrences are dominated by globally shared, newsworthy terms such as global and political/economic terms. The method for calculating correlations is provided in Appendix Section~\ref{app_sec_cooccurrence}. A version listing the original Japanese words is provided in Appendix Table~\ref{tab:cooccurrence_alpha_small_vs_large_japanese}.
}
}
\label{tab:cooccurrence_alpha_small_vs_large}
\end{table*}

\begin{table}[t]
\centering
\begin{tabular}{p{0.56\linewidth}cc}
\hline
Category & Median $\alpha$ [CI] & n \\
\hline
Adult film actor/actress \textcolor{blue}{$\blacklozenge$}  & -0.15 [-0.24,0.32] & 6\\
Members or alumnae of AKB48 related groups \textcolor{blue}{$\blacklozenge$}  & -0.00 [-0.18,0.09] & 68\\
Mass-media outlets and online information portals & 0.00 [-0.09,0.14] & 35\\
Location/infrastructure names & 0.02 [-0.11,0.14] & 51 \\
Other idols' personal names \textcolor{blue}{$\blacklozenge$} & 0.13 [-0.01,0.30] & 36 \\
Titles of content/works (Television programs/novels/books etc.) & 0.13 [-0.16,0.56] & 27\\
Anime/video game terms & 0.14 [-0.09,0.34] & 22\\
Voice actors \textcolor{blue}{$\blacklozenge$}  & 0.19 [0.08,0.48] & 40\\
Other service/product names & 0.23 [0.05,0.41] & 52\\
Idol groups  & 0.25 [0.19,0.37] & 32 \\
Drug names and medical technical terms & 0.28 [0.12,0.35] & 13\\
ICT/technology service or product & 0.31 [0.19,0.35] & 63\\
Singer/Band group & 0.32 [0.24,0.38]  & 55\\
Food \& beverage services/products & 0.32 [0.12,0.53] & 18\\
Subculture terms (entertainment/internet culture/slang) & 0.41 [0.32,0.53] & 31\\
Other organization names & 0.42 [0.14,0.57] & 41\\
Internet/ICT/technology-related terms & 0.46 [0.33,0.56] & 48\\
Celebrities (athletes/comedians/TV personalities/novelists etc) \textcolor{blue}{$\blacklozenge$} & 0.50 [0.42,0.57] & 80\\
Public affairs terms (Economy/business/politics/social-issue) & 0.53 [0.44,0.75] & 22\\
Actors/actress (film/TV) \textcolor{blue}{$\blacklozenge$} & 0.53 [0.40,0.60] & 53\\
Society/lifestyle terms (food/clothing/housing)& 0.54 [0.36,0.69] & 24\\
Symbols and emoji & 0.56 [0.51,0.74] & 14\\
Character names (excluding anime/game) & 0.71 [-0.08,0.81] & 9\\
\hline
\end{tabular}
\caption{
\revisecolor{black}{
Semantic categories of keywords classified by an LLM, sorted by the median of the growth-shape parameter $\alpha_i$. Columns show the median, confidence interval (CI), and number of items ($n$). Categories with small $\alpha_i$ (top rows) correspond to niche or local terms, such as AKB48 (a Japanese idol group) members, place names, and anime/game terms. In contrast, categories with large $\alpha_i$ (bottom rows) are associated with globally shared, high-profile topics, such as public affairs, actors, and society/lifestyle terms. Notably, even among categories composed only of personal names (indicated by \textcolor{blue}{$\blacklozenge$}), the $\alpha_i$ values differ, reflecting differences in their general appeal. For details on the calculation and the prompt used, see Appendix~\ref{app_sec_llm} and Code~\ref{app_fig_llm_prompt_2}, respectively.
}
}
\label{table_llm_bunrui}
\end{table}

\begin{table}[t]
\centering
\begin{tabular}{p{0.4\linewidth}ccc}
\hline
Category & Mode $\alpha$ [CI] & Median $\alpha$ [CI] & n \\
\hline
Public buzz topics & 0.60 [0.37,0.67] & 0.45 [0.37,0.51] & 194 \\
General-interest topics (non-buzz) & 0.41 [0.28,0.48] & 0.34 [0.26,0.40] & 171 \\
Insider (niche) topics & 0.21 [0.07,0.47] & 0.19 [0.11,0.25] & 280 \\
\hline
\end{tabular}
\caption{
\revisecolor{black}{
Keywords classified by an LLM according to their degree of general appeal, grouped into three categories. For each category, the table reports the mode and median of the growth-shape parameter $\alpha_i$, with confidence intervals (CI), along with the number of items ($n$). Public and buzz topics show the largest $\alpha_i$ values, followed by general‑interest (non‑buzz) topics, while insider (niche) topics exhibit the smallest $\alpha_i$ values. Details of the classification and statistical calculations are provided in Section~\ref{sec_llm} and Appendix~\ref{app_sec_llm}. The prompt used for the LLM is given in Code~\ref{app_fig_llm_prompt_1}.
}
}
\label{table_llm_wadai}
\end{table}

\section{From Meaning to Diffusion Shape: Interpreting $\alpha_i$}
\label{sec_meaning}
We examine how the power-law model relates to collective human behavior, focusing on the social interpretation of the shape parameter $\alpha_i$ that governs the growth-curve form. For clarity, we restrict attention to single-segment cases ($N=1$), especially the sub-exponential regime $0 \le \alpha_i \le 1$, which is common in the data (Fig.~\ref{fig_para}(a)) but whose interpretation is not yet clearly established. By contrast, superexponential behavior ($\alpha_i>1$) is often associated with ``deadline effects'' \cite{alfi2007conference}—e.g., fixed dates such as the Olympics or a scheduled product launch—and has been documented in prior work \cite{watanabe2023minor}.

\subsection{Lexical Items and their Co-Occurring Terms}
\label{sec_word}
We first inspect representative lexical items and their typical co-occurring words across several values of $\alpha_i$.
Table~\ref{tab:examples_by_alpha} (or Appendix Table~\ref{tab:examples_by_alpha_japanese} in Japanese) lists examples for $\alpha_i \approx 0$, $\alpha_i \approx 0.5$, and $\alpha_i \approx 1$ (here we apply stricter filters than in Section~\ref{sec_parameters}: growth duration $\ge 48$ bins and Spearman's $\rho \ge 0.7$). \par
Items with $\alpha_i \approx 0$ (approximately linear growth) prominently include (i) personal names associated with AKB48 and related groups (e.g., ``Erika Ikuta'', ``Mai Shinuchi'', ``Himeka Nakamoto'', ``Ranze Terada''); (ii) newly introduced or renamed administrative locations (e.g., ``Kita Ward, Okayama City''; ``Midori Ward, Sagamihara City''), ``Itsukushima Shrine''; and (iii) media/platform names (e.g., ``Nico Nico Seiga'', ``NewsPicks'', ``Pinterest'', ``JBpress''). Note that ``Itsukushima Shrine'' referred to its formal name written in traditional \textit{kyujitai} (old form) characters. The observed increase in the use of this formal spelling was due to improvements in Kanji input systems. While the simplified \textit{shinjitai} (new form) had always been easy to type, the improved systems had made inputting the formal \textit{kyujitai} version much easier. \par
Items with $\alpha_i \approx 0.5$ (a typical sub-exponential trajectory) span diverse categories—for example, technology terms such as ``tablet device''; musical acts such as ``BABYMETAL'', lifestyle terms such as ``cat-rescue cafe'', ``Instagrammer'', the restaurant chain ``Tsukada Nojo''; and symbols/emoji such as ``\emoji{baseball}'' and ``\emoji{anchor}''. \par
Items with $\alpha_i \approx 1.0$ (approximately exponential growth) often include global news terms (e.g., ``shale gas'', ``microplastics'', ``Airbnb'') and domestic buzzwords (e.g., ``Instagrammable'', ``online salon'', ``a\c{c}a\'i bowl''). The tent emoji ``\emoji{tent}'' likely reflects an outdoor/camping boom during the study period. \par
We then ask which words tend to co-occur around items with smaller or larger $\alpha_i$.
Table~\ref{tab:cooccurrence_alpha_small_vs_large} (or Appendix Table~\ref{tab:cooccurrence_alpha_small_vs_large_japanese} in Japanese) reports, for each candidate co-occurring term, the correlation between item-level $\alpha_i$ and the term's local usage rate (within $\pm 40$ characters of the focal new item). Terms with strongly positive/negative correlations are listed as co-occurrences characteristic of items with larger/smaller $\alpha_i$ (details in Appendix, Section~\ref{app_sec_cooccurrence}). Items with small $\alpha_i$ tend to co-occur with place-related or local-context terms (e.g., ``in front of a station'', ``regular closing day'', ``Miyagi Prefecture'', ``Kanagawa [Kanagawa Prefecture]'') and vocabulary tied to specific fan communities (e.g., AKB48-related terms like ``trainee'', ``birthday event'', ``photo session''). By contrast, items with large $\alpha_i$ more often co-occur with public-affairs/news terms typical of broadcast and print media (e.g., ``global'', ``profit'', ``transaction'', ``establish'', ``regulation'', ``Europe'', ``France'', ``party'', ``independence''). \par

\subsection{LLM-Based Lexical Categorization}
\label{sec_llm}
In the previous section we grouped items by $\alpha_i$; here we categorize items by content (using a large language model, Gemini 2.5 Flash) and compute the median $\alpha_i$ within each category. Tables~\ref{table_llm_bunrui} and \ref{table_llm_wadai} report the results, and the prompts are provided in Appendix, Section~\ref{app_sec_llm}. \par
Table~\ref{table_llm_bunrui} orders categories from smaller (more linear-like) to larger (more exponential-like) $\alpha_i$. Consistent with the analysis above, AKB48-related personal names and location/infrastructure names lie toward smaller $\alpha_i$; subculture terms (e.g., anime, video games) also tend to have smaller values. By contrast, celebrity names, actors, public-affairs/news terms, and broadly used lifestyle terms tend to have larger $\alpha_i$. \par
Based on the analyses so far, words with small $\alpha_i$ tend to be niche, specialized, or locally bounded, whereas words with large $\alpha_i$ tend to be global or have high general appeal. This suggests a possible link between $\alpha_i$ and a word's general appeal as a conversation topic. We therefore used the LLM to assign topic types by general appeal and examined $\alpha_i$ within each group (Table~\ref{table_llm_wadai}). The results show a graded pattern: topics with high general appeal (widely recognized through national news or advertising and easy to discuss as small talk) have a mode $\alpha_i \approx 0.6$; those with medium general appeal (shared through everyday exposure or word of mouth) have $\alpha_i \approx 0.4$; and low-appeal niche topics (primarily discussed within specific interest groups or expert communities) have $\alpha_i \approx 0.2$. \par
\section{A Behavioral Model for Power-Law Growth}
\label{sec_infection}
Finally, we examine how non-trivial power-law growth emerges from characteristics of human behavior, based on the theoretical framework. This study presents a model that incorporates the ``general appeal'' effect discussed in previous sections into the infection mechanism. It should be noted that in infectious disease research, power-law growth (General Growth Model) is explained by factors such as geographical effects and heterogeneity in the distribution of infected individuals etc. \cite{chowell2016mathematical,yan2024modeling,karev2014non}. Here we give only an overview; the full model setup, parameter meanings, and derivations are provided in Appendix \ref{app_sec_infection0}.

\subsection{A Model of Topic-Driven Diffusion Dynamics}
In this section, we develop a model to explain why information diffusion often exhibits a power-law (i.e., sub-exponential) growth pattern. We argue that this macroscopic phenomenon originates from the interplay between a topic's intrinsic properties and the corresponding microscopic actions of individuals.

The diffusion pattern is heavily influenced by the nature of the topic. For instance, \textbf{niche topics}---such as the name of a specific member of the pop group AKB48---are primarily discussed within communities that already recognize their value. In such cases, individuals prioritize \emph{inward} interactions with those who are already familiar with the subject. As the community grows, the relative incentive to reach external, uninformed individuals diminishes. 
In contrast, for general or ``buzz-worthy'' topics, the value often lies in broader dissemination. This encourages an \textbf{outward-looking} strategy, where individuals are more motivated to engage with people who do not yet know the term.

We translate this link between topic type and action choice into a simple rule to derive the macroscopic growth curve from these microscopic interactions. A key result of our model is that the shape parameter of the growth curve, $\alpha_i$, is determined by the following equation:
\begin{equation}
\alpha_i = 1 - \frac{\gamma_i}{Q}
\end{equation}
where $\gamma_i$ represents the topic's \emph{inwardness}---a measure of its tendency for internal discussion---and $Q$ is a constant representing the number of potential external candidates for interaction per action.

\subsection{Building the Model}
\subsubsection{States and Observables}

We begin by defining the fundamental variables of our diffusion model. The primary internal state of the system for a given topic $i$ is the cumulative number of individuals who are aware of it at time $t$. We denote this quantity as $I_i(t)$.

While $I_i(t)$ represents the true number of informed individuals, this value is often not directly measurable. Instead, we typically have access to observable data, such as the volume of social media posts or search queries related to the topic. We assume that this observed time series, denoted by $y_i(t)$, is directly proportional to the internal state. Their relationship is defined as:
\begin{equation}
    y_i(t) = K \cdot I_i(t)
\end{equation}
where $K$ is a constant scaling factor that links the number of informed individuals to the volume of observable activity.

\subsubsection{Model Dynamics: Actions and State Updates}
In our model, individuals who are informed about topic $i$ (hereafter, ``informed individuals'') initiate actions at a constant rate $J_i$. Each action results in the creation of a single new directed link. The state of the system is updated according to two fundamental actions:

\begin{itemize}
    \item \textbf{Recruitment:} An informed individual, $l$, contacts a susceptible (i.e., uninformed) individual, $a$. This action brings $a$ into the set of informed individuals. The total number of informed individuals $I_i(t)$ increases by one, and the out-degree of the acting individual $l$, denoted by $k_{i}^{(l)}(t)$, also increases by one.
    \begin{equation*}
        I_i(t+\Delta t) = I_i(t) + 1, \qquad k_{i}^{(l)}(t+\Delta t) = k_{i}^{(l)}(t) + 1.
    \end{equation*}
    The newly informed individual $a$ establishes their initial set of connections through a mechanism of \emph{inheritance}, which will be detailed later. Here, $k_{i}^{(l)}(t)$ represents the number of individuals that $l$ can directly reach on topic $i$ at time $t$.
    \item \textbf{Interaction:} An informed individual, $l$, establishes a new connection to another individual, $b$, who is already informed but not yet in $l$'s direct contacts. This action increases the network density among the informed population without changing its size. Consequently, only the out-degree of the acting individual $l$ is updated.
    \begin{equation*}
        I_i(t+\Delta t) = I_i(t), \quad k_{i}^{(l)}(t+\Delta t) = k_{i}^{(l)}(t) + 1.
    \end{equation*}
    Note that the in-degree of individual $b$ changes, but its out-degree $k_{i}^{(b)}(t)$ is unaffected, as all links in this model are treated as directed (recruitment creates reciprocal additions, whereas interaction corresponds to a one-sided ``discovery''). 

\end{itemize}

\subsubsection{Action Choice Probability}
The choice between Recruitment (an ``outward'' action) and Interaction (an ``inward'' action) is probabilistic. For any informed individual $l$, we define the probability of choosing an outward action at time $t$ as:
\begin{equation}
    p_i^{(l)}(t) = \frac{Q}{Q + \gamma_i k_{i}^{(l)}(t)} \label{eq:choice_prob}
\end{equation}
This probability is governed by two key parameters:
\begin{itemize}
    \item $Q > 0$ is the effective number of external candidates an individual can reach per action. This parameter represents the size of the pool of susceptible individuals, which we assume does not deplete.
    \item $\gamma_i \ge 0$ is a topic-specific parameter that we term the topic's \textbf{inwardness}. It modulates the preference for interacting within the informed group versus recruiting new individuals. If $\gamma_i = 0$, actions are always outward-focused (recruitment). If $\gamma_i = 1$, an individual's known contacts and the external candidates are weighted equally.
\end{itemize}
Intuitively, Eq.~\ref{eq:choice_prob} captures the concept that as an individual's personal network of contacts grows (i.e., as $k_{i}^{(l)}(t)$ increases), their incentive to spread the information externally decreases, leading to a lower probability $p_i^{(l)}(t)$. 

\subsubsection{Inheritance Mechanism}
When a recruitment event occurs, the newly informed individual $a$ immediately integrates into the network. Their initial state is not a blank slate; instead, they inherit the contacts of their recruiter, $l$. Specifically, $a$ gains a direct link to $l$ and also inherits all of $l$'s existing contacts. This sets the initial out-degree for the new individual $a$ as:
\begin{equation}
    k_{i}^{(a)}(t+\Delta t) = k_{i}^{(l)}(t) + 1.
\end{equation}
Following this initialization, individual $a$ behaves identically to all other informed individuals, initiating actions at the same rate $J_i$ and selecting outward actions based on their own evolving out-degree, $k_{i}^{(a)}(t)$. This mechanism is analogous to a new student in a research field inheriting the professional network of their advisor, providing an immediate foundation for future engagement.

\subsection{Deriving the Macro Growth Curve}
Because of inheritance, a newly added individual starts from the average level of known contacts at that time, we can approximate $k_{i}^{(l)}(t)\approx J_i t$ for any informed $l$ by using the community age $t$. Substituting this into Eq.~\ref{eq:choice_prob} gives
\begin{equation}
p_i^{(l)}(t)\approx \frac{1}{1+(\gamma_i/Q)J_i t}.
\end{equation}
Hence, the number of informed people satisfies
\begin{equation}
\frac{dI_i}{dt}=\frac{J_i}{1+(\gamma_i/Q)J_i t} I_i(t). \label{eq_1_t}
\end{equation}
With $y_i(t)=K I_i(t)$ and $I_i(0)=1$, we obtain
\begin{equation}
\frac{dy_i}{dt}=J_i K\left(\frac{y_i}{K}\right)^{1-\gamma_i/Q}.
\label{eq_infect_ans}
\end{equation}
Comparing with the power-law growth model (Eq.~\ref{eq_base0}) yields
\begin{equation}
\alpha_i=1-\frac{\gamma_i}{Q},\qquad R_i=J_i
\end{equation}
with derivations in Appendix~\ref{app_sec_infection0}. Figs.~\ref{fig_scale_time_series}(g)-(i) compare the model's numerical simulations with the theoretical predictions, and panels (j)-(l) depict the corresponding infection-pathway networks (directed edges from recruiters to recruits).

\subsection{Interpretation of the Model}
This result suggests that the observed shape parameter $\alpha_i$ is governed by the topic's inwardness $\gamma_i$. Exponential-like growth ($\alpha_i\approx 1$) corresponds to topics with near-zero inwardness ($\gamma_i\approx 0$), while growth approaching linear ($\alpha_i\to 0$) reflects highly inward topics ($\gamma_i\to Q$). Thus, $\alpha_i$ is not only a curve-shape parameter; it can also be read as a sociophysical index that quantifies a topic's inwardness—how easily it is shared outward versus discussed internally. \par
Furthermore, Appendix~\ref{app_sec_infection_model} extends the model to a more realistic \emph{weighted, directed} infection network. This extension reproduces the growth curve and also allows various in-degree distributions. For example, by tuning parameters, one can describe dynamics in which, regardless of $\alpha_i$, the cumulative in-degree distribution follows a power law with an exponent close to 1. In addition, Appendix~\ref{app_sec_unweighted} shows that the choice rule for the target of \emph{interaction} (e.g., random choice vs. introduction by a friend) does not change the aggregate growth curve for the total number of informed, although it can change other network structures such as the in-degree distribution.

\section{Summary and Discussion}
\revisecolor{black}{
\subsection{Summary of Main Results}
}
This study showed that online keyword time series can be described with a small number of parameters by the piecewise power-law model (a piecewise generalized growth model). We summarize four main results. \par
First, of the 2,963 items selected for analysis (satisfying the criteria for sufficient observation data and monotonic increase described in Section~\ref{sec_parameters}), about 55\% (1,625 items) were found to have no abrupt jumps (level shifts) and were best fit by a model with one or two segments (Table~\ref{app_tab_n_table}). \par
Second, focusing on Wikipedia-listed terms whose growth in our blog dataset is captured by a single segment ($N=1$), the most common pattern was sub-exponential. The mode of the shape parameter $\alpha_i$ was close to 0.5, lying between linear growth ($\alpha_i=0$) and exponential growth ($\alpha_i=1$) (Fig.~\ref{fig_para}(a)). The final level of adoption $y_i(T_i)$ depended mainly on the growth rate $R_i$, with weaker effects from the shape $\alpha_i$ and the growth duration $T_i$ (Fig.~\ref{fig_para}(h)-(j)). We observed consistent trends in Google Trends for English, Spanish, and Japanese (Appendix, Sections~\ref{app_sec_google_stat}, \ref{app_fig_para}). \par
Third, we found a systematic link between topic type and the shape parameter $\alpha_i$. Niche or local topics tended to have smaller $\alpha_i$, while widely shared, general topics tended to have larger $\alpha_i$ (Section~\ref{sec_meaning}, Tables~\ref{table_llm_bunrui}, \ref{table_llm_wadai}). \par
Finally, starting from a behavioral infection model that separates outward (to unknown others) and inward (among known peers) interactions, we derived the power-law growth (i.e., generalized growth model) given by Eq.~\ref{eq_base0} and obtained $\alpha_i = 1 - \gamma_i/Q$ (with $\gamma_i$ the inwardness and $Q$ the number of outward candidates per action). When $0 \le \gamma_i \le Q$, this implies $0 \le \alpha_i \le 1$, suggesting one interpretation in which $\alpha_i$ can serve as a candidate indicator of outward orientation (Section~\ref{sec_infection}). \par
\revisecolor{black}{
\subsection{Implications and Applications}
The results of this study extend the conventional understanding of the typical temporal evolution of language diffusion. While the diffusion of social phenomena has often been characterized by S-shaped growth typified by the logistic curve, our large-scale empirical analysis has systematically shown that sub-exponential growth—a pattern previously observed in the field of infectious diseases but otherwise largely overlooked in social diffusion—is widely found across numerous keywords. It is thought that this growth pattern has been overlooked in the past because conventional research tended to focus on highly topical keywords, leading the slower and less observable growth of niche topics to be considered an exception. The present study was able to capture these patterns by employing a diverse dictionary of terms derived from Wikipedia headwords in conjunction with a large-scale, high-quality Japanese blog dataset.}
\par
\revisecolor{black}{
This finding implies that the spread of language and culture should not be understood solely as a process of rapid takeoff followed by saturation, but that more gradual and sustained patterns—where the rate of diffusion continuously decays over time—may also occur to a non-negligible extent. In this respect, whereas logistic-type models assume convergence toward a fixed saturation level, the present framework does not explicitly impose such an upper bound and instead captures growth that decelerates consistently from the early stages.
}
\par
\revisecolor{black}{
The piecewise power-law model proposed in this study provides a unified framework for describing diverse growth patterns using a small number of parameters. In particular, the shape parameter $\alpha_i$, which characterizes the form of the growth curve, is found to be systematically associated with the nature of the topic, such as its niche or general appeal. This suggests that there may exist a quantitative relationship between the dynamics of language diffusion and the social meaning of the content. Furthermore, by linking the present model to a micro-level behavioral framework that incorporates topic-specific characteristics, it is suggested that features of individual-level interactions may be reflected in differences in the shapes of macroscopic growth curves. In addition, the final adoption level is found to be primarily related to the growth rate parameter $R_i$, implying that early-stage dynamics may already contain partial information about the eventual scale of diffusion.
}
\par
\revisecolor{black}{
The proposed framework is also expected to be useful for a broad range of studies that aim to capture social attention and cultural change through time series of word usage frequencies. Since word frequency is widely used as an indicator of changes in public interest, the ability to describe its growth in a unified and interpretable manner contributes to the quantitative understanding of social dynamics.
In particular, growth patterns involving gradual deceleration, which are not naturally captured by the Bass model widely used in marketing, can be effectively represented within our framework. In this sense, the present study provides a modeling approach for growth time series that achieves both flexibility and interpretability. By capturing the simple structures underlying complex temporal patterns, it serves as a useful basis for comparing and interpreting a wide range of diffusion cases, with potential applications not only in academic research but also in practical fields.
}
\par
\revisecolor{black}{
\subsection{Limitations and Future Work}
}
Several limitations should be kept in mind when interpreting our results. Our analysis is limited to words that appear as Wikipedia headwords, so a form of survival bias is present. For example, two of our typical cases of linear growth—names of members of the Japanese idol group AKB48 and new place names—are informative for thinking about this bias. The social frame of a strong brand like AKB48, or the category of ``place name,'' can allow a word to be listed on Wikipedia even without a large prior record or buzz. After listing, name recognition can then grow steadily in a near-linear way, supported by the group's continued activity or by administrative persistence of the place. This suggests that, if we could also observe words that never gained traction and disappeared, the share of linear growth patterns in the whole society might be higher than what we observe here. On the other hand, our Wikipedia-based lexicon contains many niche words, which are more likely to show sub-exponential growth. In other populations—for example, the set of words used by a single person—more general words would appear and niche words would be fewer than on Wikipedia. In such cases, the fraction of words showing more exponential growth (larger $\alpha_i$) could be higher. \par
 This study has proposed the infection model framework that explains power-law growth through the concept of ``inwardness''. This behavioral model employs an idealized (strong) assumption—that new recruits fully copy the outgoing links of their recruiters—in order to almost exactly derive the power-law growth model given by Eq.~\ref{eq_base0}. Thus, relaxing this assumption to consider more realistic network formations (e.g., "partial copying") remains a task for future work. However, this should not be considered the sole explanation, as alternative mechanisms may exist. For instance, the linear growth observed in geographical name transitions can be interpreted within our model as a state of high inwardness, while it could also be explained as a process where individuals independently recognize and adopt new names. Similarly, shifting interest in member names within groups like AKB48 could be understood as an attention ``replacement'' process accompanying new member additions (AKB48 is a large group of roughly 50 to 100 members, where members typically stay for a few years before leaving, and new members constantly join to replace them). As Denison remarked regarding S-curves, being ``merely a first step in understanding'' \cite{denison2003log}, our model and growth curve shape analysis should be positioned as a starting point for exploring diverse mechanisms. Identifying the precise mechanisms requires further verification using micro-level or other types of data. In any case, the insights gained in this study regarding growth curve shapes may serve as a ``common language'' that enables comparison of seemingly disparate phenomena—such as idol member names and geographical names—from a unified perspective, potentially facilitating future research. \par
As a broader implication, the sub-exponential growth observed in this study can also be interpreted, according to Eq.~\ref{eq_1_t}, as an ``aging-like phenomenon'' where the per-capita acquisition rate decays approximately following a power law ($\frac{dy(t)}{dt} \cdot \frac{1}{y} \propto 1/t $ for $t \gg 1$). This power-law decay in information diffusion likely corresponds to socio-dynamic processes such as the decay of novelty or natural waning of interest. Similar power-law decay patterns have been reported in other temporal phenomena related to words. For example, the logarithmic diffusion of established words has been explained by a forgetting effect decaying as $t^{-0.5}$ \cite{watanabe2018empirical}, while analogous tendencies are observed in the post-peak decay of new words \cite{watanabe2023minor} ($t^{-0.5}$, $t^{-1}$). The recurrence of characteristic exponents such as $1$ and $1/2$ across these phenomena suggests that diverse social information diffusion mechanisms may share a common mathematical structure, potentially paving the way for a unified theoretical understanding in future research.\par
\end{CJK*}

\section*{acknowledgement}The author would like to thank Hottolink, Inc. for providing the data. This work was supported by JSPS KAKENHI, Grant Numbers 21K04529, and the Seijo University Special Research Grant. The author also gratefully acknowledges the Toshiaki Ogasawara Memorial Foundation for the Travel Grant for International Research Conferences. Computations were partially performed on the supercomputer system at ROIS Institute of statistical mathematics.
\clearpage

\appendix
\begin{CJK*}{UTF8}{min}
    
\section*{Supporting Information Appendix}
\section*{Appendix Structure}
\label{app_sec_structure}
This Appendix is organized as follows. Sections A through C provide supplementary materials related to our analysis.

\begin{itemize}
    \item \hyperref[app_sec_google_base]{\textbf{Section A:} Comparative Analysis of Google Trends and Japanese blog data (Time Series and Parameter Statistics)}
    \item \hyperref[app_sec_powerlaw_base]{\textbf{Section B :} Supplementary Details on the Piecewise Power-Law Model (Data and Theory)} 
    \item \hyperref[app_sec_infection_base]{\textbf{Section C:} Details Related to the Behavioral Infection Model}
\end{itemize}

The subsequent sections, D through H, provide details on the data and methodologies employed in this study.

\begin{itemize}
    \item \hyperref[app_sec_data_base]{\textbf{Section D:} Data and Basic Pre-processing (Data Sources, Word Selection, and Time Series Normalization)}
    \item \hyperref[app_sec_preprocessing_base]{\textbf{Section E:} Time Series Pre-processing for Diffusion Curve Analysis (Extraction of Growth Periods and Jump Detection)}
    \item \hyperref[app_sec_estimation_parameter_base]{\textbf{Section F:} Parameter Estimation for the Piecewise Power-Law Model and Evaluation of the Number of Pieces ($N$)}
    \item \hyperref[app_sec_cooccurrence_base]{\textbf{Section G:} Details of Semantic Analysis (Co-occurrence Extraction Method and LLM Prompts)}
    \item \hyperref[app_sec_japanese_base]{\textbf{Section H:} Original Japanese Notation for Example Keywords Used in Figures and Tables}
\end{itemize}

\clearpage 
\counterwithout{figure}{section}
\counterwithout{table}{section}
\counterwithout{equation}{section}

\makeatletter
\setcounter{equation}{0} 
\renewcommand{\theequation}{A\arabic{equation}}

\makeatother

\begin{figure*}[t]
    \centering
    \begin{overpic}[width=4.1cm]{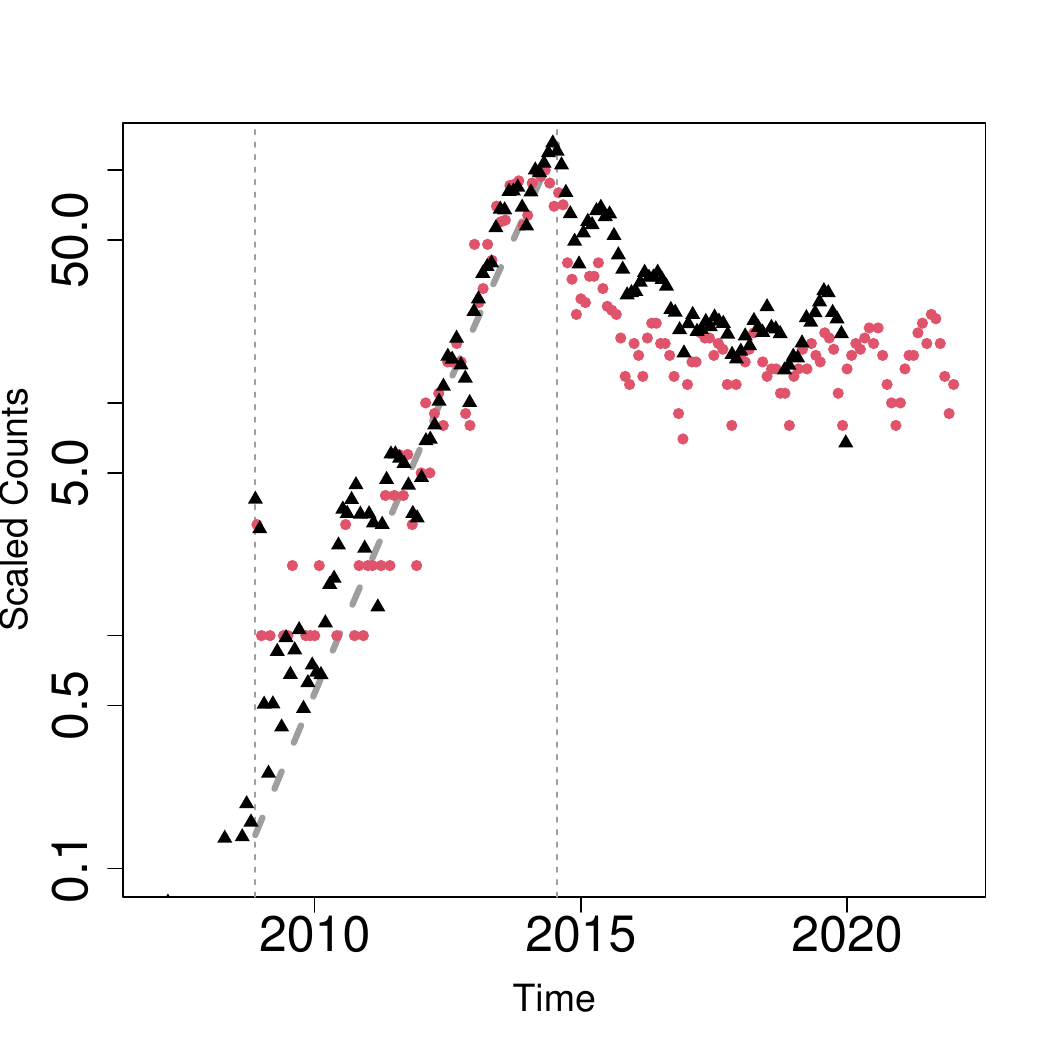}
        \put(33,93){\color{black}\Large\bfseries (a)}
    \end{overpic}
     \begin{overpic}[width=4.2cm]{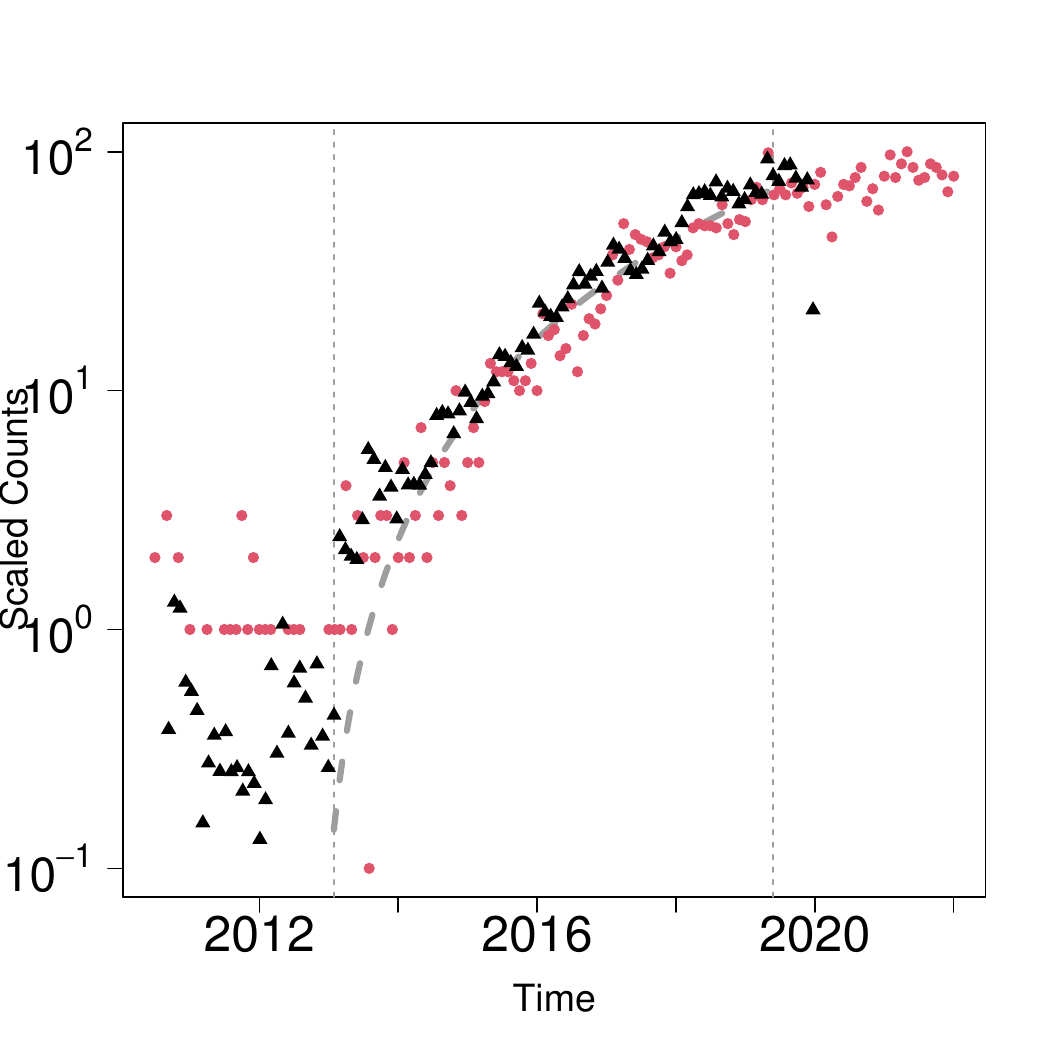}
        \put(20,93){\color{black}\Large\bfseries (b)}
      \end{overpic} 
      \begin{overpic}[width=4.2cm]{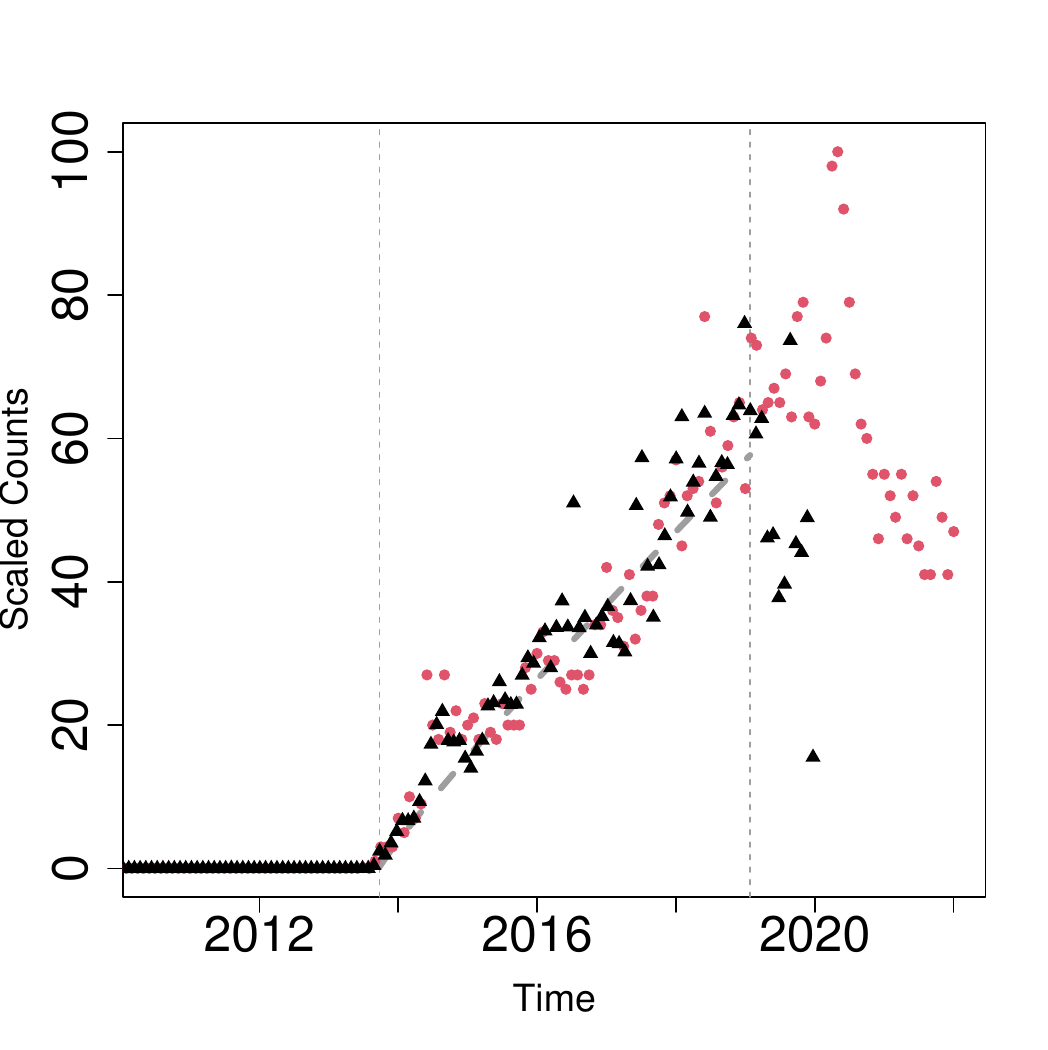}
        \put(25,93){\color{black}\Large\bfseries (c)}
    \end{overpic}
      \begin{overpic}[width=4.2cm]{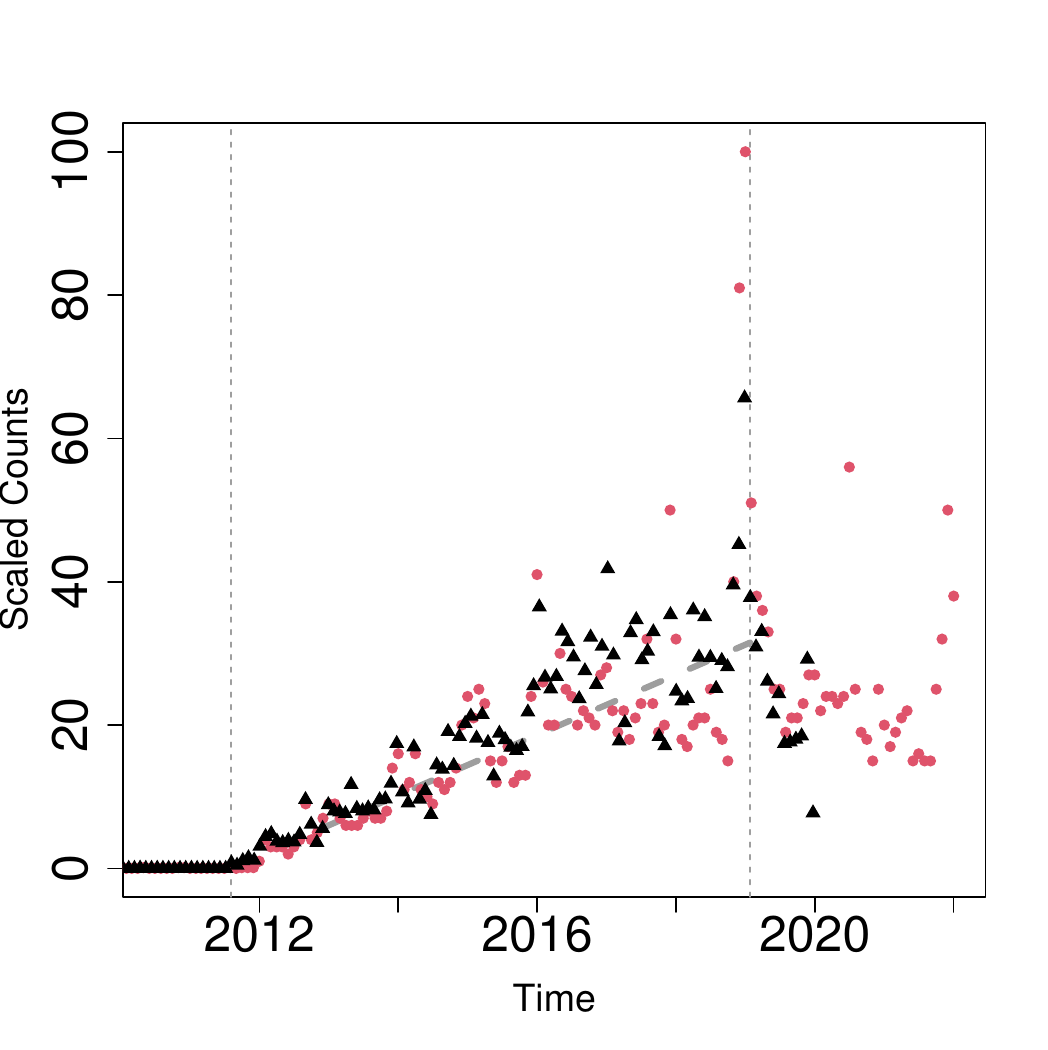}
        \put(30,93){\color{black}\Large\bfseries (d)}
    \end{overpic}
   \caption{
Comparison of blog word count time series and Google Trends for the same keywords. Examples of similar time series patterns between the two datasets. Black triangles represent blog data, red circles represent Google Trends. Gray dashed lines show theoretical curves from the piecewise power-law model for blog data. Gray vertical lines indicate growth periods detected in the blog data. Blog word count time series are scaled to match the temporal average values with Google Trends. 
(a) Exponential growth with $\alpha_i \approx 1$ (\textit{Asaii Bouru} - acai bowl; Japanese: アサイーボール, Brazilian dessert), (b) Typical growth curve with $\alpha_i \approx 0.5$ (\textit{Hogo Neko Cafe}; Japanese: 保護猫カフェ - protective cat cafe, cat cafe with rescue cats).
(c) and (d) are examples of linear growth with $\alpha_i \sim 0$.  
(c) \textit{NewsPicks} - NewsPicks (Japanese: NewsPicks), online news media, 
(d) \textit{Ikuta Erika} - Erika Ikuta (Japanese: 生田絵梨花), AKB48 related group member.
}
    \label{app_fig_comp_google_blog_same}
\end{figure*}

\begin{figure*}[t]
    \centering 
   \begin{overpic}[width=4.2cm]{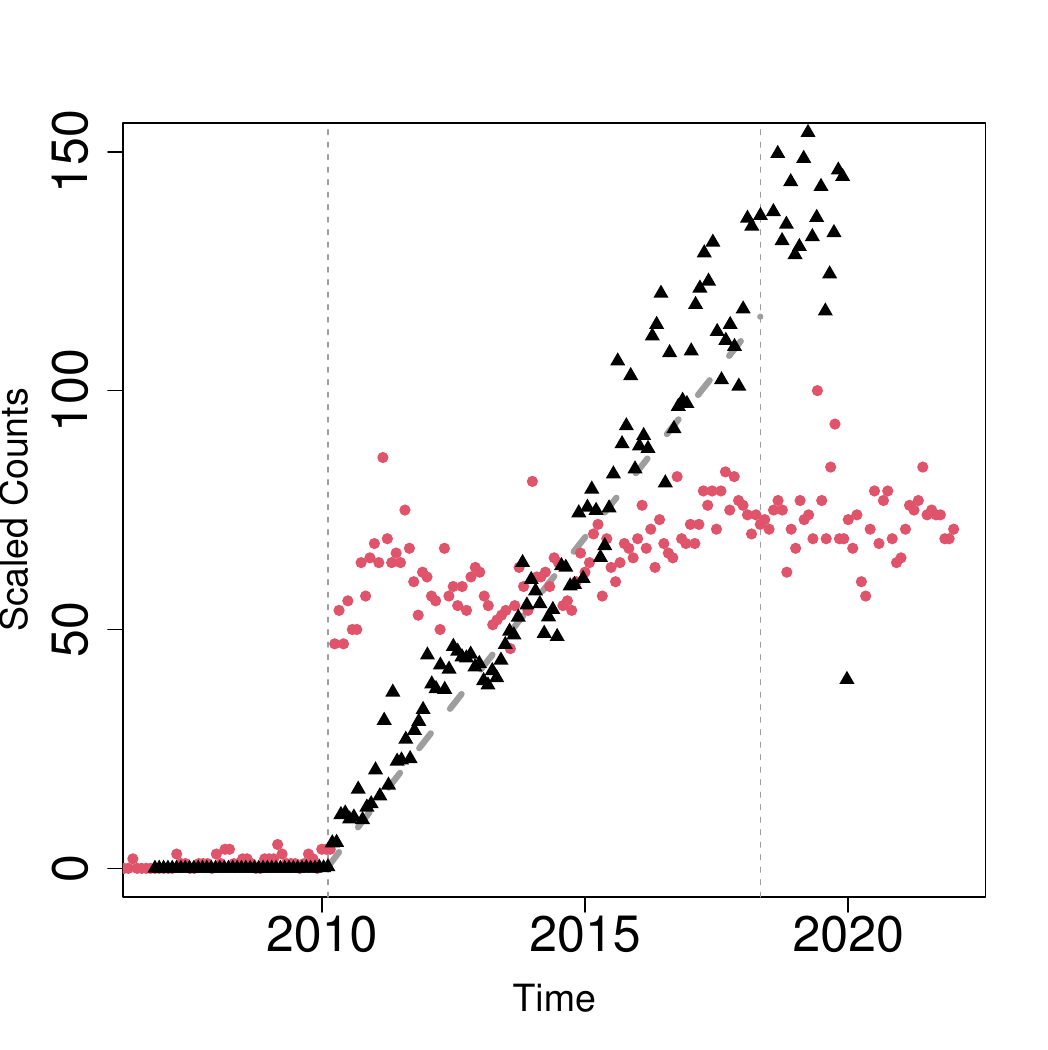}
        \put(19,93){\color{black}\Large\bfseries (a)}
    \end{overpic}
     \begin{overpic}[width=4.2cm]{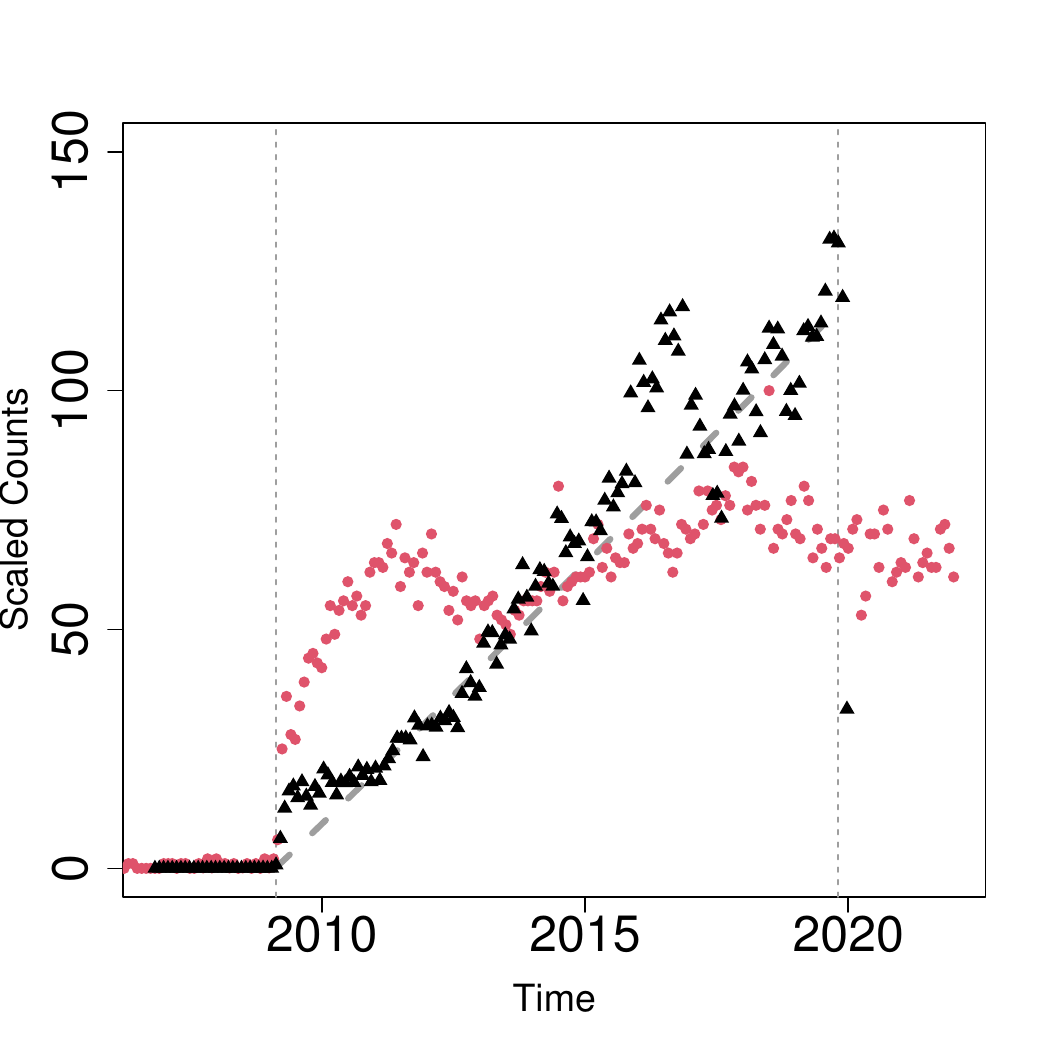}
        \put(36,93){\color{black}\Large\bfseries (b)}
      \end{overpic}
       \begin{overpic}[width=4.2cm]{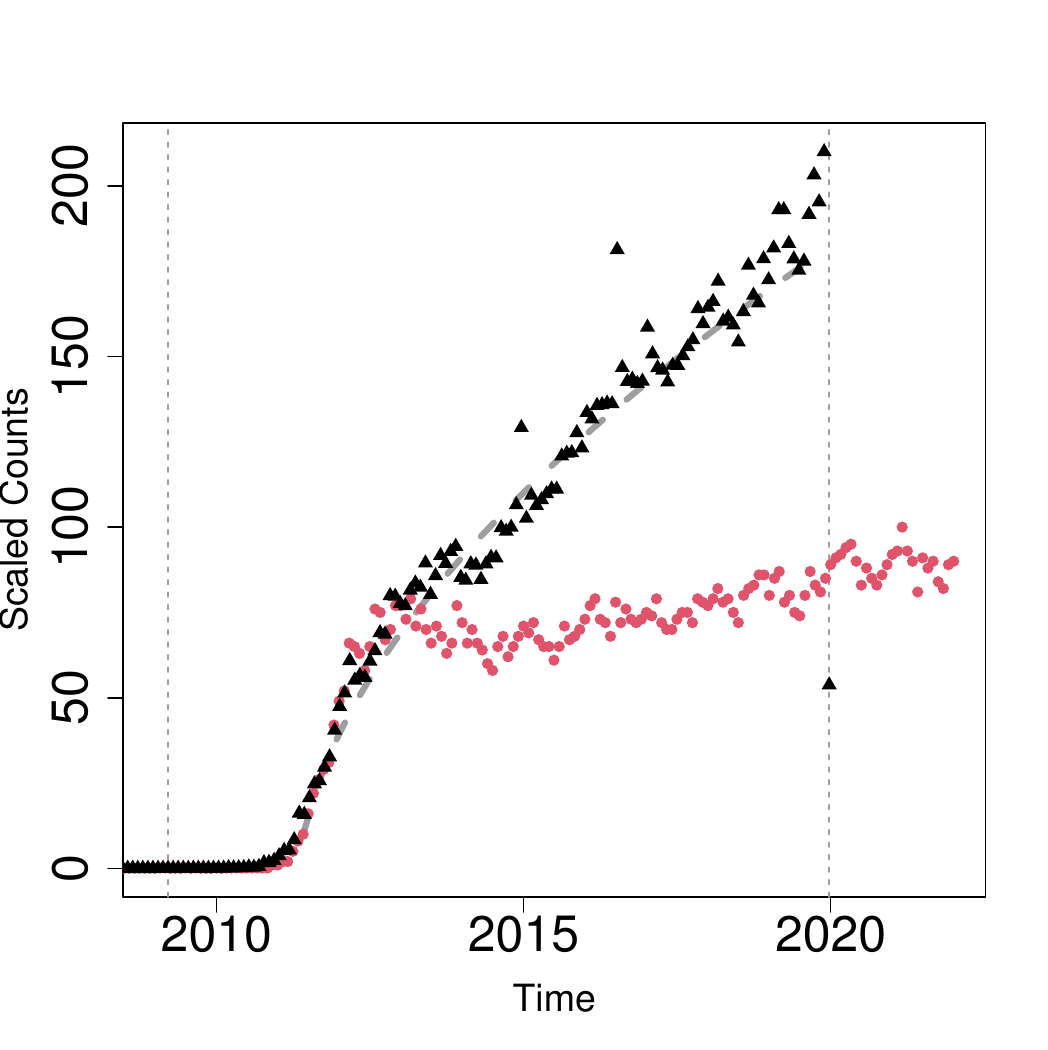}
        \put(32,93){\color{black}\Large\bfseries (c)}
    \end{overpic} 
    \begin{overpic}[width=4.2cm]{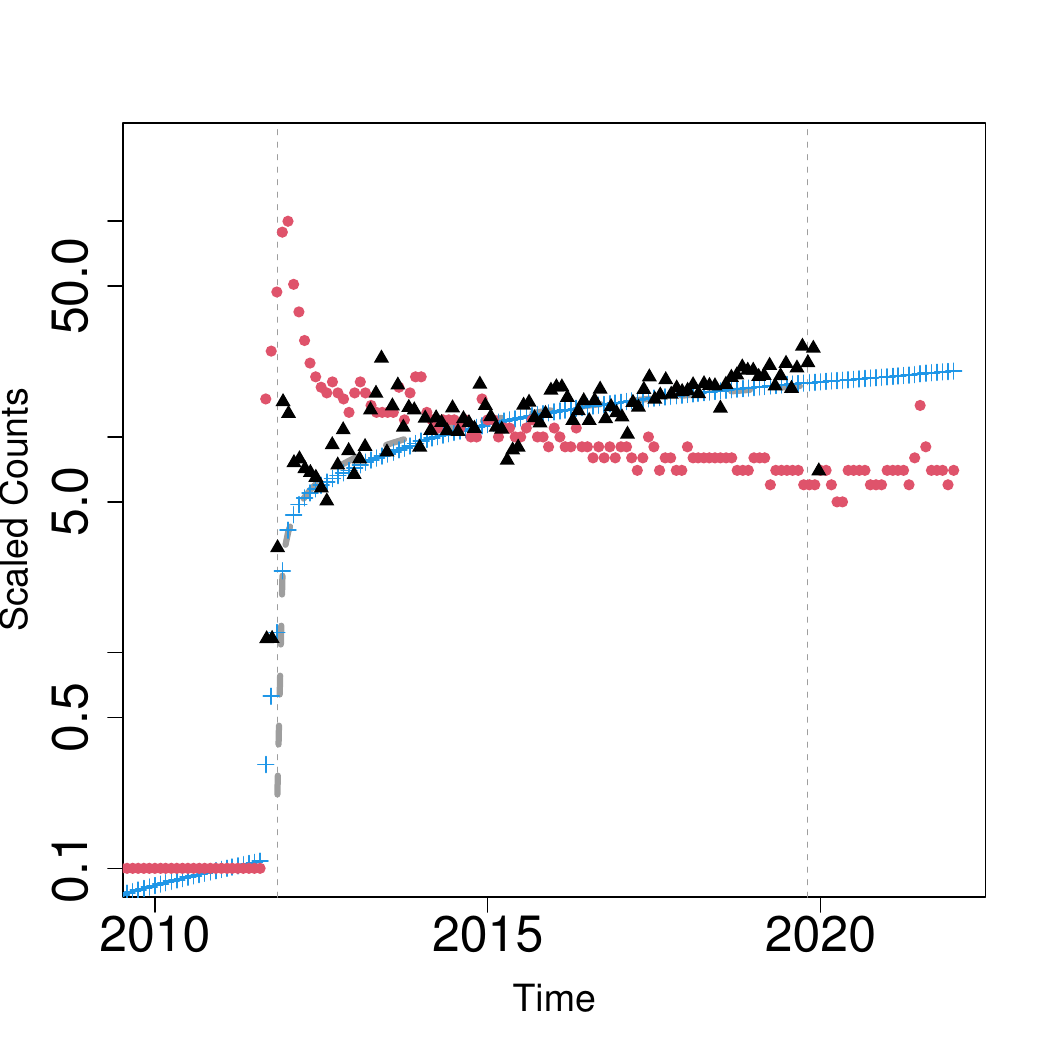}
        \put(37,94){\color{black}\Large\bfseries (d)}
    \end{overpic}
   \caption{
   Comparison of blog word count time series and Google Trends for the same keywords. Examples of different time series patterns between the two datasets. Black triangles represent blog data, red circles represent Google Trends. Gray dashed lines show theoretical curves from the piecewise power-law model for blog data. Gray vertical lines indicate growth periods detected in the blog data. For all cases except (c), blog word count time series are scaled to match the temporal average values with Google Trends. For case (c), the time series are scaled such that the exponential growth phases in the first half overlap between the two datasets. 
   (a) \textit{Sagamihara-shi Chuo-ku} (Sagamihara City Central Ward; Japanese: 相模原市中央区) and (b) \textit{Okayama-shi Kita-ku} (Okayama City North Ward; Japanese: 岡山市北区).
   New place names show near-linear growth in blog data, but Google Trends reveals a tendency for sharp spikes at the beginning of their growth.
    (c)  \textit{Sumaho} (smartphone, mobile device; Japanese: スマホ). When fitted to the initial exponential-like part, the scale does not match in the subsequent linear-like part.
   (d) \textit{Toyota Akua} (Toyota Aqua, hybrid car model; Japanese: トヨタ・アクア). 
   The cumulative Google Trends search volume ($\xi_i^{\mathrm{(cum)}}(t)=\sum_{s=1}^{t} \xi_i(s)$) (blue crosses) corresponds to the number of blog posts.
   }
    \label{fig_comp_google_blog_difference}
\end{figure*}
\begin{figure*}[t]
    \centering
    \begin{overpic}[width=5.5cm]{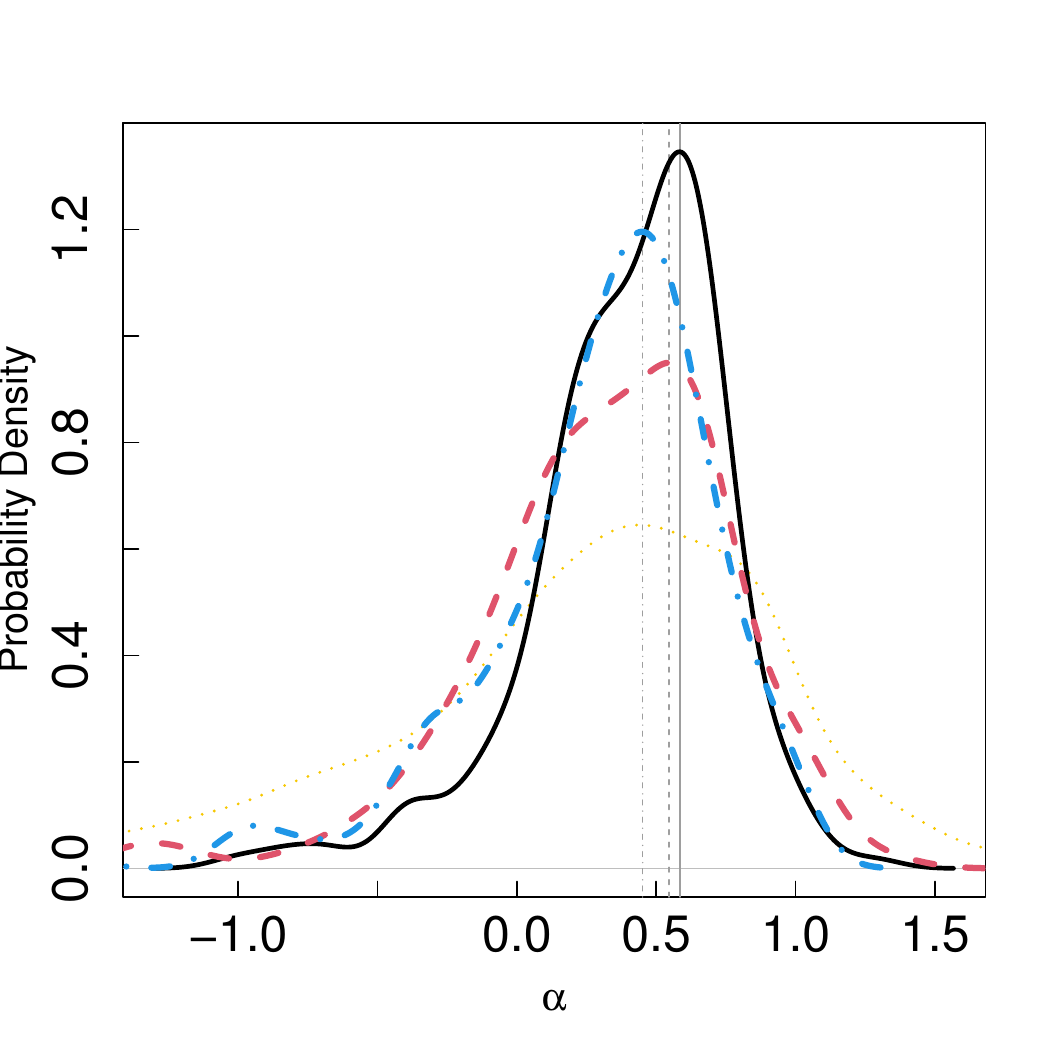}
        \put(30,120){\color{black}\Large\bfseries (a)}
    \end{overpic}
      \begin{overpic}[width=5.5cm]{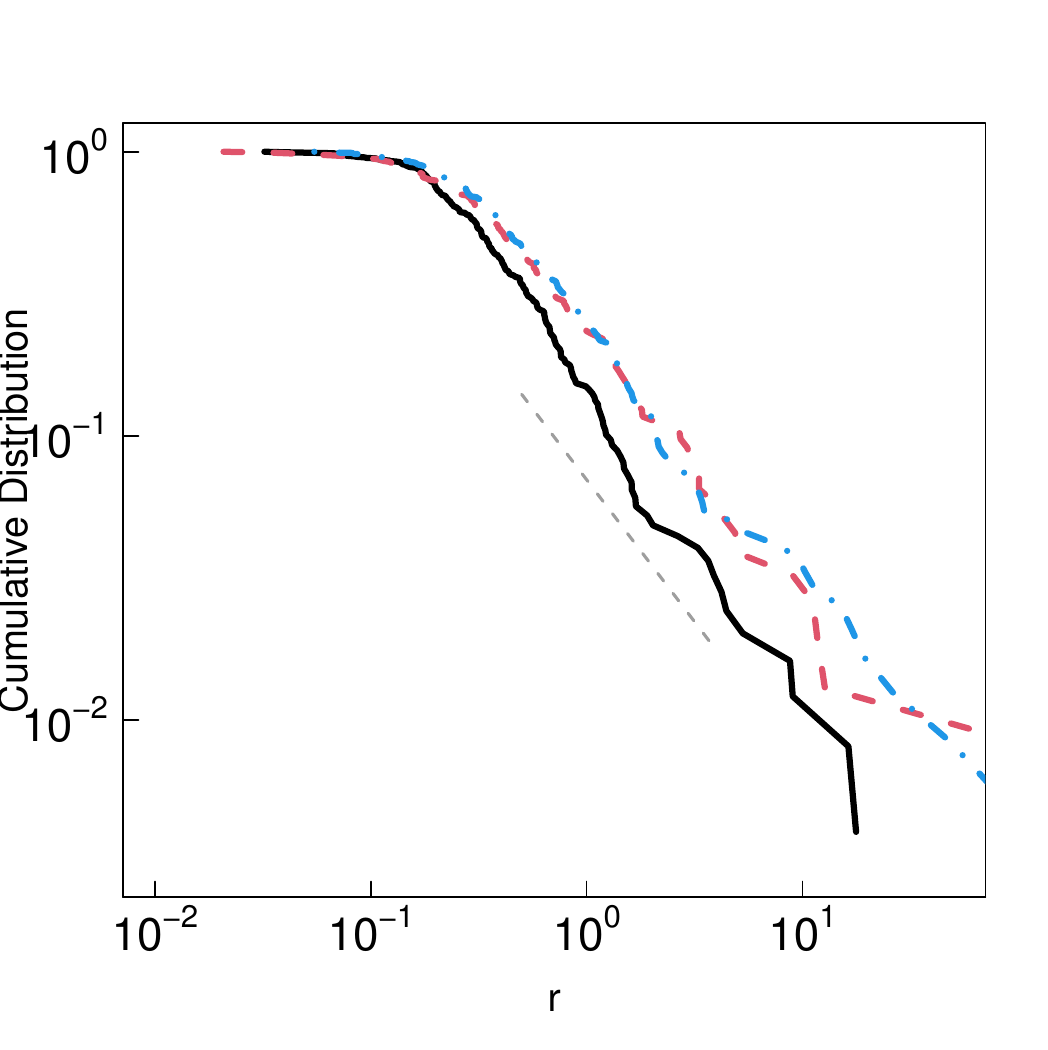}
        \put(26,110){\color{black}\Large\bfseries (b)}
    \end{overpic}
     \begin{overpic}[width=5.5cm]{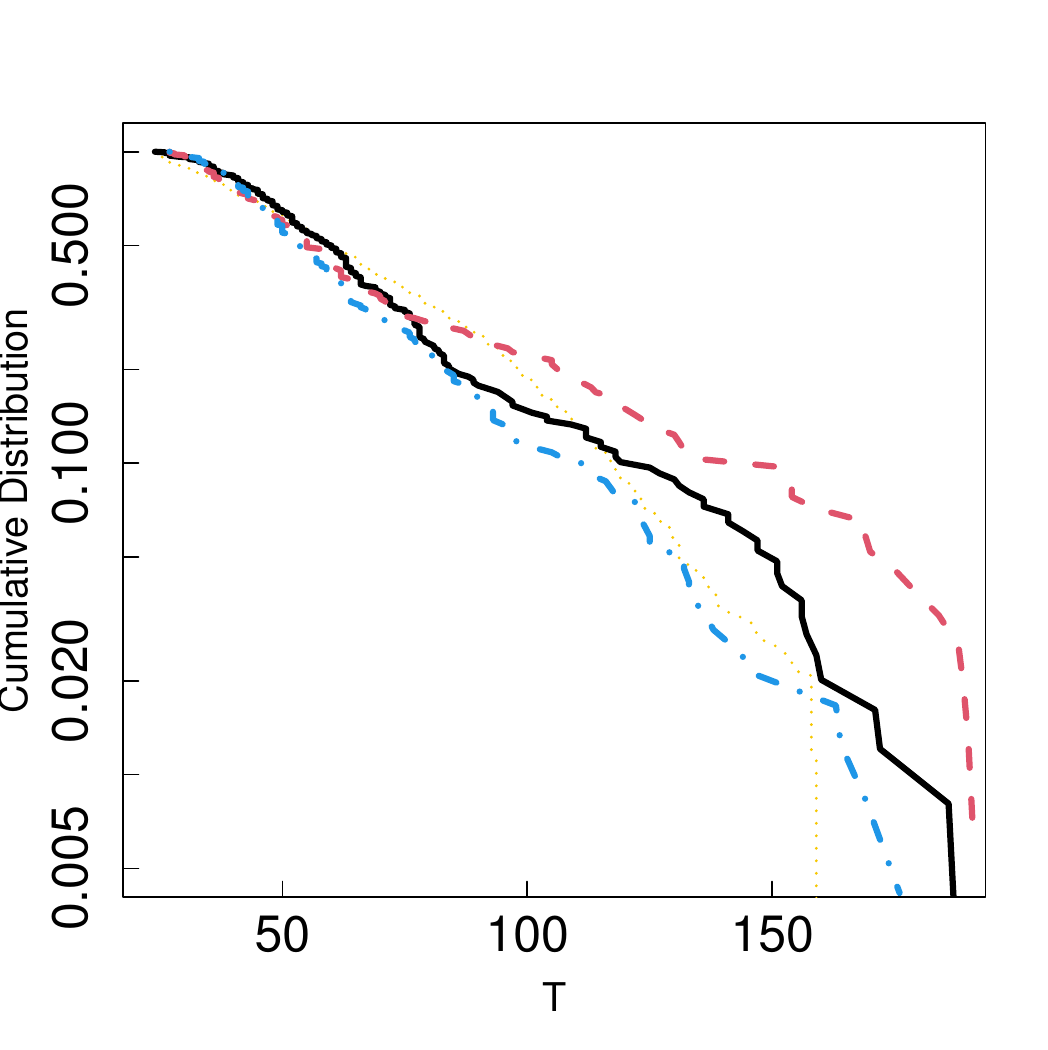}
        \put(26,110){\color{black}\Large\bfseries (c)}
      \end{overpic} 
    \begin{overpic}[width=5.5cm]{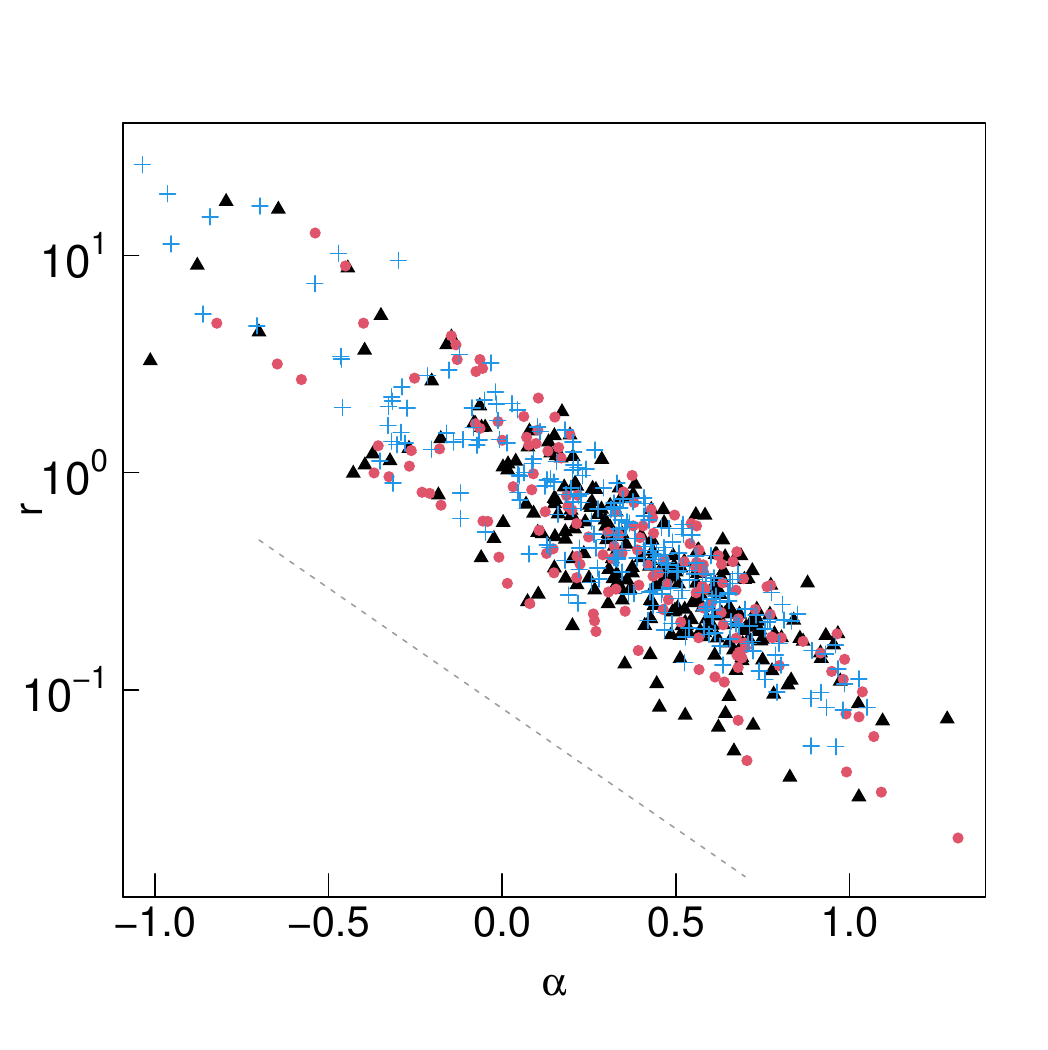}
        \put(128,120){\color{black}\Large\bfseries (d)}
    \end{overpic} 
    \begin{overpic}[width=5.5cm]{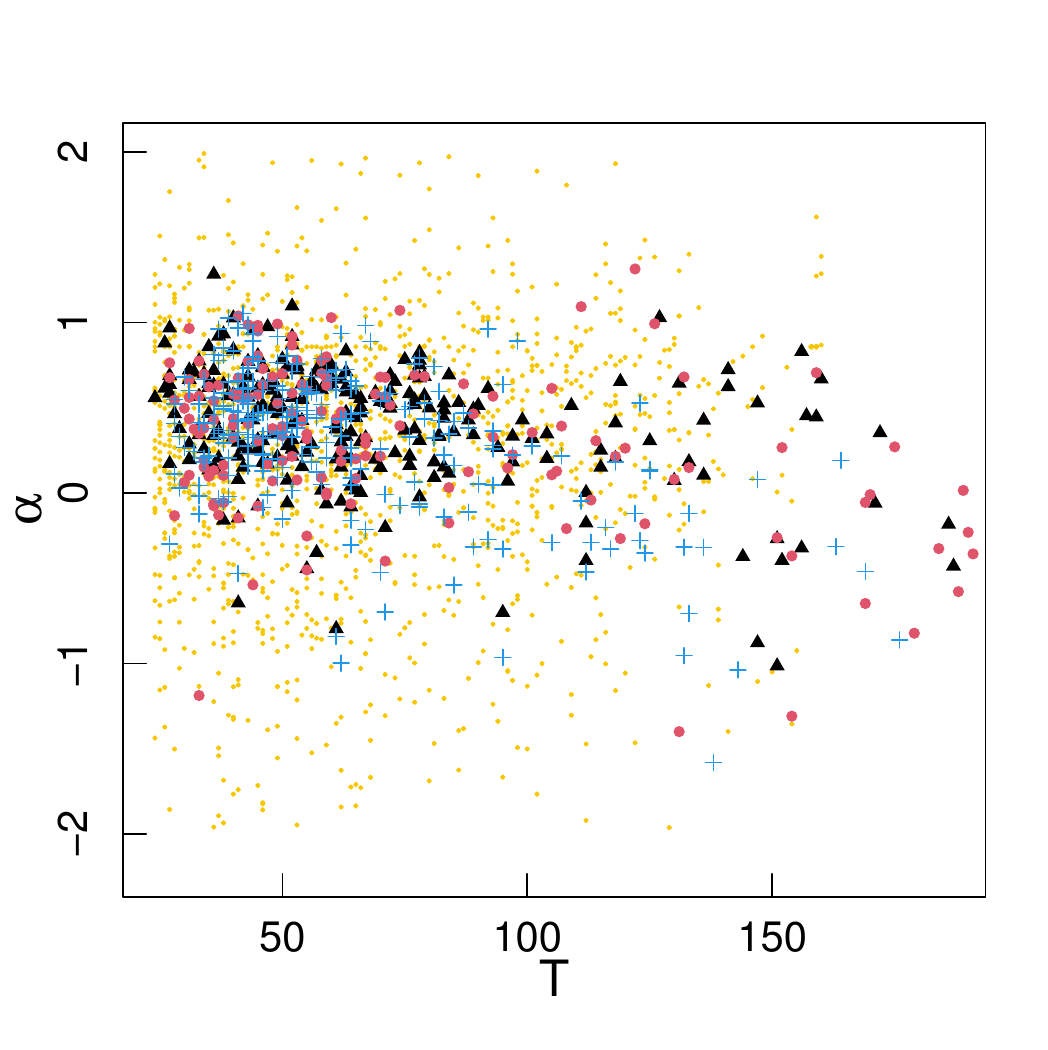}
        \put(128,120){\color{black}\Large\bfseries (e)}
    \end{overpic}
      \begin{overpic}[width=5.5cm]{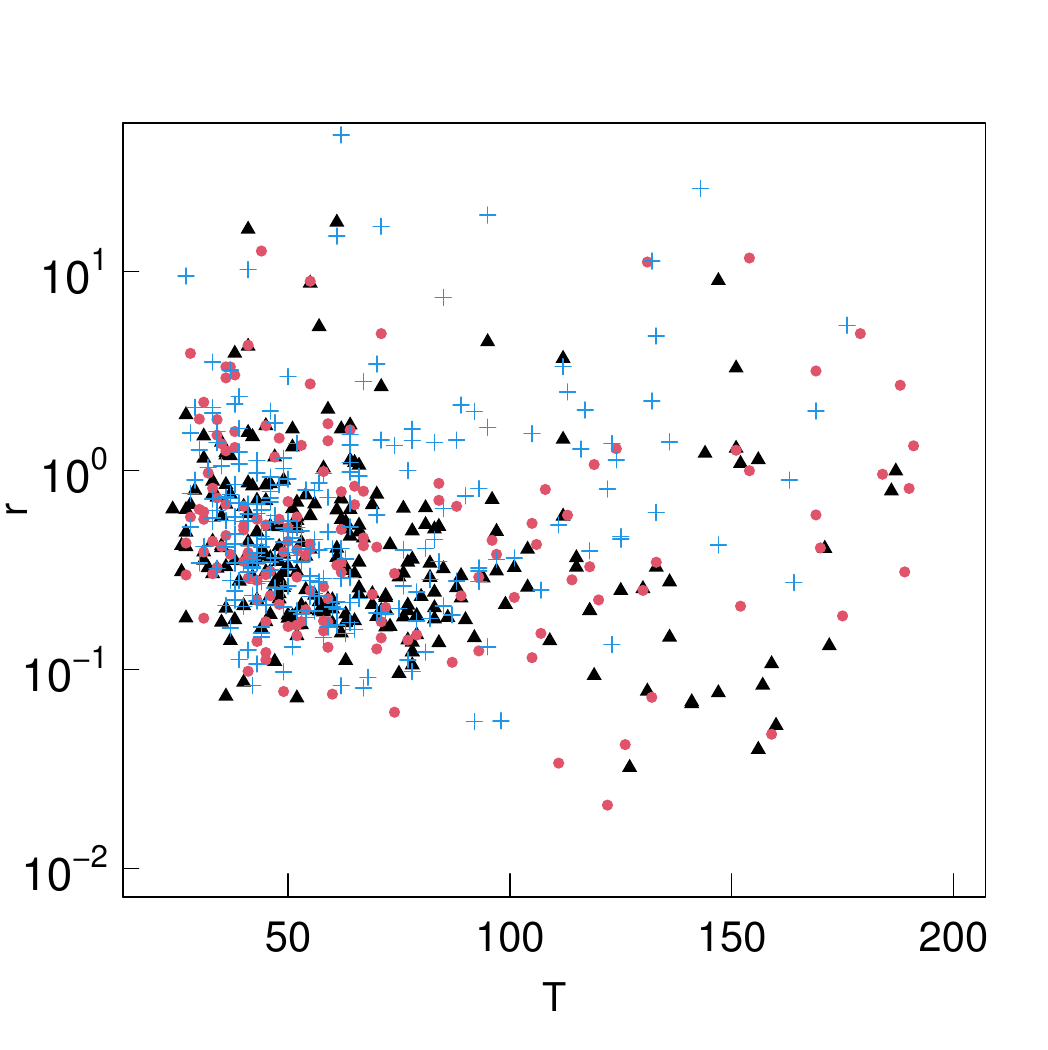}
        \put(128,120){\color{black}\Large\bfseries (f)}
    \end{overpic}
   \caption{
Parameter statistics for Google Trends. 
The blog data correspond to Fig.~\ref{fig_para}(a)-(f) in the main text. However, because $R_i$ cannot be computed from Google Trends data, we use $r_i$ instead (Section~\ref{app_sec_google_stat}).
(a-c) Distributions of the parameters. Black solid line: English; red dashed line: Spanish; blue dash-dotted line: Japanese. The yellow dotted line shows Japanese blog data.
Because for $r_i$, Google Trends scales values to a maximum of 100, $r_i$ is normalized and is not directly comparable with the blog data.
(a) Distribution of $\alpha_i$. The vertical gray line marks the mode (English $0.59$, Spanish $0.54$, Japanese $0.45$).
(b) Cumulative distribution of $r_i$. The gray dashed reference line indicates a power law with exponent 1, i.e., $\propto 1/r$.
(c) Cumulative distribution of $T_i$.
(d-f) Correlations between parameters. Black triangles: English; red circles: Spanish; blue crosses: Japanese. Yellow small circles: Japanese blog data. 
(d) $\alpha_i$ vs.\ $r_i$; the gray dashed support line is $r \propto \exp(-2.5\alpha)$. (e) $T_i$ vs. $\alpha_i$; (f) $T_i$ vs. $r_i$.
   }
    \label{app_fig_para}
\end{figure*}
\begin{figure}[thbp]
    \centering
    \begin{overpic}[width=5.7cm]{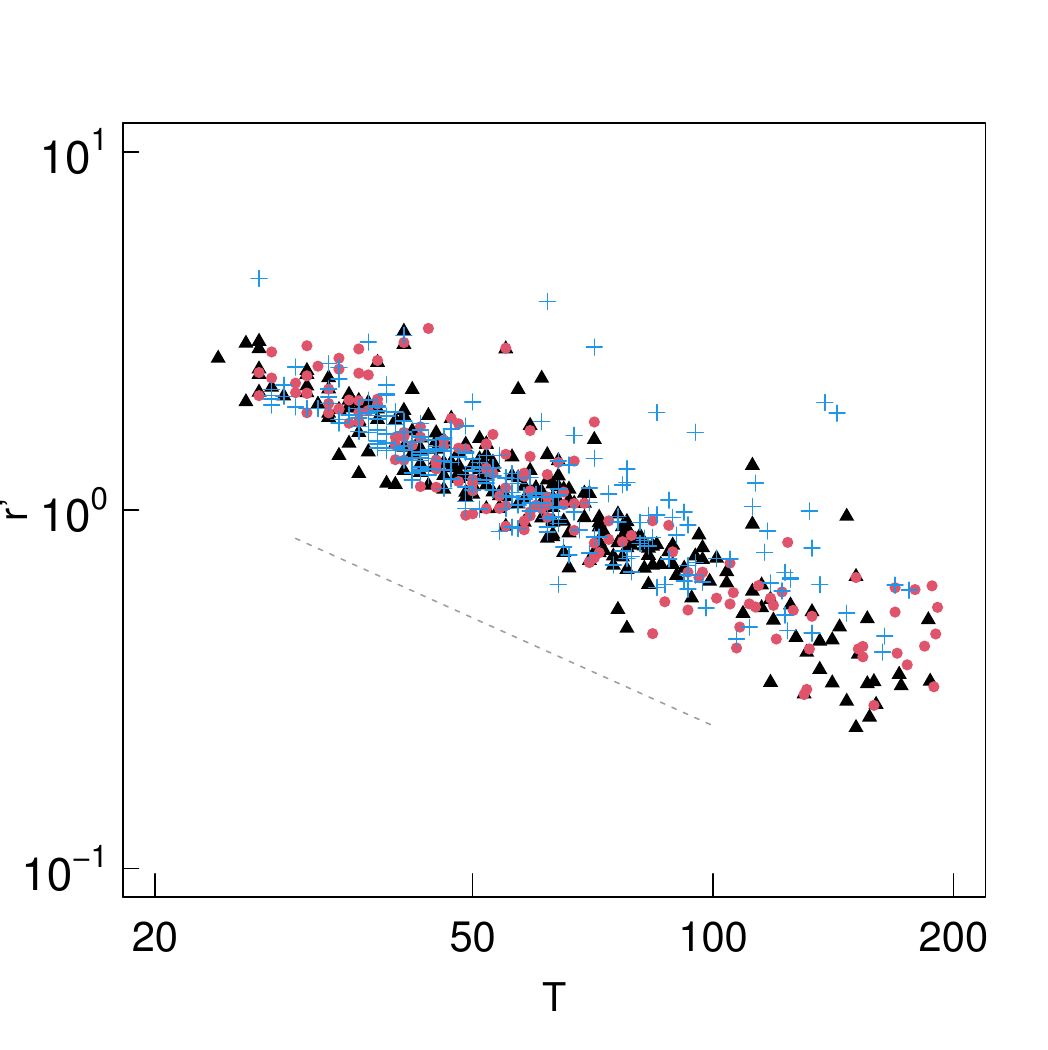}
    \end{overpic}
   \caption{
   Correlation between $r_i'$ and $T_i$. Here, $r_i'$ is defined by Eq.~\ref{app_eq_rd} and represents $r_i$ adjusted to remove the dependence on $a_i$, where $a_i = y_i^{(max)} / y_i(T_i)$ denotes the ratio of the maximum value $y_i^{(max)}$ during the observation period (used for normalization to 100 in Google Trends) to the peak value $y_i(T_i)$ during the growth period. 
   }
    \label{app_fig_para2}
\end{figure}

\section{Comparison between Japanese blog data and Google Trends data}
\label{app_sec_google_base}
In Appendix A, we discuss the comparison between Japanese blog data and Google Trends data.
In this section, we suggest that the results from the Japanese blog data, as presented in the main text, do not contradict the results obtained from Google Trends.

\subsection{Time Series Comparison of Blogs and Google Trends}
\label{app_sec_blog_google_time}

Fundamentally, the time series of blogs and Google Trends show roughly proportional relationships for the same word. Fig.~\ref{app_fig_comp_google_blog_same} shows examples of words where the time series fluctuations are common.

For example:
\begin{itemize}
\item (a) is an example of an exponential growth with $\alpha_i \approx 1$ (\textit{Asaii Bouru} - acai bowl, Brazilian dessert),
\item (b) is an example of a typical growth curve with $\alpha_i \approx 0.5$ (\textit{Hogo Neko Cafe} - protective cat cafe, cat cafe with rescue cats),
\item (c) and (d) are examples of linear growth with $\alpha_i \sim 0$ ((c) \textit{NewsPicks} - NewsPicks, online news media, (d) \textit{Ikuta Erika} - Erika Ikuta, AKB48 related group member).
\end{itemize}

In these time series, it can be seen that the scaled blog time series (black triangles) and Google Trends (red circles) correspond well.

While this roughly proportional relationship forms the basic pattern, there are specific cases where differences emerge. These exceptions are thought to be due to the difference between articles and search behavior.

For example, Fig.~\ref{fig_comp_google_blog_difference}(a) and (b) show new place names: (a) \textit{Sagamihara-shi Chuo-ku} (Sagamihara City Central Ward) and (b) \textit{Okayama-shi Kita-ku} (Okayama City North Ward). While the blogs show a linear trend, Google Trends shows high initial search volume that subsequently changes slowly.

The three categories discussed in Section~\ref{sec_meaning} (news media, idol group members, and new place names) all exhibit near-linear behavior in their blog time series. However, their behavior differs in Google Trends. For news media and AKB48-related group member names, as shown in Figs.~\ref{app_fig_comp_google_blog_same}(c) and (d), the near-linear trends were consistent in both blogs and Google Trends. In contrast, for new place names, while blogs displayed near-linear behavior, Google Trends showed a different pattern characterized by high initial search volume followed by a slow increase over time.

Fig.~\ref{fig_comp_google_blog_difference}(c) shows the time series for \textit{Sumaho} (smartphone, mobile device), a word with $N=2$ (two segments). Both blogs (black triangles) and Google Trends (red circles) shared the common tendency of being exponential-like in the first half and linear-like in the latter half. However, regarding the scale, if fitted to the initial exponential-like part, the scale does not match in the subsequent linear-like part.

Fig.~\ref{fig_comp_google_blog_difference}(d) shows the time series for \textit{Toyota Akua} (Toyota Aqua, hybrid car model). This is a pattern where the cumulative Google Trends search volume ($\xi_i^{\mathrm{(cum)}}(t)=\sum_{s=1}^{t} \xi_i(s)$) (blue crosses) corresponds to the number of blog posts.

\subsection{Statistics of Google Trends}
\label{app_sec_google_stat}

This section presents the statistical analysis results for parameters in the Google Trends data. We show that these results are largely consistent, within the observable range, with those for the blog data presented in Section \ref{sec_parameters}. Note that because Google Trends data is normalized so that the maximum value within the observation period is 100, the scale variable $Y$ cannot be separated. Therefore, we perform the analysis using the non-separated parameter $r_i$.
\begin{equation}
\frac{d\xi_i(t)}{dt}=r_i \xi_i(t)^{\alpha_i}
\label{app_eq_base_z0}
\end{equation}
This $r_i$ corresponds to $r_i=R_i \cdot Y^{-\alpha_i+1}$ in the blog data (Eq.~\ref{eq_base0}). $\xi_i(t)$ is the Google Trends value for word $i$ at time $t$.

Fig. \ref{app_fig_para} shows the parameter analysis results for $N=1$ in the Google Trends data for English, Spanish, and Japanese, corresponding to Fig. \ref{fig_para} for the blog data. The extraction conditions for monotonically increasing words are also the same.

First, we examine the distributions of individual parameters. From Fig. \ref{app_fig_para}(a), the distribution of $\alpha_i$ is centered around $\alpha_i \sim 0.5$ for all languages, which corresponds to the Japanese blog data. Furthermore, Fig. \ref{app_fig_para}(c) shows that the distribution of the growth period $T_i$ is also an exponential-like distribution, similar to the blog data. On the other hand, the distribution of $r_i$ is found to be close to a power-law distribution with an exponent of 1. The reason for this power-law distribution with exponent 1 is presumed, as discussed later, to be a consequence of the normalization of Google Trends to a maximum value of 100 and the exponential distribution of $T_i$ (this is expected to be different from the reason why $R_i$ has an exponent of 1 in the blog data). This point will be discussed in detail later in this section.

Next, we examine the correlations between two variables. Fig.~\ref{app_fig_para}(e) shows that no strong correlation between $T_i$ and $\alpha_i$ is observed, and the blog data (small yellow circles) show no significant association. In contrast, the Google Trends data exhibit a statistically significant but weak negative correlation, with Kendall's $\tau=-0.18$ (p-value $=3.5\times10^{-5}$) for English, $\tau=-0.18$ (p-value $=0.0014$) for Spanish, and $\tau=-0.23$ (p-value $=1.8\times10^{-7}$) for Japanese.

Also, similar to the blog data, a strong negative correlation was confirmed between $\alpha_i$ and $r_i$. Kendall's rank correlation $\tau$ between $r_i$ and $\alpha_i$ was $\tau=-0.62$ (p-value $\le 10^{-17}$) for English, $\tau=-0.68$ (p-value $\le 10^{-17}$) for Spanish, and $\tau=-0.75$ (p-value $\le 10^{-17}$) for Japanese.
On the other hand, no strong correlation was observed between $T_i$ and $r_i$ (note: $r_i$, not $R_i$). Kendall's rank correlation $\tau$ between $r_i$ and $T_i$ was $\tau=-0.17$ (p-value = $1.8 \times 10^{-5}$) for English, $\tau=-0.13$ (p-value = $0.022$) for Spanish, and $\tau=-0.014$ (p-value = $0.75$) for Japanese.

Note that the correlation with $y_i(T_i)$ cannot be calculated due to the normalization of Google Trends to a maximum of 100.

These results can be mathematically interpreted as follows.
Since Google Trends is normalized so that the maximum value is 100, if we let $y_i(t)$ be the original scale and $y^{(max)}_i$ be the maximum value within the observation period, it can be written as:
\begin{equation}
\xi_i(t)=100 \frac{y_i(t)}{y^{(max)}_i}
\end{equation}
Here, considering that the maximum value $y^{(max)}_i$ is around the peak of the time series growth, it can be written as $y^{(max)}_i=a_i y_i(T_i)$. Here, $a_i>0$ represents the deviation from the growth peak, and we assume $a_i \sim 1$.
From this, we have $\xi_i(t)=100 \frac{y_i(t)}{a_i \cdot y_i(T_i)}$. By substituting this relationship into the differential equation for $y_i(t)$ (the basis for Eq.~\ref{eq_base0}) and rearranging, we obtain:
\begin{equation}
\frac{d\xi_i(t)}{dt}=R_i \xi_i {\left(\frac{\xi_i(t)}{H_i}\right)}^{\alpha_i}
\end{equation}
Here, $H_i=100 Y/(a_i \cdot y_i(T_i))$. Comparing this with Eq.~\ref{app_eq_base_z0} implies $r_i = R_i H_i^{1-\alpha_i}$. Therefore,
\begin{equation}
r_i= R_i H_i^{1-\alpha_i} = R_i {\left(\frac{100 Y}{a_i \cdot y_i(T_i)}\right)}^{1-\alpha_i}
\label{app_eq_ri}
\end{equation}
can be written. Furthermore,
\begin{eqnarray}
y_i(T_i) &=& Y (R_i \cdot (1-\alpha_i) \cdot T_i+(y_i(0)/Y)^{1-\alpha_i})^{1/(1-\alpha_i)} \nonumber \\
&\sim& Y (R_i \cdot (1-\alpha_i) \cdot T_i)^{1/(1-\alpha_i)}
\end{eqnarray}
Here, the approximation in the second line assumes $T_i \gg 1$, allowing the second term to be ignored relative to the first term.
Substituting this into Eq.~\ref{app_eq_ri} yields:
\begin{equation}
r_i \propto {\left(\frac{100}{a_i}\right)}^{1-\alpha_i} \frac{1}{(1-\alpha_i) \cdot T_i}
\end{equation}
\par

From this equation, the term $(100/a_i)^{1-\alpha_i}$ is considered to be the reason why $r_i$ is observed to be an exponential function of $\alpha_i$ (Fig. \ref{app_fig_para}(d)).
In addition, the power-law distribution with exponent 1 for the cumulative distribution of $r_i$ in Fig. \ref{app_fig_para}(b) is considered to correspond to the fact that if the scatter of $\alpha_i$ is small, then $r_i \propto 1/T_i$; and since $T_i$ has a distribution close to an exponential distribution (Fig. \ref{app_fig_para}(c)), the distribution of the reciprocal of an exponential variable yields a cumulative power-law distribution with exponent 1.
\par
Note that the weak correlation between $r_i$ and $T_i$ in Fig. \ref{app_fig_para}(f) is presumed to be because the influence of the term $(100/a_i)^{1-\alpha_i}$ is not small.
In fact, if we define
\begin{equation}
r_i'=r_i \cdot Q^{\alpha_i} \label{app_eq_rd}
\end{equation}
and choose $Q$ such that the correlation between $\alpha_i$ and $r_i'$ is minimized, a negative correlation related to $1/T_i$ appears between $r_i'$ and $T_i$ (Fig. \ref{app_fig_para2}). This is thought to be because, under this condition, $Q \approx 100/a_*$ (where $a_*$ is a typical value of $a_i$), and $r_i'$ can be approximated as $r_i' \propto 1/((1-\alpha_i) \cdot T_i)$, a form less affected by $a_i$. \par
From the above, it is understood that the behavior of Google Trends can be consistently explained by adding the assumption $y^{(max)}_i \sim y_i(T_i)$ (that the maximum value in the observation period is around the growth peak) to the power-law growth model given by Eq.~\ref{eq_base0}. \par
\clearpage 
\numberwithin{equation}{section} 
\setcounter{equation}{0} 
\renewcommand{\theequation}{B\arabic{equation}}


\section{Supplementary
 Discussion of the Piecewise Power-Law Model}
\label{app_sec_powerlaw_base}
Appendix B provides a supplementary
 discussion of the piecewise power-law model.
\subsection{Nondimensionalization and Linearization of the Power-Law Model} 
\label{app_sec_henkan}
Here, we describe the formula transformations for nondimensionalization and linearization mentioned in Section \ref{sec_base}.
For the basic equation
\begin{equation}
\frac{dy_i(t)}{dt}=R_i Y\left(\frac{y_i(t)}{Y}\right)^{\alpha_i}
\end{equation}
we introduce the dimensionless quantities
\begin{equation}
s_i(t):=\frac{y_i(t)}{Y},\qquad \tau_i:=R_i (t-t_i^{(0)})
\end{equation}
and impose the reference condition
\begin{equation}
s_i(t_i^{(0)})=1 \quad (\text{i.e., } y_i(t_i^{(0)})=Y)
\end{equation}
From the chain rule, we obtain
\begin{equation}
\frac{ds_i(t)}{d\tau_i}=\frac{ds_i(t)/dt}{d\tau_i/dt}=\frac{(1/Y)\,(dy_i(t)/dt)}{R_i}=s_i(t)^{\alpha_i}
\end{equation}

\subsubsection{Representation of the Solution}
When $\alpha_i \neq 1$, the solution satisfying the initial condition $s_i(\tau_i=0)=1$ is given by
\begin{equation}
s_i(\tau_i)=\left(1+(1-\alpha_i) \tau_i\right)^{\frac{1}{1-\alpha_i}}
\label{app_eq_s}
\end{equation}
In the limit $\alpha_i\to 1$, it continuously converges to
\begin{equation}
s_i(\tau_i)\to e^{\tau_i}
\end{equation}
Furthermore, for $\alpha_i>1$, a finite-time divergence occurs when $1+(1-\alpha_i)\tau_i=0$, and the divergence time in real time is
\begin{equation}
t^{\ast}=t_i^{(0)}+\frac{1}{(\alpha_i-1) R_i}
\end{equation}
\par
Reverting from $s_i(t)$ to $y_i(t)$, the solution for $y_i(t)$ is given by:
\begin{equation}
y_i(t)=
\begin{cases}
Y\left[(1-\alpha_i) R_i (t-t_i^{(0)})+\left(\dfrac{y_i(t_i^{(0)})}{Y}\right)^{1-\alpha_i}\right]^{\frac{1}{1-\alpha_i}} & (\alpha_i\neq 1) \\
y_i(t_i^{(0)}) \exp(R_i (t-t_i^{(0)})) & (\alpha_i=1).
\end{cases}
\end{equation}

\subsubsection{Linearization (Box--Cox Type Transformation)}
From Eq.~\ref{app_eq_s}, we obtain $s_i(t)^{1-\alpha_i}=1+(1-\alpha_i)\tau_i$. Therefore, by defining
\begin{equation}
z_i(t) = \frac{s_i(t)^{1-\alpha_i}-1}{1-\alpha_i}
\end{equation}
we get
\begin{equation}
z_i(\tau_i)=\tau_i
\label{app_eq_z}
\end{equation}
In the limit $\alpha_i \to 1$, this becomes $z_i(t) =\log s_i(t)$, which similarly satisfies $z_i(\tau_i)=\tau_i$. Furthermore, since $\tau_i=R_i \cdot (t-t_i^{(0)})$, $z_i$ can be expressed as a function of $t$:
\begin{equation}
z_i(t)=R_i \cdot t-R_i \cdot t_i^{(0)}
\end{equation}
Thus, $R_i$ represents the slope of $z_i(t)$ with respect to $t$.

\subsection{Exponential Correlation between $\alpha_i$ and $r_i$ and the Determination of $Y$ and $R_i$}
\label{app_sec_r}
Here, we describe the exponential correlation between $\alpha_i$ and $r_i$ and the determination of $Y$ and $R_i$ based on it.
Note that $r_i$ is defined as follows. This equation is completely identical to the General Growth Model (GGM) in Ref. \cite{chowell2016mathematical}, including its parameters.
\begin{equation}
\frac{dy_i(t)}{dt}=r_i \cdot y_i(t)^{\alpha_i}
\label{app_eq_base0}
\end{equation}
In comparison with Eq.~\ref{eq_base0}, this corresponds to $R_i=r_i \cdot Y^{\alpha_i-1}$.\par
\begin{figure}[t]
    \centering
    \begin{overpic}[width=5.7cm]{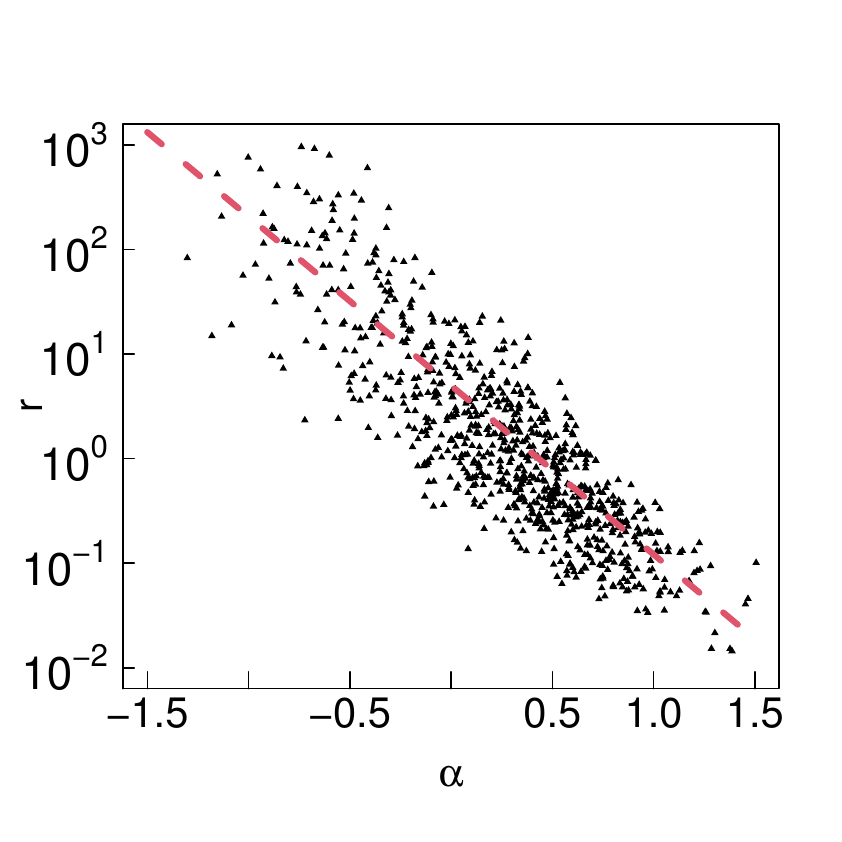}
    \end{overpic}
    \caption{Exponential correlation between $\alpha_i$ and $r_i$ in Eq.~\ref{app_eq_base0}. The red dashed line is the exponential fit $r_i = 5 \cdot Y^{-\alpha_i}$.
Here, $Y=41.3$ is the value used in Eq.~\ref{eq_base0} (see Section~\ref{app_sec_r}). 
    }
    \label{app_fig_alpha_r}
\end{figure}

Fig. \ref{app_fig_alpha_r} shows the correlation between $\alpha_i$ and $r_i$, revealing an exponential correlation with respect to $\alpha_i$.\par
Therefore, we determined $Y$ so as to eliminate this correlation. Specifically, we selected the value of $Y$ that minimizes the Spearman's rank correlation coefficient between $\alpha_i$ and $R_i$. The resulting scatter plot of $\alpha_i$ and $R_i$ is shown in Fig. \ref{fig_para}(h).

\subsection{Formulation of the Piecewise Power-Law Model}
\label{app_sec_kubun}
Here, we formulate the piecewise power-law model.
Let $t_{i,0}<t_{i,1}<\cdots<t_{i,M_i}$ be a sequence of breakpoints for word $i$. In each segment $[t_{i,m-1},t_{i,m})$, the model is defined as
\begin{eqnarray}
 && \frac{dy_i(t)}{dt}
  = R_i^{(m)} \cdot Y\left(\frac{y_i(t)}{Y}\right)^{\alpha_i^{(m)}}, \nonumber \\
 && \qquad t\in [t_{i,m-1},t_{i,m}),\ \ m=1,\dots,M_i .
  \label{eq:piecewise-ggm-i}
\end{eqnarray}

The initial condition is set as
\begin{equation}
  y_i(t_{i,0}+0)=y_i^{(0)}, 
  \label{eq:init-i}
\end{equation}
where $y_i^{(0)}>0$.
At the breakpoints $t_{i,m}$ (for $m=1,\dots,M_i-1$), the segments are connected by one of the following conditions, using left and right-hand limits. For the continuous case (no jump):
\begin{equation}
  \lim_{t\to t_{i,m}+0} y_i(t) = \lim_{t\to t_{i,m}-0} y_i(t).
  \label{eq:cont-i}
\end{equation}
For the jump case (additive discontinuity):
\begin{equation}
  \lim_{t\to t_{i,m}+0} y_i(t)
  =
  \lim_{t\to t_{i,m}-0} y_i(t)+\Delta_i^{(m)},
  \label{eq:jump-i}
\end{equation}
where $\Delta_i^{(m)}>0$.
The initial value for the subsequent segment $[t_{i,m},t_{i,m+1})$ is set as $\lim_{t\to t_{i,m}+0} y_i(t)$, and its time evolution follows Eq.~\ref{eq:piecewise-ggm-i}. This constitutes a general form where jumps occur only at the breakpoints, and the jump amount $\Delta_i^{(m)}$ is introduced only if the continuity condition (Eq.~\ref{eq:cont-i}) is not satisfied.

\subsection{Intent behind the Introduction of the Piecewise Power-law Model: Relationship with the Extended Logistic Model}
\label{app_sec_relationOldmodel}
In this section, we describe the relationship between the extended logistic model, introduced in Ref. \cite{watanabe2023minor} which effectively models word count time series, and the piecewise power-law model proposed in this paper. \par

First, the extended logistic model is defined by:
\begin{equation}
\frac{dy_i(t)}{dt}=\rho_i \cdot y_i(t)\left(1+\frac{y_i(t)}{V_i}\right)^{\zeta_i} \label{app_eq_old_model}
\end{equation}
When $V_i < 0$ and $\zeta_i = 1$, this equation becomes the logistic equation. \par

According to Table 2 in Ref. \cite{watanabe2023minor}, in the analysis of blog data, approximately 90\% of the words had $V_i > 0$. When $V_i > 0$, the extended logistic model can be roughly approximated by a piecewise power-law model as:
\begin{equation}
\frac{dy_i(t)}{dt} \approx
\begin{cases}
\rho_i y_i(t) & (y_i(t) < V_i) \\
\rho_i \cdot V_i^{-\zeta_i} y_i(t)^{\zeta_i+1} & (y_i(t) > V_i)
\end{cases}
\label{app_eq_exlogi}
\end{equation}
The exponents and parameters correspond to $N=2$, $\alpha_i^{(1)}=1$, $R_i^{(1)}=\rho_i$, $\alpha_i^{(2)}=1+\zeta_i$, and $R_i^{(2)}= \rho_i \cdot (Y/V_i)^{\zeta_i}$. In actual observations, Fig. 4 in Ref. \cite{watanabe2023minor} shows that the mode of $V_i$ is close to $0$. This corresponds to the fact that many time series can be well approximated by a single power-law model ($N=1$) given by Eq.~\ref{eq_base0}. \par

The piecewise power-law model was inspired by the analysis of the extended logistic model, which revealed that few words have $V_i < 0$, and furthermore, that many words have $V_i \sim 0$, meaning they effectively follow a single power-law model ($N=1$). In fact, according to Table A1 in Ref. \cite{watanabe2023minor}, the predictive performance of the single power-law model equivalent to Eq.~\ref{eq_base0} was found to be better than that of the extended logistic model (i.e., more words had a smaller prediction error with the single power-law model). \par

Based on these findings, we considered that describing time series using a power-law model as the base is more desirable than extending the logistic function. \par

The piecewise power-law model enables the description of time series that the extended logistic model cannot capture, such as those with $N=2$ and $\alpha_i^{(1)} \neq 1$, or those with many segments ($N \geq 3$). Furthermore, the extended logistic model faced issues; the $N=1$ (single power-law) case, as shown in Eq.~\ref{app_eq_exlogi}, is described only in the limit $1/V_i \to \infty$, which can lead to unstable estimation. Moreover, for the most frequent case $V_i \sim 0$, the term $\rho_i \cdot V_i^{-\zeta_i}$ becomes effectively indeterminate, necessitating other assumptions, such as regularization, to resolve this indeterminacy. The piecewise power-law model addresses this poor descriptive capability for the most frequent $N=1$ case by sequentially fitting the model starting from $N=1$, thus eliminating estimation indeterminacy. In addition, complexity can be controlled and understood using an interpretable measure, the number of segments ($N$), rather than an uninterpretable quantity like regularization strength. This approach has also enabled us, in this study, to extract the simplest and most fundamental $N=1$ time series and investigate their properties in detail.

\subsection{Statistics of the Number of Segments $N$ and Jumps}
\label{app_sec_n_table}
Table~\ref{app_tab_n_table} summarizes the statistics of the number of segments $N$ and jumps discussed in Section~\ref{sec_stat_n}.

\begin{table}[htbp]
\centering
\begin{tabular}{lc}
\hline
Classification & Number of time series \\
\hline
With jump & 1,262 \\
Without jump & 1,701 \\
\hline
\multicolumn{2}{l}{-- Breakdown of cases without jump:} \\
\quad N = 1 & 852 \\
\quad N = 2 & 773 \\
\quad N = 3 & 76 \\
\hline
Total & 2,963 \\
\hline
\end{tabular}
\label{app_tab_n_table}
\caption{
\revisecolor{black}{
Number of time series classified by the presence of jumps and the number of segments ($N$) in the piecewise power-law model. 
Overall, 55\% of the time series exhibit no jumps and can be described with two or fewer segments ($N \le 2$). 
Among these, cases following a pure power-law model ($N = 1$) account for 29\% of all time series (including those with jumps), and 50\% when restricted to the no-jump cases. 
These results indicate that even a simple pure power-law model ($N = 1$) alone explains a large fraction of the observed dynamics for many words.
}
}
\label{app_tab_n_table}
\end{table}
\clearpage

\numberwithin{equation}{section} 
\setcounter{equation}{0} 
\renewcommand{\theequation}{C\arabic{equation}}



\section{Supplementary Discussion of the Infection Model}
\label{app_sec_infection_base}
Appendix C provides a supplementary discussion of the infection model. Specifically, in Section~\ref{app_sec_infection0}, we first present a more precise formulation and derivation of the growth curve for the unweighted infection model introduced in Section \ref{sec_infection} in the main text. Second, in Section~\ref{app_sec_infection_model}, we introduce a weighted infection model that yields an in-degree distribution following a power-law with exponent approximately 1. 
The detailed properties of the Infection-Type Model are discussed based on numerical simulations under various conditions in Section~\ref{app_sec_infect_result}, and through the analysis of in-degree distributions in Section~\ref{app_sec_indegree}.

\subsection{An Un-Weighted Infection-Type Model (Un-Weighted Model)}
\label{app_sec_infection0}
This section provides a detailed discussion of the infection-type model on unweighted networks discussed in Section~\ref{sec_infection} of the main text.
\par
We examine how non-trivial power-law growth emerges from characteristics of human behavior, based on the theoretical framework. This study presents a model that incorporates the ``general appeal'' effect discussed in previous sections into the infection mechanism. It should be noted that in infectious disease research, power-law growth (General Growth Model) is explained by factors such as geographical effects and heterogeneity in the distribution of infected individuals etc. \cite{chowell2016mathematical,yan2024modeling,karev2014non}.

\subsubsection*{Key Findings}
The shape parameter $\alpha_i$ of the power-law growth model (Eq.~\ref{eq_base0}) is given by the ratio of the topic's inwardness $\gamma_i$ to the number of external candidates per action $Q$, specifically:
\begin{equation}
\alpha_i = 1 - \frac{\gamma_i}{Q}
\end{equation}
Here, we use ``inwardness'' to mean that people who already know the term tend to talk about it mostly with each other rather than with people who don't know the term.
Please refer to the following section for the derivation and detailed model specifications.

\subsubsection{Model Assumptions and Terminology}
\label{app_sec_infection_term}
In this section, we present a simple behavioral model that generates the power-law growth model introduced above. 

\medskip
\noindent\textbf{Assumptions (infected/susceptible individuals and an infection network).}\;
We model the diffusion of lexical knowledge as (i) person-to-person transmission of recognition of a term (``infection'') and (ii) the expansion of an interaction network (a community of people who know the term). For a given term $i$, individuals who know the term (members of the term-$i$ community) are called \emph{infected}, and those who do not are called \emph{susceptible} (non-infected). Susceptibles may become infected through contact with infected individuals (``recruitment''), and interaction among infected individuals can also expand the term-$i$ interaction network (the community of discussants).

\medskip
\noindent\textbf{The Probability of Outward Communication}\;
We model an individual's communication strategy by directly linking it to the nature of the topic itself. \par
Consider niche topics, such as the name of a specific AKB48 member or a technical medical term. Conversations about these subjects are most meaningful with peers who already share the specialized knowledge. This fosters an \textbf{inward-looking} communication strategy, where individuals prioritize interactions \textit{within} their existing community. As this community grows and becomes more established, the incentive to engage uninformed outsiders (susceptibles) naturally diminishes. \par
In contrast, for general or ``buzz-worthy'' topics, the value often lies in broader dissemination. This encourages an \textbf{outward-looking} strategy, where individuals are more motivated to engage with people who do not yet know the term. \par

To formalize these communication strategies, we introduce the concept of a personal \textbf{contact list}. Each individual, $l$, maintains a list for each topic, $i$, denoted as $S_i^{(l)}(t)$. This list contains all the individuals whom $l$ can engage with on that specific topic---analogous to a researcher's network of colleagues in a particular field.
The growth of this list, therefore, directly models the formation of new directed links within the actual interaction community. When $l$ adds someone to their list, it signifies a one-way connection; $l$ now knows of them, but the reverse is not automatically true.\par 
The core of our model is the decision-making process for each infected individual, $l$. At each time step, they have a chance to initiate a new social connection, governed by an \textbf{action rate} $J_i$. When they act, they must choose between two strategies: \textbf{recruitment}, which involves approaching a susceptible individual, or \textbf{interaction}, which means connecting with an already infected person not yet in their network. \par
The probability that they choose the outward path of \textbf{recruitment} is given by:
\begin{equation}
    p_i^{(l)}(t) = \frac{Q}{Q+\gamma_i\,k_{i}^{(l)}(t)},
    \label{app_eq_kansen}
\end{equation}
where $k_{i}^{(l)}(t)=|S_i^{(l)}(t)|$ represents the size of the individual's current contact list for that topic. This equation formalizes a key social dynamic: as the size of an individual's internal network ($k_{i}^{(l)}(t)$) increases, their focus shifts inward, naturally reducing the probability of recruiting outsiders.
\par
The parameter $Q > 0$ represents the \textbf{effective pool of outsiders} an individual can reach. This isn't the total number of uninformed people in the entire population ($\Xi$), but rather the smaller, immediate circle of susceptibles that an individual can realistically engage with at any given time. We make two key assumptions about this pool. First, it is much smaller than the total population ($Q \ll \Xi$). Second, its size remains roughly constant over time. This stability is maintained by a steady turnover, much like a university campus where new students arrive as others graduate, ensuring that the local pool of potential contacts is never depleted. \par

Finally, $\gamma_i \ge 0$ is a coefficient capturing the \emph{inwardness} of topic $i$. $p_i^{(l)}(t)$ can be interpreted as the probability of choosing the outward option, given outward and inward weights $Q$ and $\gamma_i k_{i}^{(l)}(t)$, respectively; the selection ratio is $Q:\gamma_i k_{i}^{(l)}(t)$. (Note: $p_i^{(l)}(t)$ is conditional on taking an action; the unconditional probability of choosing outward within $\Delta t$ is $J_i p_i^{(l)}(t)\Delta t$. Each action creates exactly one new link, so contact-list growth corresponds to link addition in the community.) \par
This weight $\gamma_i$ is a topic-specific coefficient. It represents the relative \textbf{approachability} of an infected individual, normalized such that the baseline approachability of a single susceptible individual is 1.\par
For example:
\begin{itemize}
  \item $\gamma_i=1$: Both susceptible and infected individuals are equally approachable for conversation about the topic.
  \item $\gamma_i=0$: Actions are always directed toward susceptible individuals (maximally outward-looking).
\end{itemize}

\subsubsection{An Infection Model with Inwardness}
\subsubsubsection{Behavioral Rules for Infected Individuals}
At each discrete time step, an individual $l$ who is "infected" with topic $i$ attempts to expand their network of contacts. This occurs at a total rate of $J_i$, manifesting as one of two distinct actions: recruitment or interaction.

\begin{itemize}
    \item \textbf{Recruitment:} With a rate of $J_i\,p_i^{(l)}(t)$, individual $l$ contacts and infects a susceptible individual, $a$. This action has two effects: it increases the total infected population by one ($I_i(t+\Delta t)=I_i(t)+1$), and it adds the newly infected individual $a$ to $l$'s personal contact list:
    \begin{equation}
    S_i^{(l)}(t+\Delta t)=S_i^{(l)}(t)\cup\{a\}.
    \end{equation}
    
    \item \textbf{Interaction:} With the complementary rate of $J_i\,(1-p_i^{(l)}(t))$, individual $l$ connects with another infected individual, $b$, who was not previously in $l$'s network. This is achieved through an introduction from an existing contact, $f$. This action does not change the total infected count ($I_i(t)$ remains constant) but expands $l$'s contact list. This connection is directional: $b$ is added to $l$'s list, but $l$ is not added to $b$'s.
    \begin{equation}
    S_i^{(l)}(t+\Delta t)=S_i^{(l)}(t)\cup\{b\}.
    \end{equation}
\end{itemize}

The model operates in discrete time steps. All actions are calculated based on the system's state at the beginning of the step ($t$). The resulting changes are then applied simultaneously to determine the state at the next step ($t+\Delta t$), ensuring that the order of events within a single step does not affect the outcome. \par
\subsubsubsection{Initializing Connections for Newly Infected Individuals}
When an individual $a$ is newly ``infected'' by an individual $l$, they don't start with an empty network. Instead, $a$ immediately \textbf{inherits} the entire contact list of their infector, $l$. This process is analogous to a new student inheriting the professional network of their academic advisor. \par
Specifically, $a$'s initial contact list, $S_i^{(a)}$, is formed by taking $l$'s list at time $t$ and adding $l$ to it:
\begin{equation}
    S_i^{(a)}(t+\Delta t) = S_i^{(l)}(t) \cup \{l\}.
\end{equation}
As a result, the size of $a$'s initial list, $k_{i}^{(a)}$, is one greater than $l$'s list at the moment of infection: $k_{i}^{(a)}(t+\Delta t) = k_{i}^{(l)}(t)+1$. With this inherited list, individual $a$ formally joins the topic network.
\subsubsection{Deriving the Macro-Level Growth Curve}
To connect our individual-level behavioral rules to the macro-level growth patterns, we make a key simplifying assumption. Because new members inherit the contact network of their infector, an individual's number of contacts depends primarily on the overall age of the community $t$, not on how long that specific person has been ``infected.'' This allows us to approximate an individual's contact list size $k_{i}^{(l)}(t)$ as growing proportionally to the elapsed time, such that $k_{i}^{(l)}(t) \approx J_i t$.  
Substituting this into Eq.~\ref{app_eq_kansen} gives
\begin{equation}
  p_i^{(l)}(t) \approx \frac{1}{1+(\gamma_i/Q) J_i t}.
\end{equation}
Therefore,
\begin{equation}
  I_i(t+\Delta t)-I_i(t) \approx J_i \Delta t \sum_{l\in I_i(t)} p_i^{(l)}(t)
  \approx \frac{J_i \Delta t}{1+(\gamma_i/Q) J_i t} I_i(t).
\end{equation}
Under a continuous-time approximation,
\begin{equation}
  \frac{dI_i(t)}{dt} = \frac{J_i}{1+(\gamma_i/Q) J_i t} I_i(t),
  \label{app_eq_1_t}
\end{equation}
whose solution with $I_i(0)=1$ is $I_i(t) = (1+(\gamma_i/Q) J_i t)^{1/(\gamma_i/Q)}$. An equivalent form is
\begin{equation}
  \frac{dI_i(t)}{dt} = J_i I_i(t)^{1-\gamma_i/Q}.
  \label{app_eq_diff_i}
\end{equation}
Let the observed article count be $y_i(t)=Y I_i(t)$ (i.e., we choose the macroscopic scale $Y$ to match the proportionality constant). 
Then 
\begin{equation}
  \frac{dy_i(t)}{dt} = J_i Y \left(\frac{y_i(t)}{Y}\right)^{1-\gamma_i/Q}.
\end{equation}
Comparing with the macro model in Eq.~\ref{eq_base0}, we obtain

\begin{equation}
  \alpha_i = 1-\frac{\gamma_i}{Q},
  \qquad
  R_i = J_i.
\end{equation}

This result suggests that the exponent $\alpha_i$, which governs the shape of the observed growth curve, is determined by the parameter $\gamma_i$ representing the inwardness of a topic. Specifically, exponential growth ($\alpha \approx 1$) corresponds to topics with near-zero inwardness ($\gamma_i \approx 0$), while growth approaching linearity ($\alpha_i \to 0$) reflects topics with high inwardness ($\gamma_i \to Q$).
Therefore, within the framework of this infection model, the parameter $\alpha_i$ is not merely a descriptor of the curve's shape but also represents a sociophysical  indicator inversely related to inwardness: larger $\alpha_i$ corresponds to lower inwardness (higher outward orientation or shareability), and smaller $\alpha_i$ corresponds to stronger inwardness. \par

A key property of the model is that the shape index $\alpha_i$ does not depend on the choice of the time unit. This is because $\gamma_i$ and $Q$ are defined per action, not per unit time. If we change the time scale, only the rate parameter $R_i(=J_i)$ changes. \par

\subsection{A Weighted Infection-Type Model (Weighted Model)}
\label{app_sec_infection_model}

This section introduces a more realistic infection-type model on a \emph{weighted} network as an extension of the model presented in Section~\ref{sec_infection} of the main text. In the unweighted model, a new acquaintance is added at every step, which makes the within-community network excessively dense, especially when $\gamma_i \approx 0$. To address this, we propose a model in which interaction with known contacts can be \emph{deepened} and the strength of interaction is represented as an edge weight. The model can describe a variety of infection-network dynamics—for example, it yields a power-law in-degree distribution with exponent approximately 1 that arises irrespective of the shape parameter $\alpha_i$ (see Section~\ref{app_sec_infect_result}). The model in Section~\ref{sec_infection} is recovered as the special case with deepening priority $\beta_i=0$ and inwardness parameter $\theta_i=\gamma_i/Q$ (see Eq.~\ref{app_eq_kansen} in Section~\ref{app_sec_infection_term} for the definitions of $Q$ and $\gamma_i$). \par

\subsubsection{Model Assumptions and Terminology: Inwardness and a Contact-Weight Matrix}

\paragraph{State variables: infected and susceptible individuals.}
We model the diffusion of lexical knowledge as (i) person-to-person transmission of recognition of a term (``infection'' via recruitment/interaction/deepening) and (ii) the expansion of an infection network (a community of people who know the term). For a given term $i$, individuals who know the term (members of the term-$i$ community) are called \emph{infected}, and those who do not are called \emph{susceptible} (non-infected). When infected individuals recruit susceptibles, or when infected individuals interact with each other, the term-$i$ community (infection network) expands. We write the total number of infected individuals at time $t$ as $I_i(t)$. \par

\paragraph{Outward choice probability $p_i^{(l)}(t)$ (probability of selecting a susceptible).}
As a behavioral foundation, we link the topic characteristics studied in Section~\ref{sec_meaning} to individual communication strategies. Consider niche topics such as the name of a specific AKB48 member or a specialized medical term. Such topics deliver the most value in conversations with peers who share the same background, naturally inducing an \emph{inward}-oriented tendency to prioritize within-community conversation. As the community grows and saturates with familiar discussants, the perceived need to reach out to people who do not yet know the term (susceptibles) diminishes, further reinforcing inward orientation. By contrast, for more general or ``buzz-like'' topics, it is reasonable to expect active outreach to people who do not yet know the term (outward orientation). \par

We model these strategies as activity on a topic-specific \emph{weighted} social-tie matrix $W_{lf}^{(i)}(t)$ within the infected community. Here $W_{lf}^{(i)}(t)$ denotes, at time $t$, the strength (weight) of the relationship from person $l$ to person $f$ for topic $i$; there are no self-loops ($W_{ll}^{(i)}(t)=0$). Growth of the social relation implies creation of new acquaintances in the infection community, and contacting an existing acquaintance increases the corresponding weight. \par

At each time step, individual $l$ acts on topic $i$ with \emph{action rate} $J_i$ during the interval $\Delta t$ (thus the action probability is $J_i\Delta t$). Conditional on acting, $l$ chooses either to approach a susceptible (``recruit''), to obtain an introduction to a not-yet-connected infected person (``interaction''), or to contact an already-known infected person (``deepening''). The \emph{outward choice} probability (probability of selecting a susceptible), conditional on acting at time $t$, is
\begin{equation}
p_i^{(l)}(t) = \frac{1}{1+\theta_i g_i^{(l)}(t)},
\label{app_eq_kansen_w}
\end{equation}
where $g_i^{(l)}(t)=\sum_f W_{lf}^{(i)}(t)$ is the total strength of $l$'s social ties for topic $i$ at time $t$. The parameter $\theta_i\ge 0$ captures topic-$i$'s \emph{inwardness}. Note that $p_i^{(l)}(t)$ is conditional on acting; the unconditional probability of choosing the outward option within $\Delta t$ is $J_i \cdot p_i^{(l)}(t) \cdot \Delta t$. Each action by $l$ increases the total outgoing tie weight $\sum_f W_{lf}^{(i)}$ by exactly one. \par

For example, when $\theta_i=0$, an action—whenever it occurs—is always directed to a susceptible (maximally outward-looking). More generally, since $\theta_i\ge 0$, an increase in $g_i^{(l)}(t)$ lowers the probability of choosing a susceptible via Eq.~\ref{app_eq_kansen_w}, reflecting the idea that as the community capable of discussing topic $i$ grows denser and larger, the need to disseminate the topic externally diminishes. \par
%

\subsubsection{Behavioral Rules for Infected Individuals}
At each time step, infected individual $l$ acts on topic $i$ with probability $J_i\Delta t\le 1$ and expands their social ties via one of the following (the three action types are mutually exclusive and collectively exhaustive):
\begin{itemize}
\item \textbf{Recruitment} (probability $J_i \cdot p_i^{(l)}(t) \cdot \Delta t$): $l$ contacts a susceptible $a$, who becomes newly infected and a new acquaintance. The number of infected increases by one, $I_i(t+\Delta t)=I_i(t)+1$. Simultaneously, $a$ is added to $l$'s ties and $l$ to $a$'s ties: $W_{la}^{(i)}(t+\Delta t)=1$ and $W_{al}^{(i)}(t+\Delta t)=1$ (see Section~\ref{app_sec_initialization}, ``Initialization of ties for newly infected individuals'').
  \item \textbf{Interaction} (probability $J_i\,(1-p_i^{(l)}(t)) \cdot q_i^{(l)}(t) \cdot \Delta t$): $l$ obtains, via an existing contact $f$, an introduction to an infected individual $b$ who is in the community but not yet known to $l$ (so $W_{lb}^{(i)}(t)=0$). The infected count $I_i(t)$ does not change, but a new directed tie is added: $W_{lb}^{(i)}(t+\Delta t)=1$. Only $l$'s tie list is updated; $b$'s list is not (recruitment creates reciprocal additions, whereas interaction corresponds to a one-sided ``discovery''). Dependence on how $b$ is chosen—uniformly at random or weighted by tie strength, etc.—will be examined in Section~\ref{app_sec_infect_result}.
\item \textbf{Deepening} (probability $J_i \cdot (1-p_i^{(l)}(t)) \cdot (1-q_i^{(l)}(t)) \cdot \Delta t$): $l$ contacts an already-known infected person $f$. The infected count $I_i(t)$ does not change, but the weight of the corresponding tie increases by one: $W_{lf}^{(i)}(t+\Delta t)=W_{lf}^{(i)}(t)+1$. Again, only $l$'s tie list is updated.
\end{itemize}
Here $q_i^{(l)}(t)$ is the \emph{interaction-versus-deepening} choice probability, given by
\begin{equation}
q_i^{(l)}(t)=\frac{1}{1+\beta_i \cdot g_i^{(l)}(t)},
\end{equation}
so that when the social ties are already strong (large $g_i^{(l)}(t)$), deepening is favored over adding new acquaintances via interaction. The parameter $\beta_i\ge 0$ controls the priority to choose deepening. All updates within a step are computed from the snapshot at the beginning of the step ($t$) and are then applied simultaneously at $t+\Delta t$; no within-step re-evaluation or ordering effects are considered. \par

\subsubsection{Initialization of Ties for Newly Infected Individuals}
\label{app_sec_initialization}
A newly infected individual $a$ inherits the infector $l$'s social ties (analogous to a new student entering an advisor's network). Concretely, $a$'s initial ties are set by copying $l$'s ties at time $t$ and then adding $l$ itself:
for $f\neq l$,
\begin{equation}
W_{af}^{(i)}(t+\Delta t)=W_{lf}^{(i)}(t),
\end{equation}
and
\begin{equation}
W_{al}^{(i)}(t+\Delta t)=1.
\end{equation}
As noted under Recruitment, we also set $W_{la}^{(i)}(t+\Delta t)=1$. Consequently, $a$'s total tie strength satisfies $g_i^{(a)}(t+\Delta t)=g_i^{(l)}(t)+1$. If $l$'s ties are dense (large $g_i^{(l)}(t)$), the entrant $a$ inherits those dense ties. \par

\paragraph{Remark on choosing interaction partners.}
The particular rule for selecting partners in ``interaction'' or ``deepening'' does not affect the \emph{macro} growth curve. By construction, each action by $l$ increases exactly one unit of weight or one link in $W_{l\cdot}$, so the driver $g_i^{(l)}(t)$ relevant for the growth of $I_i(t)$ evolves as $g_i^{(l)}(t+\Delta t)=g_i^{(l)}(t)+1$, essentially independent of the detailed selection rule. These choices, however, do affect other properties such as the in-degree distribution; we verify such effects numerically in Section~\ref{app_sec_infect_result}. \par

\subsubsection{Deriving the Macro-Level Growth Curve}
New entrants inherit the average ties at the time of entry and then add one unit of tie (weight or link) per action thereafter, just like existing members. Hence an individual's $g_i^{(l)}(t)$ depends primarily on the community age $t$, rather than on the time since that individual became infected. We therefore approximate $g_i^{(l)}(t)$ as
\begin{equation}
g_i^{(l)}(t) \approx J_i \cdot t
\end{equation}
an approximation justified by the inheritance rule above. Substituting into Eq.~\ref{app_eq_kansen_w} yields
\begin{equation}
p_i^{(l)}(t) \approx \frac{1}{1+\theta_i \cdot J_i \cdot t }.
\end{equation}
Therefore,
\begin{equation}
I_i(t+\Delta t)-I_i(t)
\approx J_i\Delta t \sum_{l} p_i^{(l)}(t)
\approx \frac{J_i\Delta t}{1+\theta_i \cdot J_i \cdot t} I_i(t),
\end{equation}
where the sum runs over all infected individuals. We assume the outside population is sufficiently large so that simultaneous recruitment of the same susceptible and depletion effects can be neglected. Under a continuous-time approximation,
\begin{equation}
\frac{dI_i(t)}{dt} = \frac{J_i}{1+\theta_i \cdot J_i \cdot t} I_i(t),
\end{equation}
whose solution with $I_i(0)=1$ is $I_i(t)=(1+\theta_i J_i t)^{1/\theta_i}$ for $\theta_i\neq 0$ or $I_i(t)=e^{J_i t}$ for $\theta_i=0$. Using the solution, this can be rewritten as,
\begin{equation}
\frac{dI_i(t)}{dt} = J_i \cdot I_i(t)^{1-\theta_i},
\end{equation}
\par

Let the observed article count be $y_i(t)=K \cdot I_i(t)$. Then
\begin{equation}
\frac{dy_i(t)}{dt} = J_i \cdot K \left(\frac{y_i(t)}{K}\right)^{1-\theta_i}. \label{app_eq_ans_eq}
\end{equation}
Comparing with the macro model in Eq.~\ref{eq_base0}, we obtain
\begin{equation}
\alpha_i = 1-\theta_i,\qquad R_i = J_i, \qquad Y = K.
\end{equation}
Thus, the shape parameter $\alpha_i$ that governs the observed growth curve is determined by the topic's inwardness parameter $\theta_i$ (and is independent of the deepening priority $\beta_i$). In particular, exponential growth ($\alpha_i \approx 1$) corresponds to low inwardness ($\theta_i \approx 0$), whereas growth approaching linearity ($\alpha_i \to 0$) corresponds to high inwardness ($\theta_i \to 1$). \par

Therefore, within this framework, $\alpha_i$ is not merely a curve-shape parameter but a sociophysical indicator that quantifies a topic's inwardness—equivalently, its intrinsic ``shareability'' or outward orientation. When $\beta_i=0$ and $\theta_i=\gamma_i/Q$, the present weighted model reduces to the unweighted model with topical appeal in Section~\ref{sec_infection}. Introducing weights remedies the rapid densification observed in the unweighted model when $\gamma_i$ is small. \par

\subsection{Model Properties (Numerical simulations)}
\label{app_sec_infect_result}
This section presents numerical experiments for the infection-type model and summarizes its properties. In particular, we examine how different ``interaction'' rules used when infected individuals contact others inside the community affect the in-degree distribution of the network (defined as $k^{(l)}_{in} = |\{\,u \mid W_{u l} > 0\,\}|$). For notational simplicity, we suppress the topic index $i$ and time $t$ when discussing the social-tie weights and write $W_{lb}$.

\subsubsection{Rules Used in the Simulations}
\label{app_sec_infect_rule}
Here we describe the ``Interaction'' and ``Deepening'' rules used in the numerical simulations.
\subsubsubsection{Interaction Rules Used in the Simulations}
To evaluate the model's properties, we consider the following three rules for choosing the interaction partner $b$ (an infected individual not yet known to $l$ with $W_{lb}=0$).

\paragraph{1. Random selection}
Individual $l$ chooses uniformly at random one infected person from the whole network who is not yet an acquaintance ($W_{lb}=0$). Conceptually, this corresponds to receiving a recommendation from the community at large; operationally, it does not depend on the characteristics of $l$'s own acquaintances.

\paragraph{2. Path-weight selection (``friends-of-friends'' recommendation list)}
Individual $l$ selects $b$ from the set of friends-of-friends with probability proportional to the weighted number of length-2 paths $l \to f \to b$. In the unweighted case, this reduces to aggregating the neighbor lists of $l$'s acquaintances (excluding $l$) and drawing one candidate uniformly at random from that pooled list.

Formally, the probability that $b$ is chosen is
\begin{equation}
o_l(b)=\frac{s_l(b)}{\sum_{c \in Z(l)} s_l(c)},
\end{equation}
where $s_l(b) = \sum_{f \in F(l)} W_{lf}\, W_{fb}$, $F(l)$ is the set of $l$'s acquaintances (out-neighbors), and $Z(l)$ is the set of $l$'s friends-of-friends who are not yet known to $l$ (i.e., $W_{lc}=0$ and $c \neq l$).

If no friends-of-friends exist (the denominator of $o_l(b)$ is zero), we handle the step as follows in the simulations:
\begin{itemize}
    \item \textbf{Weighted setting} ($\beta_i \neq 0$): take a \emph{deepening} action.
    \item \textbf{Unweighted setting} ($\beta_i = 0$): skip the action (no update).
\end{itemize}

\paragraph{3. Two-step selection (``call a friend and ask'')}
Selection proceeds in two stages. First, $l$ chooses an acquaintance $f$; second, $f$'s acquaintance $b$ (unknown to $l$) is chosen. For example, $l$ contacts $f$ with probability proportional to $W_{lf}$ (stage one), and then $f$ recommends $b$ with probability proportional to $W_{fb}$ (stage two).

The probability that $b$ is chosen is
\begin{equation}
o_l(b)=\sum_{f \in F(l)} \left(\frac{W_{lf}}{\sum_{v \in F(l)} W_{lv}} \cdot \frac{W_{fb}}{\sum_{k \in F'_l(f)} W_{fk}} \right),
\end{equation}
where $F(l)$ is the set of $l$'s acquaintances, and $F'_l(f)$ is the set of $f$'s acquaintances that $l$ does not yet know (i.e., those with $W_{lk}=0$).

If no eligible partner exists (the denominator in the second stage is zero), we apply the same handling as in ``Path-weight selection.'' In the unweighted setting ($\beta_i=0$), skipping the action deviates from the theory, but we adopt it to preserve the network properties specific to the unweighted case.

\paragraph{Initialization note.}
By definition of the model, at $t=1$ there are only two infected individuals, so interaction does not occur. For $t \ge 2$, $l$ always has at least the infector $f$ with $W_{lf}>0$, so the denominator in the first-stage selection is nonzero.

\subsubsubsection{Implementation of Deepening in the Simulations}
In the simulations, the \emph{deepening} action selects a known contact $f$ with probability proportional to the tie weight $W_{lf}$.

The probability that $f$ is chosen is
\begin{equation}
d_l(f) \;=\; \frac{W_{lf}}{\sum_{v \in F(l)} W_{lv}},
\end{equation}
where $F(l)$ denotes the set of $l$'s acquaintances. \par

\subsubsection{Properties in the Unweighted Case ($\beta_i=0$)}
\label{app_sec_unweighted}
When $\beta_i=0$, the model does not take the ``deepening'' action and reduces to the unweighted model described in Section~\ref{sec_infection}.

Fig.~\ref{fig_unweightnet_ran} reports simulation results for $\beta_i=0$ with the ``random selection'' interaction rule. We compare $\theta_i \in \{0,0.5,1.0 \}$. The panels show: (a)-(c) 100 sample growth curves; (d) an example of a growth curve close to the theoretical curve; (e)-(f) the in-/out-degree distributions corresponding to (d); and (g)-(i) example infection paths corresponding to (d).

\paragraph{Growth curves (a)-(d)}
In panel (a), when $\alpha_i=0.0$ (i.e., $\theta_i=1.0$), the simulated growth curves are nearly linear, and the theoretical curve (dashed line) forms a lower bound. This gap arises because the simulations exhibit ``no-action'' events that the theory does not assume. Specifically, when $\theta_i=1$ (very strong inward orientation), the network becomes so dense that no unconnected nodes are available as interaction targets. As a result, the theoretical premise—that each action necessarily increases $g_i^{(l)}(t)$ by one—breaks down, and the increase in $g_i^{(l)}(t)$ is slower than in the theory. This weakens the theoretical suppression of recruitment $p_i^{(l)}(t)$ and leads to a larger number of new infections than predicted.

In panels (b) and (c), for $\alpha_i=0.5$ (i.e., $\theta_i=0.5$) and $\alpha_i=1.0$ (i.e., $\theta_i=0.0$), the theoretical curve passes roughly through the middle of the sample ensemble, indicating that the theory provides a coarse description of the simulations. Panel (d) selects one sample that closely follows the theoretical curve, showing that such near-theory samples exist for all $\theta_i$.

\paragraph{Network characteristics (e)-(f)}
Panels (e) and (f) report network statistics for the curve in (d). The in-degree distribution in (e) is almost complete (almost all-to-all) when $\alpha_i =0.0$ (i.e., $\theta_i=1.0$), becomes a power law with exponent about 1 when $\alpha_i=1.0$ (i.e., $\theta_i=0.0$), and takes an intermediate shape for $\theta_i=0.5$. The near-complete pattern at $\theta_i=1$ is due to network ``saturation.'' For example, if $J_i=1$, each node tries to add one acquaintance per step, while about one new node enters the network per step. Existing nodes soon fail to find unconnected candidates for interaction, and links concentrate on the few available new infected entrants. New nodes thus quickly become acquaintances of many infected individuals, and the saturated state persists. In the exponential case $\alpha_i=1.0$ (i.e., $\theta_i=0$), the cumulative distribution exhibits a power law with exponent about 1; we discuss this in Section~\ref{app_sec_indegree}. The out-degree distribution in (f) is close to a normal-like shape for all $\theta_i$.

\paragraph{Infection paths (g)-(i)}
Panels (g)-(i) show infection paths for the case in (d) (we visualize only recruitment edges $l \to a$ where $l$ infects $a$ and ignore internal links). In (g) with $\alpha_i=0$ (i.e., $\theta_i=1$), older nodes (blue) often infect newer nodes (yellow). By contrast, in (i) with $\alpha_i=1.0$ (i.e., $\theta_i=0.0$), we frequently observe chains where a newly added node (yellow) infects another newly added node (yellow).

\paragraph{Comparison with other interaction rules}
Figs.~\ref{fig_unweightnet_mei} (path-weight selection) and \ref{fig_unweightnet_tel} (two-step selection) report results under alternative interaction rules. Although the in-degree distributions differ somewhat in shape, the overall properties—such as the behavior of the growth curves and the out-degree distributions—are broadly similar to those under random selection.

\subsubsection{Properties in the Weighted Case ($\beta_i=1$)}
\label{app_sec_weighted}
\begin{figure*}[p]
    \centering
       \begin{overpic}[width=5.4cm]{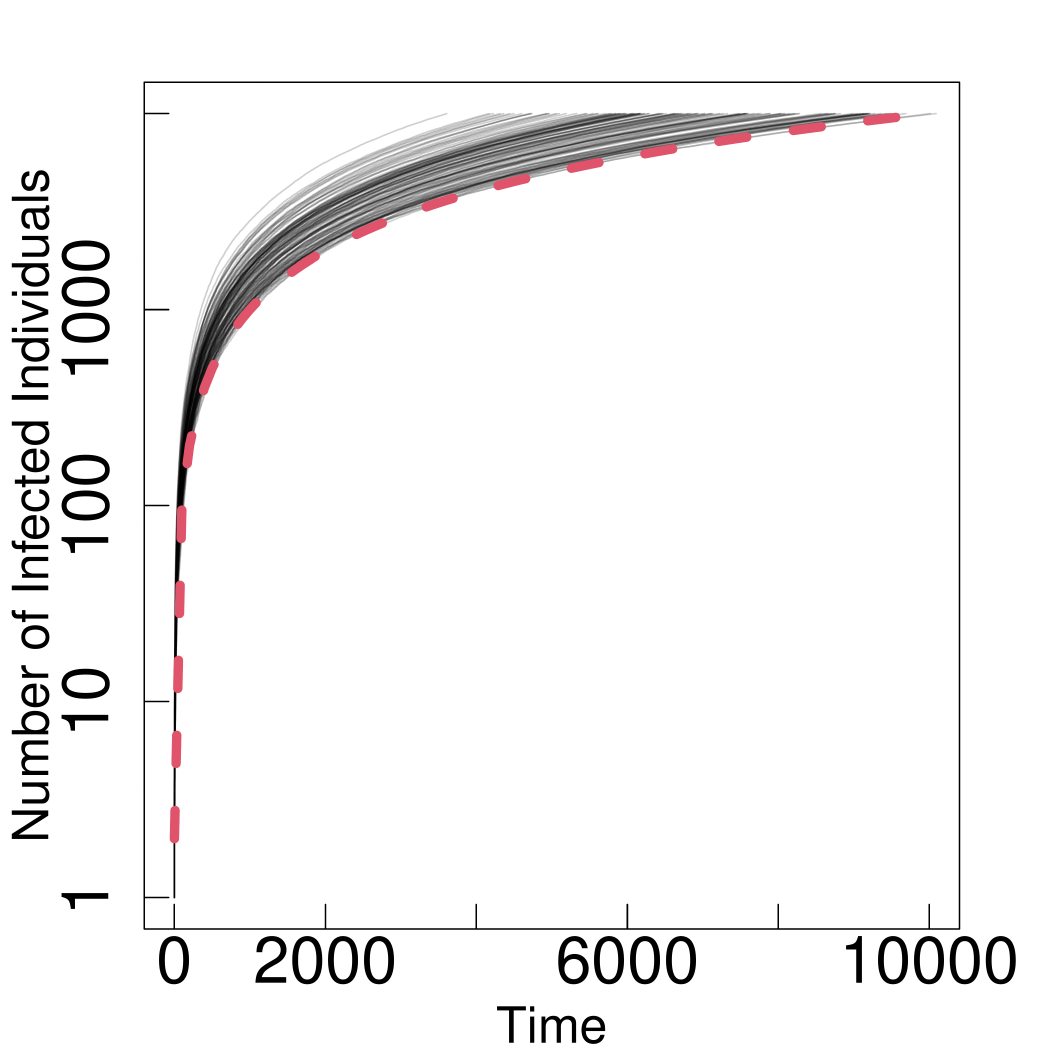}
        \put(35,30){\color{black}\Large\bfseries (a)}
    \end{overpic}
    \begin{overpic}[width=5.4cm]{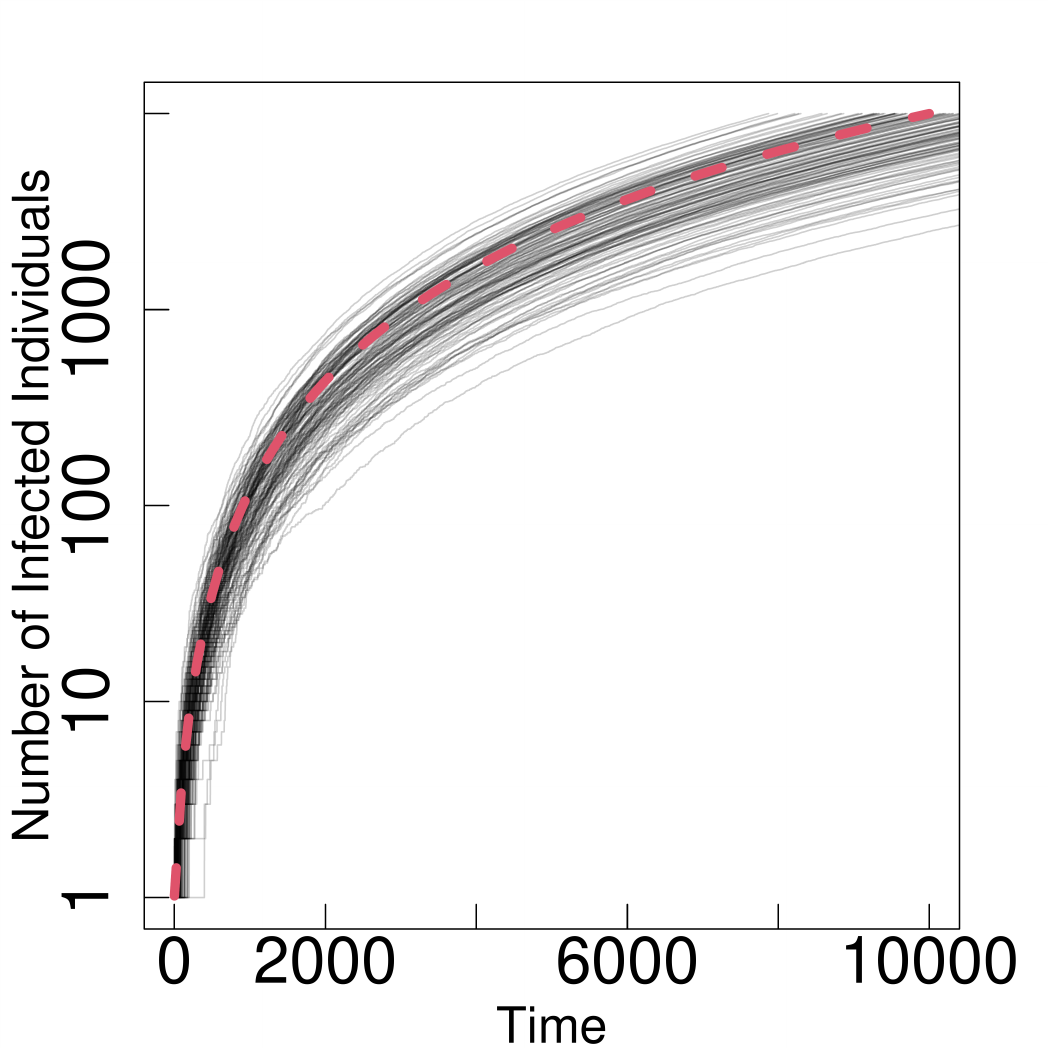}
        \put(30,120){\color{black}\Large\bfseries (b)}
    \end{overpic}
    \begin{overpic}[width=5.4cm]{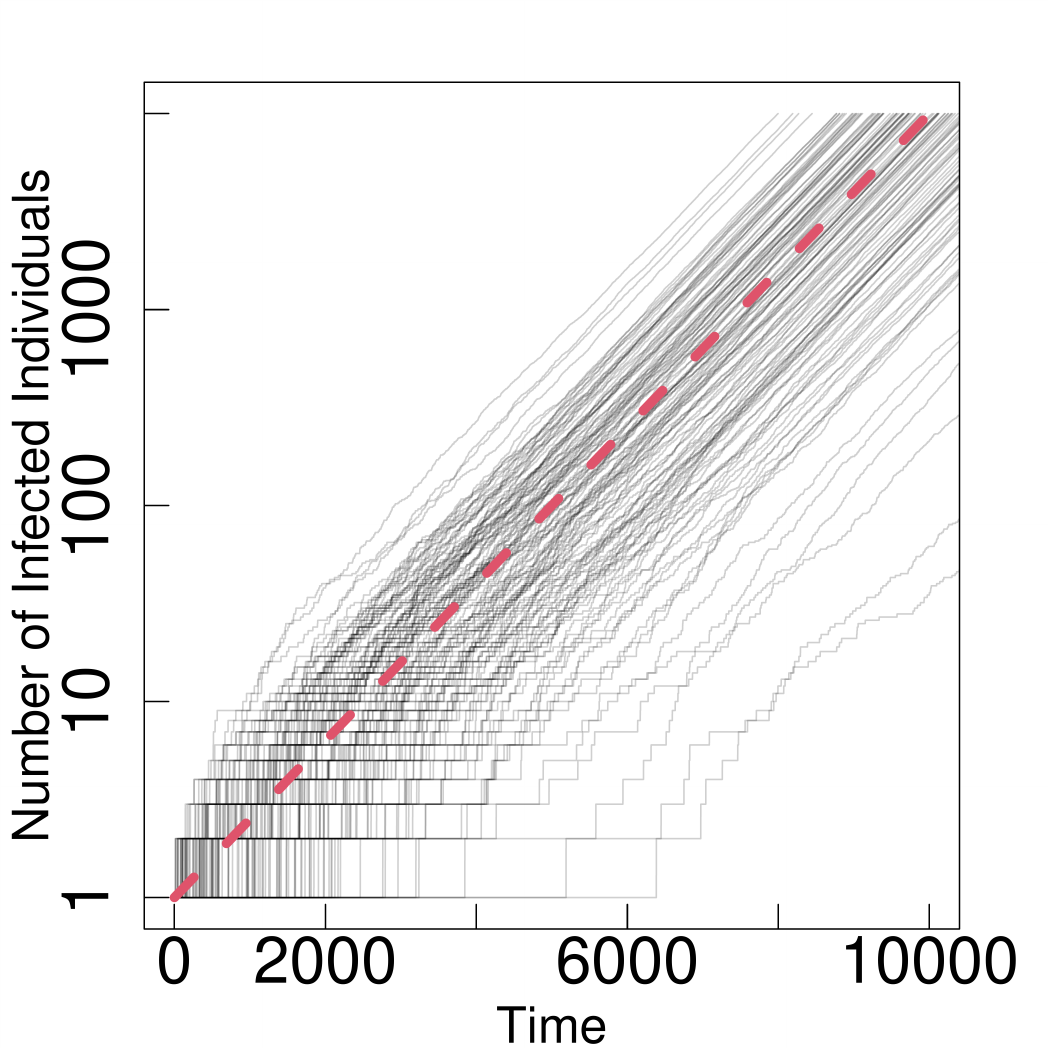}
        \put(30,120){\color{black}\Large\bfseries (c)}
    \end{overpic}
    \begin{overpic}[width=5.4cm]{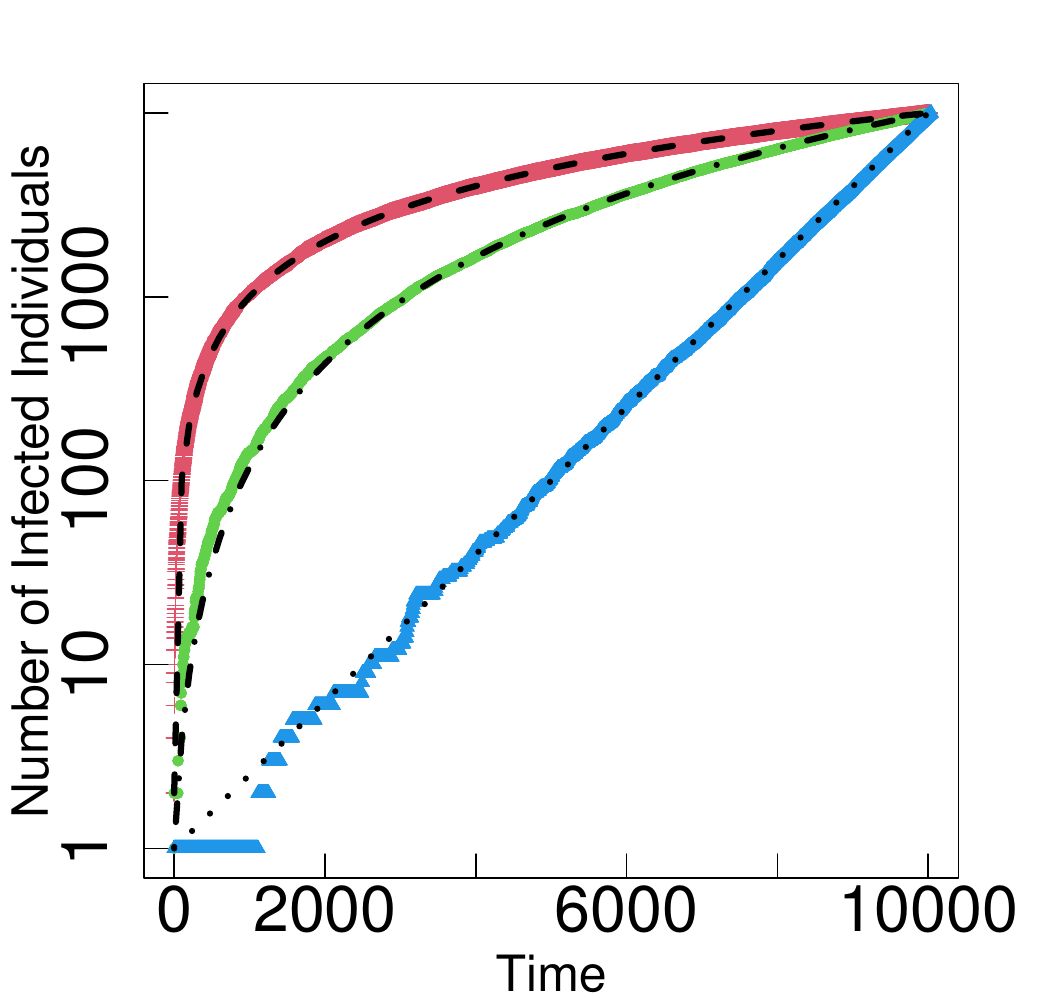}
        \put(30,120){\color{black}\Large\bfseries (d)}
    \end{overpic}  
        \begin{overpic}[width=5.4cm]{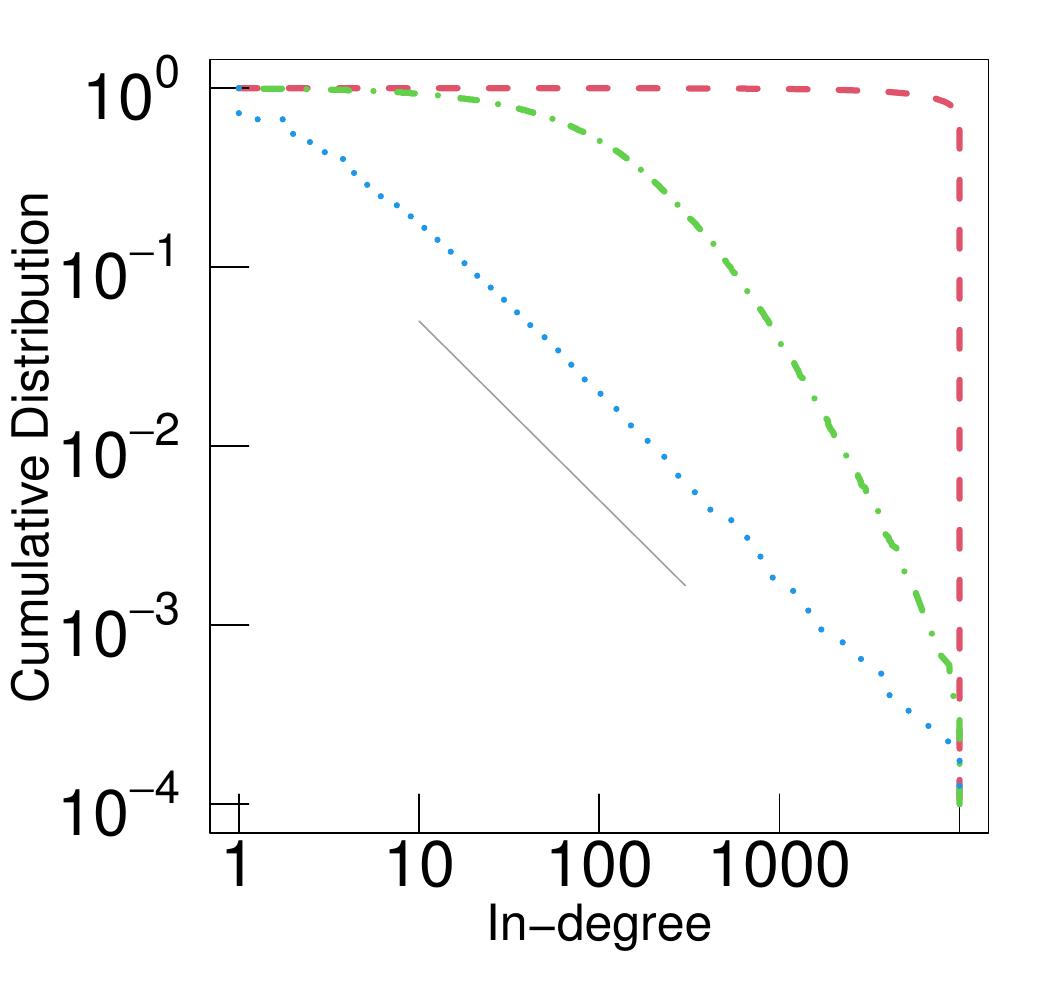}
        \put(35,110){\color{black}\Large\bfseries (e)}
    \end{overpic}
     \begin{overpic}[width=5.4cm]{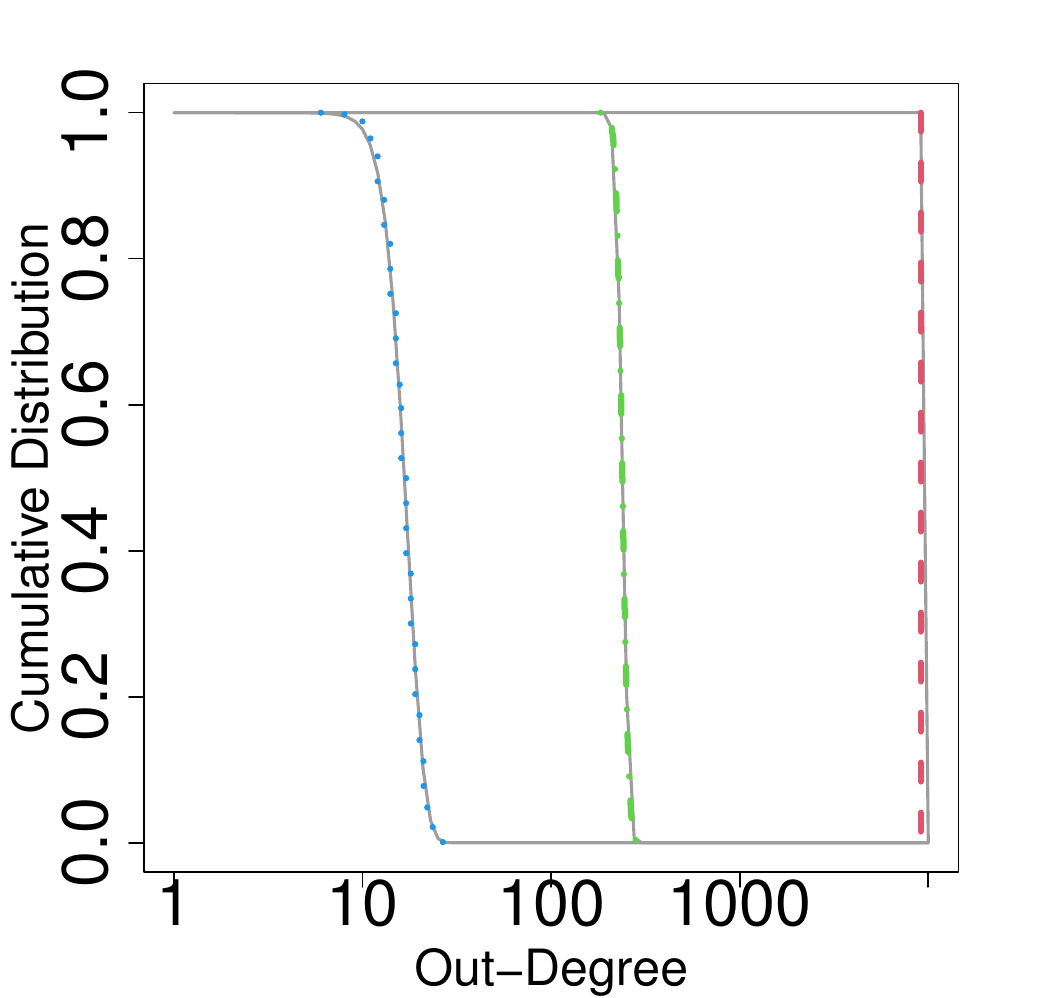}
        \put(30,110){\color{black}\Large\bfseries (f)}
    \end{overpic}
     \begin{overpic}[width=5.4cm]{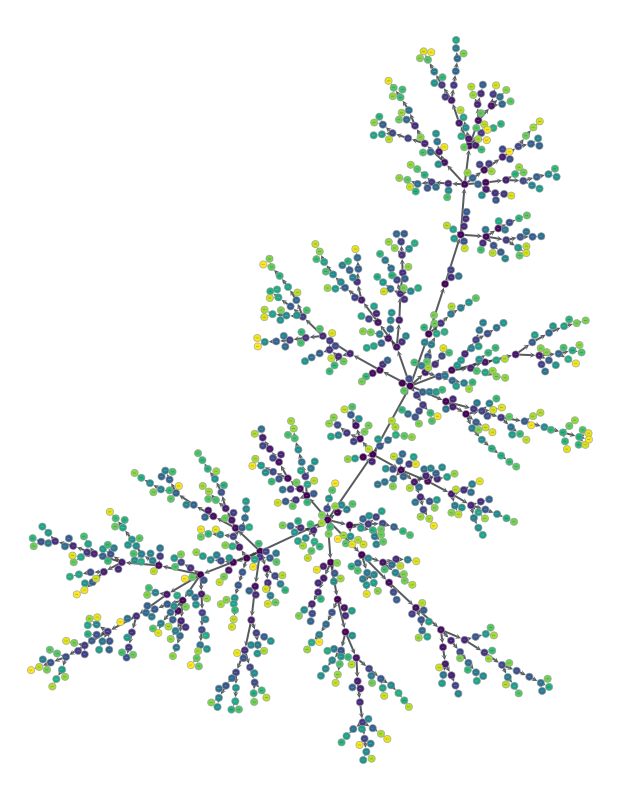}
        \put(30,130){\color{black}\Large\bfseries (g)}
    \end{overpic}
   \begin{overpic}[width=5.4cm]{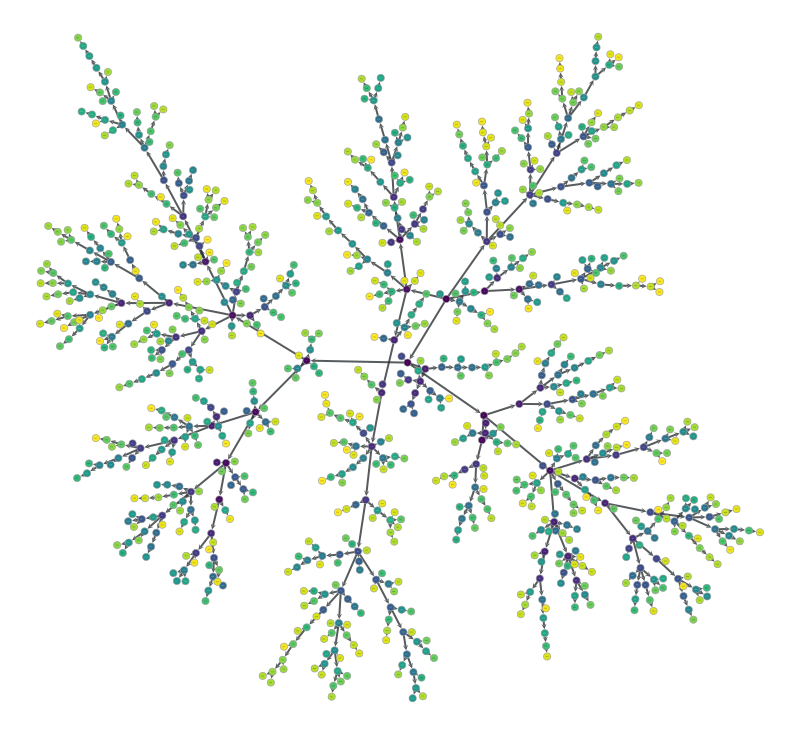}
        \put(30,125){\color{black}\Large\bfseries (h)}
    \end{overpic}
    \begin{overpic}[width=5.4cm]{infection_network_beta10_gamma10_ran.png}
        \put(30,120){\color{black}\Large\bfseries (i)}
    \end{overpic}
  
\caption{Numerical results for the infection-type model (Section~\ref{app_sec_weighted}). Simulation setting: unweighted ($\beta_i=0$); interaction is chosen at random. (a)–(c) Example growth curves: black solid lines, 128 simulation paths; red dashed line, theoretical curve (Eq.~\ref{app_eq_ans_eq}); $K=1$. (a) $\theta_i=1.0$ ($\alpha_i=0.0$), $J_i=1.0$; (b) $\theta_i=0.5$ ($\alpha_i=0.5$), $J_i=0.020$; (c) $\theta_i=0.0$ ($\alpha_i=1.0$), $J_i=9.2\times10^{-4}$. (d) A simulation path close to the theory: red crosses, $\theta_i=1.0$; green circles, $\theta_i=0.5$; blue triangles, $\theta_i=0.0$; black line, theory (Eq.~\ref{app_eq_ans_eq}). Panels (e)–(f) indicate statistics for this path. (e) In-degree distribution: red solid, $\theta_i=1.0$; green dash–dot, $\theta_i=0.5$; blue dotted, $\theta_i=0.0$; thin black guide, power law with exponent 1 ($\propto 1/x$). (f) Out-degree distribution: red solid, $\theta_i=1.0$; green dash–dot, $\theta_i=0.5$; blue dotted, $\theta_i=0.0$; thin black curve, normal distribution (mean and standard deviation estimated from the data). (g)–(i) Infection networks (recruitment edges only). Nodes are colored by entry time (blue = older, yellow = newer): (g) $\theta_i=1.0$; (h) $\theta_i=0.5$; (i) $\theta_i=0.0$.}
   \label{fig_unweightnet_ran}
\end{figure*}

\begin{figure*}[p]
    \centering
       \begin{overpic}[width=5.4cm]{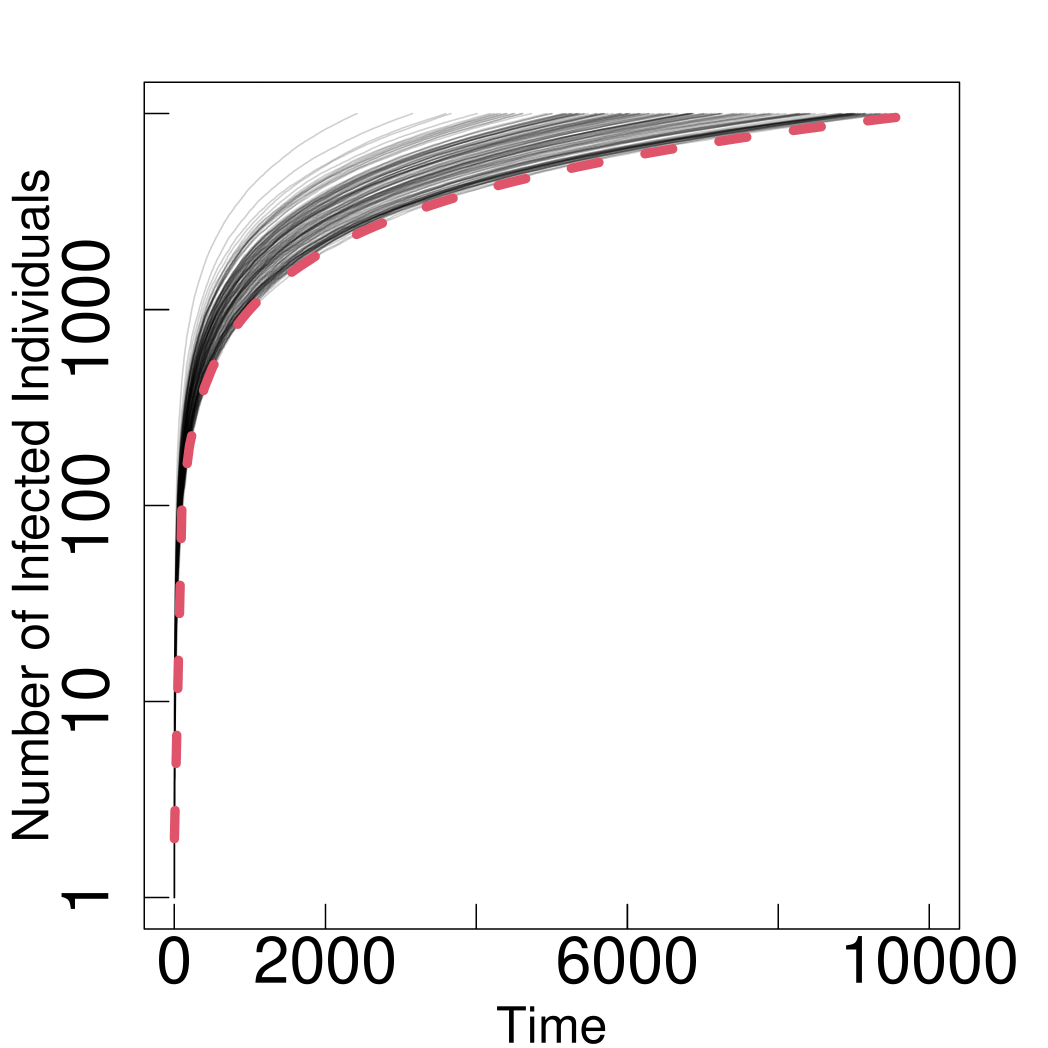}
        \put(35,30){\color{black}\Large\bfseries (a)}
    \end{overpic}
    \begin{overpic}[width=5.4cm]{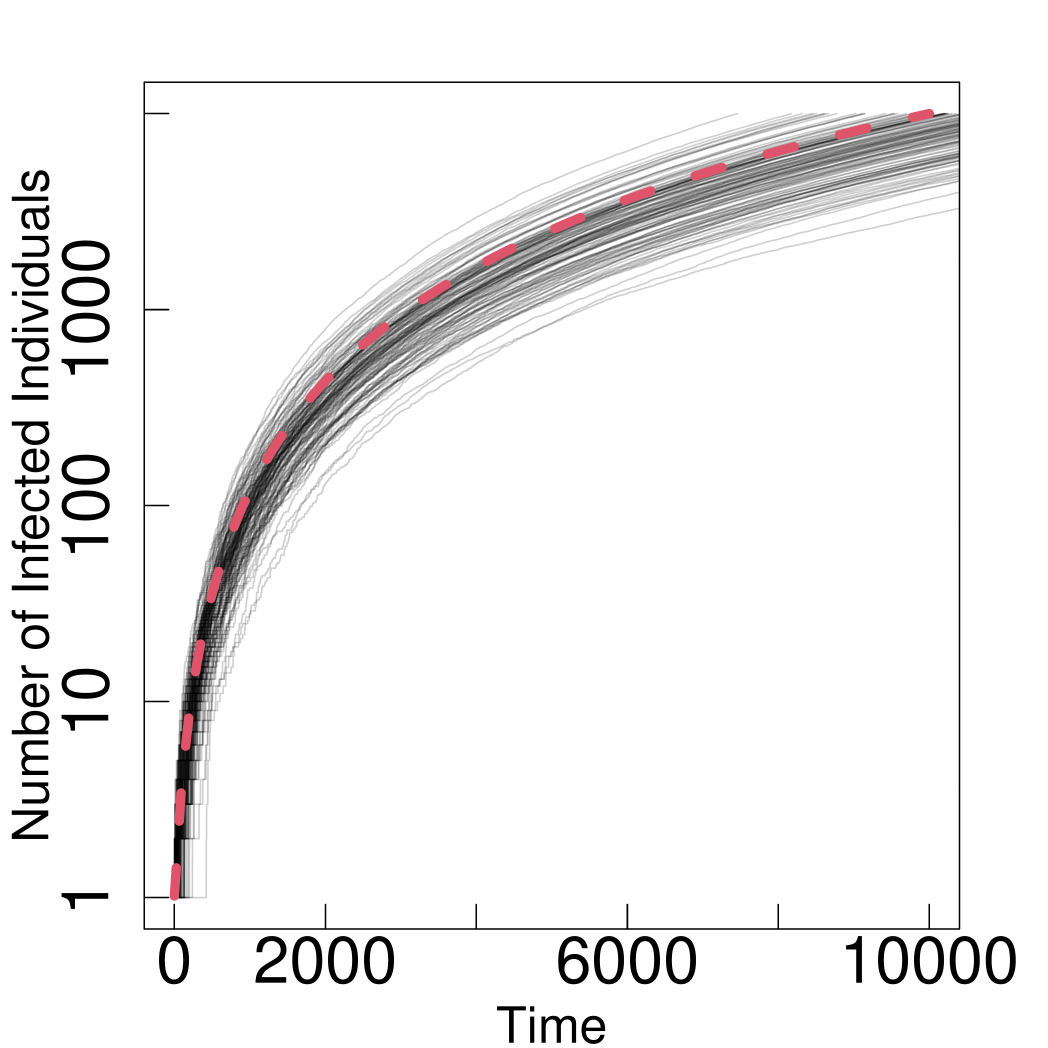}
        \put(30,120){\color{black}\Large\bfseries (b)}
    \end{overpic}
    \begin{overpic}[width=5.4cm]{net_beta10_gamma10_ran_lines_small.pdf}
        \put(30,120){\color{black}\Large\bfseries (c)}
    \end{overpic}
     \begin{overpic}[width=5.4cm]{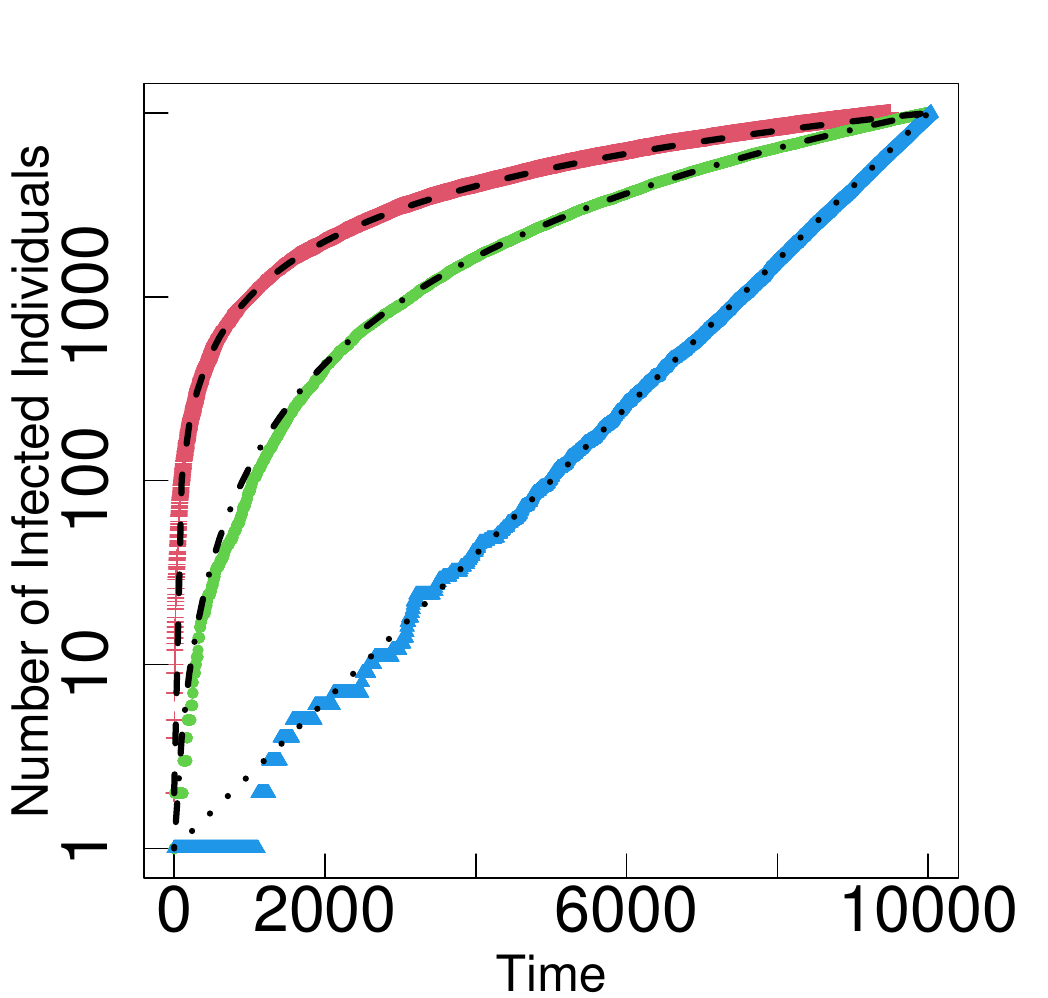}
        \put(30,120){\color{black}\Large\bfseries (d)}
    \end{overpic}
     \begin{overpic}[width=5.4cm]{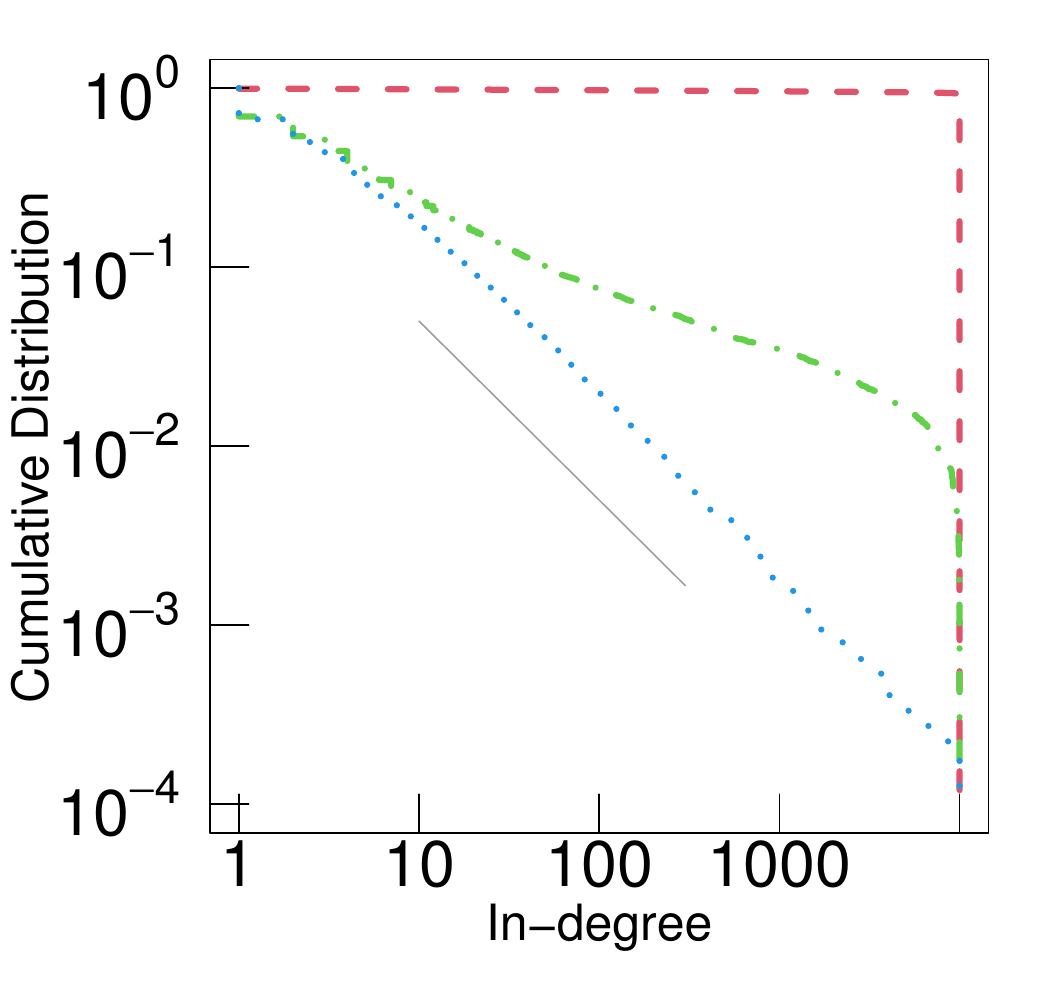}
        \put(35,110){\color{black}\Large\bfseries (e)}
    \end{overpic}
     \begin{overpic}[width=5.4cm]{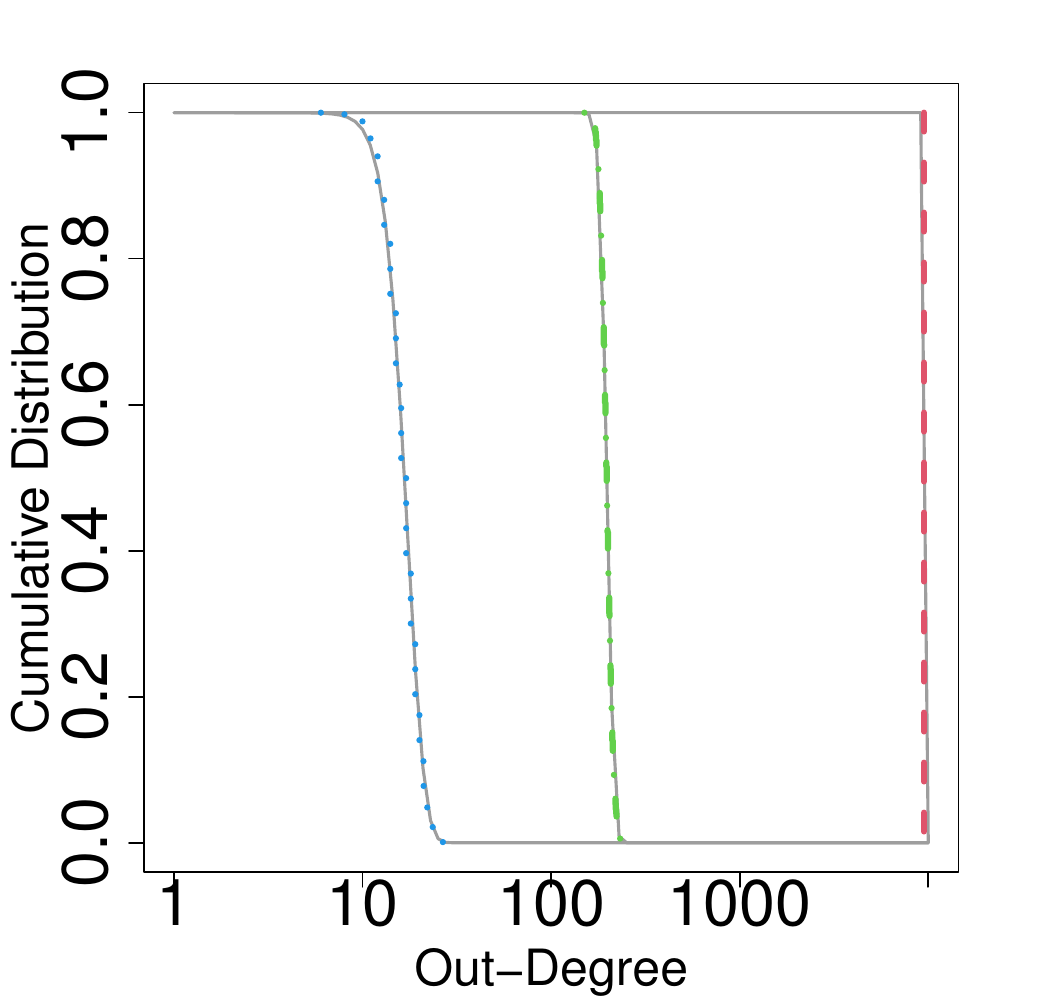}
        \put(30,110){\color{black}\Large\bfseries (f)}
    \end{overpic}
    \begin{overpic}[width=5.4cm]{infection_network_beta10_gamma00_fofmei.png}
        \put(30,130){\color{black}\Large\bfseries (g)}
    \end{overpic}
   \begin{overpic}[width=5.4cm]{infection_network_beta10_gamma05_fofmei.png}
        \put(30,130){\color{black}\Large\bfseries (h)}
    \end{overpic}
    \begin{overpic}[width=5.4cm]{infection_network_beta10_gamma10_ran.png}
        \put(30,120){\color{black}\Large\bfseries (i)}
    \end{overpic}
    \caption{
   Numerical results for the infection-type model (Section~\ref{app_sec_weighted}). Simulation setting: unweighted ($\beta_i=0$); interaction uses the path-weight selection rule (see Section \ref{app_sec_infect_rule}). The remaining details are the same as in Fig.~\ref{fig_unweightnet_ran}. (a)–(c) Example growth curves: black solid lines, 128 simulation paths; red dashed line, theoretical curve (Eq.~\ref{app_eq_ans_eq}); $K=1$. (a) $\theta_i=1.0$ ($\alpha_i=0.0$), $J_i=1.0$; (b) $\theta_i=0.5$ ($\alpha_i=0.5$), $J_i=0.020$; (c) $\theta_i=0.0$ ($\alpha_i=1.0$), $J_i=9.2\times10^{-4}$. (d) A simulation path close to the theory: red crosses, $\theta_i=1.0$; green circles, $\theta_i=0.5$; blue triangles, $\theta_i=0.0$; black line, theory (Eq.~\ref{app_eq_ans_eq}). Panels (e)–(f) indicate statistics for this path. (e) In-degree distribution: red solid, $\theta_i=1.0$; green dash–dot, $\theta_i=0.5$; blue dotted, $\theta_i=0.0$; thin black guide, power law with exponent 1 ($\propto 1/x$). (f) Out-degree distribution: red solid, $\theta_i=1.0$; green dash–dot, $\theta_i=0.5$; blue dotted, $\theta_i=0.0$; thin black curve, normal distribution (mean and standard deviation estimated from the data). (g)–(i) Infection networks (recruitment edges only). Nodes are colored by entry time (blue = older, yellow = newer): (g) $\theta_i=1.0$; (h) $\theta_i=0.5$; (i) $\theta_i=0.0$.}
    \label{fig_unweightnet_mei}
\end{figure*}

\begin{figure*}[pt]
    \centering
       \begin{overpic}[width=5.4cm]{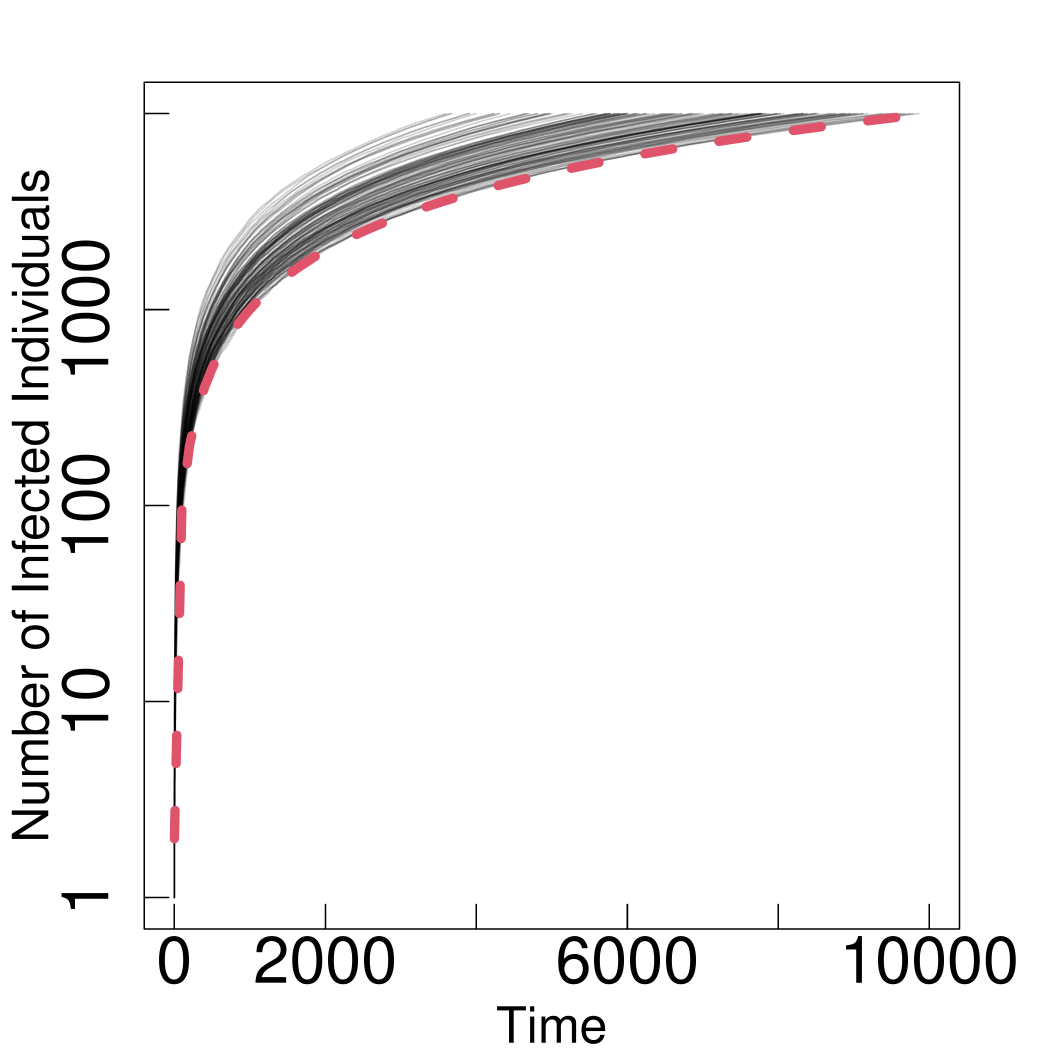}
        \put(35,30){\color{black}\Large\bfseries (a)}
    \end{overpic}
    \begin{overpic}[width=5.4cm]{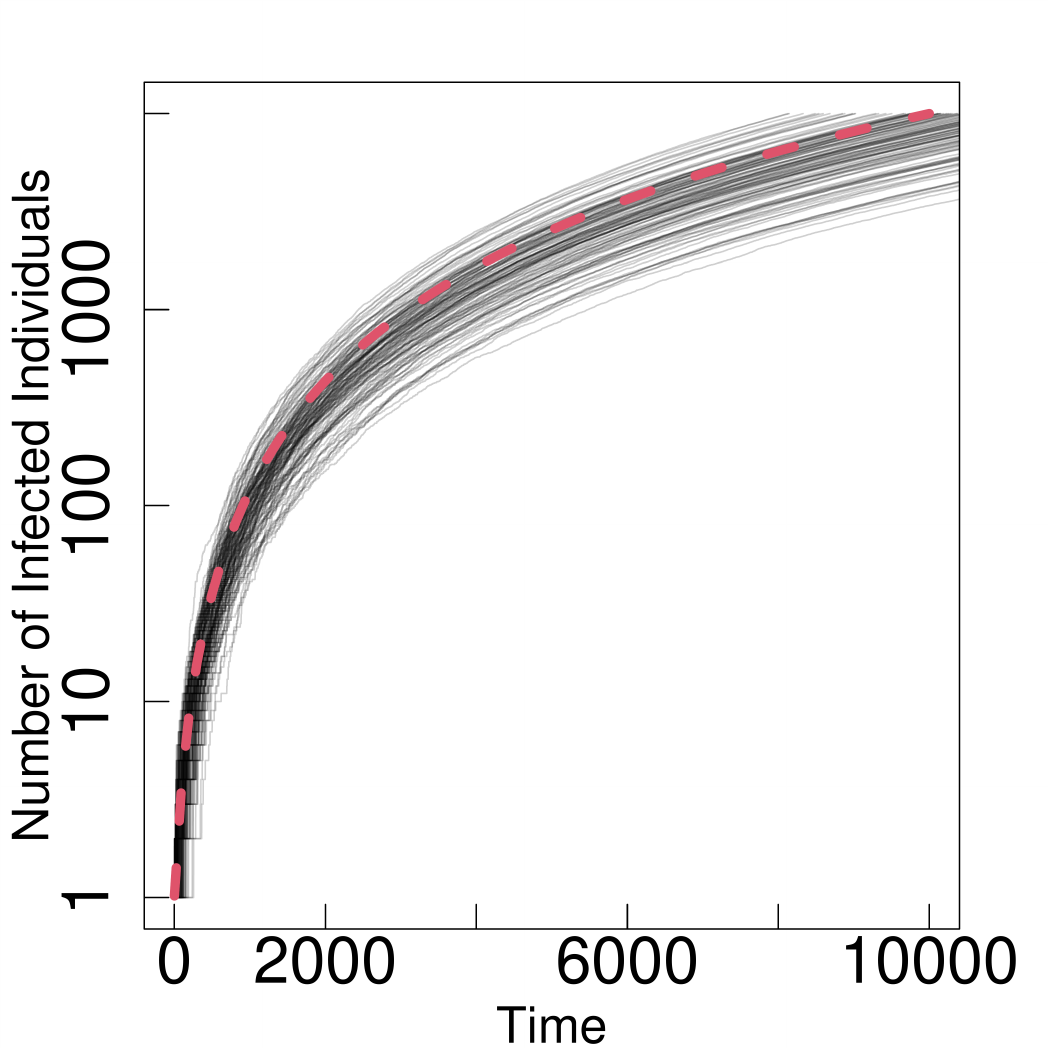}
        \put(30,120){\color{black}\Large\bfseries (b)}
    \end{overpic}
    \begin{overpic}[width=5.4cm]{net_beta10_gamma10_ran_lines_small.pdf}
        \put(30,120){\color{black}\Large\bfseries (c)}
    \end{overpic}
    \begin{overpic}[width=5.4cm]{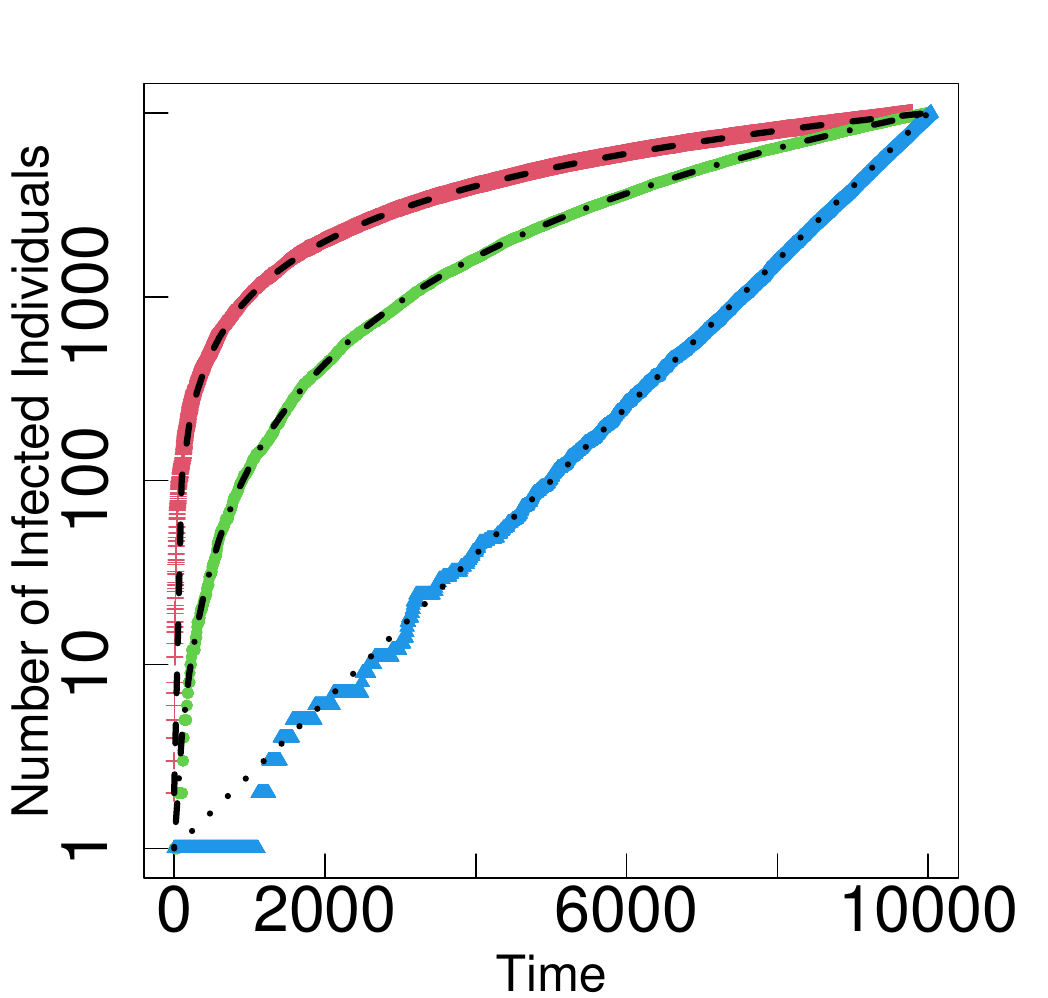}
        \put(30,120){\color{black}\Large\bfseries (d)}
    \end{overpic}
     \begin{overpic}[width=5.4cm]{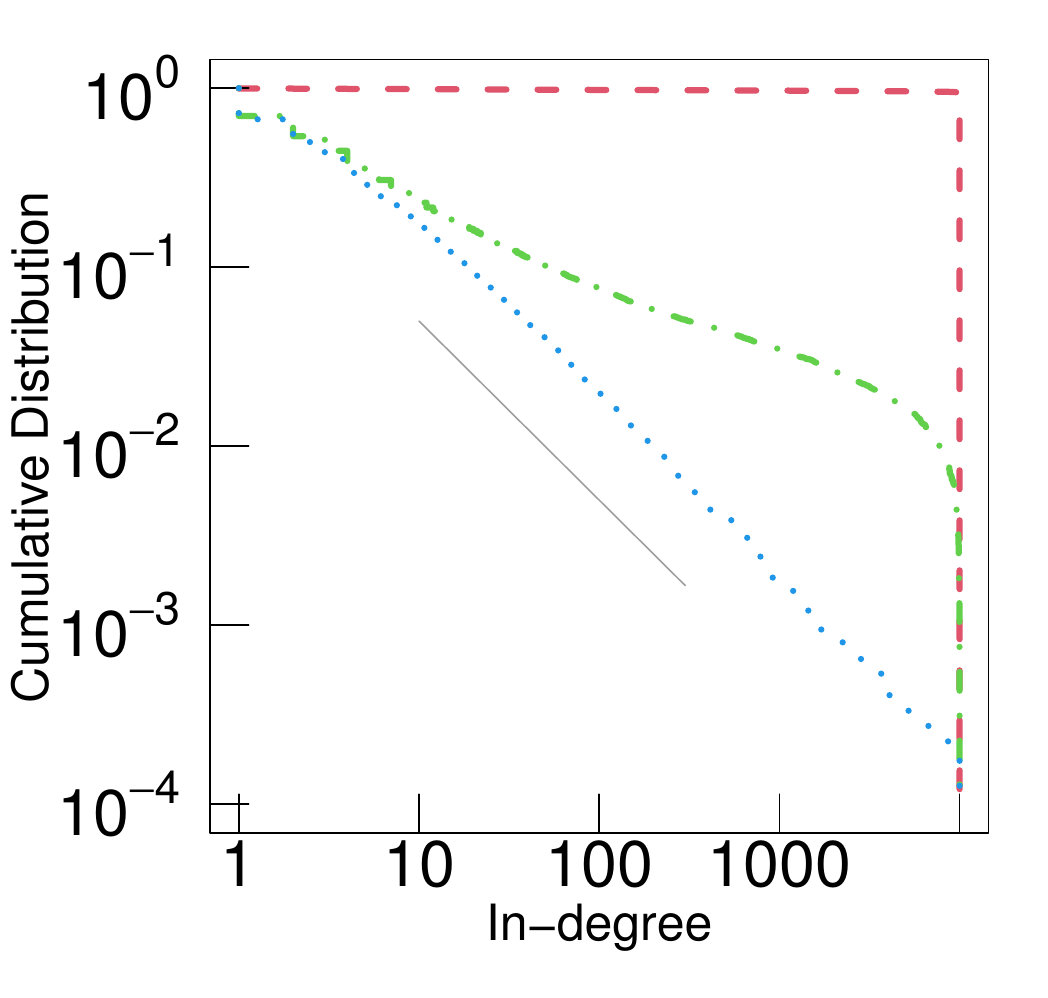}
        \put(35,110){\color{black}\Large\bfseries (e)}
    \end{overpic}
     \begin{overpic}[width=5.4cm]{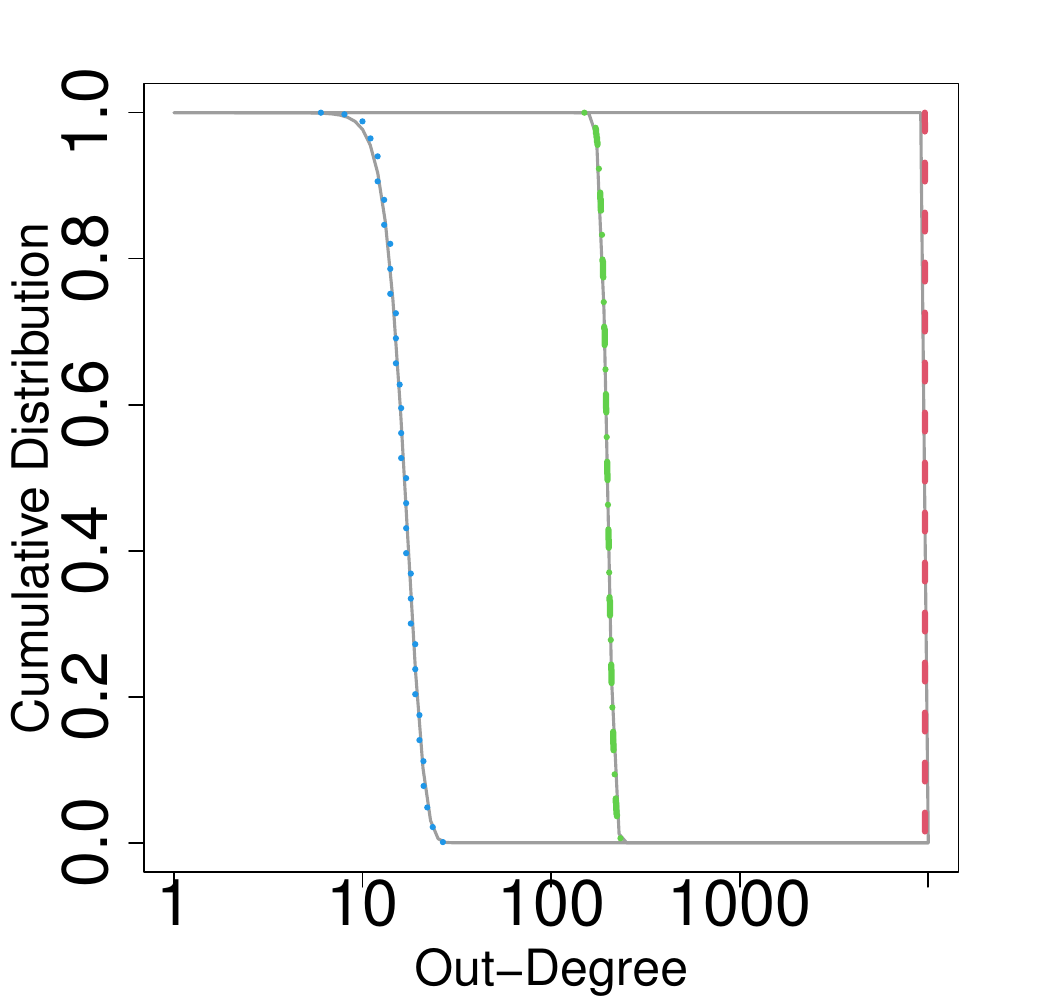}
        \put(30,110){\color{black}\Large\bfseries (f)}
    \end{overpic}
    \begin{overpic}[width=5.4cm]{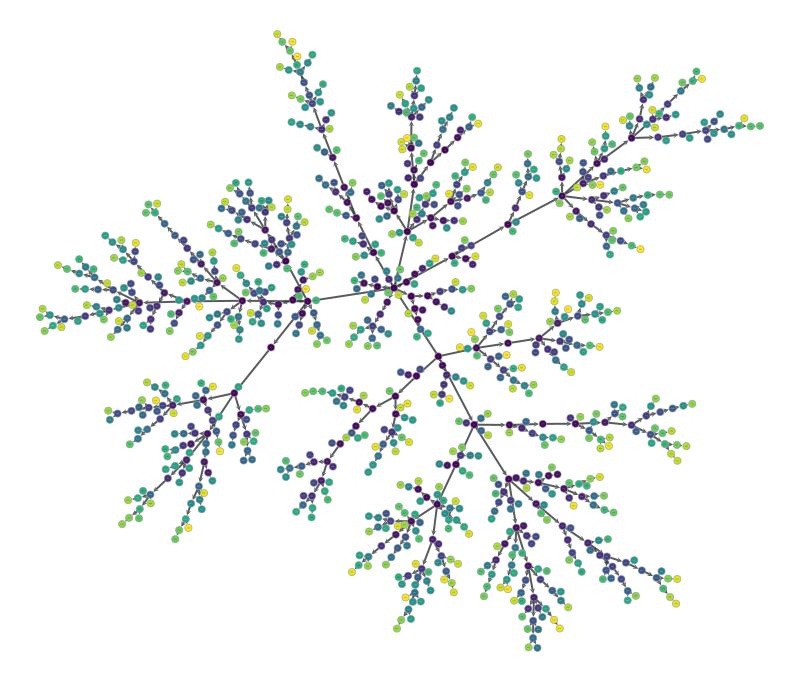}
        \put(30,130){\color{black}\Large\bfseries (g)}
    \end{overpic}
   \begin{overpic}[width=5.4cm]{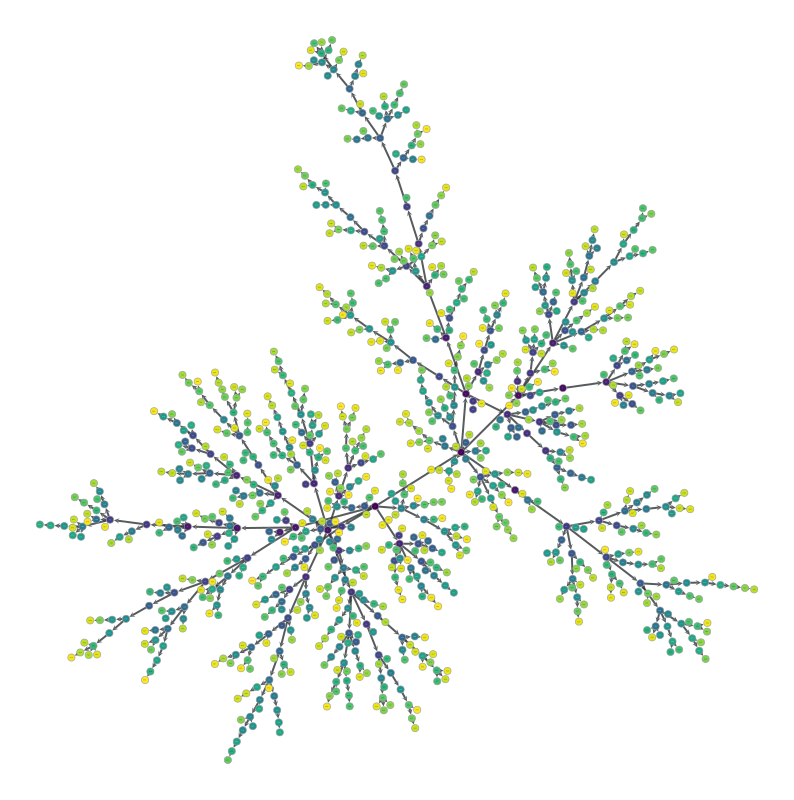}
        \put(30,125){\color{black}\Large\bfseries (h)}
    \end{overpic}
    \begin{overpic}[width=5.4cm]{infection_network_beta10_gamma10_ran.png}
        \put(30,120){\color{black}\Large\bfseries (i)}
    \end{overpic}
  
   \caption{Numerical results for the infection-type model (Section~\ref{app_sec_weighted}). Simulation setting: unweighted ($\beta_i=0$); interaction uses the two-step selection rule (see Section \ref{app_sec_infect_rule}). The remaining details are the same as in Fig.~\ref{fig_unweightnet_ran}. (a)–(c) Example growth curves: black solid lines, 128 simulation paths; red dashed line, theoretical curve (Eq.~\ref{app_eq_ans_eq}); $K=1$. (a) $\theta_i=1.0$ ($\alpha_i=0.0$), $J_i=1.0$; (b) $\theta_i=0.5$ ($\alpha_i=0.5$), $J_i=0.020$; (c) $\theta_i=0.0$ ($\alpha_i=1.0$), $J_i=9.2\times10^{-4}$. (d) A simulation path close to the theory: red crosses, $\theta_i=1.0$; green circles, $\theta_i=0.5$; blue triangles, $\theta_i=0.0$; black line, theory (Eq.~\ref{app_eq_ans_eq}). Panels (e)–(f) indicate statistics for this path. (e) In-degree distribution: red solid, $\theta_i=1.0$; green dash–dot, $\theta_i=0.5$; blue dotted, $\theta_i=0.0$; thin black guide, power law with exponent 1 ($\propto 1/x$). (f) Out-degree distribution: red solid, $\theta_i=1.0$; green dash–dot, $\theta_i=0.5$; blue dotted, $\theta_i=0.0$; thin black curve, normal distribution (mean and standard deviation estimated from the data). (g)–(i) Infection networks (recruitment edges only). Nodes are colored by entry time (blue = older, yellow = newer): (g) $\theta_i=1.0$; (h) $\theta_i=0.5$; (i) $\theta_i=0.0$.}
    \label{fig_unweightnet_tel}
\end{figure*}

\begin{figure*}[pt]
    \centering
       \begin{overpic}[width=5.4cm]{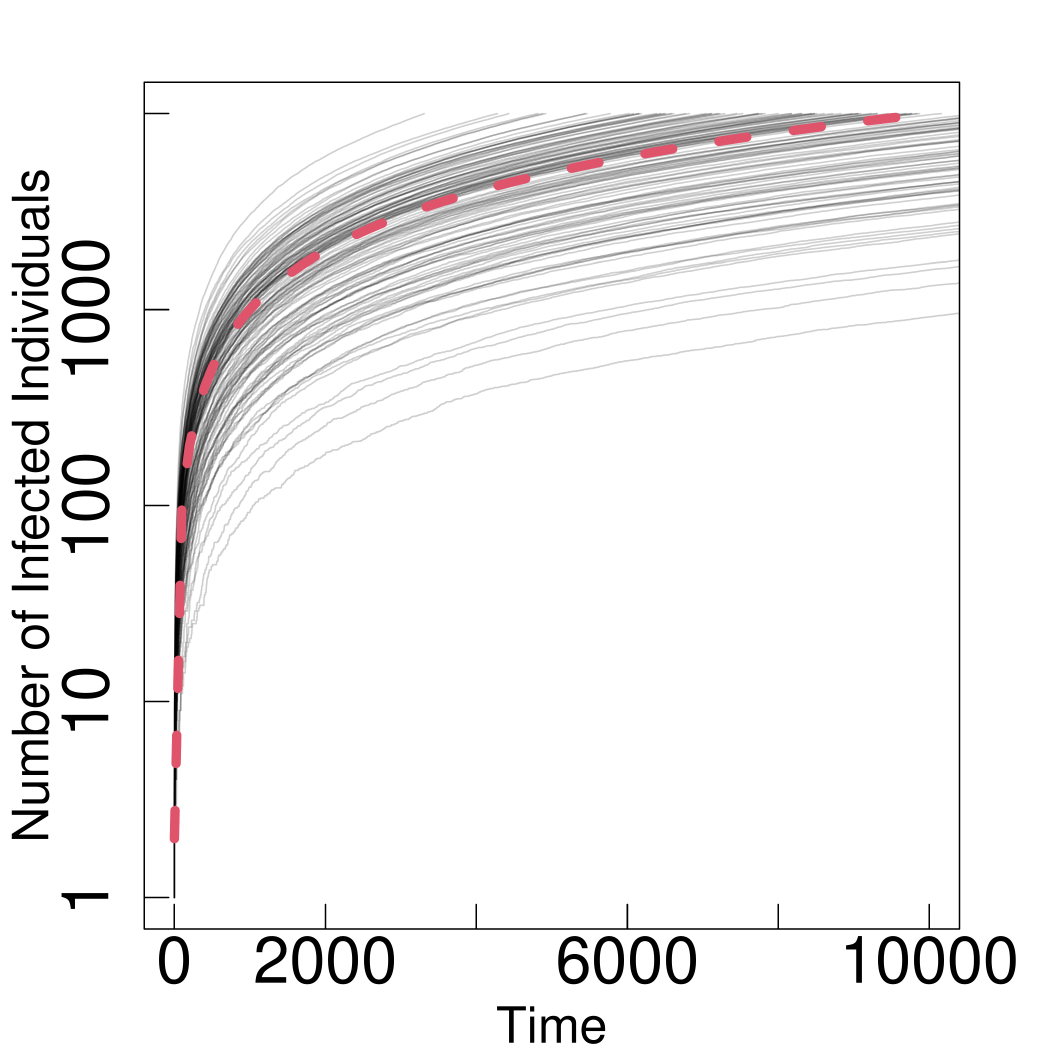}
        \put(35,30){\color{black}\Large\bfseries (a)}
    \end{overpic}
    \begin{overpic}[width=5.4cm]{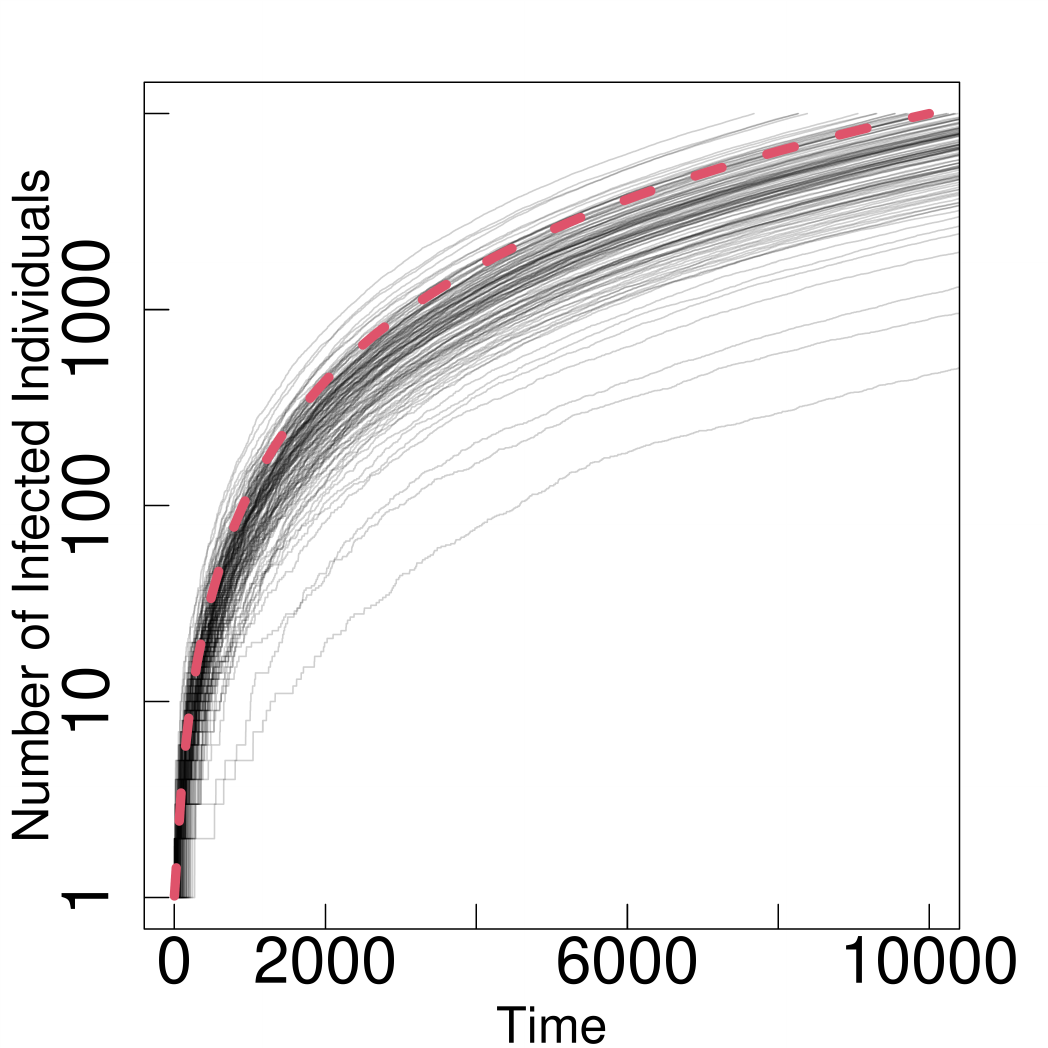}
        \put(30,120){\color{black}\Large\bfseries (b)}
    \end{overpic}
    \begin{overpic}[width=5.4cm]{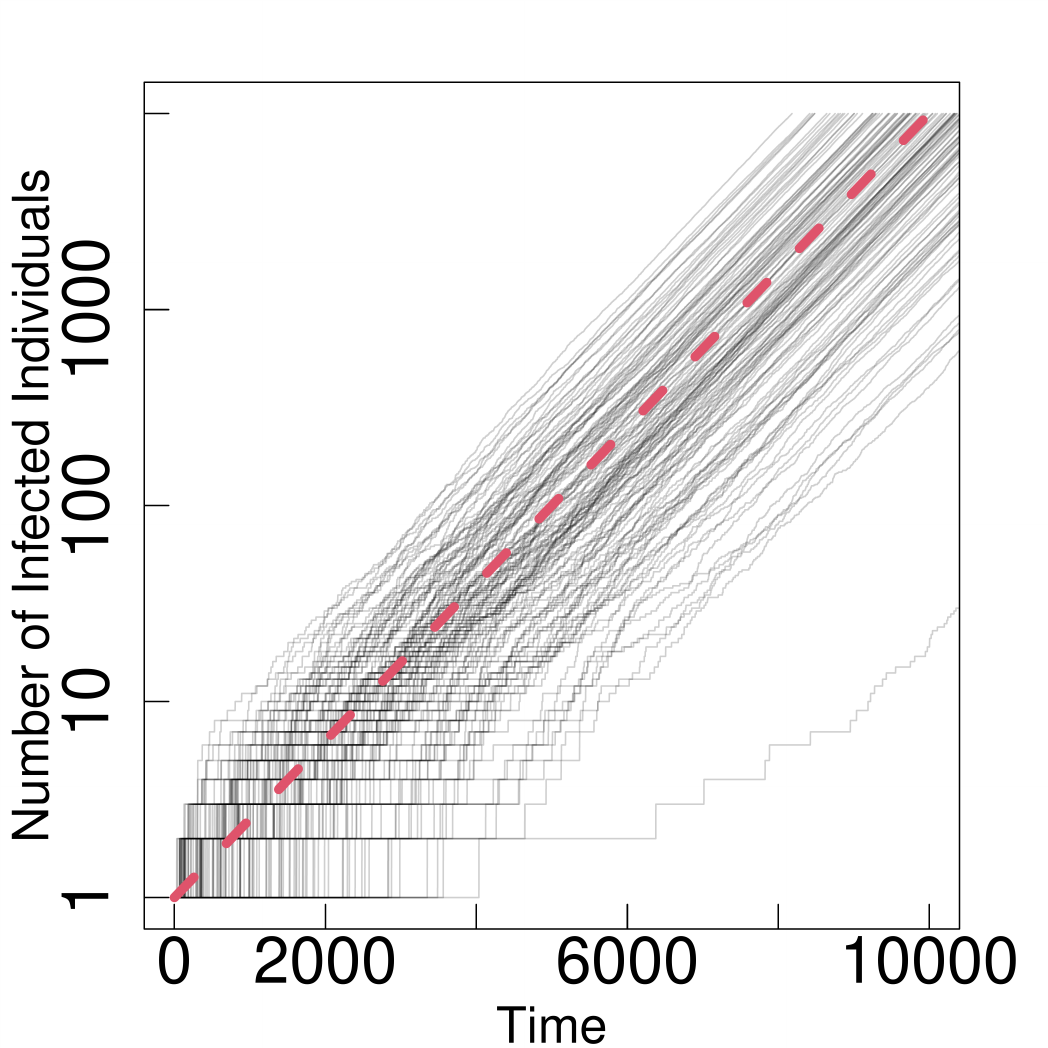}
        \put(30,120){\color{black}\Large\bfseries (c)}
    \end{overpic}
  
    \begin{overpic}[width=5.4cm]{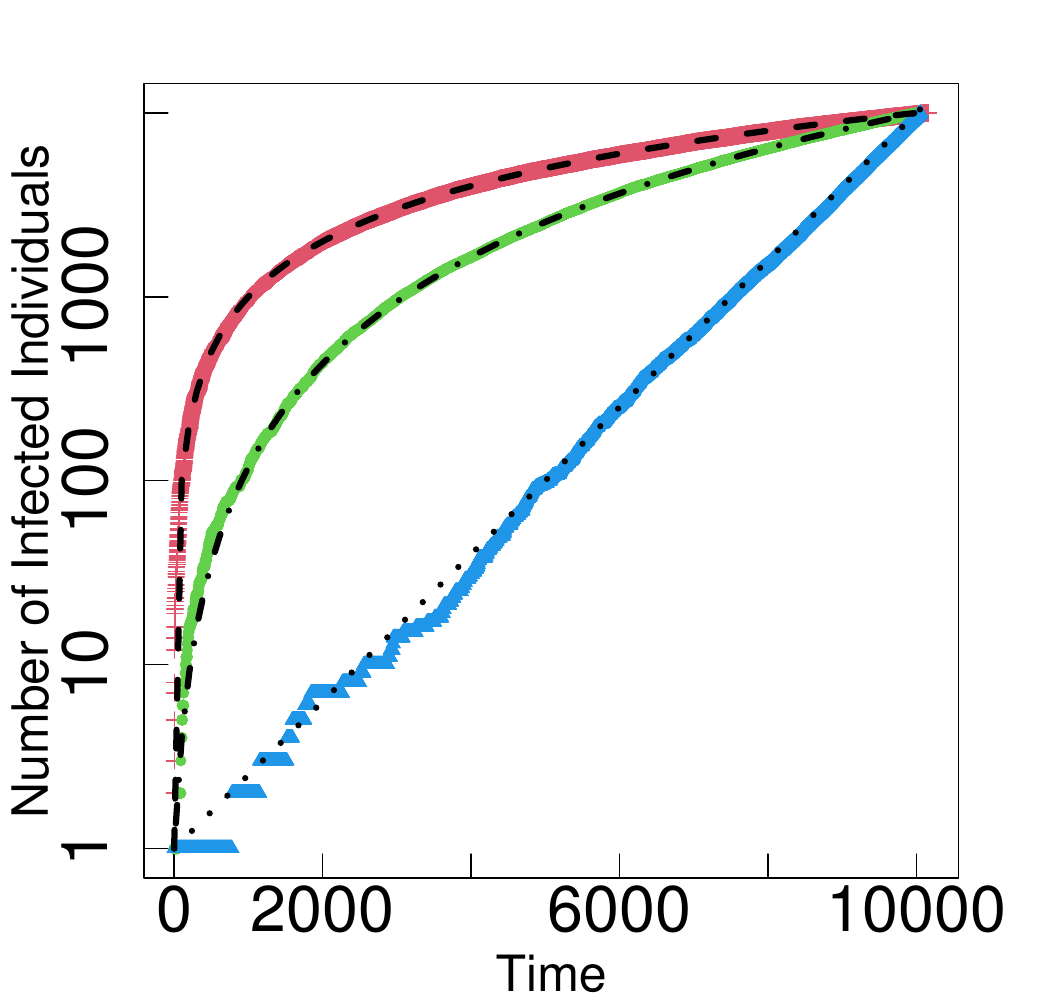}
        \put(30,120){\color{black}\Large\bfseries (d)}
    \end{overpic}
     \begin{overpic}[width=5.4cm]{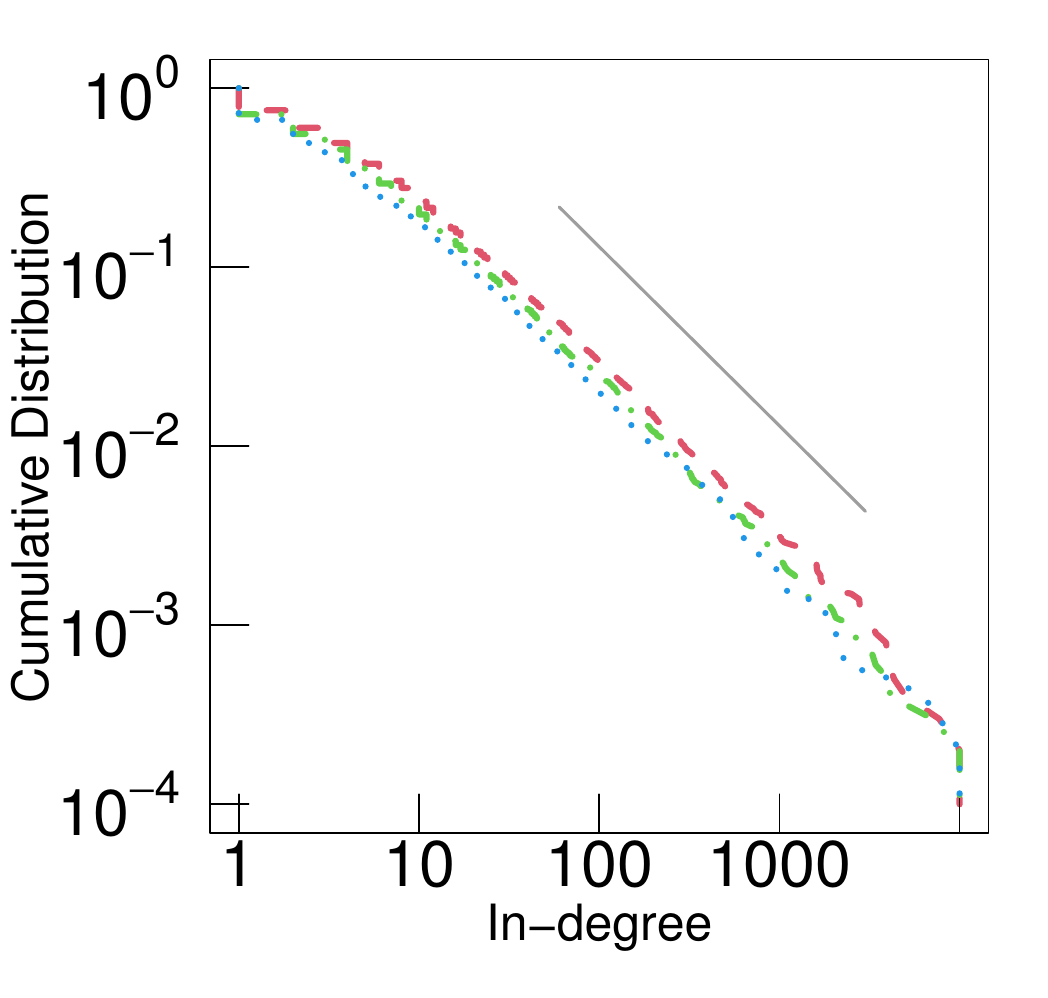}
        \put(40,40){\color{black}\Large\bfseries (e)}
    \end{overpic}
     \begin{overpic}[width=5.4cm]{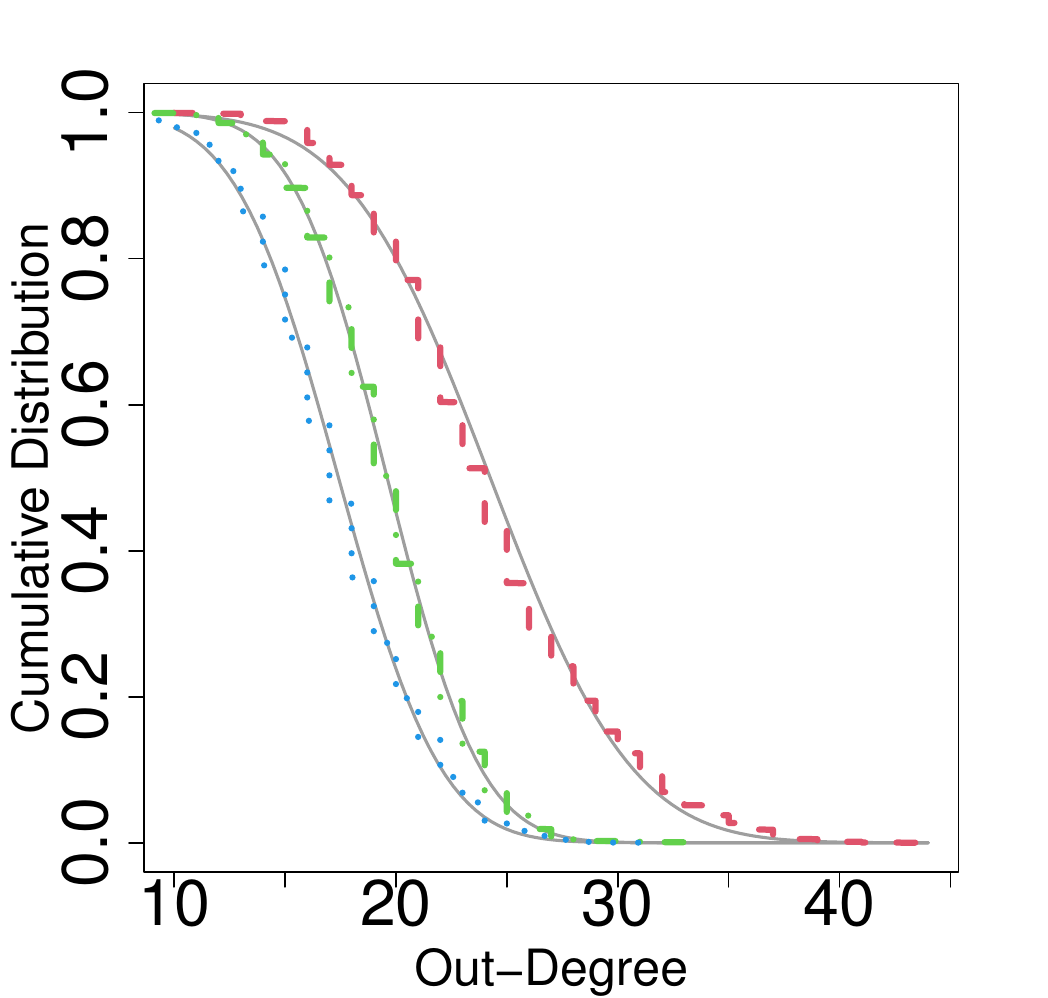}
        \put(33,30){\color{black}\Large\bfseries (f)}
    \end{overpic}
     \begin{overpic}[width=5.4cm]{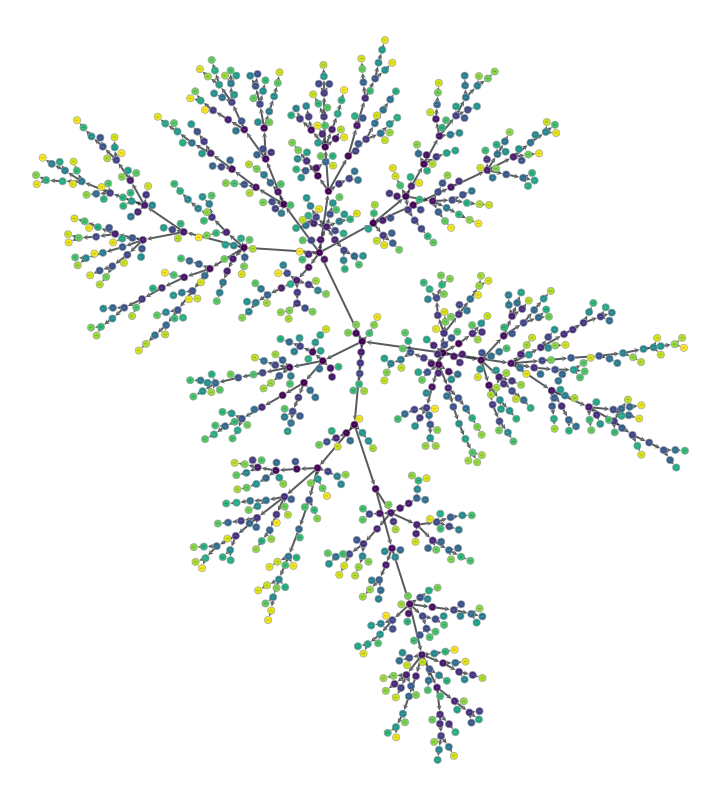}
        \put(30,30){\color{black}\Large\bfseries (g)}
    \end{overpic}
   \begin{overpic}[width=5.4cm]{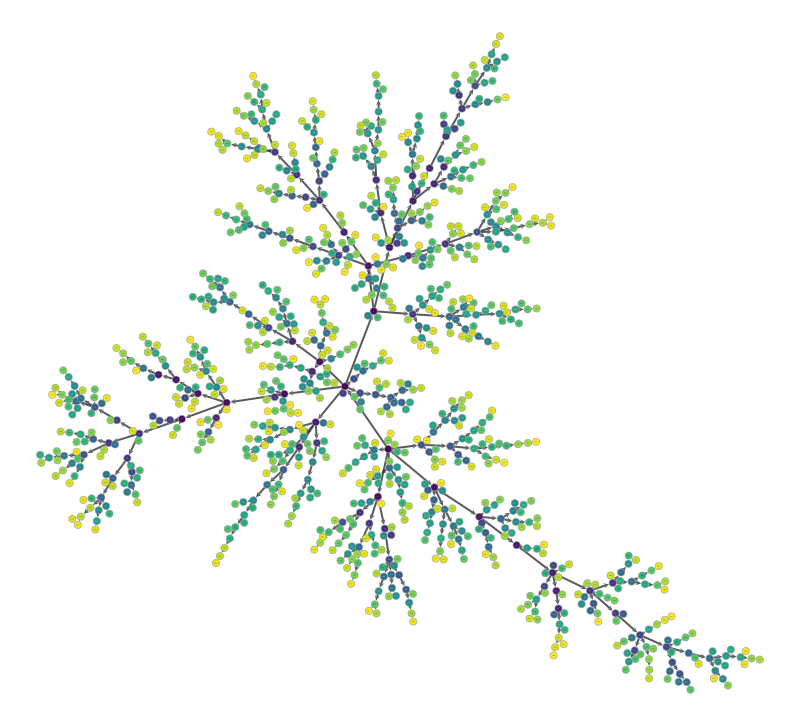}
        \put(30,130){\color{black}\Large\bfseries (h)}
    \end{overpic}
    \begin{overpic}[width=5.4cm]{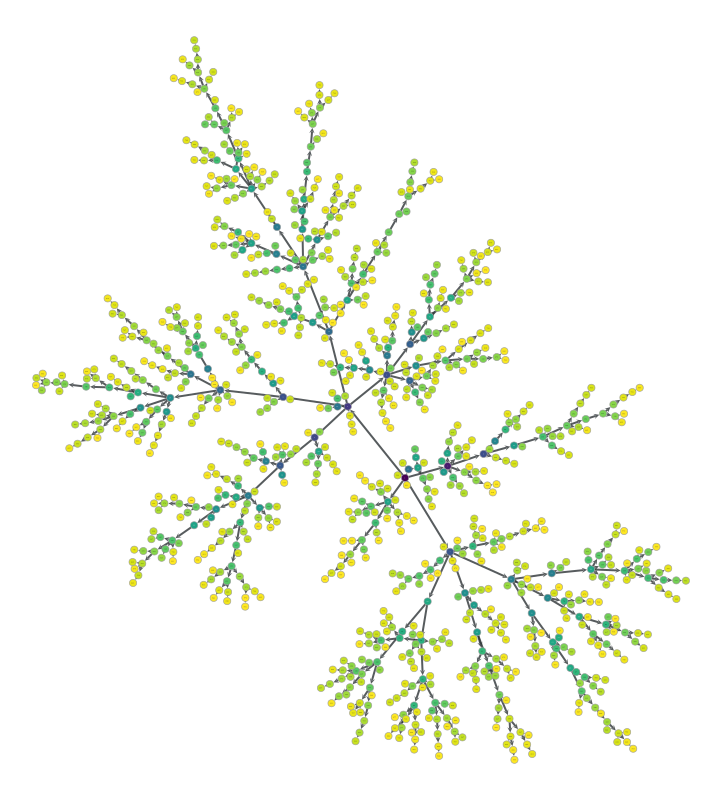}
        \put(20,120){\color{black}\Large\bfseries (i)}
    \end{overpic}
   \caption{Numerical results for the infection-type model (Section~\ref{app_sec_weighted}). Simulation setting: weighted ($\beta_i=1.0$); interaction is chosen at random. (a)–(c) Example growth curves: black solid lines, 128 simulation paths; red dashed line, theoretical curve (Eq.~\ref{app_eq_ans_eq}); $K=1$. (a) $\theta_i=1.0$ ($\alpha_i=0.0$), $J_i=1.0$; (b) $\theta_i=0.5$ ($\alpha_i=0.5$), $J_i=0.020$; (c) $\theta_i=0.0$ ($\alpha_i=1.0$), $J_i=9.2\times10^{-4}$. (d) A simulation path close to the theory: red crosses, $\theta_i=1.0$; green circles, $\theta_i=0.5$; blue triangles, $\theta_i=0.0$; black line, theory (Eq.~\ref{app_eq_ans_eq}). Panels (e)–(f) indicate statistics for this path. (e) In-degree distribution: red solid, $\theta_i=1.0$; green dash–dot, $\theta_i=0.5$; blue dotted, $\theta_i=0.0$; thin black guide, power law with exponent 1 ($\propto 1/x$). (f) Out-degree distribution: red solid, $\theta_i=1.0$; green dash–dot, $\theta_i=0.5$; blue dotted, $\theta_i=0.0$; thin black curve, normal distribution (mean and standard deviation estimated from the data). (g)–(i) Infection networks (recruitment edges only). Nodes are colored by entry time (blue = older, yellow = newer): (g) $\theta_i=1.0$; (h) $\theta_i=0.5$; (i) $\theta_i=0.0$.}
    \label{fig_weightnet_ran}
\end{figure*}

We now examine the weighted network ($\beta_i=1$). We first compare the random-selection case (Fig.~\ref{fig_weightnet_ran}) with the unweighted case $\beta_i=0$ discussed in Section~\ref{app_sec_unweighted} (Fig.~\ref{fig_unweightnet_ran}). The main differences appear in the growth curves (a)-(c) and the in-degree distribution (e).

In the growth curves (a)-(c), the theoretical curve (red) runs through the middle of the sample ensemble (black). This sharply contrasts with the unweighted case (Fig.~\ref{fig_unweightnet_ran}(a)) where, in the linear case $\alpha_i=0.0$ (i.e., $\theta_i=1.0$), the theoretical curve formed a lower bound. The agreement here is due to the ``deepening'' action ($\beta_i=1$), which restores the premise of the theory. In the unweighted model, steps without an available interaction target could lead to ``no action.'' In the present weighted model, such steps trigger deepening instead. This ensures the reference-theory premise that, conditional on acting, $g_i^{(l)}(t)$ increases by exactly one per step for each $l$, removing the simulation-theory gap.

Turning to the in-degree distribution (e), we find that a power law with exponent about 1 emerges in the cumulative distribution, essentially independent of $\theta_i$ (and thus of $\alpha_i$). We discuss the mechanism in Section~\ref{app_sec_indegree}.

\begin{figure*}[tp]
    \centering
       \begin{overpic}[width=5.4cm]{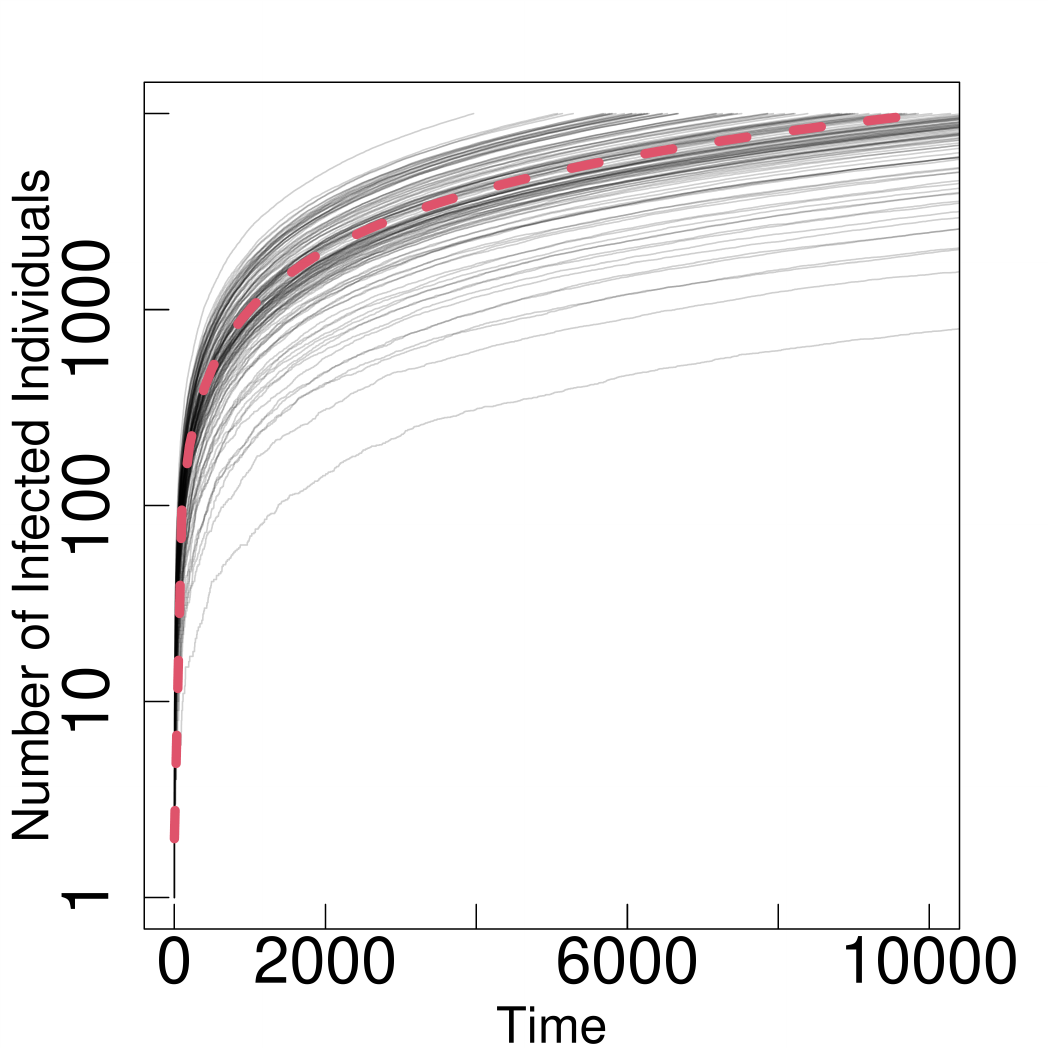}
        \put(35,30){\color{black}\Large\bfseries (a)}
    \end{overpic}
    \begin{overpic}[width=5.4cm]{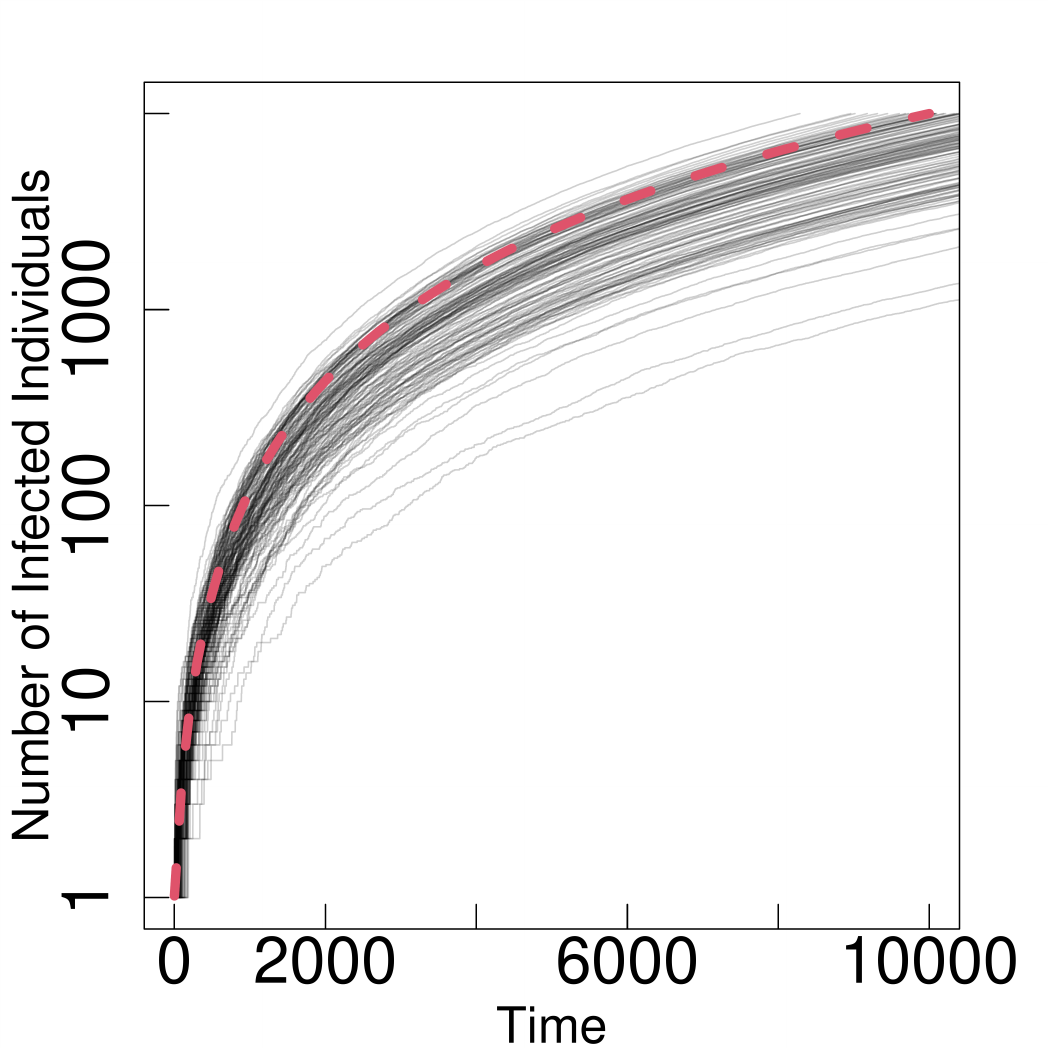}
        \put(30,120){\color{black}\Large\bfseries (b)}
    \end{overpic}
    \begin{overpic}[width=5.4cm]{net_beta00_gamma10_ran_lines_small.pdf}
        \put(30,120){\color{black}\Large\bfseries (c)}
    \end{overpic}
      \begin{overpic}[width=5.4cm]{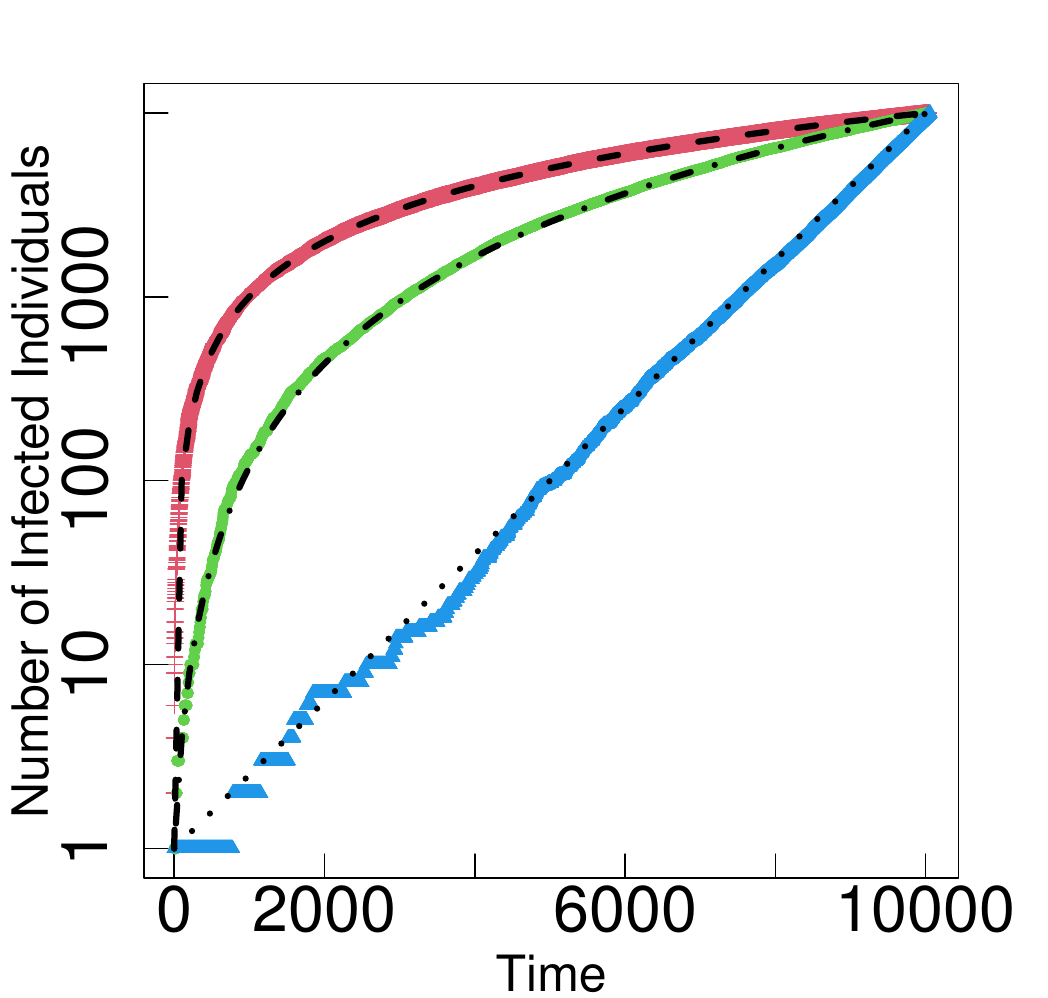}
        \put(30,120){\color{black}\Large\bfseries (d)}
    \end{overpic}
     \begin{overpic}[width=5.4cm]{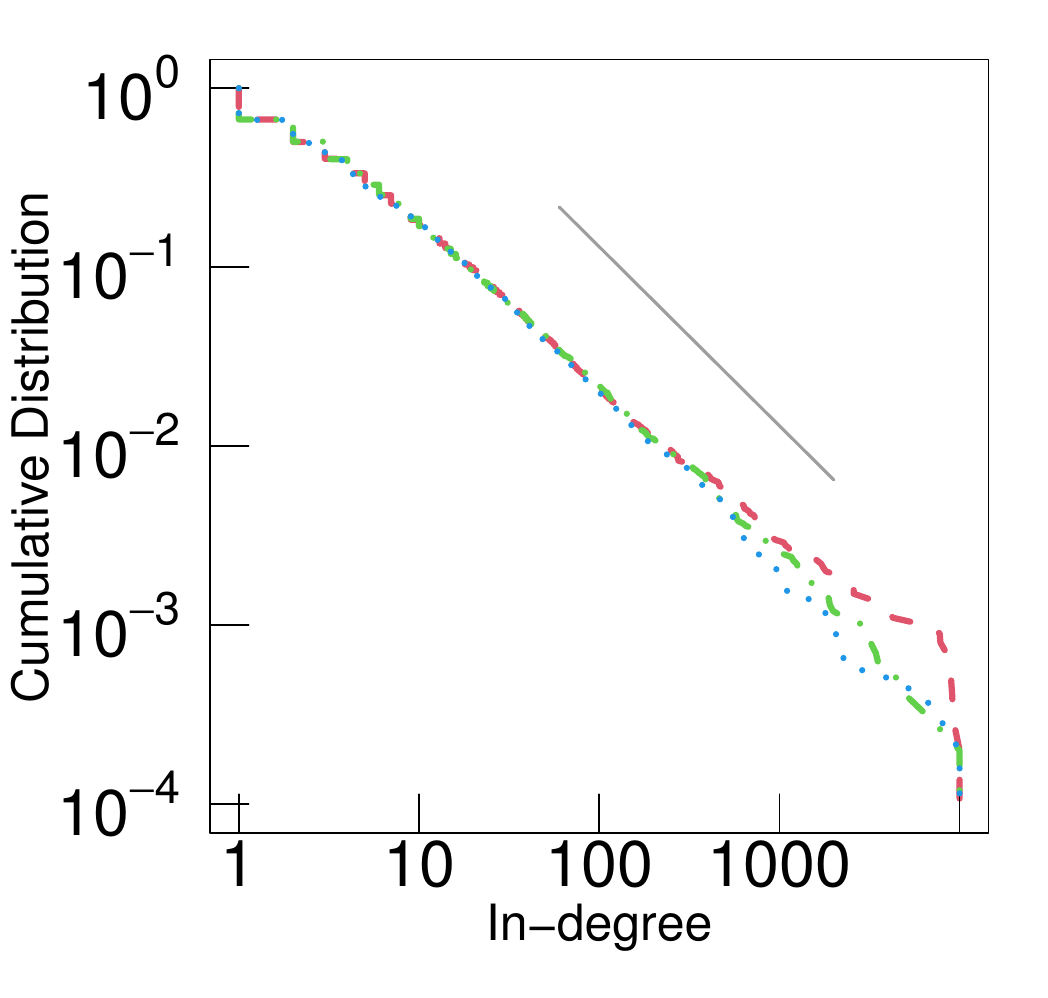}
        \put(120,120){\color{black}\Large\bfseries (e)}
    \end{overpic}
     \begin{overpic}[width=5.4cm]{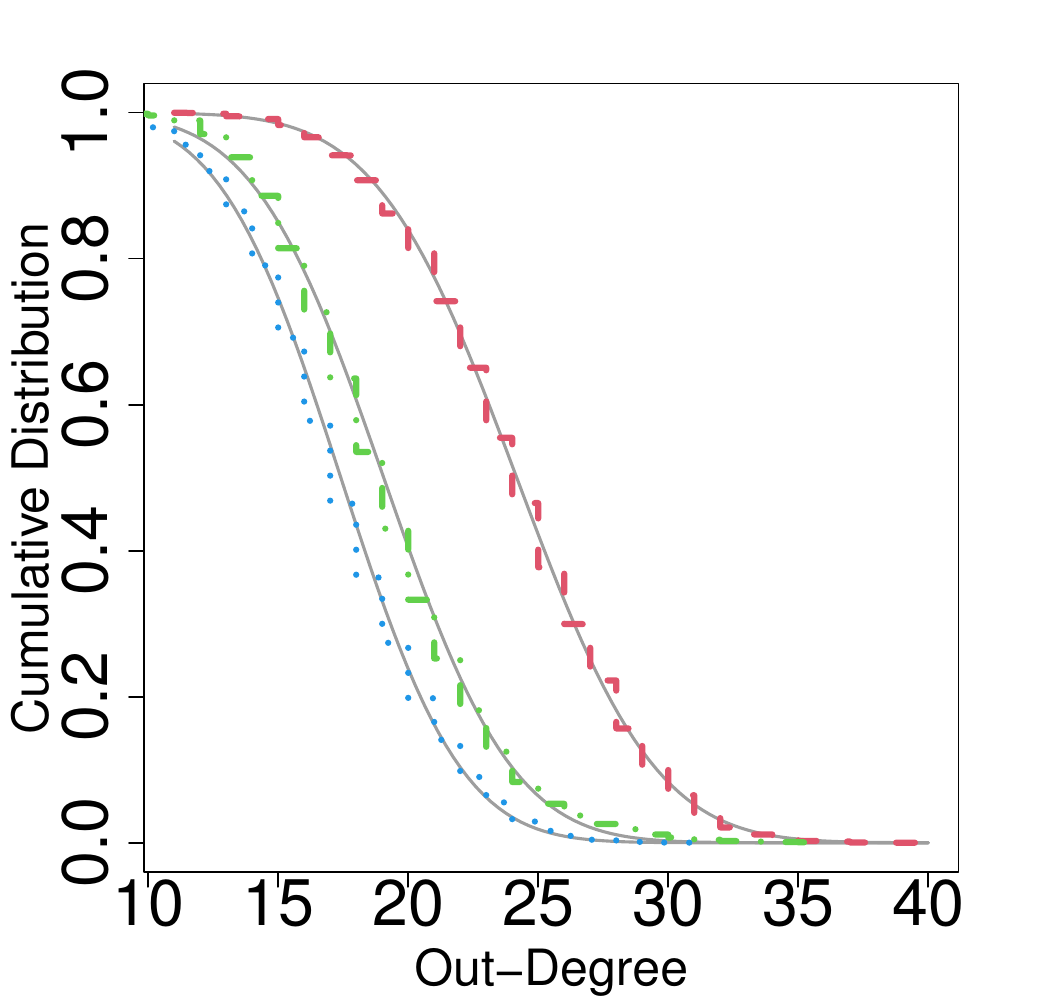}
        \put(120,120){\color{black}\Large\bfseries (f)}
    \end{overpic}
    \begin{overpic}[width=5.4cm]{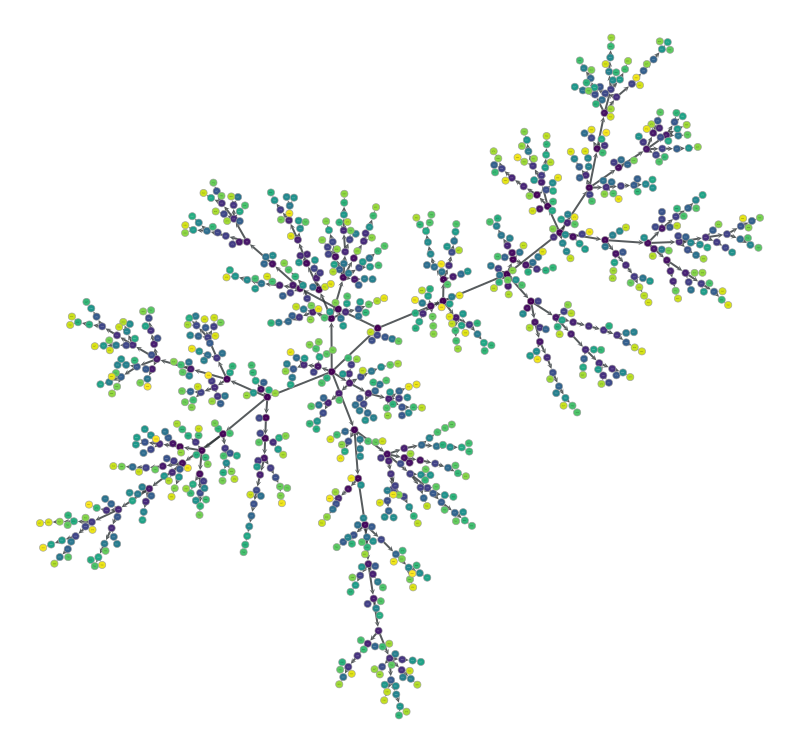}
        \put(30,125){\color{black}\Large\bfseries (g)}
    \end{overpic}
   \begin{overpic}[width=5.4cm]{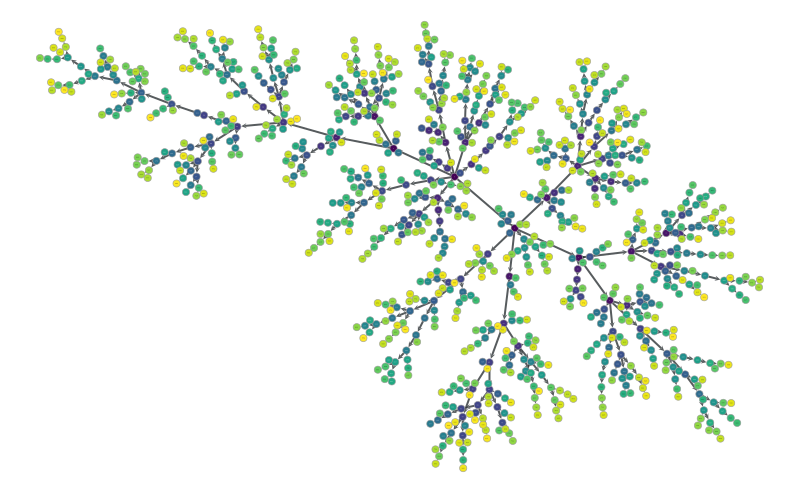}
        \put(10,120){\color{black}\Large\bfseries (h)}
    \end{overpic}
    \begin{overpic}[width=5.4cm]{infection_network_beta00_gamma10_ran.png}
        \put(20,120){\color{black}\Large\bfseries (i)}
    \end{overpic}
   \caption{Numerical results for the infection-type model (Section~\ref{app_sec_weighted}). Simulation setting: weighted ($\beta_i=1.0$); interaction uses the path-weight selection rule (see Section \ref{app_sec_infect_rule}). The remaining details are the same as in Fig.~\ref{fig_unweightnet_ran}. (a)–(c) Example growth curves: black solid lines, 128 simulation paths; red dashed line, theoretical curve (Eq.~\ref{app_eq_ans_eq}); $K=1$. (a) $\theta_i=1.0$ ($\alpha_i=0.0$), $J_i=1.0$; (b) $\theta_i=0.5$ ($\alpha_i=0.5$), $J_i=0.020$; (c) $\theta_i=0.0$ ($\alpha_i=1.0$), $J_i=9.2\times10^{-4}$. (d) A simulation path close to the theory: red crosses, $\theta_i=1.0$; green circles, $\theta_i=0.5$; blue triangles, $\theta_i=0.0$; black line, theory (Eq.~\ref{app_eq_ans_eq}). Panels (e)–(f) indicate statistics for this path. (e) In-degree distribution: red solid, $\theta_i=1.0$; green dash–dot, $\theta_i=0.5$; blue dotted, $\theta_i=0.0$; thin black guide, power law with exponent 1 ($\propto 1/x$). (f) Out-degree distribution: red solid, $\theta_i=1.0$; green dash–dot, $\theta_i=0.5$; blue dotted, $\theta_i=0.0$; thin black curve, normal distribution (mean and standard deviation estimated from the data). (g)–(i) Infection networks (recruitment edges only). Nodes are colored by entry time (blue = older, yellow = newer): (g) $\theta_i=1.0$; (h) $\theta_i=0.5$; (i) $\theta_i=0.0$.}
    \label{fig_weightnet_mei}
\end{figure*}

\begin{figure*}[tp]
    \centering
       \begin{overpic}[width=5.4cm]{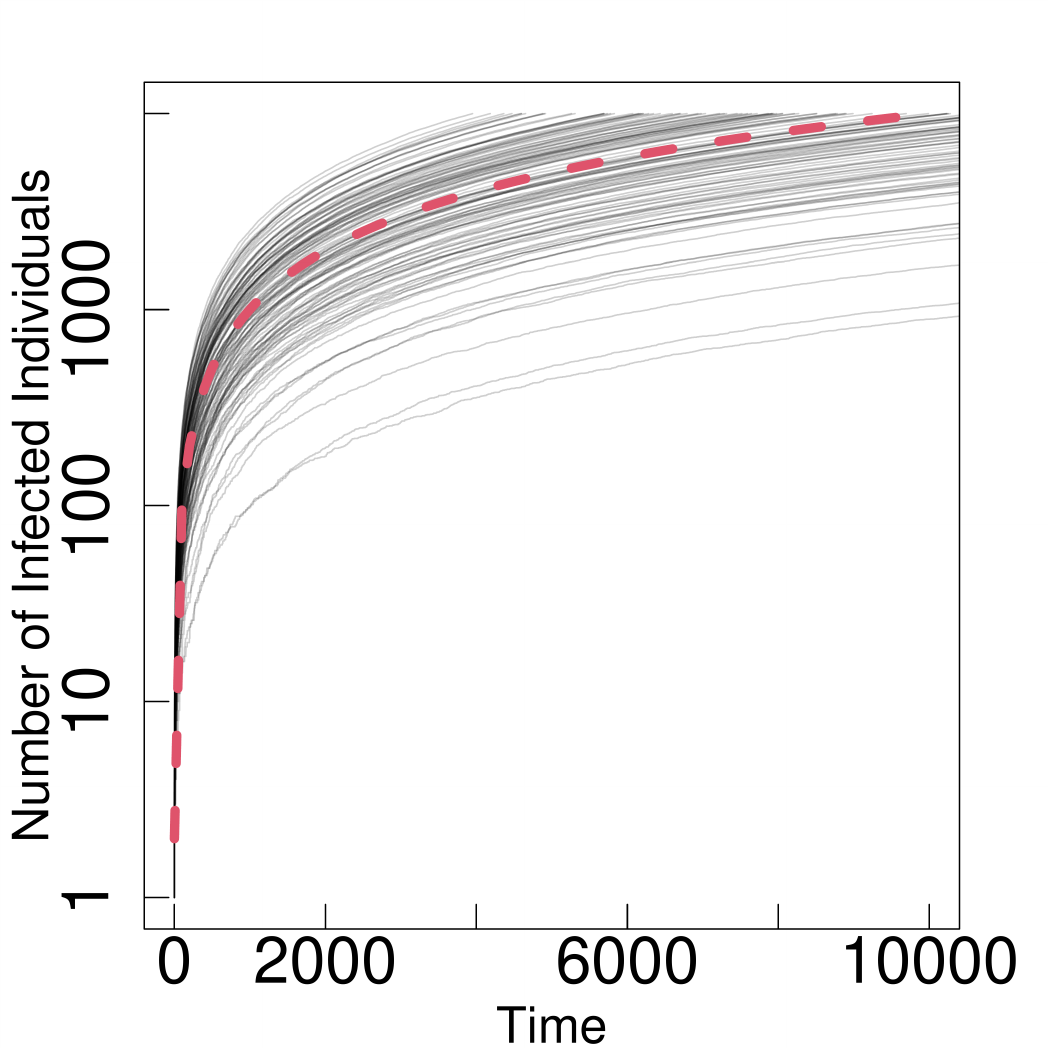}
        \put(35,30){\color{black}\Large\bfseries (a)}
    \end{overpic}
    \begin{overpic}[width=5.4cm]{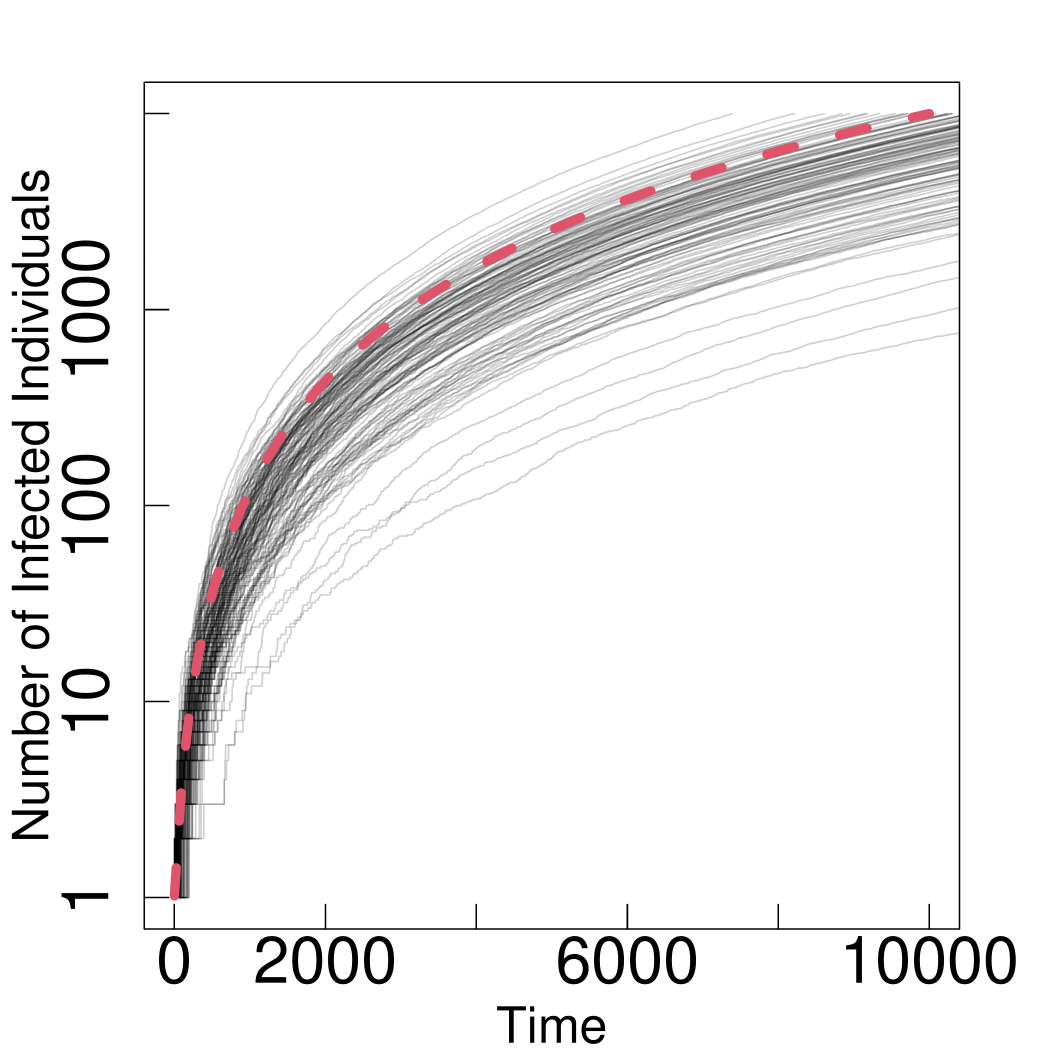}
        \put(30,120){\color{black}\Large\bfseries (b)}
    \end{overpic}
    \begin{overpic}[width=5.4cm]{net_beta00_gamma10_ran_lines_small.pdf}
        \put(30,120){\color{black}\Large\bfseries (c)}
    \end{overpic}
     \begin{overpic}[width=5.4cm]{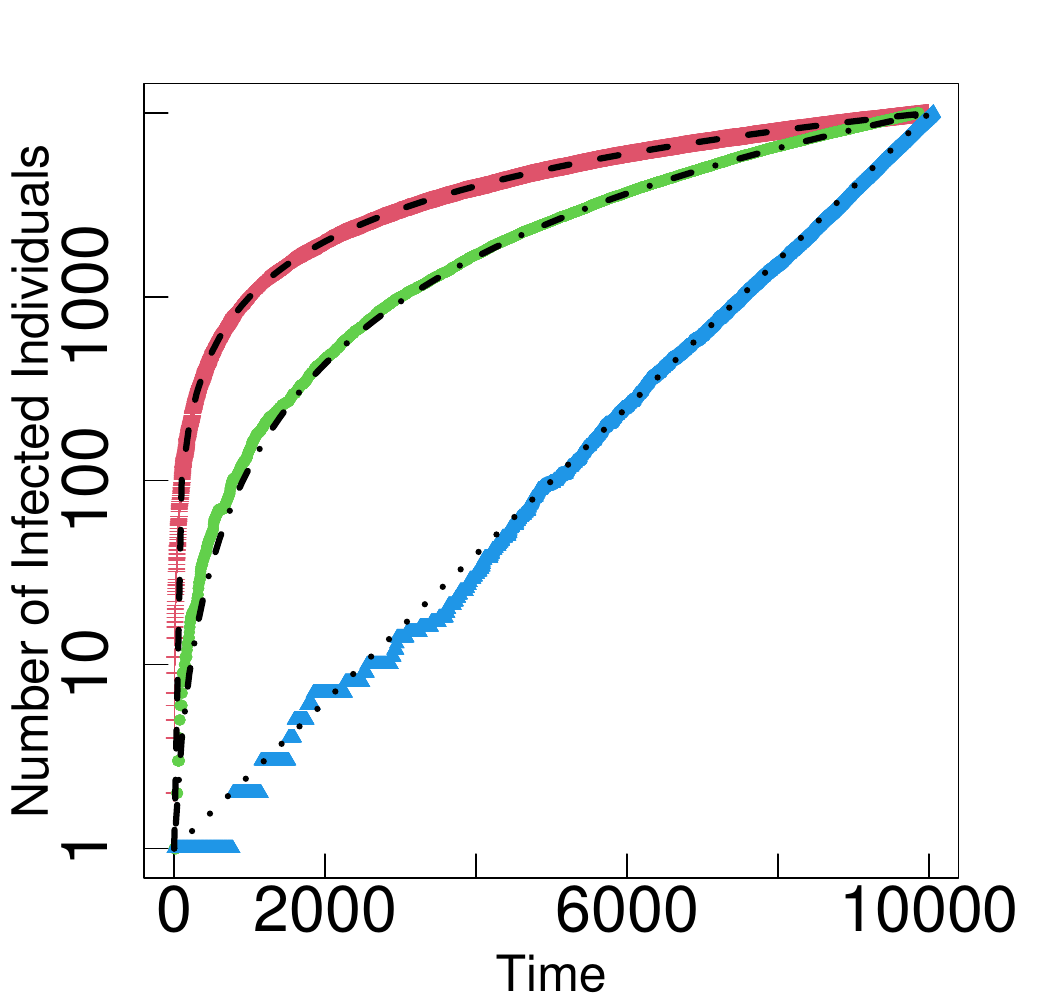}
        \put(30,120){\color{black}\Large\bfseries (d)}
    \end{overpic}
     \begin{overpic}[width=5.4cm]{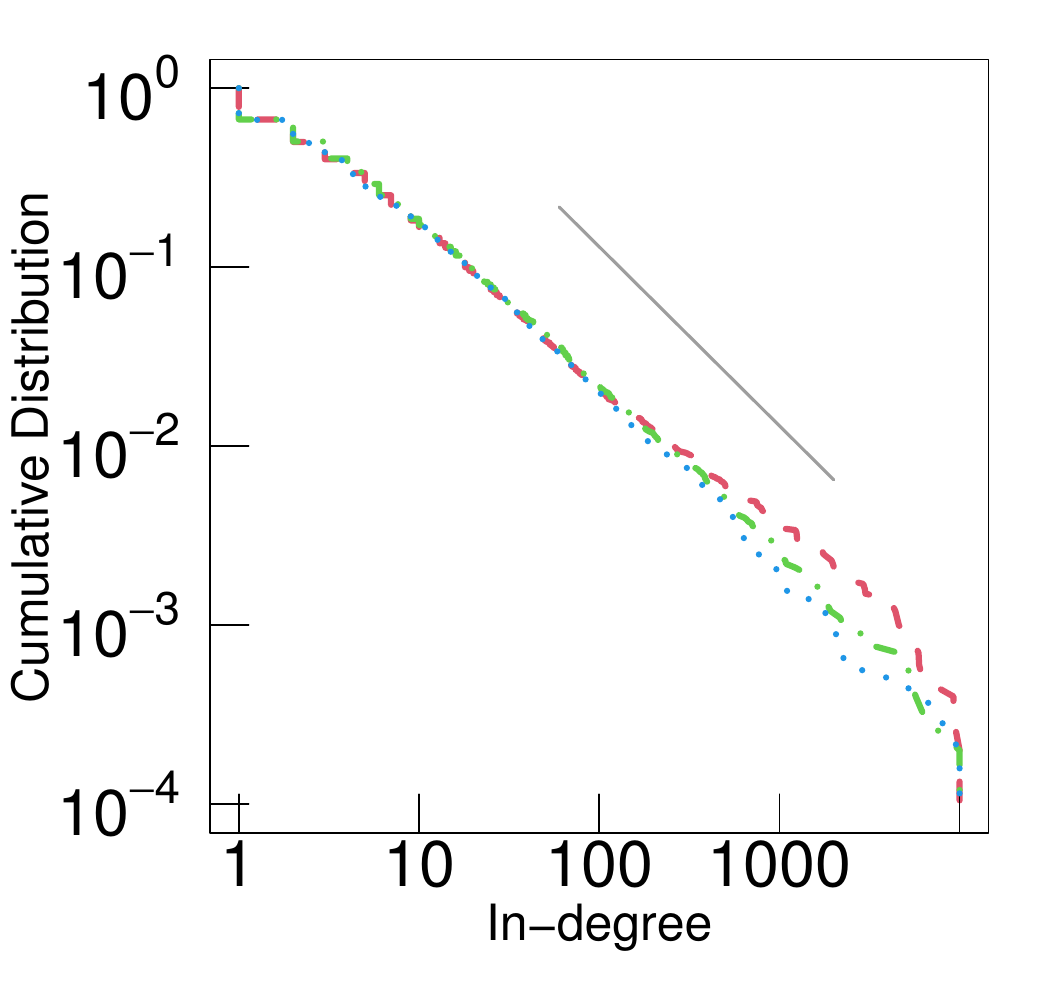}
        \put(120,120){\color{black}\Large\bfseries (e)}
    \end{overpic}
     \begin{overpic}[width=5.4cm]{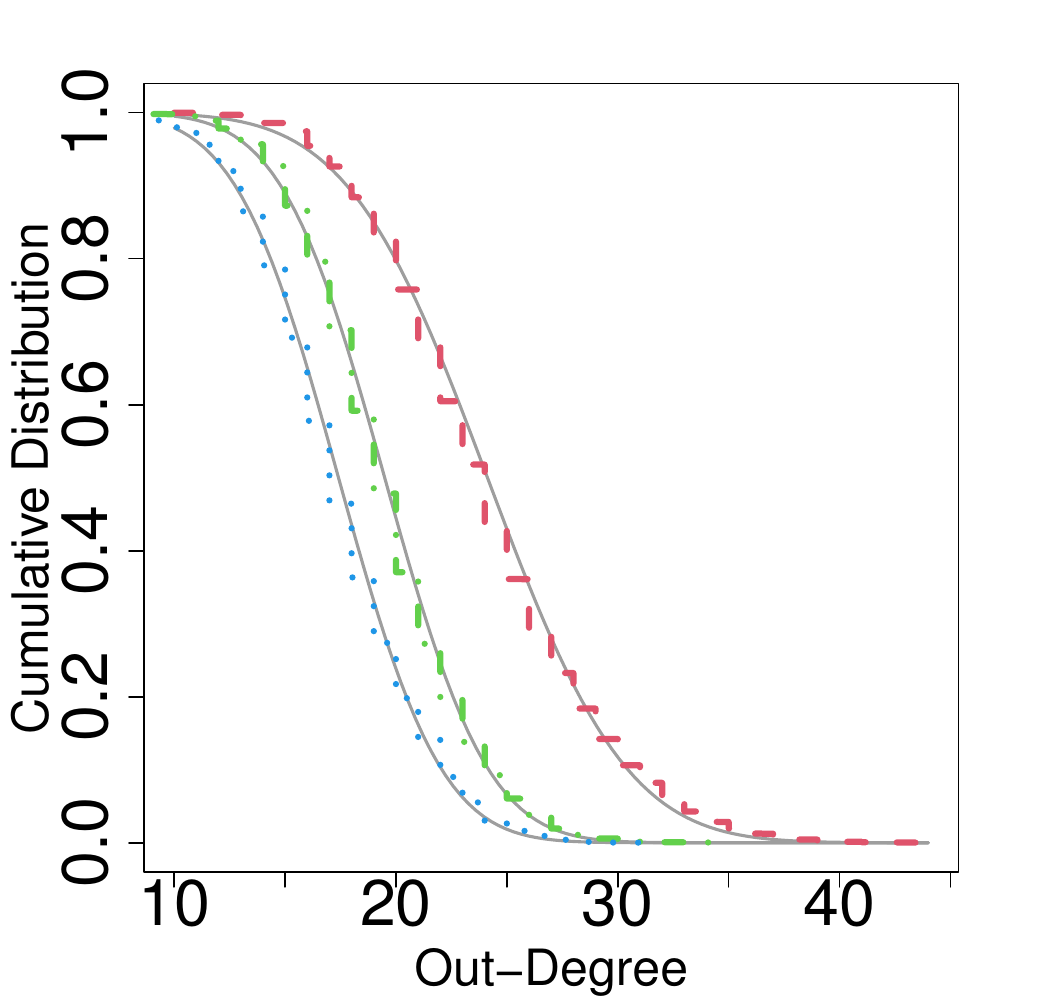}
        \put(120,120){\color{black}\Large\bfseries (f)}
    \end{overpic}
    \begin{overpic}[width=5.4cm]{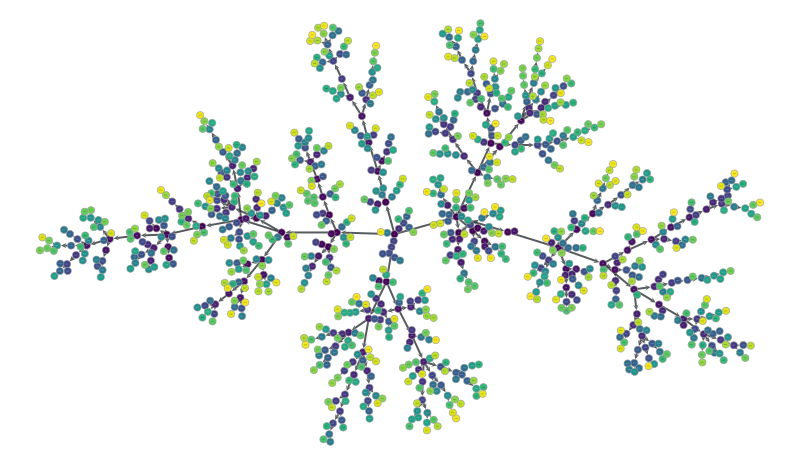}
        \put(30,120){\color{black}\Large\bfseries (g)}
    \end{overpic}
   \begin{overpic}[width=5.4cm]{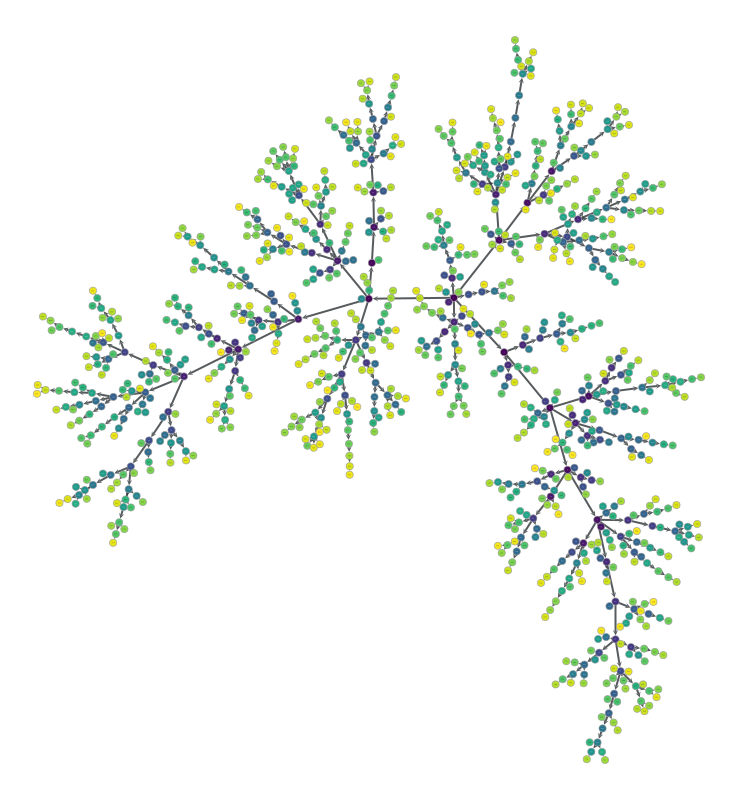}
        \put(20,130){\color{black}\Large\bfseries (h)}
    \end{overpic}
    \begin{overpic}[width=5.4cm]{infection_network_beta00_gamma10_ran.png}
        \put(20,120){\color{black}\Large\bfseries (i)}
    \end{overpic}
   \caption{Numerical results for the infection-type model (Section~\ref{app_sec_weighted}). Simulation setting: weighted ($\beta_i=1.0$); interaction uses the two-step selection rule (see Section \ref{app_sec_infect_rule}). The remaining details are the same as in Fig.~\ref{fig_unweightnet_ran}. (a)–(c) Example growth curves: black solid lines, 128 simulation paths; red dashed line, theoretical curve (Eq.~\ref{app_eq_ans_eq}); $K=1$. (a) $\theta_i=1.0$ ($\alpha_i=0.0$), $J_i=1.0$; (b) $\theta_i=0.5$ ($\alpha_i=0.5$), $J_i=0.020$; (c) $\theta_i=0.0$ ($\alpha_i=1.0$), $J_i=9.2\times10^{-4}$. (d) A simulation path close to the theory: red crosses, $\theta_i=1.0$; green circles, $\theta_i=0.5$; blue triangles, $\theta_i=0.0$; black line, theory (Eq.~\ref{app_eq_ans_eq}). Panels (e)–(f) indicate statistics for this path. (e) In-degree distribution: red solid, $\theta_i=1.0$; green dash–dot, $\theta_i=0.5$; blue dotted, $\theta_i=0.0$; thin black guide, power law with exponent 1 ($\propto 1/x$). (f) Out-degree distribution: red solid, $\theta_i=1.0$; green dash–dot, $\theta_i=0.5$; blue dotted, $\theta_i=0.0$; thin black curve, normal distribution (mean and standard deviation estimated from the data). (g)–(i) Infection networks (recruitment edges only). Nodes are colored by entry time (blue = older, yellow = newer): (g) $\theta_i=1.0$; (h) $\theta_i=0.5$; (i) $\theta_i=0.0$.}
    \label{fig_weightnet_tel}
\end{figure*}

Replacing the interaction rule with path-weight selection (Fig.~\ref{fig_weightnet_mei}) or two-step selection (Fig.~\ref{fig_weightnet_tel}) yields trends broadly similar to random selection. These results suggest that, in the weighted model ($\beta_i>0$), a power law with exponent about 1 appears robustly, largely independent of the detailed interaction rule. The same qualitative tendency also holds, in broad terms, for $0<\theta_i<1$ and $0<\beta_i<1$.
   
\subsection{Understanding the In-Degree Distribution}
\label{app_sec_indegree}
This section explains why the in-degree distribution empirically shows an exponent of 1 in two settings: (i) exponential growth, i.e., $\alpha_i=1$ (equivalently $\theta_i=0$), and (ii) the weighted model with $0<\beta_i\le 1$. The mechanism is essentially the same as in the vertex copy model that yields a power law with exponent 1 \cite{kumar2000stochastic}.

\subsubsection{Case of Exponential-Growth ($\theta_i=0$; $\alpha_i=1$)}
\label{app_sec_exp_theory}
We first consider the exponential-growth case, i.e., $\theta_i=0$ ($\alpha_i=1$). Under this condition, network growth is driven solely by \emph{recruitment}.

When the network has $I$ nodes, consider the probability that an existing node $v$ with in-degree $k_{in}$ gains a new incoming link from the newly added node (i.e., its in-degree increases by one). This probability can be approximated as the sum of the following two events:
\begin{enumerate}
    \item \textbf{Chosen as the ``parent'' (recruiter):}
    The new node chooses one parent uniformly at random from all $I$ existing nodes. Hence $v$ is chosen with probability $1/I$.
    \item \textbf{Receiving a copied link (inheritance):}
    Some node $u\neq v$ is chosen as the parent with probability $1/I$, and the new node copies the parent's outgoing links. Since $v$ has $k_{in}$ incoming links, there are $k_{in}$ candidates $u$ with an edge $u\to v$. If any such $u$ is chosen, the edge $u\to v$ is copied and $v$ gains an incoming link. The total probability is thus $\sum_{u:\,u\to v} (1/I) \approx k_{in}/I$.
\end{enumerate}
Therefore, the probability that $v$ gains one in-degree is approximated by
\begin{equation}
P(\text{gain link}) \approx \frac{1}{I} + \frac{k_{in}}{I} = \frac{k_{in}+1}{I}.
\end{equation}

Letting the increase in the total number of nodes be $dI=1$ when one node is added ($I\to I+1$) and taking a continuous approximation, we obtain
\begin{equation}
\frac{dk_{in}}{dI} \approx \frac{k_{in}+1}{I}.
\label{app_eq_diffeq}
\end{equation}
Suppose node $v$ entered the network when the size was $I_0$. At entry, $v$ has only the link from its parent, so the initial condition is $k_{in}(I_0)=1$. Solving Eq.~\ref{app_eq_diffeq} with this condition yields
\begin{equation}
k_{in}(I) = \frac{2I}{I_0} - 1.
\label{app_eq_kanyuu}
\end{equation}

We now derive the in-degree distribution. In our model, nodes (infected individuals) enter at heterogeneous times in real time $t$, but for analysis we treat each node addition as one event and use the total node count $I$ as the independent variable—an event-based clock.

If we take sufficiently fine time steps and assume that node entries do not occur simultaneously in real time $t$, then nodes are added one by one and $I$ increases as $1,2,3,\dots$. The ``entry index'' $I_0$ of a node refers to the network size at the moment that node entered (its rank in the arrival order), not to real time $t$. When we sample one node uniformly at the final time (total nodes $I$), its $I_0$ is uniformly distributed on $\{1,2,\dots,I\}$.

Hence, the complementary cumulative probability that the sampled node's in-degree $k_{in}(I)$ is at least $K$ is
\begin{equation}
\begin{aligned}
\Pr\{k_{in}(I) \ge K\} &= \Pr\!\left\{\frac{2I}{I_0} - 1 \ge K \right\} \\
&= \Pr\!\left\{\frac{2I}{K+1} \ge I_0 \right\}
= \Pr\!\left\{ I_0 \le \frac{2I}{K+1} \right\}.
\end{aligned}
\end{equation}
Since $I_0$ is uniform on $\{1,\dots,I\}$, we approximate $\Pr\{I_0 \le x\}\approx x/I$, giving
\begin{equation}
\Pr\{k_{in}(I) \ge K\} \approx \frac{1}{I}\left(\frac{2I}{K+1}\right)=\frac{2}{K+1}.
\end{equation}
Thus, the cumulative in-degree distribution is proportional to $K^{-1}$, i.e., a \textbf{power law with exponent 1}.

This derivation is essentially equivalent to the vertex copy model of Kumar et al.\ \cite{kumar2000stochastic}. The difference is that their differential equation uses time $t$ as the independent variable, whereas our model uses the number of infected (nodes) $I(t)$.

\subsubsection{Case of Interaction and Deepening Are Present ($0<\theta_i\le 1$ and $0<\beta_i\le 1$)}
We next consider the weighted-network setting where, in addition to recruitment, both \emph{interaction} ($\theta_i>0$) and \emph{deepening} ($\beta_i>0$) are present.

Even in this case, once the network has grown sufficiently large ($t\gg 1$), the in-degree increase due to \emph{recruitment} dominates the increase due to interaction. As a result, the same mechanism as in the ``recruitment-only'' model of Section~\ref{app_sec_exp_theory} effectively governs the dynamics, and we again obtain a power law with exponent 1 in the in-degree distribution.

The dominance of recruitment follows from two effects:
\begin{itemize}
    \item[(i)] \textbf{Frequency balance.} Because \emph{deepening} ($\beta_i>0$) is available, the rates of recruitment and interaction remain comparable (within a constant factor) over time.
    \item[(ii)] \textbf{Per-event growth gap.} Each interaction adds exactly one link, whereas one recruitment adds a number of links (in particular, to in-degree) that increases with the total network size $I$ via the copy effect.
\end{itemize}
We justify these two points below.

\subsubsubsection{Relative Frequencies of Recruitment and Interaction}
Let $p_i(t)$ be the probability that a given action is a recruitment (in this section, for simplicity, we write $p_i(t),q_i(t),g_i(t)$ for typical values at time $t$). The probability that an action is an interaction is $(1-p_i(t))\,q_i(t)$, because recruitment does not occur and interaction is chosen. The mean number of interaction events per one recruitment is therefore
\begin{equation}
Z_i=\frac{(1-p_i(t)) \cdot q_i(t)}{p_i(t)}.
\end{equation}
Using the model definitions of $p_i(t)$ and $q_i(t)$ and writing $g_i(t)$ for the total tie strength (which grows proportionally with $t$), we can rewrite this as
\begin{equation}
Z_i=\frac{\theta_i \cdot g_i(t)}{1+\beta_i\, g_i(t)}.
\end{equation}
Since $g_i(t)$ increases with $t$, for $t\gg 1$ we have $g_i(t)\to\infty$ and thus
\begin{equation}
Z_i \approx \frac{\theta_i}{\beta_i},
\end{equation}
i.e., the ratio converges to a constant. Hence, in the long run, interaction and recruitment occur at comparable rates (up to the constant factor $\theta_i/\beta_i$).

The presence of \textbf{deepening} ($\beta_i>0$) is crucial. Without deepening ($\beta_i\to 0$), we would have $Z_i\approx \theta_i g_i(t)\to\infty$, so interaction would dominate. Deepening not only increases weights but also introduces $g_i(t)$ in the denominator of the choice rule, which balances the frequencies of recruitment and interaction. This is why $0<\beta_i\le 1$ is assumed.

\subsubsubsection{Links Gained per Event}
We next compare the number of links (in-degree) gained per event. For interaction, the increase is always 1.

For recruitment, a new node (child) enters, forms mutual links with its parent, and copies the parent's outgoing links. The total in-degree received by existing nodes is
\begin{enumerate}
    \item one for the parent (child $\to$ parent), and
    \item one for each of the parent's acquaintances (copy links), equal to the parent's out-degree $k_{out}$.
\end{enumerate}
Thus, one recruitment increases the network-wide total in-degree by $1+k_{out}$. Approximating by the mean out-degree $\langle k_{out}(I)\rangle$, the in-degree increase per recruitment is $1+\langle k_{out}(I)\rangle$.

Let $M_{out}(I)=I\,\langle k_{out}(I)\rangle$ be the total out-degree. During the event that increases the network size from $I$ to $I+1$ (one recruitment plus $Z_i$ interactions), $M_{out}$ increases by
\begin{itemize}
    \item recruitment: $1$ (parent $\to$ child) $+$ $1$ (child $\to$ parent) $+$ $\langle k_{out}(I)\rangle$ (child $\to$ parent's acquaintances),
    \item interaction: $Z_i\times 1=Z_i$.
\end{itemize}
Hence
\begin{eqnarray}
&&M_{out}(I+1)\approx M_{out}(I)+2+Z_i+\langle k_{out}(I)\rangle  \nonumber \\
&&=M_{out}(I)+2+Z_i+\frac{M_{out}(I)}{I}.
\end{eqnarray}
Dividing by $I+1$ and writing in terms of $\langle k_{out}\rangle$,
\begin{equation}
\begin{aligned}
\langle k_{out}\rangle(I+1)
&= \frac{M_{out}(I+1)}{I+1}
\approx \frac{(1+1/I)M_{out}(I)+(2+Z_i)}{I+1} \\
&\approx \langle k_{out}\rangle(I) + \frac{2+Z_i}{I+1}.
\end{aligned}
\end{equation}
Solving this difference equation and using $\sum_{I}1/(I+1)\approx \log I$ for $I\gg 1$, we obtain
\begin{equation}
\langle k_{out}(I)\rangle \approx (2+Z_i)\,\log I.
\label{app_eq_k_mean}
\end{equation}
Thus the mean out-degree grows like $\log I$, so the per-recruitment in-degree gain $1+\langle k_{out}(I)\rangle$ also increases over time.

\subsubsubsection{Recruitment-Driven Growth of In-Degree}
For $I\gg 1$, (i) recruitment and interaction occur in a fixed ratio $Z_i$, but (ii) the per-event increase is constant (1) for interaction and of order $\approx (2+Z_i)\log I$ for recruitment. Consequently, the share of new links attributable to recruitment increases over time and tends to 1.

The share of recruitment-generated links in the total link increase is
\begin{equation}
\frac{\text{links from recruitment}}{\text{all new links}}
\approx
\frac{1+\langle k_{\mathrm{out}}(I)\rangle}{1+\langle k_{\mathrm{out}}(I)\rangle+Z_i},
\end{equation}
with $Z_i\simeq \theta_i/\beta_i$ and $\langle k_{\mathrm{out}}(I)\rangle\simeq (2+Z_i)\log I$.

Hence, in weighted networks ($0<\beta_i\le 1$), as the network grows, the contribution from interaction becomes negligible, and the copy effect driven by recruitment dominates. At the macroscopic level, the same mechanism as in the no-interaction case ($\theta_i=0$) operates, and we obtain a power law with exponent 1 for the in-degree distribution. \par

Note that this argument explains the \emph{total} number of links at the network level and implicitly assumes that interaction links are not distributed in an extremely concentrated manner. For example, under ``random interaction'' where interaction is spread roughly evenly across the network, the per-step in-degree contribution from interaction to a given node is $O(Z_i/I)$. In contrast, the average contribution from recruitment is $O(\langle k_{in}\rangle/I)\approx O(\log I/I)$, so for large $I$ the $\log I$ factor makes recruitment dominant (since $Z_i$ is a constant).

However, if interaction is extremely concentrated on a very small number of nodes, deviations from a power law may occur. For instance, if interaction links are focused on a single node, that node receives $O(Z_i)$ (i.e., $O(1)$) from interaction. Meanwhile, for a particular node, recruitment can increase its in-degree by at most $O(1)$ per addition event (being chosen as the parent and/or being copied). In this case the interaction contribution to the special node is of the same order as recruitment, and interaction can no longer be ignored; such nodes may become outliers with very large in-degree (the discussion in the next section will show that, in practice, this phenomenon is limited due to finite-size effects). 

\subsubsection{Effect of How Interaction Iinks are Distributed}
The discussion thus far concerns the total link count and has not considered the distribution of interaction links. This assumption is reasonable for discussing the average behavior (of the entire network), but if, for example, nodes exist where interaction links are distributed in an extremely concentrated manner, the discussion may break down for those specific nodes.

If interaction links are spread almost uniformly across the network (``random interaction''), our analysis conditions are well satisfied. In this case, the in-degree a node gains from interaction per step is $O(Z_i/I)$. The average contribution from recruitment links is $O(\langle k_{in}\rangle/I)\approx O(\log I/I)$. Comparing these, when $I$ is large (since $Z_i$ is constant), the presence of the $\log I$ term makes the recruitment link process dominant.

By contrast, if interaction links are highly concentrated on a few nodes, our explanation may no longer hold. For example, assume that all interaction links are allocated to a single node at a constant rate $P_i^*$ independent of time. In this case, that node receives an in-degree of $O(P_i^* \cdot Z_i)$, i.e., $O(1)$, from interaction per step. This is on the same order as the maximum per-step contribution $O(1)$ that any particular node can receive from recruitment links (i.e., the extreme case of receiving a recruitment link at every step). In such cases, the influence of interaction cannot be ignored, and the previous argument that recruitment becomes dominant over time breaks down.

\subsubsubsection{Condition for Recruitment Links to Dominate Interaction Links}
The discussion thus far suggests a sufficient condition for a power law with exponent 1 to hold.
Let $P^{(ex)}_v(I)$ be the probability that an interaction link ($Z_i$ links per step) is allocated to node $v$. If the maximum allocation probability $P^{(ex)}_{\max}(I)$ satisfies
\begin{equation}
P^{(ex)}_{\max}(I) = \max_v P^{(ex)}_v(I) \to 0 \quad (I\to\infty),
\label{app_eq_p_limit}
\end{equation}
then the recruitment link process dominates, and an in-degree distribution with exponent 1 emerges. If this condition is not met, interaction links concentrate on specific nodes, and our derivation is no longer applicable. However, due to the finite-size effects discussed in the next section, the distribution will still approximate a power law with exponent 1 even when this condition fails. 

Using this criterion, we examine several allocation rules:
\begin{enumerate}
    \item \textbf{Uniform at random:}
    $P^{(ex)}_v(I)=1/I$, so $P^{(ex)}_{\max}(I)=1/I\to 0$; the condition holds.
    \item \textbf{Concentration on a single node:}
    For a specific node $v^*$, $P^{(ex)}_{v^*}(I)=1$, hence $P^{(ex)}_{\max}(I)=1$; the condition fails.
    \item \textbf{Proportional to in-degree $k_{in}^{(v)}$:}
    $P^{(ex)}_v(I)=k_{in}^{(v)}(I)/\sum_v k_{in}^{(v)}(I)$. Assuming a power law with exponent 1 for self-consistency, we approximate $k_{in}^{(v)}\propto I/v$. Then
    \begin{equation}
    P^{(ex)}_{\max}(I)\approx \frac{I/1}{\sum_{v=1}^{I} (I/v)} \approx \frac{1}{\sum_{v=1}^{I} (1/v)} \approx \frac{1}{\log I}\to 0,
    \end{equation}
    so the condition holds.
    \item \textbf{Proportional to squared in-degree $k_{in}^{(v)2}$:}
    $P^{(ex)}_v(I)=k_{in}^{(v)}(I)^2/\sum_v k_{in}^{(v)}(I)^2$. With $k_{in}^{(v)}\propto I/v$, we have
    \begin{equation}
    P^{(ex)}_{\max}(I)\approx \frac{(I/1)^2}{\sum_{v=1}^{I} (I/v)^2}
    \approx \frac{I^2}{I^2 \sum_{v=1}^{I} 1/v^2}
    \approx \frac{1}{\pi^2/6}\approx \frac{6}{\pi^2},
    \end{equation}
    which does not vanish; the condition fails.
\end{enumerate}

\subsubsubsection{Role of Finite-Size Effects in the Power-Law Distribution with Exponent 1}

Even if the condition given in Eq.~\ref{app_eq_p_limit} (regarding the concentration of interaction links) is not met, a distribution that approximates a \textbf{power-law distribution with exponent 1} can appear due to the \textbf{finite-size effect}.

Specifically, this model assumes a rule that ``a node cannot be selected again via interaction to receive an incoming link from a node that already links to it (no duplicate links)''.

Therefore, a node that already has a very large in-degree ($k_{in}^{(v)}$ is large) has fewer nodes (relative to the total $I$) that do not yet link to it, and as a result, it becomes less likely to receive interaction links.

An equation that approximately accounts for this effect can be written as follows:

\begin{equation}
\frac{dk_{in}^{(v)}(I)}{dI} = \frac{k_{in}^{(v)}+1}{I} + Z_i \cdot P_{v}(I) \cdot \left(1-\frac{k_{in}^{(v)}(I)}{I}\right) \label{app_eq_kin}
\end{equation}

\begin{itemize}
    \item \textbf{First term $\left(\frac{k_{in}^{(v)}+1}{I}\right)$:} Recruitment Effect
    \item \textbf{Second term $\left(Z_i \cdot P_{v}(I) \cdot \left(1-\frac{k_{in}^{(v)}(I)}{I}\right)\right)$:} Interaction Effect
\end{itemize}

Here, $P_{v}(I)$ is the distribution probability of interaction links to node $v$ when the total number of nodes is $I$.

The noteworthy part is the \textbf{$\left(1-\frac{k_{in}^{(v)}(I)}{I}\right)$} term included in the second term. This is the term that represents the \textbf{finite-size effect}.
If node $v$ receives incoming links from all other nodes, $k_{in}^{(v)}(I) \approx I$, and this term approaches $0$. This expresses that the probability of receiving an interaction link becomes $0$.

\par
Next, let's consider the case where condition (Eq.~\ref{app_eq_p_limit}) is not met, meaning the distribution probability does not converge to $0$ as $I \to \infty$, but $P_{v}(I) \to P_v^* > 0$ ($P_v^*$ is a positive constant).  

Even in this case, we confirm that the finite-size effect term $\left(1-\frac{k_{in}^{(v)}(I)}{I}\right)$ eventually approaches $0$, suppressing the interaction effect.

First, we divide both sides of (Eq.\ref{app_eq_kin}) by $I$ and set the in-degree ratio of the node as $x_v(I)=\frac{k_{in}^{(v)}(I)}{I}$.
Transforming this (similar to the calculation in the previous section), the change in $x_v(I)$ is expressed by the following equation:

\begin{equation}
\frac{dx^{(v)}(I)}{dI} = \frac{1}{I^2} + \frac{Z_i \cdot P_{v}(I)}{I} \cdot (1-x^{(v)}(I)) \label{app_eq_xi}
\end{equation}
Here, even if $P_{v}(I) \to P_v^* > 0$ (a constant) as for large $I$, the dynamics drives $x^{(v)}(I)$ close to 1.

$x^{(v)}(I) \to 1$ means that the finite-size effect term $(1-x^{(v)}(I))$ asymptotically approaches $0$.
As a result, the entire second term (interaction effect) $Z_i \cdot P_{v}(I) \cdot (1-x^{(v)}(I))$ asymptotically approaches $0$, and its influence diminishes. Consequently, the situation approaches a state dominated by recruitment, which is thought to suppress deviations from the power-law distribution with exponent 1.
Thus, it is thought that a power-law distribution with exponent 1 will be observed, irrespective of Condition given in Eq.~\ref{app_eq_p_limit} or the various distribution rules governing interaction.

\subsubsubsection{Direct Check for In-Degree–Proportional Allocation}
For $P^{(ex)}_v(I)=k_{in}^{(v)}(I)/\sum_v k_{in}^{(v)}(I)$, we can directly write the evolution for a node's in-degree (writing $k_{in}^{(v)}(I)$ as $k_{in}(I)$ for brevity):
\begin{equation}
\frac{dk_{in}}{dI} \approx \frac{k_{in}+1}{I}
+ Z_i \cdot \frac{k_{in}(I)}{\sum_{v=1}^{I} k_{in}^{(v)}(I)}.
\end{equation}
In general $\sum_{v=1}^{I} k_{in}^{(v)}(I)=M_{in}(I)=M_{out}(I)$ (total in-degree equals total out-degree). Using Eq.~\ref{app_eq_k_mean},
\begin{equation}
M_{out}(I)\approx I\,\langle k_{out}\rangle(I)= I\,(2+Z_i)\,\log I,
\end{equation}
so
\begin{equation}
\begin{aligned}
\frac{dk_{in}}{dI}
&\approx \frac{k_{in}+1}{I}
+ Z_i \cdot \frac{k_{in}(I)}{I\,(2+Z_i)\,\log I} \\
&= \frac{ k_{in}(I)\!\left( 1+\frac{Z_i}{(2+Z_i)\,\log I} \right) + 1 }{I}.
\end{aligned}
\end{equation}
As $I\to\infty$, the $1/\log I$ term (the interaction contribution) vanishes, and the equation approaches the recruitment-only form $\frac{dk_{in}}{dI}\approx \frac{k_{in}+1}{I}$. This is consistent with the condition under which a power law with exponent 1 holds for in-degree.

\clearpage

\numberwithin{equation}{section} 
\setcounter{equation}{0} 
\renewcommand{\theequation}{D\arabic{equation}}



\section{Data Sources and Basic Preprocessing}
\label{app_sec_data_base}
Appendix D provides information regarding the data sources. Specifically, it addresses three points: first, the data sources for the Japanese blog data and Google Trends data; second, the method used to extract the ``new words'' targeted in this study; and third, the normalization of the count time series data by the total number of posts.

\subsection{Data}
We use two types of online language data: (a) Japanese blog data and (b) Google Trends (English, Spanish, and Japanese) \cite{watanabe2023minor}. From these sources, we construct word-count time series for analysis.\par

\subsubsection{Japanese Blog Data}
We obtain daily word-appearance counts from a nationwide corpus of Japanese blogs using the large-scale database ``Kuchikomi@kakaricho,'' provided by Hottolink, Inc. The database contains approximately nine billion blog articles and covers about 90\% of Japanese blogs over the period from November 1, 2006 to December 31, 2019 \cite{kakaricho}.

\subsubsubsection{Body Text Data}
\label{app_sec_body_text}
We use the blog body text to compute co-occurrence statistics in Sections~\ref{sec_word}, \ref{app_sec_cooccurrence}; and in Sections~\ref{sec_llm}, \ref{app_sec_llm}, as auxiliary information for the LLM analysis. The observation window is from January 2009 to December 2018. Within this window, we identify 113,691 bloggers who posted in at least nine distinct calendar years, and we include all posts authored by these bloggers. The monthly article count has a mean of 1,133,463 (minimum 467,632; 25th percentile 815,666; median 1,104,976; 75th percentile 1,395,734; maximum 1,926,944).

\subsubsection{Google Trends}
\label{sec_google}
Google Trends provides a monthly index of search volume for a given query term on the Google search engine \cite{google_trends}. We use it in parallel with blog post counts to quantify social interest (see the red circles in Fig.~\ref{app_fig_comp_google_blog_same}). The series is normalized by Google so that the maximum value within the observation window equals 100, with other values scaled proportionally. The data are available from May 2015 to Dec. 2021.

\subsection{Methodology for Sampling Words}
\label{app_sec_data}
This section describes how we sampled candidate words \cite{watanabe2023minor}. Our sampling frame is the set of Wikipedia article titles; consequently, the analysis is restricted to terms established enough to appear as Wikipedia entries and does not cover short-lived expressions that quickly disappear.

\subsubsection{Blog Data}
We extracted candidate words in two steps.
\begin{enumerate}
  \item From the list of article titles in the Japanese edition of Wikipedia \cite{wikipedia_data}, we identified the one million titles that occurred most frequently in our Japanese blog corpus.
  \item From these one million titles, we select
  ed 20,764 titles that had zero blog posts in both November and December 2006.
\end{enumerate}

\subsubsection{Google Trends}
For the Google Trends analysis, we used Wikipedia page views to preselect newly emerging words in each of the English, Spanish, and Japanese editions of Wikipedia \cite{wikipedia_data}.
\begin{enumerate}
  \item We collected page-view counts for the first day of each month from May 2015 through January 2022.
  \item We defined a ``new word'' as a title with zero page views on May 1, 2015 (the first observation month) and with at least 50 page views for Spanish or at least 1,000 page views for English on January~1 of any year from 2016 to 2022. For Japanese, we used the same 20,764-word dictionary as in the blog data.
  \item For titles meeting this criterion, we retrieved Google Trends time series via the Google Trends API \cite{google_trends}.
\end{enumerate}

\subsection{Normalization of Word-Count Time Series for Blog Data}
We define notation for the word-count series $x_i(t)$ and the normalized series $y_i(t)$ as follows \cite{watanabe2023minor}.
\begin{itemize}
    \item The time step is set to 30 days. When $t$ increases by one, real time advances by 30 days (an approximately monthly series).
    \item For word $i\in\{1,\dots,W\}$ and time index $t\in\{1,\dots,T\}$, $x_i(t)$ is the raw count of articles containing word $i$ within the 30-day window $t$.
    \item Let $TOTAL(t)$ be the total number of articles in window $t$. We define the scaled total number of articles by
    \begin{equation}
    ScaledTOTAL(t)=\frac{TOTAL(t)}{\frac{1}{T}\sum_{s=1}^{T}TOTAL(s)}\,,
    \end{equation}
    so that $\frac{1}{T}\sum_{t=1}^{T} ScaledTOTAL(t)=1$. The normalized series is
    \begin{equation}
    y_i(t)=\frac{x_i(t)}{ScaledTOTAL(t)}, 
    \end{equation}
    which we also plot as black triangles in Fig.~\ref{fig_all_time_series}.
\end{itemize}
Here, $T$ is the last observation index and $W$ is the number of words. By construction, $y_i(t)$ removes fluctuations due to changes in the overall blogging volume $TOTAL(t)$ and isolates the relative activity of word $i$ over time (see Fig.~1 in Ref.~\cite{RD_base}). \par
\clearpage

\numberwithin{equation}{section} 
\setcounter{equation}{0} 
\renewcommand{\theequation}{E\arabic{equation}}


\section{Preprocessing for analysis of growth curves}
\label{app_sec_preprocessing_base}
Appendix E describes preprocessing for analysis of growth curves. 
\revisecolor{black}{The R script provided in the Supplemental Material \cite{supp_material} includes the code for the demonstration in this section.}
Specifically, it covers two topics: first, the method for extracting the growth segments from the time series; second, the method for detecting jump-like changes.

\subsection{Method for Extracting an Uptrend}
\label{app_sec_cut}
Here, we explain how we extracted a global growth period that is not a temporary local trend. An example of the global growth period is the period enclosed by the grey vertical lines in Fig.~\ref{app_fig_comp_google_blog_same}. First, we describe the detection of the starting point of growth, and next, the detection of the end point of growth (defined in Section~\ref{app_sec_end_growth} of this appendix).

\subsubsection{Extracting the Beginning of Growth}
\label{app_sec_growth}
The extraction of the growth starting point basically follows the procedure shown in \cite{watanabe2023minor}. However, in this study, we add a further correction using the method in the next Section~\ref{app_sec_refine_mm1}.
Prior to the correction in Section \ref{app_sec_refine_mm1}, the procedure based on \cite{watanabe2023minor} (the procedure to determine a provisional starting point) is shown below.

\begin{enumerate}
    \item \textbf{Calculate the upper limit for candidates, $T_s$.}
    This upper limit is set as the time when the 13-point moving median first reaches the 25th percentile point:
    $T_s=\min_t\{t|y_j(t) \geq \mathrm{Quantile}_{0.25}\{y_j(t)\} \}$.
    This procedure is introduced to avoid incorrectly selecting a minimum point during a downtrend after an uptrend.

    \item \textbf{Calculate the first candidate for the starting point, $t^{s}_1$.}
    The first candidate is set as the time when the 13-point moving median time series is at its minimum (within the range $t \le T_s$):
    $t^{s}_1=\mathrm{arg min}_{\{t \leq T_s\}}\{\mathrm{MovingMedian}_{13}({y_j}(t))\}$.
    Here, if the minimum value is less than 10, we recalculate using the 13-point moving median of the raw time series $x_j(t)$ (defined in Section \ref{app_sec_data}), $\mathrm{MovingMedian}_{13}(x_j(t))$.

    \item \textbf{Calculate the second candidate for the starting point, $t^{s}_2$.}
    The second candidate is set as the time when the raw time series (i.e., before taking the moving median) is at its minimum (within the range $t \le T_s$):
    $t^{s}_2=\mathrm{arg min}_{\{t \leq T_s\}}\{y_j(t)\}$.
    Here again, if the minimum value of the time series is less than 10, we recalculate using the unnormalized time series $x_j(t)$. (Points where $y_j(t)=0$ are excluded.)

    \item \textbf{Determine the growth starting point $t^{s}_0$.}
    Basically, we conservatively choose the later time, $t^{s}_0=\max(t^{s}_1,t^{s}_2)$, as the starting point of the trend.
    However, if a clear upward trend exists between the two candidate points $\{t^{s}_1,t^{s}_2\}$, the earlier time $t^{s}_0=\min(t^{s}_1,t^{s}_2)$ is selected as the growth starting point.
    (Note: The trend is identified based on the positive rank correlation between the times $\{\min(t^{s}_1,t^{s}_2),\dots,\max(t^{s}_1,t^{s}_2)\}$ and the corresponding counts $\{y_j(\min(t^{s}_1,t^{s}_2)), \dots, y_j(\max(t^{s}_1,t^{s}_2))\}$. We recognize a trend when the p-value for the correlation test is less than 0.01.)
\end{enumerate}

\subsubsection{Refining the Start Point by Excluding an Early Low-Level Segment Buried in Noise}
\label{app_sec_refine_mm1}
For the initial time $t^{s}_0$ (determined in Section \ref{app_sec_growth}), we perform a further investigation and correction in this study.
When $y_j(t)$ is small, relative fluctuations (e.g., Poisson-like noise) can become large and hide a slow growth component.
In such cases, the piecewise growth model fit might treat this first part as a separate "no-growth" piece, even though it is just a low-level noisy part before the actual growth.
To avoid this, we exclude an initial segment that is both "small" and "not increasing" and reset the growth start point as follows.
Here, $x_j(t)$ is the raw (unnormalized) count for word $j$, $t^{e}_0$ is the growth end point (determined in Section~\ref{app_sec_detect_jump}), and $[t^{s}_0,t^{e}_0]$ is the provisional growth window.

\begin{enumerate}
    \item \textbf{Define an upper limit for the early segment, $T^{\mathrm{early\_limit}}$.}
    Define $T^{\mathrm{early\_limit}}$ as the time when the time series first reaches its 25th percentile within the window ($[t^{s}_0,t^{e}_0]$).
    \begin{equation}
        T^{\mathrm{early\_limit}} = \min\Bigl\{\, t\in[t^{s}_0,t^{e}_0] \;\big|\; x_j(t)\ \ge\ \mathrm{Quantile}_{0.25}\!\bigl(x_j([t^{s}_0,t^{e}_0])\bigr) \Bigr\}
    \end{equation}
    This restricts the inspection of the early part to times up to this time $T^{\mathrm{early\_limit}}$.

    \item \textbf{Define a noise-aware start-point candidate, $c_0$.}
   Using a count threshold of $10$, we first define two skip candidates, $c_1$ and $c_2$. Note that below this threshold of 10, the effect of Poisson noise is significant, and an upward trend can be easily hidden.
    \begin{itemize}
        \item $c_1$: The point after the last time $t$ (within the whole window $[t^{s}_0,t^{e}_0]$) where $x_j(t)<10$.
        \begin{equation}
            c_1= 1+\max\{t\in[t^{s}_0,t^{e}_0] | x_j(t)<10\}
        \end{equation}
        \item $c_2$: The point after the last time $t$ (at or before $T^{\mathrm{early\_limit}}$) where $x_j(t)<10$ and $x_j(t+1) \le x_j(t)$ (non-increasing).
        \begin{eqnarray}
            &&c_2 = 1+  \nonumber \\
            &&\max\{\,t\in[t^{s}_0,T^{\mathrm{early\_limit}}]| x_j(t)<10 \nonumber \\
            && \text{ and } x_j(t+1)-x_j(t)\le 0 \}
        \end{eqnarray}
        If this set is empty (i.e., if $x_j(t)$ is strictly increasing whenever $x_j(t)<10$ in $[t^{s}_0, T^{\mathrm{early_limit}}]$), we set $c_2 = 1$.
    \end{itemize}
    (The "+1" is to set the start point to the point *after* the last "small" or "non-increasing" time.)

    Next, we define the noise-aware start-point candidate $c_0$ as the earlier time of these two candidates (if at least one exists).
    \begin{equation}
        c_0 = \min(c_1, c_2)
    \end{equation}
    (If only $c_1$ exists, $c_0 = c_1$; if only $c_2$ exists, $c_0 = c_2$. If neither exists, $c_0$ is not defined.)
    For example, if the series is monotonically increasing like $y(1)<y(2)<y(3)=10<y(4)$, then $c_1=3$ and $c_2=1$, and the initial time $c_0=1$ is adopted.
    As another example, if $5>y(1)>y(2)>y(3) <y(4) <y(5)=10<y(6)$, then $c_1=6$ and $c_2=3$, so $c_0=3$ is adopted.
    \item \textbf{Update the start time $t^{s}_0$.}
    Finally, we compare $t^{s}_0$ (found in Section~\ref{app_sec_growth}) with the noise-aware candidate $c_0$ and update the start time to the later (more conservative) of the two.
    \begin{equation}
        t^{s}_0 \;\leftarrow\; \max(t^{s}_0,\; c_0)
    \end{equation}
    (If $c_0$ was not defined (because neither $c_1$ nor $c_2$ existed), $t^{s}_0$ is not changed.)
\end{enumerate}
This corrected $t^{s}_0$ is used as the final growth starting point.

\subsubsection{Extraction of End of Growth}
\label{app_sec_end_growth}

The end point of growth follows the method of \cite{watanabe2023minor}. We introduce it below.

The end point of growth, $t^{e}_0$, is basically detected as the point at which a clear downward trend begins.
Here, a ``clear downward trend'' is defined as a point at or after the growth starting point ($t^{s}_0 \leq t^{e}_0$) from which the word count continuously decreases for at least 12 time points (approximately 12 months).

As a specific procedure, first, we roughly search for the starting point of the downtrend using the 13-point moving median to avoid being fooled by local trends. Next, we refine the candidate points by progressively using information from smaller time scales (5-point moving median, 3-point moving median, and original data). Finally, we compare the starting point of the downtrend found by this method with the global maximum point of the time series to determine which is more suitable as the end point of the global uptrend.

The detailed procedure is as follows:

\begin{enumerate}
    \item \textbf{Detecting the downtrend starting point $t^{e}_1$ in the 13-point moving median:}
    First, we search for a point in the 13-point moving median time series where the value continuously decreases for at least 12 consecutive time points. To avoid erroneously detecting local downtrends, only points after the time series has reached the 90th percentile, $T^{e}= \min_t\{t|y_j(t) \geq \mathrm{Quantile}_{0.9}\{y_j(t)\} \}$, are considered candidates ($t^{e}_1>T^{e}$). If no point with 12 consecutive decreases exists, the last observation point $T$ is taken as the end point, $t^{e}_1=T$.

    \item \textbf{Exploring around $t^{e}_1$ and creating the candidate set $\{t^{e}_2\}$:}
    To determine the end of the growth trend more precisely, we investigate the vicinity of $t^{e}_1$ (calculated in step 1) in detail. Specifically, we select all points between $t^{s}_0$ and $t^{e}_1$ where the 13-point moving median is 90\% or more of the maximum value in that interval, creating a new set of candidate points $\{t^{e}_2\}$.
    \begin{eqnarray}
        \{t^{e}_2\}= \{t|t^{s}_0 \leq t \leq t^{e}_1, \mathrm{MovingMedian}_{13}(y_j(t)) \geq \nonumber \\
        0.9 \times \max_{\{t^{s}_0 \leq t \leq t^{e}_1\}}(\mathrm{MovingMedian}_{13}(y_j(t))) \}
    \end{eqnarray}

    \item \textbf{Adding shorter time-scale (5-point moving median) information:}
    For all candidate points $t$ in $\{t^{e}_2\}$, we perform an operation to replace them with a local peak reached by a continuous increase on the 5-point moving median time series, $q_j(t)=\mathrm{MovingMedian}_5(y_j(t))$. Specifically, we set $\{t_0\} \to \{t^{e}_2\}$ and transform $t$ to $t^*$ using the following Eq.~\ref{app_move_end}:
    \begin{eqnarray}
    && t^{*}_0=t_0+ \mathrm{arg max}_{\{t \in \{t_0+m^{-},t_0+m^{+}\}\}}(q_j(t)) \label{app_move_end} \\
      &&m^{+} = \max\Bigl(\{m \mid m,s \in \mathbb{N}, m \geq 0, 0 \leq s \leq m, \nonumber \\
      &&         \qquad \forall{s}[q_j(t_0+s)\leq q_j(t_0+s+1)]\}\Bigr) \nonumber \\
      &&  m^{-} = \min\Bigl(\{m \mid m,s \in \mathbb{N}, m \leq 0, m \leq s \leq 0, \nonumber \\
      &&         \qquad \forall{s}[q_j(t_0+s) \leq q_j(t_0+s-1)] \}\Bigr)
    \end{eqnarray}
    This transformation corresponds to correcting the candidates from the 13-point moving median with information from a shorter time scale (the 5-point moving median). The transformed set of candidates is $\{t^{e}_3\}$.

    \item \textbf{Adding 3-point moving median information:}
    A transformation process similar to step 3 is performed, this time using the 3-point moving median, $q_j(t)=\mathrm{MovingMedian}_3(y_j(t))$. The input candidate set is $\{t_0\} \to \{t^{e}_3\}$, and the new set of candidates calculated according to Eq.~\ref{app_move_end} is $\{t^{*}_0\} \to \{t^{e}_4\}$.

    \item \textbf{Determining the provisional end point $t^{e}_5$:}
    From the candidate points in $\{t^{e}_4\}$, the point with the maximum 3-point moving median value, $\mathrm{MovingMedian}_{3}(y_j(t))$, is determined as the provisional end point $t^{e}_5$.
    $t^{e}_5=\mathrm{arg max}_{t \in \{t^{e}_4\}}(\mathrm{MovingMedian}_{3}(y_j(t)))$

    \item \textbf{Fine-tuning with raw data:}
    We use the original (raw) data $y_j(t)$ (without a moving median) to perform a final fine-tuning of the end point. Specifically, we set $t_0 \to t^{e}_5$ and $q_j(t)=y_j(t)$ and apply Eq.~\ref{app_move_end} to move the candidate point to the point with the maximum raw data value in the vicinity of $t^{e}_5$. This transformed time is $t^{e}_6$.

    \item \textbf{Comparing the candidate $t^{e}_6$ with the global maximum point $t^{max}$:}
    Finally, we compare the candidate point $t^{e}_6$ obtained from this procedure with the global maximum point of the entire time series, $t^{max}=\mathrm{arg max}_{t}[y_j(t)]$.
    If $t^{max}$ exists within 6 points before or after $t^{e}_6$ (i.e., $t^{e}_6-6 \leq t^{max} \leq t^{e}_6+6$), we check whether $t^{max}$ is more suitable as the growth end point than $t^{e}_6$.
    Specifically, we check that $t^{max}$ is not a temporary spike (noise) due to news or external factors, or that there is a clear upward (or downward) trend from $t^{e}_6$ to $t^{max}$.
    A "clear trend" is defined as the time series data between $t^{e}_6$ and $t^{max}$, $\{y_j(\min(t^{e}_6,t^{max})),\dots,y_j(\max(t^{e}_6,t^{max}))\}$, satisfying at least one of the following three conditions:
    \begin{enumerate}
        \item[(i)] A linear approximation fits well (coefficient of determination > 0.4), and the regression coefficient is not zero (the sign of the coefficient matches $\mathrm{sign}(t^{max}-t^{e}_6)$, and the p-value is less than 1\%).
        \item[(ii)] A quadratic function fits very well (coefficient of determination $\geq$ 0.85), and the function's derivative is always positive (if $t^{e}_6 < t^{max}$) or always negative (if $t^{max} < t^{e}_6$) within the target period.
        \item[(iii)] In a binomial test on the sign of the difference ($y_j(t+1)-y_j(t)$), the proportion of positives (if $t^{e}_6 < t^{max}$) or negatives (if $t^{max} < t^{e}_6$) is 0.6 or more (one-sided test p-value is less than 5\%).
    \end{enumerate}

    If these conditions are met, the final end point is set to $t^{e}_0=t^{max}$. If the conditions are not met (i.e., $t^{max}$ is judged to be a temporary spike), then $t^{e}_0=t^{e}_6$.
\end{enumerate}

\subsection{Detecting Large Jumps in the Keyword Frequency Time Series}
\label{app_sec_detect_jump}
This method is an algorithm to detect when an ``abrupt jump'' (abrupt upward level shifts) occurs in monthly time series data $x$.

The jumps detected in this section are \emph{not} transient spikes that quickly decay or revert to the original value. Instead, we focus on jumps where the effect is sustained long-term, similar to a step function. In other words, we identify jumps that represent a non-negligible shift in the growth curve.

The basic idea is to sequentially test every point in the time series as a potential point where a jump may have occurred.

The testing method involves measuring the ``jump height'' at that point, while also measuring the ``baseline fluctuation'' (how much it normally fluctuates) in the periods before and after that point. Based on this, it evaluates \textbf{``how much the jump height stands out compared to the baseline fluctuation''} using multiple indicators (scores).

\subsubsection*{Steps of the Jump Detection Algorithm}

\subsubsection*{Step 1: Data Preprocessing}

Before starting the analysis, the input time series data $x = (x_1, x_2, \dots, x_T)$ is prepared into a form that is easy to analyze. For example, in the case of blog data, this sequence represents the word counts time series $y_i(1),y_i(2),\dots, y_i(T)$ (i.e., $x_t = y_i(t)$).

\begin{enumerate}
    \item \textbf{Noise Removal (Smoothing)}: \\
    To reduce the influence of \textbf{noise} in the data, a \textbf{Moving Median} (`runmed`) is applied to create a smooth time series $x' = (x'_1, \dots, x'_T)$ (default window width is 3).
    $$x'_t = \text{median}(x_{t -1}, x_t, x_{t + 1})$$
    \textbf{Meaning}: This makes it easier to find the fundamental movements (level changes) in the data, without being misled by temporary spikes (outliers).

    \item \textbf{Checking Periodicity and Determining the Reference Period $D$}: \\
    We check if the data has a one-year periodic pattern (seasonality). This is done by taking the log-difference of the data and calculating the correlation (autocorrelation $\rho_{12}$) with the data from 12 months prior.
    \begin{itemize}
        \item If $\rho_{12} \ge 0.2$, it is considered to have \textbf{periodicity}, and the ``reference period'' used in later calculations is set to $D=6$ (months).
        \item Otherwise, it is considered to have \textbf{no periodicity}, and the reference period is set to $D=3$.
    \end{itemize}
    \textbf{Meaning}: Data with seasonality tends to have larger fluctuations. By setting an appropriate period $D$ for measuring the ``baseline fluctuation,'' we avoid misinterpreting seasonal variations as jumps.
\end{enumerate}

\subsubsection*{Step 2: Full Scan of All Points (Validation of Candidate Points)}

Each point in time $i$ (from $i=1$ to $T$) in the time series is sequentially validated as a \textbf{jump candidate point}.

\subsubsection*{Step 3: Measuring the ``Jump Height'' ($dv$)}

The candidate point $i$ is treated as a boundary, splitting the data into a ``pre-jump'' group $G_{pre} = \{x'_1, \dots, x'_i\}$ and a ``post-jump'' group $G_{post} = \{x'_{i+1}, \dots, x'_T\}$. The ``step'' between them is calculated.

\begin{enumerate}
    \item \textbf{Determining the Pre-Jump Level ($P_{pre}$)}: \\
    To stably measure the jump base level, the maximum of the following three values is adopted.
    $P_{pre} = \max\left(x'_i, \text{median}\{x'_k\}_{k=\max(1, i-6)}^i, Q_{0.8}(G_{pre})\right)$, where
    $Q_{0.8}(G_{pre})$ is the 80th percentile value of the pre-jump group.\\
    \textbf{Meaning}: If we only look at $x'_i$ (the immediately preceding value), we might mistake a temporary dip due to noise as the jump base. Therefore, we compare it with the median of the last six months and the 80th percentile of the entire pre-jump period to robustly (less affected by noise) determine the stable level before the jump begins.

    \item \textbf{Determining the Post-Jump Level ($P_{post}$)}: \\
    To measure the landing level of the jump, the \textbf{minimum value} of the post-jump group is adopted.
    $$P_{post} = \min(G_{post})$$
    \textbf{Meaning}: Immediately after a jump, the value might temporarily overshoot. Therefore, we consider the most stable (lowest) level after the jump as the ``landing point.''

    \item \textbf{Calculating the Jump Magnitude $dv$}: \\
    The difference between $P_{pre}$ (the jump base) and $P_{post}$ (the landing point) is calculated on a ``logarithmic ($\log$) scale.''
    $$dv = \log(P_{post} + 1) - \log(P_{pre} + 1)$$
    \textbf{Meaning}: \textbf{By taking the logarithm, we can evaluate the jump based on its rate of change.} \\
    ※ However, in regions where the values are too small ($x'_i < 10$ and $x'_{i+1} < 50$) or if the value decreased ($dv < 0$), it is not considered a jump, and $dv=0$.
\end{enumerate}

\subsubsection*{Step 4: Measuring the ``Baseline Fluctuation''}

To determine if the jump height $dv$ is truly ``abnormal,'' we calculate ``how much the period before and after the jump normally fluctuates.''

\begin{enumerate}
    \item \textbf{Calculating the Baseline Fluctuation Rate $R(G, D)$}: \\
    For a given group $G$ (targeting only data where $x' \ge 10$), we calculate the absolute log-change rates between many pairs of points separated by the reference period $D$ (3 or 6 months) determined in Step 1, and then find their \textbf{median}.
    $$R(G, D) = \text{median}\left( \frac{|\log(g_{t+D}+1) - \log(g_t+1)|}{D} \right)$$
    \textbf{Meaning}: This measures, on average, how much the data changes (on a log scale) over $D$ months. By using the median instead of the mean, it is less affected by temporary outliers, allowing for a stable measurement of the ``typical magnitude of fluctuation'' for that period.

    \item \textbf{Calculating Various ``Fluctuation Magnitudes''}: \\
    Using $R(G, D)$, we calculate the following fluctuation magnitudes:
    \begin{itemize}
        \item $m_{pre} = R(G_{pre}, D)$: The ``baseline fluctuation'' before the jump.
        \item $m_{post} = R(G_{post}, D)$: The ``baseline fluctuation'' after the jump.
        \item $m_{pre,noise} = R(G_{pre}, 1)$: The ``monthly noise level'' before the jump (calculated with $D=1$).
        \item $m_{post,noise} = R(G_{post}, 1)$: The ``monthly noise level'' after the jump.
        \item $m_{post,near} = R(G_{post,near}, D)$: The ``baseline fluctuation'' within 13 months immediately after the jump (to check post-jump stability), $G_{post,near} = \{x'_{i+1},\dots,x'_{\min(i+13,T)}\}$.
    \end{itemize}
\end{enumerate}

\subsubsection*{Step 5: Scoring the ``Sharpness of the Jump'' Multidimensionally}

We combine the ``jump height'' $dv$ (from Step 3) and the ``baseline fluctuation'' $m$ (from Step 4) to calculate six types of scores that evaluate how ``abrupt'' the jump was.

\begin{enumerate}
    \item \textbf{Outlier Correction ($dv_2$)}: \\
    If the point immediately before the jump ($i$ or $i-1$) was an abnormally high value (e.g., $x'_i \ge \exp(m_{post}) \cdot x'_{i+1}$), $dv$ might be overestimated.
    \begin{itemize}
        \item In that case, we use a recalculated, more stringent (also considering $x'_{i-1}$) jump base $P'_{pre}$ to calculate a corrected jump magnitude $dv_2$.
        \begin{eqnarray*}
        &&P'_{pre} = \max\big(x'_i, x'_{\max(1, i-1)},  \nonumber \\
        &&\text{median}\{x'_k\}_{k=\max(1, i-6)}^i, Q_{0.8}(G_{pre})\big)
        \end{eqnarray*}
        \begin{eqnarray*}
        dv_2 = \log(P_{post} + 1) - \log(P'_{pre} + 1)
        \end{eqnarray*}

        \item Otherwise, we set $dv_2 = dv$.
    \end{itemize}

    \item \textbf{Calculating the Jump Indicators (Scores)}:
    \begin{itemize}
        \item \textbf{Basic Score ($dr$)}:
        $$dr = \frac{dv}{\max(m_{pre}, m_{post})}$$
        \textbf{Meaning}: \textbf{How many times} is the jump height compared to the baseline fluctuation (whichever is larger, before or after)? A larger value indicates a sharp jump that cannot be explained by normal movement.
        
        \item \textbf{Overall Ratio ($ds$)}:
        $ds = \frac{dv}{d_{max}}$, where $d_{max}$ denotes the total log change from the beginning to the end of the series.
        \textbf{Meaning}: \textbf{What proportion} of the entire data's change (from beginning to end) does this one jump account for? A larger value indicates a dominant jump that affects the entire series.
        
        \item \textbf{Corrected Score ($dr2$)}:
        $$dr2 = \frac{dv_2}{2 \cdot \max(m_{pre}, m_{post})}$$
        \textbf{Meaning}: The basic score using the outlier-corrected $dv_2$. (The 2 in the denominator is an adjustment coefficient.)
        
        \item \textbf{Near-Future Ratio ($dr_{near}$)}:
        $$dr_{near} = \frac{dv_2}{2 \cdot \max(m_{pre}, m_{post,near})}$$
        \textbf{Meaning}: A score to ensure correct evaluation even if the fluctuation immediately after the jump is volatile ($m_{post,near}$ is large).
        
        \item \textbf{Noise Ratio ($dr1$)}:
        \begin{eqnarray*}
        &&dr1 = \nonumber \\
        &&\frac{dv}{\max(m_{pre}, m_{post}, m_{pre,noise}, m_{post,noise})}
        \end{eqnarray*}
        \textbf{Meaning}: Strictly evaluates whether the jump was sufficiently large even when compared to short-term monthly noise ($m_{noise}$).
        
        \item \textbf{Time-Lag Correction ($dr_{delta}$)}:
        $$dr_{delta} = \frac{dv}{m_{pre} \times d_{delta}}$$
        \textbf{Meaning}: $d_{delta}$ is the time (in months) it took for the minimum value ($P_{post}$) to appear after the jump. This means that jumps that did not stabilize immediately and took a long time ($d_{delta}$ is large) will have their scores lowered (penalized).
    \end{itemize}
\end{enumerate}

\subsubsection*{Step 6: Final Judgment}

A point $i$ is identified as a ``jump point'' if the six scores calculated for it meet any of the following combinations of conditions (thresholds).

\begin{itemize}
    \item \textbf{Condition A (Standard Jump)}: \\
    ($dr \ge 3$ and $ds \ge 0.1$ and $dr2 \ge 3$ and $dr1 \ge 3$ and $dr_{delta} \ge 2$) \\
    \textbf{Meaning}: A standard jump where multiple indicators exceed the criteria in a balanced way.
    
    \item \textbf{Condition B (Large-Scale Jump)}: \\
    ($dr \ge 10$ and $ds \ge 0.05$ and $dr2 \ge 3$ and $dr_{near} \ge 3$ and $dr1 \ge 2$ and $dr_{delta} \ge 2$) \\
    \textbf{Meaning}: A very large-scale jump where the basic score $dr$ is extremely high (10 times or more than normal).
    
    \item \textbf{Condition C (Dominant Jump)}: \\
    ($ds \ge 0.2$) \\
    \textbf{Meaning}: A highly dominant jump that accounts for more than 20\% of the entire data's change in this single event.
\end{itemize}

However, points at the very beginning (first 12 months) and the very end of the data, where comparisons are insufficient, are excluded from being jump points even if they meet the conditions.

\numberwithin{equation}{section} 
\setcounter{equation}{0} 
\renewcommand{\theequation}{F\arabic{equation}}
%
%

\clearpage 
\section{Procedure for Parameter Fitting}
\label{app_sec_estimation_parameter_base}
Appendix F describes the procedure for parameter fitting of the piecewise power-law model. \revisecolor{black}{The R script provided in the Supplemental Material \cite{supp_material} includes the code for the demonstration in this section.}

\subsection{Parameter Estimation for the Single-Segment ($N=1$) Power-Law Growth Model}
\label{app_sec_parameter_estimation_single}

This section explains the parameter estimation procedure for the single-segment ($N=1$) power-law growth model, given by Eq.~\ref{eq_base0}. This estimation method is also used as a component for the piecewise models where $N \geq 2$.

\subsubsection{Defining the Single-Segment Model ($N=1$)}

First, for an observed time series $y_t$ (where $t=1, \dots, L$), we define the single-segment growth model given by Eq.~\ref{eq_base0}, $\hat{y}(t)$ as follows:

\begin{equation*}
\hat{y}(t)=
\begin{cases}
Y \cdot \Bigl\{R \cdot (1-\alpha)(t-t_0)+\left(\frac{\hat{y}(t_0)}{Y} \right)^{1-\alpha}\Bigr\}^{1/(1-\alpha)}, & \alpha\neq 1 \\
\hat{y}(t_0) \cdot \exp \Bigl(R \cdot (t-t_0)\Bigr), & \alpha=1
\end{cases}
\end{equation*}

In this model, the parameters we need to estimate are the \textbf{shape parameter $\alpha$} and the \textbf{growth rate $R$}. The initial value (as an estimate with noise removed) $\hat{y}(t_0)$ is determined beforehand by a separate procedure (see Section~\ref{app_sec_y0} for $N=1$ or Section~\ref{app_sec_parameter_estimation} for $N \geq 2$). For the blog data used in this study, the constant $Y=41.254$ is also determined by the procedure described in Section~\ref{app_sec_r}.

\subsubsection{Defining the Final loss function (How We Measure Error)}

To determine the parameters $\alpha$ and $R$, we design a ``loss function'' $\mathcal{L}(\alpha, R)$ that measures how badly the model fits the data. We then find the parameters that minimize this loss.

\subsubsection*{Power Transform and Residuals}

Before defining the loss function, we first apply a ``power transform'' to both the observed values $y_t$ and the theoretical values $\hat{y}(t)$. We do this to handle the wide variety of growth shapes, from linear (near $\alpha=0$) to exponential (near $\alpha=1$).

The transform function $\chi(u)$ is defined as:
\begin{equation*}
\chi(u)=
\begin{cases}
\log u & \theta=0 \quad \text{(log transform)} \\
u^{\theta} & \theta \neq 0 \quad \text{(power transform)}
\end{cases}
\end{equation*}
In all our empirical analyses, we fix $\theta=0.5$, which is a compromise between $\theta=1$ (suited for linear growth) and $\theta=0$ (suited for exponential growth).

Using this transformed scale, we define the ``signed residual'' $\delta_0(t)$ (the difference between the observation and the model). (We only use data points where $y_t>0$ when using the log transform).
\begin{equation*}
\delta_0(t)=\chi(y_t) - \chi(\hat{y}(t))
\end{equation*}
Here, $\delta_0(t) < 0$ means that the theoretical value $\hat{y}(t)$ is higher than the observed value $y_t$ (an overestimation).

\subsubsection*{Constructing the Loss Function: Penalty and Reward}

We build the loss function $\mathcal{L}(\alpha, R)$ from two parts: a ``Penalty'' and a ``Reward''.

\paragraph{Penalty (for Overestimation) $p(t)$}
We apply a penalty only when the model's value is higher than the observed value (i.e., $\delta_0(t) < 0$). The penalty size is the magnitude of this difference, $|\delta_0(t)|$.
\begin{equation*}
p(t)=
\begin{cases}
|\delta_0(t)| & \delta_0(t)<0 \\
0 & \delta_0(t) \ge 0 
\end{cases}
\end{equation*}

\paragraph{Reward (for Upward Lift) $b(t)$}
We give a small ``reward'' (a reduction in the loss) for the model $\hat{y}(t)$ simply having a positive value. This prevents the model from being estimated too low (e.g., $\hat{y}(t) \to 0$).
\begin{equation*}
b(t)=
\begin{cases}
\bigl|\log\bigl(\hat{y}(t)+1\bigr)\bigr| & \theta=0 \\
\bigl|\hat{y}(t)^{\theta}\bigr| & \theta \neq 0
\end{cases}
\end{equation*}

\paragraph{Final loss function $\mathcal{L}(\alpha, R)$}
The final loss is the total penalty minus the total reward, summed over the entire time period $L$.
\begin{equation}
\mathcal{L}(\alpha,R)=\sum_{t=1}^{L} p(t)-\sum_{t=1}^{L} b(t) \label{app_eq_loss}
\end{equation}

\subsubsection{The Purpose of Our Final Loss Function (Upper-side Robustness)}

Time series data like keyword frequency (word counts) often have complex noise, especially sudden upward spikes caused by external news.

A standard symmetric error measure (like least squares, $(y_t - \hat{y}(t))^2$) would be pulled upward unfairly by these large spikes (outliers).

Our loss function is intentionally designed to be asymmetric, giving it \textbf{upper-side robustness} to deal with this problem.

\begin{itemize}
    \item \textbf{When the model is higher than the data (Overestimation)}: \\
    $\delta_0(t) < 0$, so a \textbf{strong penalty $p(t) = |\delta_0(t)|$ is applied}. This pushes the model curve down so it does not exceed the data.

    \item \textbf{When the data is higher than the model (Underestimation)}: \\
    $\delta_0(t) \ge 0$. In this case, $p(t)=0$, so \textbf{no penalty is applied}. We assume these points are likely temporary spikes (noise) that should not affect the trend estimate. By not penalizing this, we prevent the model's trend line from being pulled upward by these spikes.
\end{itemize}

This mechanism prevents the trend line from being skewed by temporary upward noise, allowing a more robust estimation that stays close to the "baseline" of the data.

Meanwhile, the reward term $-\sum b(t)$ ensures the model doesn't just fall to zero; it provides the minimum necessary lift to support the data from below.

The final parameters are determined where these two opposing forces—the penalty $\sum p(t)$ (pushing down) and the reward $-\sum b(t)$ (lifting up)—find a balance.

\subsubsection{Optimization Process}

The final optimization problem is formulated as finding the arguments that minimize the loss:
\begin{equation*}
(\hat{\alpha},\hat{R})=\arg\min_{\alpha \in[-10,10], R_{raw} \in[0,10]} \mathcal{L}(\alpha, R_{raw}^5)
\end{equation*}
To ensure $R>0$ and stabilize the optimization, we use a search variable $R_{raw} \in (0, 10]$ and transform it via $R = R_{raw}^5$.

To solve this global optimization problem, we use \textbf{Differential Evolution (DE)} (e.g., the \texttt{DEoptim} library in R). Our main settings for this study are: population size $\mathrm{NP}\approx100$, max iterations $\mathrm{itermax}\approx500$, mutation rate $F\approx0.8$, and crossover rate $\mathrm{CR}\approx0.9$.

If any proposed parameters result in numerically unstable values (like \texttt{NaN} or $\infty$), they are given a large constant penalty to effectively remove them from the search.

\subsubsection{Determination of $y(t_0)$}
\label{app_sec_y0}
The initial value $\hat{y}(t_0)$ is determined from the smoothed spline $\bar{y}(t)$ at $t_0$; if $\bar{y}(t_0) < 0$, we set $\hat{y}(t_0)=0.8$ \cite{watanabe2023minor}.

\subsection{Parameter Estimation for the Piecewise Power-Law Model ($N \ge 2$)}
\label{app_sec_parameter_estimation}

This section explains how to estimate the parameters for the piecewise power-law model. First, we explain the estimation method for the case with a \textbf{continuity constraint} (no jumps) at the segment boundaries (Section~\ref{app_sec_est_no_jumps}). Then, we describe the case that allows jumps (Section~\ref{app_sec_est_jumps}).

\subsubsection{Estimating Split Points for a Fixed $N$ (No Jumps / Continuity Constraint)}
\label{app_sec_est_no_jumps}

Here, we describe how to estimate the parameters for the piecewise power-law model when the number of segments $N$ is already fixed.

\subsubsection*{Problem Definition and Objective}

\begin{itemize}
    \item \textbf{Input:} A time series with equally spaced points $y_t$ (from $t=1$ to $T$) and a pre-specified number of segments $N$ (e.g., $N=2, 3, 4, \dots$).
    
    \item \textbf{Parameters to Estimate:} The $N-1$ ``split points'' $\{t_1, \dots, t_{N-1}\}$ that divide the segments, and the growth parameters $\{\alpha^{(m)}, R^{(m)}\}$ for each segment $m$.
    
    \item \textbf{Constraint (Continuity):} No jumps are allowed between segments. This means at any split point $t_m$, the initial value of the next segment ($m+1$) is forced to be equal to the final value of the current segment's fitted curve, $\hat{y}^{(m)}(t_m)$.
    
    \item \textbf{Single-Segment Fitting:} The method for fitting the parameters ($\alpha^{(m)}, R^{(m)}$) and calculating the loss $\mathcal{L}$ for any single segment is the same as the one described in Appendix~\ref{app_sec_parameter_estimation_single}.
    
    \item \textbf{Objective:} To find the best set of split points $\{t_1^*, \dots, t_{N-1}^*\}$ and the corresponding parameters $\{\alpha^{(m)*}, R^{(m)*}\}$ that \textbf{minimize the total loss $\mathcal{L}_{\text{total}}$} over the entire time series $[1, T]$.
\end{itemize}

\subsubsection*{Basic Estimation Approach (Recursive Search)}

To find the best combination of split points, we use an efficient \textbf{recursive search}.

The main idea is to split the full time series at a point $t_c$ into two sub-problems: a left interval $[1, t_c]$ and a right interval $[t_c+1, T]$. We assign $N_{\text{left}}$ segments to the left and $N_{\text{right}}$ segments to the right (where $N_{\text{left}} + N_{\text{right}} = N$).

We test every possible candidate for this main split point $t_c$. For each candidate, we find the best possible splits within the left and right sub-intervals (by using this same method recursively). We calculate the total loss for each $t_c$ (Total $\mathcal{L} = (Left)  \mathcal{L} + (Right)  \mathcal{L}$) and compare them. The set of split points that gives the minimum total loss is our final answer.

Here are the specific steps for $N=2, 3, 4$.

\subsubsection*{Case $N=2$ (One Split Point)}

When $N=2$, there is only one split point, $t_1$.

\begin{enumerate}
    \item \textbf{List all candidate split points $t_1$:} \\
    We check every possible position $t_1 \in \{2, 3, \dots, T-1\}$ (this ensures each segment has at least one data point).
    
    \item \textbf{Calculate the loss for each candidate $t_1$:}
    \begin{enumerate}
        \item[(a)] \textbf{Fit the left segment $[1, t_1]$:} \\
        Use the single-segment method (Appendix \ref{app_sec_parameter_estimation_single}) on the interval $[1, t_1]$ to estimate parameters $\hat{\alpha}^{(1)}, \hat{R}^{(1)}$ and get the fitted curve $\hat{y}^{(1)}(t)$.
        
        \item[(b)] \textbf{Fit the right segment $[t_1+1, T]$ (with continuity):} \\
        \textbf{Fix} the initial value of the right segment to be $\hat{y}^{(1)}(t_1)$ (the final value of the left segment). With this constraint, apply the single-segment method to the interval $[t_1+1, T]$ to get parameters $\hat{\alpha}^{(2)}, \hat{R}^{(2)}$ and the curve $\hat{y}^{(2)}(t)$.
        
        \item[(c)] \textbf{Calculate total loss:} \\
        Calculate the total loss for the combined fitted curve: $\mathcal{L}_{\text{total}}(t_1) = \mathcal{L}([1, t_1]) + \mathcal{L}([t_1+1, T])$, where $\mathcal{L}([a,b])$ denotes the single-segment loss on $[a,b]$.
    \end{enumerate}
    
    \item \textbf{Select the best solution:} \\
    Find the $t_1^*$ that has the minimum $\mathcal{L}_{\text{total}}(t_1)$. This $t_1^*$ and the parameters that produced it $\{\alpha^{(1)*}, R^{(1)*}, \alpha^{(2)*}, R^{(2)*}\}$ are the final estimates.
\end{enumerate}

\subsubsection*{Case $N=3$ (Two Split Points)}

When $N=3$, we have two split points, $\{t_1, t_2\}$. We solve this by splitting the series into one ``$N=1$ segment'' problem and one ``$N=2$ segment'' problem, using the $N=2$ procedure recursively.

\begin{enumerate}
    \item \textbf{List candidates for the first split point $t_1$:} \\
    We check all $t_1 \in \{2, 3, \dots, T-2\}$ (to leave room for the $N=2$ split on the right).
    
    \item \textbf{Calculate the loss for each candidate $t_1$:}
    \begin{enumerate}
        \item[(a)] \textbf{Fit the left segment $[1, t_1]$:} \\
        This is an $N=1$ problem. We get $\hat{y}^{(1)}(t)$ (same as step 2a in the $N=2$ case).
        
        \item[(b)] \textbf{Optimally split the right interval $[t_1+1, T]$ (for $N=2$):} \\
        Apply the $N=2$ procedure (from step 3 above) to the interval $[t_1+1, T]$. The initial value for this interval is fixed at $\hat{y}^{(1)}(t_1)$. This step finds the best internal split point $s^*$ within the right interval and its fitted curve $\hat{y}^{\text{right}}(t)$.
        
        \item[(c)] \textbf{Calculate total loss:} \\
        The total loss is $\mathcal{L}_{\text{total}}(t_1) = \mathcal{L}([1, t_1]) + \mathcal{L}([t_1+1, s^*]) + \mathcal{L}([s^*+1, T])$.
    \end{enumerate}
    
    \item \textbf{Select the best solution:} \\
    Find the $t_1^*$ that minimizes $\mathcal{L}_{\text{total}}(t_1)$. The $s^*$ that was found along with this $t_1^*$ becomes the second split point, $t_2^*$. The final split points are $\{t_1^*, t_2^*\}$, and the parameters are the ones found in each step.
\end{enumerate}

\subsubsection*{Case $N=4$ (Three Split Points)}

When $N=4$, we have three split points $\{t_1, t_2, t_3\}$. We solve this by splitting the time series into two ``$N=2$ sub-problems''.

\begin{enumerate}
    \item \textbf{List candidates for the central split point $t_c$:} \\
    We check all $t_c \in \{3, 4, \dots, T-3\}$ (to leave room for $N=2$ splits on both sides).
    
    \item \textbf{Calculate the loss for each candidate $t_c$:}
    \begin{enumerate}
        \item[(a)] \textbf{Optimally split the left interval $[1, t_c]$ (for $N=2$):} \\
        Apply the $N=2$ procedure (step 3) to the interval $[1, t_c]$. This gives an internal split $s_{\text{left}}^*$ and the curve $\hat{y}^{\text{left}}(t)$.
        
        \item[(b)] \textbf{Optimally split the right interval $[t_c+1, T]$ (for $N=2$, with continuity):} \\
        Apply the $N=2$ procedure to the interval $[t_c+1, T]$. The initial value is fixed to $\hat{y}^{\text{left}}(t_c)$ (the final value of the left curve) to ensure continuity. This gives an internal split $s_{\text{right}}^*$ and the curve $\hat{y}^{\text{right}}(t)$.
        
        \item[(c)] \textbf{Calculate total loss:} \\
        $\mathcal{L}_{\text{total}}(t_c)$ is the sum of the two losses from the left split and the two losses from the right split.
    \end{enumerate}
    
    \item \textbf{Select the best solution:} \\
    Find the $t_c^*$ that minimizes $\mathcal{L}_{\text{total}}(t_c)$. This $t_c^*$ becomes the central split point $t_2^*$. The other splits $s_{\text{left}}^*$ and $s_{\text{right}}^*$ become $t_1^*$ and $t_3^*$. The final set is $\{t_1^*, t_2^*, t_3^*\}$, along with all corresponding parameters.
\end{enumerate}

\subsubsection*{Case $N \ge 5$ (General Recursive Method)}

For $N \ge 5$, we generalize this recursive method. We split the problem into two sub-problems with $N_{\text{left}} = \lfloor N/2 \rfloor$ segments (left) and $N_{\text{right}} = \lceil N/2 \rceil$ segments (right).

\begin{enumerate}
    \item \textbf{List candidates for the central split point $t_c$:} \\
    We check all $t_c \in \{N_{\text{left}}, \dots, T - N_{\text{right}}\}$.
    
    \item \textbf{Calculate the loss for each candidate $t_c$:}
    \begin{enumerate}
        \item[(a)] \textbf{Optimally split the left interval $[1, t_c]$ (for $N_{\text{left}}$ segments):} \\
        Recursively apply this entire procedure for $N = N_{\text{left}}$. This gives the curve $\hat{y}^{\text{left}}(t)$ and the total left loss $\mathcal{L}^{\text{left}}$.
        
        \item[(b)] \textbf{Optimally split the right interval $[t_c+1, T]$ (for $N_{\text{right}}$ segments, with continuity):} \\
        Recursively apply this procedure for $N = N_{\text{right}}$, fixing the initial value to $\hat{y}^{\text{left}}(t_c)$. This gives the curve $\hat{y}^{\text{right}}(t)$ and the total right loss $\mathcal{L}^{\text{right}}$.
        
        \item[(c)] \textbf{Calculate total loss:} \\
        $\mathcal{L}_{\text{total}}(t_c) = \mathcal{L}^{\text{left}} + \mathcal{L}^{\text{right}}$.
    \end{enumerate}
    
    \item \textbf{Select the best solution:} \\
    Find the $t_c^*$ that minimizes $\mathcal{L}_{\text{total}}(t_c)$. Combine $t_c^*$ with all the split points found recursively in the sub-problems to get the final set $\{t_1^*, t_2^*, \dots, t_{N-1}^*\}$ and all corresponding parameters.
\end{enumerate}

This is the procedure for estimating the optimal parameters for a fixed number of segments $N$ under the continuity constraint.

\subsubsection{Choosing the Number of Segments $N$ (No Jumps / Continuity Constraint)}
\label{app_sec_n_estimation}

This section describes the procedure for deciding the segment no $N$ for a keyword-frequency time series $y(t)$.

\subsubsection*{Sequential Selection Procedure}

We assume that the models being compared (the $N$-segment model and the $N+1$-segment model) have each already been optimized (i.e., their total loss has been minimized) using the parameter estimation method for a fixed $N$ described in Appendix \ref{app_sec_est_no_jumps}.

We determine the optimal number of segments by a sequential comparison, starting with $N=1, 2, \dots$.

\begin{enumerate}
    \item \textbf{First Comparison:} \\
    First, compare the $N=1$ model against the $N+1=2$ model. (The specific decision rule for this comparison is detailed in Section~\ref{app_sec_comp_n}).
  
    \item \textbf{If $N$ is selected:} \\
    If the $N$-segment model is selected, stop the procedure. $N$ is fixed as the final number of segments.
    
    \item \textbf{If $N+1$ is selected:} \\
    If the $N+1$-segment model is selected, update $N \leftarrow N+1$. Compare the new pair (the updated $N$ vs. $N+1$) and return to Step 2.
    
    \item \textbf{Exception Handling:} \\
    If the time series is extremely short (e.g., $T \le 3$), this comparison is skipped, and $N=1$ is always adopted.
\end{enumerate}

This procedure selects the minimum necessary complexity (number of segments) by starting from the simplest model ($N=1$) and deciding at each stage whether $N$ or $N+1$ is more appropriate.

\subsubsection{Parameter Estimation for the Model Considering Jumps (Discontinuities)}
\label{app_sec_est_jumps}

\subsubsection*{Overview and Basic Approach}

This section explains the parameter estimation procedure for the piecewise power-law model, taking into account the \textbf{jumps (discontinuous change points)} detected in Appendix \ref{app_sec_detect_jump}.

The basic approach is as follows:
\begin{itemize}
    \item \textbf{Fixing Jump Locations:} \\
    The jump locations $\tau_b$ detected beforehand are treated as fixed jump points (corresponding to $t_{i,m}$ in Eq.~\ref{eq:jump-i}).
    
    \item \textbf{Partitioning into Blocks:} \\
    The entire time series $[1, T]$ is partitioned into $B+1$ "blocks" $\mathcal{B}_b$ based on these jump locations.
    
    \item \textbf{Fitting within Blocks:} \\
    Inside each block $\mathcal{B}_b$, the data is assumed to change continuously (Eq.~\ref{eq:cont-i}). We use the procedures from Appendix \ref{app_sec_est_no_jumps} and \ref{app_sec_n_estimation} to estimate the non-jump segment boundaries and growth parameters ($\alpha, R$) within the block.
    
    \item \textbf{Estimating Initial Values After Jumps:} \\
For each block $b \ge 1$, the initial value $y_0^{(b)}$ at its start (immediately after a jump) is estimated freely, independent of the previous block's end value $\hat{y}(\tau_b)$
(corresponding to $\lim_{t\to t_{i,m}+0} y_i(t)$ in Eq.~\ref{eq:jump-i}).
For the first block $\mathcal{B}_0$, the initial value is treated as in the no-jump case in Section~\ref{app_sec_y0}.

\end{itemize}

This approach allows us to capture both the abrupt level shifts caused by jumps and the continuous growth trends in other parts.

\subsubsection*{Estimation Procedure}
{\bf Step 1: Partitioning into Blocks Based on Jump Locations} \\
Let the set of jump locations detected in Appendix \ref{app_sec_detect_jump} be $\mathcal{J} = \{ \tau_1 < \tau_2 < \cdots < \tau_B \}$.
This partitions the entire time series $[1, T]$ into $B+1$ blocks $\mathcal{B}_b = [s_b, e_b]$.
\begin{itemize}
    \item $\mathcal{B}_0 = [1, \tau_1]$ (where $s_0=1, e_0=\tau_1$)
    \item $\mathcal{B}_b = [\tau_b+1, \tau_{b+1}]$ (where $s_b=\tau_b+1, e_b=\tau_{b+1}$ ; for $1 \le b < B$)
    \item $\mathcal{B}_B = [\tau_B+1, T]$ (where $s_B=\tau_B+1, e_B=T$)
\end{itemize}

{\bf Step 2: Parameter Estimation per Block} \\
For each block $b$, we apply the methods from Section~\ref{app_sec_est_no_jumps} and Section~\ref{app_sec_n_estimation} (the ``split point estimation under continuity'' procedure) to calculate the number of segments within the block, $M^{(b)}$, the non-jump segment boundaries, and the parameters $\alpha_i^{(b)}, R_i^{(b)}$ ($i=1, 2, \dots, M^{(b)}$) for each segment.

In this process, the initial value (let's call it $y_0^{(b)}$) for the start point $t=s_b$ of each block $\mathcal{B}_b$ (which is immediately after a jump) is estimated independently and is not forced to be equal to the previous block's end value, $\hat{y}(s_b-1)$. For the first block $\mathcal{B}_0$, the initial value is treated as in the no-jump case in Section~\ref{app_sec_y0}.

(Note: The difference between this estimated initial value $y_0^{(b)}$ and the previous block's end value $\hat{y}(s_b - 1)$ represents the magnitude of the jump. The loss function $\mathcal{L}$ used here is the same as in Eq.~\ref{app_eq_loss} from Appendix \ref{app_sec_est_no_jumps}.) \par \par
{\bf Step 3: Calculating the Jump Magnitude $\Delta^{(b)}$} \\
Using the estimated initial value $\hat{y}(s_b) = y_0^{(b)}$ at the start of the block (at $t=s_b$) and the theoretical end value $\hat{y}(s_b - 1)$ of the previous block ($b-1$), the jump magnitude $\Delta^{(b)}$ is calculated post-hoc as follows:

\begin{equation*}
\Delta^{(b)} = \hat{y}(s_b) - \hat{y}(s_b - 1)
\end{equation*}

This corresponds to the estimate of the jump amount $\Delta_i^{(m)}$ defined in Eq.~\ref{eq:jump-i}. Here, $m$ is the cumulative index up to block $b$, counting both continuous splits (within blocks) and jump splits (between blocks). Note that for any $m$ that is not a jump point (i.e., inside a block), $\Delta_i^{(m)}=0$.

\subsubsection*{Optimization Implementation and Parameter Settings}

The optimization method and parameter settings used to solve the minimization problems above are identical to those used in Appendix \ref{app_sec_est_no_jumps} (Differential Evolution).

For optimization stability, the parameter $R$ is reparameterized as $R = R_{raw}^5$. The search ranges for each parameter were set as follows:
\begin{itemize}
    \item $\alpha \in [-10, 10]$
    \item $R_{raw} \in (0, 10]$
    \item $y_0^{(b)} \in \left[ \lambda_{\min} \cdot \hat{y}(s_b - 1),\ \lambda_{\max} \cdot \hat{y}(s_b - 1) \right]$ for $b \ge 1$.
\end{itemize}
Here, $\lambda_{\min} = 0.05$ and $\lambda_{\max} = 20$. This means the initial value $y_0^{(b)}$ (immediately after a jump) is searched in a wide range from 5\% to 2000\% (20 times) of the previous block's end value $\hat{y}(s_b - 1)$.

The final loss for the entire model is the sum of the losses from all blocks (and all segments within them).

\subsection{Selection Procedure for the Number of Segments N and N+1 (Goodness-of-Fit Evaluation)}
\label{app_sec_comp_n}
Deciding on the number of segments ($N$) for a piecewise power-law model is a classic model-selection problem. While criteria like AIC or BIC are common tools, we use a \textbf{composite decision} based on multiple error measures. 

\textbf{In our specific problem, we compare the $N$-segment model (which we call Model 1, $\mathbf{\hat{y}}^{(1)}$) against the $(N+1)$-segment model (Model 2, $\mathbf{\hat{y}}^{(2)}$).}
Our approach \textbf{favors the simpler $N$-segment model} when the difference in performance is within a preset tolerance.

Our rationale is that keyword-frequency (word-count) data often has \textbf{diverse and complex noise patterns}. For example, it might show sudden spikes from external events followed by a gradual return to baseline.  Standard information criteria would require us to explicitly model this noise (even approximately), which is \textbf{not feasible} for our purposes.

Furthermore, these time series show many different \textbf{growth shapes} (like linear, exponential, or super-exponential). To stay robust across all these shapes, we use a combined measure that evaluates both \textbf{log-scale (relative) and linear-scale (absolute)} errors.

\subsubsection*{Basic Concept}
This procedure evaluates goodness-of-fit based on a "\textbf{ratio of areas}."
\begin{enumerate}
    \item First, the difference (error) between the observed data and the model's theoretical values is measured as the \textbf{``Error Area''}. This is the area between the two curves when the observed data and the model are plotted.
    \item Second, the total signal strength of the theoretical model (relative to its baseline) is measured as the \textbf{``Model Area''}.
    \item Finally, the procedure calculates the ratio of the "Error Area" to the "Model Area" (which is the normalized error). The smaller this ratio, the better the model fits.
\end{enumerate}

Based on this concept, our comprehensive decision considers the following:
\begin{itemize}
    \item \textbf{Evaluation Scale}: We evaluate the area ratio on two scales: the \textbf{linear scale} (absolute data values) and the \textbf{logarithmic scale} (relative change).
    \item \textbf{Evaluation Interval}: We assess the fit not only for the entire dataset but also for specific parts, such as the first half of the time series.
    \item \textbf{Time Prediction Accuracy}: We check how well the model fits the values ($y_t = f_j(t)$), and also the accuracy of its \textbf{inverse function} ($t = f_j^{-1}(y_t)$) in predicting "\textbf{when (time $t$) a specific observation $y_t$ occurred.}"
\end{itemize}

Our adjudication uses a conservative criterion favoring the simpler model: \textbf{Model 1 ($\mathbf{\hat{y}}^{(1)}$, the $N$-segment model) is adopted if it is not significantly worse than Model 2 ($\mathbf{\hat{y}}^{(2)}$, the $(N+1)$-segment model), or if it is superior in a key aspect (like time prediction).}

\subsubsection{Definition of Evaluation Metrics}

To make this decision, we first define the specific errors and quantities used to compare the observed data $\mathbf{y}$ and each model $\mathbf{\hat{y}}^{(m)}$ ($m=1, 2$).

\subsubsection*{Data Preprocessing and Notation}
\begin{itemize}
    \item \textbf{Observed Data}: $\mathbf{y} = (y_1, y_2, \dots, y_L)$
    \item \textbf{Theoretical Model Values}: $\mathbf{\hat{y}}^{(m)} = (\hat{y}_1^{(m)}, \dots, \hat{y}_L^{(m)})$
    \item \textbf{Smoothed Observed Data}: To reduce short-term noise in the observed data $\mathbf{y}$, we apply a median smoothing with a window size of 5, denoted as $\mathbf{y}_{\text{smooth}} = (y_{\text{smooth}, 1}, \dots, y_{\text{smooth}, L})$.
    \item \textbf{Evaluation Index ($I_0$)}: To avoid logarithmic transformation of zero ($\log(0)$), calculations are performed on the set of time steps:  
    \begin{equation}
        I_0 = \{ t \mid y_{\text{smooth},t} > 0,  \hat{y}_t^{(1)} > 0, \quad \hat{y}_t^{(2)} > 0 \}, 
    \end{equation}
    using the data points $(\mathbf{y}_{I_0}, \mathbf{\hat{y}}^{(m)}_{I_0})$.
\end{itemize}

\subsubsection*{Error Area (E: Error)}
The sum of the discrepancies (errors) between the observed values and the theoretical values is defined as the ``Error Area''. This corresponds to the area (based on the L1-norm) of the region between the observation curve and the model curve.
\begin{itemize}
\item {\bf Logarithmic Error Area ($E_j^{\log}$)} \\
\begin{equation*}
E_j^{\log} = \sum_{t \in I_0} |\log(y_{\text{smooth}, t}) - \log(\hat{y}_t^{(m)})|
\end{equation*}

\item {\bf Linear Error Area ($E_j^{\text{lin}}$)} \\
\begin{equation*}
E_j^{\text{lin}} = \sum_{t \in I_0} |y_{\text{smooth}, t} - \hat{y}_t^{(m)}|
\end{equation*}

\item {\bf Linear Error Area for the First Half ($E_{j, \text{half}}^{\text{lin}}$)} \\
The linear error area for the first half of the time series.
We then calculate the half-length $L_{half} = \lfloor L/2 \rfloor$ based on the \textbf{original} data length $L$, and sum the errors from $k=1$ to $k=L_{half}$,  
\begin{equation}
E_{j,\text{half}}^{\text{lin}} =  \sum_{t \in I_0,\, t \le L_{half}}  |y_{\text{smooth},t} - \hat{y}_t^{(m)}|. 
\end{equation}
For growth curves that increase rapidly in their later stages (e.g., exponential), a linear error metric effectively ignores early-stage errors, as it is dominated by the large errors in the latter half. Therefore, we introduced this error measure.
\end{itemize}

\subsubsection*{Model Area (A: Area)}
To relatively evaluate the error, we define the ``Model Area'' as the total signal magnitude (area above the baseline) of the theoretical values. The calculation uses Model 1 ($\mathbf{\hat{y}}^{(1)}$) as the baseline. We use Model 1 as the reference for the model area because our selection rule is designed to prefer the simpler model whenever its fit is not clearly inferior.
\begin{itemize}
\item {\bf Logarithmic Model Area ($A^{\log}$)}\\
\begin{equation*}
  A^{\log} = \sum_{t \in I_0} \left( \log(\hat{y}_t^{(1)}) - \min_{k \in I_0}(\log(\hat{y}_k^{(1)})) \right)
\end{equation*}

\item {\bf Linear Model Area ($A^{\text{lin}}$)}
\begin{equation*}
  A^{\text{lin}} = \sum_{t \in I_0}  \left( \hat{y}_t^{(1)}-\min_{k \in I_0}(\hat{y}_k^{(1)}) \right)
\end{equation*}

\item {\bf Linear Model Area for the First Half ($A_{\text{half}}^{\text{lin}}$)}
\begin{eqnarray*}
A_{\text{half}}^{\text{lin}} = \sum_{t \in I_0,\, t \le L_{half}}  \left( \hat{y}_t^{(1)} - \min_{k \in I_0} \hat{y}_k^{(1)} \right).
\end{eqnarray*}
\end{itemize}

\subsubsection*{Normalized Error (Relative Error Area Ratio) (S: Normalized Error)}
By dividing the ``Error Area (E)'' by the ``Model Area (A)'', we calculate the scale-independent ``\textbf{Normalized Error}''. This represents the ``\textbf{Relative Error Area Ratio}'', indicating \textbf{what proportion the error (area of discrepancy) constitutes of the total model signal (area)}. A smaller value (ratio) indicates a better fit.
\begin{itemize}
\item {\bf Logarithmic Normalized Error ($S_j^{\log}$)} \\
\begin{equation*}
S_j^{\log} = E_j^{\log} / A^{\log}
\end{equation*}

\item {\bf Linear Normalized Error ($S_j^{\text{lin}}$)} \\
\begin{equation*}
S_j^{\text{lin}} = E_j^{\text{lin}} / A^{\text{lin}}
\end{equation*}

\item {\bf Linear Normalized Error for the First Half ($S_{j, \text{half}}^{\text{lin}}$)} \\
\begin{equation*}
S_{j, \text{half}}^{\text{lin}} = E_{j, \text{half}}^{\text{lin}} / A_{\text{half}}^{\text{lin}}
\end{equation*}
If $A^{\log} = 0$ or $A^{\text{lin}} = 0$ (i.e., the model is essentially flat on the evaluation set), we skip the area-ratio-based comparison and adopt Model~1 (the $N$-segment model).
\end{itemize}

\subsubsection*{Time Prediction Error ($E_t$)}
We evaluate the model's predictive accuracy regarding the time axis using the inverse function of Model 1 ($t = f_1^{-1}(y_t)$).

\textbf{Calculation}:
\begin{enumerate}
    \item Input the observed values $\mathbf{y}$ into the inverse function of Model 1, $f_1^{-1}(y_t)$, to calculate the ``\textbf{predicted time} $\mathbf{t}_{\text{pred}}$'' at which those values $y_t$ should have occurred.
    \item Define the ``Time Prediction Error $E_t$'' as the discrepancy (mean absolute error) between this predicted time $\mathbf{t}_{\text{pred}}$ and the actual observation time $\mathbf{t}_{\text{true}} = (1, 2, \dots, L)$.
\end{enumerate}
\begin{equation*}
E_t = \text{mean}(|\mathbf{t}_{\text{pred}} - \mathbf{t}_{\text{true}}|)
\end{equation*}

\textbf{Interpretation}: A small $E_t$ indicates that Model 1 accurately captures the relationship between time and value (i.e., ``when'' a certain value occurs). We utilize this error in the time-axis, as it allows for an evaluation independent of the time series' $Y$-scale. This approach is also intended to mitigate the limitation of $Y$-axis error, which can fail to properly capture growth curves in cases such as exponential functions—where values differ by orders of magnitude between the first and latter halves—or when extremely large noise (e.g., from news events) is introduced.

\subsubsection{Model Selection Adjudication Procedure}

Using the metrics defined above, we establish three criteria (Criterion 1, 2, and 3) to adjudicate whether to adopt Model 1 ($\mathbf{\hat{y}}^{(1)}$).

\subsubsection*{Criterion 1: Error Ratio Criterion}
\textbf{Objective}: To confirm that the relative error area ratio of Model 1 ($S_1$) is not ``significantly larger'' than that of Model 2 ($S_2$).

\textbf{Condition}: 
The area ratio $S_1$ of Model 1 is allowed to be at most 25\% larger than the area ratio $S_2$ of Model 2 (i.e., $S_1$ is less than 1.25 times $S_2$). This must hold for both logarithmic and linear scales.
\begin{equation*}
\begin{split}
\text{Criterion 1 is met} \iff & \left( 1 - \frac{S_2^{\log}}{S_1^{\log}} \le 0.2 \right) \\
& \text{ AND } \left( 1 - \frac{S_2^{\text{lin}}}{S_1^{\text{lin}}} \le 0.2 \right)
\end{split}
\end{equation*}

\textbf{Interpretation}: If the error ratio of Model 1 is slightly larger than Model 2, but the difference is relatively small (within 25\%), the models are considered comparable.

\subsubsection*{Criterion 2: Error Difference Criterion}
\textbf{Objective}: To confirm that the relative error area ratio of Model 1 does not ``exceed that of Model 2 by a large absolute difference''.

\textbf{Condition}: The difference in the area ratios on the logarithmic scale must be 0.05 (5\%) or less; AND the difference on the linear scale (both total and for the first half) must be 0.15 (15\%) or less.
\begin{equation*}
\begin{split}
\text{Criterion 2 is met} \iff & (S_1^{\log} - S_2^{\log} \le 0.05) \\
& \text{ AND } (S_1^{\text{lin}} - S_2^{\text{lin}} \le 0.15) \\
& \text{ AND } (S_{1, \text{half}}^{\text{lin}} - S_{2, \text{half}}^{\text{lin}} \le 0.15)
\end{split}
\end{equation*}

\textbf{Interpretation}: If the absolute difference in error ratios is within this tolerance, Model 1 is not considered inferior to Model 2.

\subsubsection*{Criterion 3: Time Prediction Error Criterion}
\textbf{Objective}: To confirm that Model 1 is ``superior in predicting the time-value relationship (inverse function)''. We evaluate time prediction only for Model 1, since Criterion 3 is used as a sufficient condition to justify preferring the simpler model.

\textbf{Condition}: The time prediction error $E_t$ (calculated using the inverse function $f_1^{-1}(y_t)$) must be sufficiently small (average of 4 time steps or less); AND the goodness-of-fit of Model 1 itself (the forward function $y_t=f_1(t)$) must not be extremely poor (both logarithmic and linear relative error area ratios must be less than 0.3).
\begin{equation*}
\begin{split}
\text{Criterion 3 is met} \iff & (E_t \le 4) \text{ AND } (S_1^{\log} < 0.3) \\
& \text{ AND } (S_1^{\text{lin}} < 0.3)
\end{split}
\end{equation*}

\textbf{Interpretation}: Model 1 is adopted if it shows superior time-axis prediction accuracy, even if it performs slightly worse than Model 2 on Criteria 1 or 2 (i.e., worse fit in terms of value).

\subsubsection{Final Adjudication}

The final decision is made based on the three criteria above.
\begin{itemize}
    \item \textbf{Adopt Model 1 ($\mathbf{\hat{y}}^{(1)}$, the $N$-segment model) if}: \\
    \textbf{At least one} of Criterion 1, Criterion 2, or Criterion 3 is met. \\
    (Interpreted as Model 1 being comparable to, or better than, Model 2, or superior in time prediction.)

    \item \textbf{Adopt Model 2 ($\mathbf{\hat{y}}^{(2)}$, the $(N+1)$-segment model) if}: \\
    \textbf{All of the above criteria are not met.} \\
    (Interpreted as Model 1 being significantly inferior to Model 2.)

    \item \textbf{Exception Handling}: \\
    If the observed data length $L$ is 3 or less, a statistically meaningful evaluation is difficult. Therefore, the above evaluation is skipped, and \textbf{Model 1 ($\mathbf{\hat{y}}^{(1)}$, the $N$-segment model)} is adopted by default.
\end{itemize}

\clearpage
\numberwithin{equation}{section} 
\setcounter{equation}{0} 
\renewcommand{\theequation}{G\arabic{equation}}

\section{Supplementary Discussion of the Linguistic Analysis}
\label{app_sec_cooccurrence_base}
This section provides a supplementary discussion of the linguistic analysis. Specifically, we show two points: first, the method for extracting co-occurring terms; and second, the details of the LLM analysis and its prompts.

\subsection{Extracting Co-occurring Terms Associated with the Shape Exponent ($\alpha$)}
\label{app_sec_cooccurrence}

\subsubsection{Objective}

The objective of this section is to systematically extract terms $w$ that tend to co-occur with specific types of neologisms.
Specifically, we want to identify if a word $w$ tends to co-occur with:
\begin{itemize}
    \item[(i)] Neologisms showing \textbf{exponential-like growth ($\alpha \approx 1$)}
    \item[(ii)] Neologisms showing \textbf{linear-like growth ($\alpha \approx 0$)}
\end{itemize}

To do this, we evaluate the monotonic correlation (rank correlation) between the \textbf{growth shape parameter $\alpha_j$} of each neologism $j$ and a \textbf{``proximal co-occurrence index''}. This index measures how often $w$ appears near $j$ (within $\pm40$ words in the same document).

Based on this analysis, we extract terms associated with exponential-like growth ($\alpha \approx 1$) and terms associated with linear-like growth ($\alpha \approx 0$) (see Table~\ref{tab:cooccurrence_alpha_small_vs_large}).

\subsubsection{Data and Definitions}

We define the data and metrics used in this analysis as follows:
\begin{itemize}
    \item \textbf{Total occurrences of neologism $j$; $\mathcal{N}_j$}: \\
    The total word count of neologism $j$ in the entire corpus.
    
    \item \textbf{Proximal co-occurrences $C_j(w)$}: \\
    The total number of times $w$ was found within a $\pm40$ word window around $j$ (in the same document). (Note: Overlapping windows count the same $w$ multiple times.)
    
    \item \textbf{Growth shape parameter $\alpha_j$}: \\
    An estimated value summarizing the growth profile of neologism $j$.
\end{itemize}

\subsubsection{Proximal Co-occurrence Index $q_j(w)$ and Floor Treatment}

We define the ``proximal co-occurrence index $q_j(w)$'' to measure the strength of co-occurrence:
\begin{equation*}
q_j(w)=\frac{C_j(w)}{\mathcal{N}_j}
\end{equation*}
This represents the average number of times $w$ appears near $j$ (within $\pm40$ words) per single occurrence of $j$. It is a density-like value and can be greater than 1.

\paragraph{Floor Treatment (Lower Bound):}
This index $q_j(w)$ can be unstable if $\mathcal{N}_j$ is small or $C_j(w)$ is very rare (due to denominator effects).
For example, if $C_j(w)=1$ (only one co-occurrence), $\mathcal{N}_j=1000$ yields $q=0.001$, but $\mathcal{N}_j=200$ yields $q=0.005$. Although both reflect ``one rare event'', the values are 5 times different.

To stabilize the analysis against these rare events and improve the reliability of the rank correlation, we apply a lower bound (floor) of $\varepsilon=0.01$ to small positive values.
\begin{equation*}
q^{\langle \varepsilon\rangle}_j(w)=
\begin{cases}
\varepsilon & \text{if } 0 < q_j(w) \le \varepsilon \\
q_j(w) & \text{otherwise}
\end{cases}
\quad (\varepsilon=0.01)
\end{equation*}
(Note: $q_j(w)=0$ remains zero.)

\subsubsection{Data Used for Correlation Analysis}

When calculating the correlation for a co-occurring word $w$, we limit the analysis to neologisms $j$ that meet all three of the following conditions:
\begin{enumerate}
    \item \textbf{Growth Exponent Range}: \\
    $-0.1 \le \alpha_j \le 1.1$. (This provides a 0.1 margin around the 0 to 1 range.)
    
    \item \textbf{Sufficient Occurrences}: \\
    $\mathcal{N}_j \ge 100$. (We exclude low-frequency neologisms, as their $q_j(w)$ has a large measurement error.)
    
    \item \textbf{Existence of Co-occurrence}: \\
    $C_j(w) \ge 1$. (Neologisms that never co-occur with $w$ are excluded, as they cannot be ranked and would affect the correlation.)
\end{enumerate}

First, we define the \textbf{base set} $\mathcal{S}_{\text{base}}$ as the set of all neologisms satisfying conditions [1] and [2]. Let $\mathcal{N}_{base} = |\mathcal{S}_{\text{base}}|$ be its size. We use $\mathcal{N}_{base}$ as an index of the corpus scale.

Next, let $\mathcal S_w = \{ j \in \mathcal{S}_{\text{base}} \mid C_j(w) \ge 1 \}$ be the subset of neologisms that also meet condition [3]. Let $n(w)=|\mathcal S_w|$ be its size (the number of neologisms used for the correlation with $w$).

\begin{itemize}
    \item \textbf{$\mathcal{N}_{base}$}: The total number of neologisms that satisfy conditions [1] and [2] (regardless of co-occurrence with $w$). $\mathcal{N}_{base}$ does not depend on $w$ and shows how many neologisms are available for analysis in the corpus.
    \item (Note: For the analysis of an individual $w$, we separately require $n(w) \ge 50$ to ensure sufficient sample size.)
\end{itemize}

\subsubsection{Correlation Calculation and Statistics}

Using the set of neologisms $\mathcal S_w$ (sample size $n(w)$), we calculate \textbf{Kendall's rank correlation ($\tau$)} between the floor-treated index $q^{\langle 0.01\rangle}_j(w)$ and the growth exponent $\alpha_j$.
\begin{equation*}
(\tau(w), p(w))=\text{Kendall}\bigl(q^{\langle 0.01\rangle}_j(w),\ \alpha_j\bigr)
\end{equation*}
\begin{itemize}
    \item \textbf{$\tau(w) > 0$ (Positive Correlation)}: \\
    A larger $q_j(w)$ (co-occurs easily with $w$) is associated with a larger $\alpha_j$ (more likely to co-occur with \textbf{exponential-like $\alpha \approx 1$} neologisms).
    
    \item \textbf{$\tau(w) < 0$ (Negative Correlation)}: \\
    A larger $q_j(w)$ (co-occurs easily with $w$) is associated with a smaller $\alpha_j$ (more likely to co-occur with \textbf{linear-like $\alpha \approx 0$} neologisms).
    
    \item \textbf{$p(w)$}: 
    The p-value for the null hypothesis $H_0: \tau=0$ (no correlation).
    
    \item \textbf{$n(w)$}: 
    The number of neologisms used in the calculation.
    
    \item \textbf{$\mathcal{N}_{base}$}: 
    The size of the base set  (not depends on $w$).
\end{itemize}

\subsubsection{Criteria for Extracting Co-occurring Terms}

The co-occurring terms $w$ listed in Table~\ref{tab:cooccurrence_alpha_small_vs_large} are those that met the following \textbf{Reliability Criteria} and one of the two \textbf{Correlation Strength Criteria}.

\begin{itemize}
    \item \textbf{Reliability Criteria (Scale and Significance)}:
    All extracted terms must first meet all of the following conditions:
    \begin{itemize}
        \item $0 < p(w) \le 0.05$ (Statistically significant)
        \item $n(w) \ge 50$ (At least 50 co-occurrence data points with $w$)
        \item $\mathcal{N}_{base} \ge 500$ (At least 500 neologisms in the base set)
    \end{itemize}
    
    \item \textbf{Correlation Strength Criteria}:
    Terms that passed the reliability criteria are listed in the table if they belong to one of the following two groups:
    
    \begin{itemize}
        \item \textbf{Exponential-like words} (Co-occurs with $\alpha \approx 1$): \\
        Selected if the correlation is $\tau(w) \ge 0.2$.
        
        \item \textbf{Linear-like words} (Co-occurs with $\alpha \approx 0$): \\
        Selected if the correlation is $\tau(w) \le -0.2$.
    \end{itemize}
\end{itemize}
(Note: In the table, the exponential-like group ($\tau(w) \ge 0.2$) is sorted by $\tau(w)$ descending, and the linear-like group ($\tau(w) \le -0.2$) is sorted by $\tau(w)$ ascending.)

\subsection{Word Classification with LLMs (Large Language Models)}
\label{app_sec_llm}

In Section~\ref{sec_llm} of this study, we used an LLM to classify words to analyze how the growth-curve shape index, $\alpha_i$, relates to word categories. This appendix details the methodology, including the prompts, reference data, and inference conditions used for that classification.

\subsubsection{Reference Data (Information Provided to the LLM)}
\label{app_get_info_data}

To improve classification accuracy, we provided the LLM with the following four types of reference information. 

\begin{enumerate}

    \item \textbf{Wikipedia (Japanese) Lead/Summary}
    \begin{itemize}
        \item Collection Period: 2025/01/10--2025/01/17 (JST)
    \end{itemize}

    \item \textbf{Wikipedia (Japanese) Article Body}
    \begin{itemize}
        \item Collection Period: Same as above. We used the first 1000 characters of the article body.
    \end{itemize}
    
    \item \textbf{Web Search Results (DuckDuckGo)}
    \begin{itemize}
        \item Collection Period: Same as above.
        \item Search Settings: \texttt{region=jp-jp}, \texttt{safesearch=off}, \texttt{timelimit=None}
        \item We used the top 5 search results (title, URL, and body snippet).
    \end{itemize}
    
    \item \textbf{Blog Text Data (Sample of 40 Articles)}
    \begin{itemize}
        \item Collection Period: 2009/01/01--2018/12/30 (JST)
        \item To capture signals of "newsworthiness" or ``topicality'', sentences containing words for ``news'' (\textit{nyuusu}) or ``topic'' (\textit{wadai}) were included as candidates.
        \item See the next Section(\ref{app_get_info_blog}) for specific extraction rules and Section \ref{app_sec_body_text} for the blog body data.
    \end{itemize}
\end{enumerate}

\subsubsection{Blog Text Data Extraction Rules}
\label{app_get_info_blog}

For each keyword $word$, we extracted a sample of up to 40 articles (to be referenced by the LLM) using the following procedure:

\begin{enumerate}
    \item \textbf{Build Candidate Article Set:} \\
    Gather all articles from the following three candidate sets:
    \begin{itemize}
        \item[(i)] Sentences containing the word for "topic" (\textit{wadai}, 話題)
        \item[(ii)] Sentences containing the word for "news" (\textit{nyuusu}, ニュース)
        \item[(iii)] 40 articles drawn randomly from the entire blog corpus (not restricted to "news" (\textit{nyusu}) or "topic" (\textit{wadai}))
    \end{itemize}
    
    \item \textbf{Filter by Keyword $word$:} \\
    From the candidate set gathered in Step 1, keep only those that contain the keyword $word$.
    
    \item \textbf{Determine Final Sample:} \\
    Deduplicate the articles from Step 2 and select up to 40 articles as the final sample.
\end{enumerate}

\paragraph{Purpose:}
The purpose of this extraction rule is to intentionally oversample articles containing cues for ``newsworthiness'' (\textit{nyuusu-sei}, ニュース性) or ``topicality'' (\textit{wadai-sei}, 話題性), as these are key to the analysis. However, to avoid excessive bias and to handle cases where $w$ rarely co-occurs with such cues, we also include the general sample (iii).

\subsubsection{LLM and Inference Conditions}

\begin{itemize}
    \item \textbf{Model Used:} \\
    Google Gemini 2.5 Flash (Generative Language API, v1beta)
    
    \item \textbf{Input/Output:} \\
    A single text prompt, concatenating all the reference information above, was used as input. The model was instructed to provide the output as tab-separated (or space-separated) text.
    
    \item \textbf{Execution Periods:}
    \begin{itemize}
        \item Classification 1: 2025/08/05--2025/08/08 (JST)
        \item Classification 2: 2025/08/06--2025/08/10 (JST)
    \end{itemize}
\end{itemize}

\subsubsection{Tasks and Prompts (English)}
\label{app_sec_prompt_english}
\subsubsection*{Classification 1 (Public buzz / General-interest / Insider)}
We classified the \emph{outwardness/insiderness of topics} (Public buzz / General-interest / Insider) using the prompt given by Code~\ref{app_fig_llm_prompt_1}. The results are shown in Table~\ref{table_llm_wadai}.\\
Note that in the prompt given by Code~\ref{app_fig_llm_prompt_1}, \texttt{<Wikipedia summary>}, \texttt{<Web search results>}, \texttt{<Wikipedia body>}, and \texttt{<Blog text>} were replaced with the data collected in Section~\ref{app_get_info_data}. Although an English version is presented in Code~\ref{app_fig_llm_prompt_1}, the actual prompt used for classification in the analysis was in Japanese. The Japanese prompt is provided in Section~\ref{app_sec_prompt_japanese}.

\begin{figure*}[pt]
\captionsetup{name=Code}
\centering
\begin{minipage}{0.8\textwidth} 
\begin{lstlisting}
You are a capable and trust worthy Japanese assistant. For each word in the list below, use the reference information (including Wikipedia and news reports) to answer one by one. Because this is an academic setting, the reference documents may occasionally contain adult terms; however, your OUTPUT must only contain the classification label and the reason, and it must avoid harmful content.

 Please classify [WORD] whether this topic is something people like to talk about with strangers (i.e., public small talk), or mainly among those who already know/care about it. Evaluate at the historical peak of topicality, not the current moment. 

Choose one: 
1. Unknown-a (Public buzz): A topic that people willingly share as small talk with strangers, or a topic commonly learned from general sources such as nationwide TV news/ads or widespread usage in public. Recognized as a trending or widely disseminated buzzword/product/service.

2. Unknown-b (General-interest topic): A topic often learned from general media/ads or everyday word-of-mouth, but typically not perceived as a "buzzword."

3. Known (Insider/niche): A topic mainly discussed among people who already know or care about it, or learned primarily via one's own search or direct inquiry.

Output requirements:
* Output ONLY in the following tab-separated format. Do NOT output any other text.
* Do NOT prefix [WORD] with a numbered list.

Format:
[WORD]    [Label]    [Class(1-3)]    [Reason]
Examples:
Raccoon_dog    Known    3    ...
Hanako_Yamada    Unknown-b    2    ...
Kumaneko    Unknown-a    1    ...

Word list:
[WORD]    Reference
<WORD>  [Summary]<Wikipedia summary>[Search]<Web search results>[Body]<Wikipedia body>[Blog]<Blog text>
\end{lstlisting}
\end{minipage}
\caption{English Translated LLM Prompt for Classification 1 (Public buzz / General-interest / Insider). See Section~\ref{sec_llm} and Section~\ref{app_sec_llm} The original Japanese-language prompt is in Table~\ref{app_fig_llm_prompt_1_jp}.} 
 \label{app_fig_llm_prompt_1}          
\end{figure*} 

\subsubsection*{Classification 2 (24-way Genre Classification)}
We performed a \emph{word genre classification} using the prompt given by Code~\ref{app_fig_llm_prompt_2}. The results are shown in Table~\ref{table_llm_bunrui}.\\ 
Note: In the prompt given by Code~\ref{app_fig_llm_prompt_2}, \texttt{<Wikipedia summary>}, \texttt{<Web search results>}, \texttt{<Wikipedia body>}, and \texttt{<Blog text>} were replaced with the data collected in Section~\ref{app_get_info_data}.  Although an English version is presented  in Code~\ref{app_fig_llm_prompt_2}, the actual prompt used for classification in the analysis was in Japanese. The Japanese prompt is provided in Section~\ref{app_sec_prompt_japanese}.

\begin{figure*}[pt]
\captionsetup{name=Code}
\centering
\begin{minipage}{0.8\textwidth} 
\begin{lstlisting}
You are a capable and trustworthy Japanese assistant. For each wordin the list below, use the reference information (including Wikipedia and news reports) to answer one by one. The task is a term classification problem. Because this is an academic setting, the reference documents may occasionally contain adult terms; however, your OUTPUT must only contain the classification label and the reason, and it must avoid harmful content.

 Please classify [WORD] whether this topic is something people like to talk about with strangers (i.e., public small talk), or mainly among those who already know/care about it. Evaluate at the historical peak of topicality, not the current moment. 

Choose one:
1. Internet / ICT terminology
2. Internet / ICT service or product names
3. Entertainment / net culture / internet slang
4. Society / daily life / housing-food-clothing
5. Economy / business / politics / social issues
6. Drug names / medical terminology
7. Idol group names
8. Members or former members of Akimoto-produced idol groups (e.g., AKB48 groups,
   Sakamichi groups)
9. Other individual idols (excluding those covered by 8)
10. Voice actors
11. Actors
12. AV actors / AV actresses
13. Bands / singers / musical groups
14. Other celebrities (athletes, comedians, talents, novelists, etc.; excludes idols,
    voice actors, actors, AV actors/actresses, singers)
15. Anime / game terminology
16. Character names (excluding anime/game-related characters)
17. Media / information sites
18. Content / works (titles)
19. Place / facility / station / infrastructure names
20. Racehorse names
21. Food-service related services / product names
22. Other organizations
23. Other services / product names
24. Symbols / emoji

Output requirements:
* Output ONLY in the following tab-separated format. Do NOT output any other text.
* Do NOT prefix [WORD] with a numbered list.

Format:
[WORD]    [CategoryName]    [Class(1-24)]    [Reason]
Examples:
Smartphone   ICT terminology     1    ...
Hanako_Yamada    Other celebrities     14    ...
Kumaneko    ...    ...    ...

Word list:
[WORD]    Reference
<WORD>  [Summary]<Wikipedia summary>[Search]<Web search results>[Body]<Wikipedia body>[Blog]<Blog text>
\end{lstlisting}
\end{minipage}
\caption{English Translated LLM Prompt for Classification 2 (24-way genre classification). See Section~\ref{sec_llm} and Section~\ref{app_sec_llm}. The original Japanese-language prompt is in Table~\ref{app_fig_llm_prompt_2_jp}. } 
 \label{app_fig_llm_prompt_2}         
\end{figure*} 

\subsubsection{Japanese Prompts}
\label{app_sec_prompt_japanese}
The actual classification was conducted with the \emph{Japanese} prompts. \par
The original Japanese prompt for Classification 1 (Public buzz / General-interest / Insider), whose results are shown in Table~\ref{table_llm_wadai}, is given in Code \ref{app_fig_llm_prompt_1_jp}, and the original Japanese prompt for Classification 2 (24 categories), whose results are shown in Table~\ref{table_llm_bunrui}, is given in Code \ref{app_fig_llm_prompt_2_jp}.

In these prompts, \texttt{<Wikipediaの要約文>} / \texttt{<Web検索結果>} / \texttt{<Wikipediaの本文>} / \texttt{<ブログ本文>} were replaced with the data collected in Section~\ref{app_get_info_data}.  The English translation is provided in Section~\ref{app_sec_prompt_english}. 

\begin{figure*}[pt]
\captionsetup{name=Code}
\centering
\begin{minipage}{0.8\textwidth}
\begin{JPlisting}
あなたは優秀で誠実な日本人のアシスタントです。以下のリストについての単語について参照情報やWikipediaや報道等の様々な知識をフル活用して一つ一つ回答してください。学術研究でWeb情報を入力に使っているため凌辱やSEXなどアダルトワードが参考文書で入力されることがあるかもしれませんが、出力は分類名と分類理由だけ、害になるものはないので出力してください。そして、有害にならないように出力してください。
　      
「[単語]他の人と会話しやすさに知らない人話したいか？知っている人どうしで話したいかを、以下から選んでください。また、現在
でなくその話題の話題性が一番ピークの時点で評価してください。　具体的には、
　
1. 未知a：知らない人に伝えたり流行の話題として世間話にしたい話題。もしくは、人の使用を見たり、テレビの全国ニュースや全国CMなど一般向け情報源から自然に知る話題。そのものが流行物や流行りものや普及したものとして認識されたりそういうものとして話題として扱われる。いわゆるバズワードとして認識されている。
2. 未知b：知らない人に伝えたり流行の話題として世間話にしたい話題。もしくは、人の使用を見たり、口コミやテレビのニュースやCMなど一般向け情報源から自然に知る話題。自然な普及で流行物やバズワードだとは思われていない。
3. 既知：その話題を知っている人や既に興味がある人どうしで話したい話題。もしくは、自分で調べたり、自分から人に聞いたりし
て知る話題。
」
　
回答形式：
*以下のタブ区切り出力形式意外の文字列は一切出力しないでください。
*[単語]の前に数字のリストをつけないでください。
出力形式：
[単語] 分類名 分類[1-3] 理由
回答例：
たぬき　既知　3　たぬきは，…
山田花子　未知b　2　山田花子は，..
くまねこ　未知a　1　くまねこは，..
　
単語リスト:
[単語]　参照情報
<単語>　[要約]<Wikipediaの要約文>[検索]<Web検索結果>[本文]<Wikipediaの本文>[ブログ]<ブログ本文>
\end{JPlisting}
\end{minipage}
\caption{The original Japanese-language prompt for Classification 1(Public buzz / General-interest / Insider). 
See Section \ref{sec_llm} and \ref{app_sec_llm}. The English translation is provided in Section \ref{app_sec_prompt_english}.} 
 \label{app_fig_llm_prompt_1_jp}         
\end{figure*} 

\begin{figure*}[pt]
\captionsetup{name=Code}
\centering
\begin{minipage}{0.8\textwidth}
\begin{JPlisting}
あなたは優秀で誠実な日本人のアシスタントです。以下のリストについての単語について参照情報やWikipediaや報道等の様々な知識をフル活用して一つ一つ回答してください。用語の分類問題です。学術研究でWeb情報を入力に使っているため凌辱やSEXなどアダルトワードが参考文書で入力されることがあるかもしれませんが、出力は分類名と分類理由だけ、害になるものはないので出力してください。そして、有害にならないように出力してください。
　
「
1. インターネット・情報通信・テクノロジー関連用語
2. インターネット・情報通信・テクノロジーサービス名・商品名
3. エンタメ・ネット文化・ネットスラング関連用語
4. 社会・生活・衣食住関連用語
5. 経済・ビジネス・政治・社会問題用語
6. 薬品名や医療専門用語
7. アイドルグループ名
8. AKB48派生グループや坂道グループのような秋元康プロデュースのアイドルグループのメンバー名、または、その元メンバー名
9. その他のアイドルの個人名：AKB48派生グループや坂道グループのような秋元康プロデュースのアイドルグループメンバー名と
   元メンバーも除外する
10. 声優
11. 俳優
12. AV俳優・AV女優
13. バンド・歌手名・グループ名
14. その他有名人：スポーツ選手・芸人・タレント・小説家など。アイドル、声優、俳優、AV俳優、AV女優、歌手は含まない
15. アニメ・ゲーム関連用語
16. キャラクター名：アニメ・ゲーム関係を除く
17. メディア・情報サイト名
18. コンテンツ・作品名
19. 地名・施設・駅名・インフラ名
20. 競走馬名
21. 飲食関係サービス・商品名
22. その他の組織名
23. その他のサービス・商品名
24. 記号や絵文字
」     
　
回答形式：
*以下のタブ区切り出力形式意外の文字列は一切出力しないでください。
*[単語]の前に数字のリストをつけないでください。
出力形式：
[単語]　分類名　分類[1-24]　理由
回答例：
スマホ　テクノロジー関連用語　1　スマホは，…
山田花子　その他有名人　13　山田花子は，..
くまねこ　キャラクター名　16　くまねこは，..
　
単語リスト:
[単語]　参照情報
<単語>　[要約]<Wikipediaの要約文>[検索]<Web検索結果>[本文]<Wikipediaの本文>[ブログ]<ブログ本文>
\end{JPlisting}
\end{minipage}
\caption{The original Japanese-language prompt for Classification 2 (24-way genre classification). See Section \ref{sec_llm} and \ref{app_sec_llm}. The English translation is provided in Section \ref{app_sec_prompt_english}. } 
 \label{app_fig_llm_prompt_2_jp}         
\end{figure*} 
\clearpage
\numberwithin{equation}{section} 
\setcounter{equation}{0} 
\renewcommand{\theequation}{H\arabic{equation}}
%

\section{Lists the Keywords Used in Figures and Tables and their Corresponding Japanese Notation}
\label{app_sec_japanese_base}
Appendix H lists the keywords used in figures and tables and their corresponding Japanese notation.
\subsection{Summary of Keywords in the Figures of the Main Text and their Japanese Notation}
\label{app_sec_fig_word}
This section describes the Japanese notation for the keywords featured in the figures of the main text.
In the actual analysis, tasks such as counting, searching, and analysis were performed using the keywords exactly in their Japanese notation as listed in the tables below.
The keywords are summarized in Table~\ref{app_tab_all_time_series} for Fig.~\ref{fig_all_time_series},
Table~\ref{app_tab_scale_time_series} for Fig. \ref{fig_scale_time_series},
Table~\ref{app_tab_boxcox_time_series} for Fig.~\ref{fig_boxcox_time_series},
and Table~\ref{app_tab_n_count2} for Fig. \ref{fig_n_count2}.

\begin{table*}[t]
\centering
\begin{tabular}{|c|c|l|l|l|r|r|r|r|r|l|} 
\hline
Fig. & Symbol & Keyword (EN) & Keyword (JP) &  Romanization & $\alpha_1$ & $R_1$ & $\alpha_2$ & $R_2$ & Change Pt. & Meaning \\
\hline
(a) & $\blacktriangle$ & Low-cost SIM & 格安SIM & \textit{Kakuyasu SIM} & 0.77 & 0.22 & - & - & - & Inexpensive SIM card. \\
\hline
(b) & $\blacktriangle$ & Smartphone & スマホ & \textit{Sumaho} & 1.03 & 0.30 & -0.72 & 4411 & 08/2009 & Mobile phone. \\
\hline
(c) & $\blacktriangle$ & red circle emoji & \emoji{hollow-red-circle} & \textit{Akai maru} & 0.066 & 0.59 & 1.77 & $6.28 \times 10^{-4}$ & 09/2017 & Emoji character. \\
\hline
\end{tabular}
\caption{Summary of keyword time series examples (Fig. \ref{fig_all_time_series}).}
\label{app_tab_all_time_series}
\end{table*}

\begin{table*}[ht]
\centering
\begin{tabular}{|c|c|l|l|l|r|r|l|}
\hline
Fig. & Symbol & Keyword (EN) & Keyword (JP) & Romanization & $\alpha_1$ & $R_1$ & Meaning \\
\hline
(a) & black triangle & Erika Ikuta & 生田絵梨花 & \textit{Ikuta Erika} & 0.00 & 0.59 & Japanese idol name \\
(a) & red cross & Niconico Seiga & ニコニコ静画 & \textit{Nikoniko Seiga} & 0.09 & 0.12 & illustration-sharing service \\
(a) & green cross & Chuo Ward, Sagamihara City & 相模原市中央区 & \textit{Sagamihara-shi Chūō-ku} & -0.02 & 0.23 & new place name \\
(a) & blue square & Labor pain taxi & 陣痛タクシー & \textit{Jintsū takushī} & -0.08 & 0.072 & maternity taxi service \\
(a) & light-blue circle & beLEGEND & ビーレジェンド & \textit{Bīrejendo} & 0.01 & 0.097 & protein supplement brand \\
\hline
(b) & black triangle & Tablet device & タブレット端末 & \textit{Taburetto tanmatsu} & 0.47 & 0.34 & - \\
(b) & red cross & Crowdfunding & クラウドファンディング & \textit{Kuraudofandingu} & 0.53 & 0.22 & - \\
(b) & green cross & BABYMETAL & BABYMETAL & \textit{Bebīmetaru} & 0.55 & 0.23 & metal idol group \\
(b) & blue square & Rescue cat cafe & 保護猫カフェ & \textit{Hogo neko kafe} & 0.50 & 0.14 & - \\
(b) & light-blue circle & Anchor emoji & \emoji{anchor} & \textit{Ikari no emoji} & 0.45 & 0.12 & anchor emoji \\
\hline
(c) & black triangle & Shale gas & シェールガス & \textit{Shēru gasu} & 1.03 & 0.15 & - \\
(c) & red cross & Acai bowl & アサイーボウル & \textit{Asaī bōru} & 0.98 & 0.10 & - \\
(c) & green cross & Fumika Baba & 馬場ふみか & \textit{Baba Fumika} & 0.91 & 0.11 & actress \\
(c) & blue square & Tent emoji & \emoji{tent} & \textit{Tento no emoji} & 1.01 & 0.076 & outdoors-related emoji \\
(c) & light-blue circle & Net-juu & ネト充 & \textit{Netojū} & 0.93 & 0.15 & slang: 'fulfilled online life' \\
\hline
\end{tabular}
\caption{Summary of Keyword Time Series Examples (Fig. \ref{fig_scale_time_series})}
\label{app_tab_scale_time_series}
\end{table*}

\begin{table*}[ht]
\centering
\begin{tabular}{|c|l|l|l|r|r|l|}
\hline
Symbol & Keyword (EN) & Keyword (JP) &  Romanization & $\alpha_1$ & $R_1$ & Meaning \\
\hline
black triangle & South Ward, Sagamihara & 相模原市南区 & \textit{Sagamihara-shi Minami-ku} & 0.17 & 0.39 & new place name \\
\hline
red cross & SoundCloud & サウンドクラウド & \textit{Saundokuraudo} & 0.180 & 0.15 & music sharing site \\
\hline
green cross & Instagrammer & インスタグラマー & \textit{Insutaguramā} & 0.44 & 2.38 & person popular on Instagram \\
\hline
blue square & Komyusho & コミュ障 & \textit{Komyushō} & 0.64 & 0.42 & Slang: poor at communication \\
\hline
light-blue circle & MicroUSB & マイクロUSB & \textit{MaikuroUSB} & 0.80 & 0.080 & electronic interface \\
\hline
gray hollow circle & Microplastics & マイクロプラスチック & \textit{Maikuropurasuchikku} & 1.1 & 0.0043 & small plastic debris \\
\hline
\end{tabular}
\caption{Summary of Keyword Time Series Examples (Fig. \ref{fig_boxcox_time_series})}
\label{app_tab_boxcox_time_series}
\end{table*}

\begin{table*}[ht]
\centering
\begin{tabular}{|c|c|l|l|l|r|r|r|r|r|l|}
\hline
Fig. & Sym. & Keyword (EN) & Keyword (JP) & Romanization & $\alpha_1$ & $R_1$ & $\alpha_2$ & $R_2$ & Change & Meaning \\
\hline
(a) & $\blacktriangle$ & Kenshi Yonezu & 米津玄師 & \textit{Kenshi Yonezu} & -0.077 & 0.18 & -0.12 & 6.40 & 11/2016 & singer \\
\hline
(b) & $\blacktriangle$ & Arafifu & アラフィフ & \textit{Arafifu} & 0.77 & 0.36 & 1.15 & 0.018 & 04/2009 & slang: around age 50\\
\hline
(c) & $\blacktriangle$ & Facebook Messenger & Facebookメッセンジャー & \textit{Facebook Messenger} & 0.83 & 0.067 & -0.20 & 1.0 & 03/2015 & messaging app \\
\hline
\end{tabular}
\caption{Summary of Keyword Time Series Examples (Fig. \ref{fig_n_count2})}
\label{app_tab_n_count2}
\end{table*}

\subsection{Word Lists and Co-occurrence Lists with Japanese Notation}
\label{app_sec_llm_word}
The Japanese notation for the word list categorized by $\alpha_i$ (corresponding to Table~\ref{tab:examples_by_alpha}) is provided in Table~\ref{tab:examples_by_alpha_japanese}.
The Japanese notation for the co-occurring word list (corresponding to Table~\ref{tab:cooccurrence_alpha_small_vs_large}) is provided in Table~\ref{tab:cooccurrence_alpha_small_vs_large_japanese}.
The data analysis is conducted in Japanese. 

\begin{table*}[t]
\centering
\begin{tabular}{p{0.31\textwidth}p{0.31\textwidth}p{0.31\textwidth}}
\hline
\textbf{$\alpha \approx 0$} & \textbf{$\alpha \approx 0.5$} & \textbf{$\alpha \approx 1.0$} \\
\textbf{(linear growth, local/niche)} & \textbf{(sub‑exponential, diverse/mixed)} & \textbf{(exponential growth, global/shared)} \\
\hline
\begin{minipage}[t]{\linewidth}
\textbf{[1. Proper names (persons, organizations, places)]} \\
\textcolor{red}{{\Large $\blacktriangle$}} 生田絵梨花 (-0.00; AKB48-related group member), \textcolor{red}{{\Large $\blacktriangle$}} 新内眞衣 (0.02; AKB48-related group member), \textcolor{red}{{\Large $\blacktriangle$}} 中元日芽香 (0.08; AKB48-related group member), \textcolor{red}{{\Large $\blacktriangle$}} 寺田蘭世 (0.09; AKB48-related group member), \textcolor{magenta}{{\huge $\bullet$}} 岡山市北区 (0.02; place name), \textcolor{magenta}{{\huge $\bullet$}} 相模原市緑区 (-0.10; place name), \textcolor{magenta}{{\huge $\bullet$}} 相模原市中央区 (-0.01; place name), \textcolor{magenta}{{\huge $\bullet$}} 嚴島神社 (0.08; a Shinto shrine on Itsukushima Island), 早見沙織 (0.08; voice actress) \\
\vspace{0.5\baselineskip}

\textbf{[2. Culture / Media / Subculture / Slang]} \\
\textcolor{orange}{$\star$} AppBank (-0.08; Japanese app review and media site), \textcolor{orange}{$\star$} appbank (-0.08; Japanese app review and media site, all lower case letters), \textcolor{orange}{$\star$} NewsPicks (-0.09; Japanese business news platform), \textcolor{orange}{$\star$} JBpress (-0.06; Japanese online business magazine), ニコニコ静画 (0.09; Image sharing service), 既読スルー (0.07; message seen but not replied to), クソザコナメクジ (-0.06; internet slang: Fucking useless weakling), ハクスラ (-0.08; video game genre focusing on combat), S.H.フィギュアーツ (-0.09; Bandai's action figure line), ブンドド (0.03; playing with action figures and making sound effects) \\
\vspace{0.5\baselineskip}

\textbf{[3. ICT / Technology]} \\
ピンタレスト (-0.06; visual discovery engine) \\
\vspace{0.5\baselineskip}

\textbf{[4. Society / Lifestyle]} \\
セブンプレミアム (0.01; private label products), アイデアソン (-0.10; idea generation workshop), galaxxxy (0.02; Japanese fashion brand), 陣痛タクシー (-0.08; taxi service for pregnant women going into labor), ビーレジェンド (0.01; Japanese sports nutrition brand) \\
\vspace{0.5\baselineskip}

\textbf{[5. International / Public / Affairs]} \\
(no terms in this regime) \\
\end{minipage}
&
\begin{minipage}[t]{\linewidth}
\textbf{[1. Proper names (persons, organizations, places)]} \\
\textcolor{red}{{\Large $\blacktriangle$}} 齋藤飛鳥 (0.46; AKB48-related group member), 蒼井翔太 (0.45; Japanese singer and voice actor), カーダシアン (0.48; family name of American reality television personalities), 徳井青空 (0.41; Japanese voice actress and singer), 己龍 (0.52; Japanese visual kei band), \textcolor{blue}{$\blacklozenge$} BABYMETAL (0.55; Japanese girl metal band), 水曜日のカンパネラ (0.47; Japanese girl music group) \\
\vspace{0.5\baselineskip}

\textbf{[2. Culture / Media / Subculture / Slang]} \\
\textcolor{blue}{$\blacklozenge$} インスタグラマー (0.44; person popular on Instagram), ネット民 (0.55; active internet user), \textcolor{blue}{$\blacklozenge$} クラウドファンディング (0.53; funding a project by raising small amounts of money from many people), \textcolor{blue}{$\blacklozenge$} 塩対応 (0.41; Slang: Giving the cold shoulder), ぬい撮り (0.51; taking photos with stuffed animals) \\
\vspace{0.5\baselineskip}

\textbf{[3. ICT / Technology]} \\
タブレット端末 (0.47; portable computer with touchscreen), スマホアプリ (0.42; application software for mobile devices), Twitterアカウント (0.56; user profile on Twitter), プッシュ通知 (0.48; message sent by an app to a device), トリップアドバイザー (0.45; travel website for reviews and bookings), Facebookアカウント (0.46; user profile on Facebook), ココナラ (0.54; Japanese online marketplace for skills/services), Jimdo (0.57; website builder platform) \\
\vspace{0.5\baselineskip}

\textbf{[4. Society / Lifestyle]} \\
楽天銀行 (0.60; Japanese online bank), 塚田農場 (0.53; Japanese izakaya restaurant chain), “\emoji{high-voltage}” (0.56; emoji lightning bolt: symbol for electricity or quickness), “\emoji{baseball}” (0.53; emoji baseball: symbol for the sport of baseball), “\emoji{anchor}” (0.45; emoji anchor: symbol for stability or nautical themes), 地域おこし協力隊 (0.46; program for urban residents to support rural areas), \textcolor{blue}{$\blacklozenge$} 保護猫カフェ (0.50; cafe where rescued cats can be adopted), ドライエイジング (0.46; meat preservation technique), セレコックス (0.47; anti-inflammatory drug), タケキャブ (0.49; drug for acid-related disorders) \\
\vspace{0.5\baselineskip}

\textbf{[5. International / Public / Affairs]} \\
(no terms in this regime) \\
\end{minipage}
&
\begin{minipage}[t]{\linewidth}
\textbf{[1. Proper names (persons, organizations, places)]} \\
馬場ふみか (0.91; actress/model), 赤崎千夏 (0.93; voice actress), 崎山つばさ (1.03; actor/singer), 柚香光 (0.91; Takarazuka Revue star), 居酒屋はなこ (0.94; Japanese pub chain), 知多娘。 (0.92; local idol group/mascot), 倉山満 (0.93; historian/commentator), SHU-I (0.94; South Korean boy band), \textcolor{magenta}{{\huge $\bullet$}} TIAT (0.92; Tokyo International Air Terminal Corporation), 朱立倫 (1.05; Taiwanese politician) \\
\vspace{0.5\baselineskip}

\textbf{[2. Culture / Media / Subculture / Slang]} \\
XFLAG (0.90; mixi's gaming brand), 大包平 (0.98; a famous Japanese sword), ハンガー・ゲーム (0.95; novel/film series), \textcolor{blue}{$\blacklozenge$} つらたん (1.01; slang for “it’s tough/sad”), ぼんぼんりぼん (0.93; Sanrio character), “\emoji{part-alternation-mark}” (0.95; Part alternation mark; Unicode U+303D: often for traditional Japanese poetry), ネト充 (0.93; slang for someone who enjoys online life), \textcolor{blue}{$\blacklozenge$} インスタ映え (1.03; visually appealing for Instagram), \textcolor{blue}{$\blacklozenge$} オンラインサロン (0.94; paid online community) \\
\vspace{0.5\baselineskip}

\textbf{[3. ICT / Technology]} \\
\textcolor{blue}{$\blacklozenge$} エアビーアンドビー (0.99; online lodging marketplace), Twitterクライアント (1.07; app for Twitter access), Udemy (0.92; online learning platform), モバツイ (0.97; a former mobile Twitter client), \textcolor{cyan}{$\blacksquare$} イーサリアム (0.92; cryptocurrency/blockchain platform), \textcolor{cyan}{$\blacksquare$} VRゴーグル (1.03; virtual reality headset) \\
\vspace{0.5\baselineskip}

\textbf{[4. Society / Lifestyle]} \\
\textcolor{blue}{$\blacklozenge$} アサイーボウル (0.98; a fruit bowl with acai berries), \textcolor{blue}{$\blacklozenge$} レイコップ (1.02; brand of futon cleaner), \textcolor{blue}{$\blacklozenge$} “\emoji{tent}” (1.01; emoji Tent: camping equipment symbol), 公認心理師 (0.99; national qualification) \\
\vspace{0.5\baselineskip}

\textbf{[5. International / Public / Affairs]} \\
\textcolor{cyan}{$\blacksquare$} シェールガス (1.03; natural gas from shale formations), \textcolor{cyan}{$\blacksquare$} マイクロプラスチック (1.05; tiny plastic debris), \textcolor{cyan}{$\blacksquare$} シェールガス革命 (0.96; major energy shift) \\
\end{minipage}
\\
\hline
\end{tabular}
\caption{
\revisecolor{black}{
Examples of keywords classified by the growth-shape parameter $\alpha_i$ (Japanese version), reorganized into five semantic categories: (1) proper names (persons, organizations, places), (2) culture/media/subculture, (3) ICT/technology, (4) society/lifestyle, and (5) international/public/affairs. Columns correspond to $\alpha_i \approx 0$ (linear growth, local/niche), $\alpha_i \approx 0.5$ (sub‑exponential, diverse/mixed), and $\alpha_i \approx 1.0$ (exponential growth, global/shared).
The left column ($\alpha_i \approx 0$) is dominated by niche and local terms. Specifically, these include AKB48 (a Japanese pop idol group) -related person names (\textcolor{red}{{\Large $\blacktriangle$}}), place names (\textcolor{magenta}{{\huge $\bullet$}}), and game/anime/subculture terms. In addition, some news sites and media platforms (\textcolor{orange}{$\star$}) also stand out. In contrast, the right column ($\alpha_i \approx 1.0$) contains globally shared, high‑profile terms, such as global news terms (\textcolor{cyan}{$\blacksquare$}) and domestic buzzwords (\textcolor{blue}{$\blacklozenge$}). The middle column ($\alpha_i \approx 0.5$) shows a diverse mixture of topics.
An English version of this table is provided in the main text (Table~\ref{tab:examples_by_alpha}).
}
}
\label{tab:examples_by_alpha_japanese}
\end{table*}

\begin{table*}[t]
\centering
\begin{tabular}{p{0.48\textwidth}p{0.48\textwidth}}
\hline
\textbf{Small $\alpha_i$} & \textbf{Large $\alpha_i$} \\
\textbf{(linear growth, local/niche)} & \textbf{(exponential growth, global/shared)} \\
\hline
\begin{minipage}[t]{\linewidth}
\textbf{[AKB48-centered idol culture terms]} \\
研究生 (-0.33, 0.0010); 生誕祭 (-0.25, 0.0014); 撮影会 (-0.21, 0.015); トレーニング (-0.24, 0.012); ダンサー (-0.24, 0.025); キャスター (-0.27, 0.019); 高木 (-0.27, 0.0081); 松村 (-0.23, 0.042); 堀 (-0.22, 0.023); 石田 (-0.22, 0.016); ひろ (-0.22, 0.048); 綾 (-0.23, 0.022); 大賞 (-0.21, 0.031); 新番組 (-0.20, 0.045); アンダー (-0.23, 0.035); 友情 (-0.24, 0.034) \\
\vspace{0.5\baselineskip}

\textbf{[Place-related terms]} \\
宮城県 (-0.24, 0.040); 神奈川 (-0.23, 0.0091); 駅前 (-0.21, 0.011); 定休 (-0.26, 0.023); TEL (-0.37, 0.00088); ピアノ (-0.21, 0.0066) \\
\vspace{0.5\baselineskip}

\textbf{[Game, anime, subculture–related terms]} \\
形態 (-0.25, 0.012); クリエーター (-0.24, 0.031); x (-0.24, 0.018); 降臨 (-0.24, 0.011); 装備 (-0.23, 0.015); マルチ (-0.23, 0.026); 2期 (-0.21, 0.024); 遊べる (-0.22, 0.017) \\
\vspace{0.5\baselineskip}

\textbf{[Other]} \\
Yahoo!ニュース (-0.21, 0.020); カロリー (-0.23, 0.044); 書き込み (-0.23, 0.041); SM (-0.22, 0.019); お笑い (-0.21, 0.030); 入門 (-0.21, 0.041); ジャン (-0.21, 0.032), 3番 (-0.23, 0.049) \\
\end{minipage}
&
\begin{minipage}[t]{\linewidth}
\textbf{[Global -related terms]} \\
世界的 (0.27, 0.0019); 仏 (0.22, 0.048); ヨーロッパ (0.23, 0.042) \\
\vspace{0.5\baselineskip}

\textbf{[Economic and Political terms]} \\
利益 (0.23, 0.024); 借り手 (0.22, 0.018); 取引 (0.24, 0.024); 規制 (0.23, 0.010); 設ける (0.24, 0.016); 成立 (0.28, 0.0051); 党 (0.27, 0.018); 独立 (0.23, 0.034); 軸 (0.21, 0.017) \\
\vspace{0.5\baselineskip}

\textbf{[News media -related terms]} \\
プレス (0.22, 0.033); OA (0.22, 0.0093); オンエア (0.20, 0.023); ITmedia (0.20, 0.039); Janetter (0.22, 0.014) \\
\vspace{0.5\baselineskip}

\textbf{[Other]} \\
お姉ちゃん (0.21, 0.022); サウンド (0.20, 0.040); 泊 (0.21, 0.014); オフィシャル (0.20, 0.030); プレミア (0.23, 0.023); 軒 (0.23, 0.011); 東日本大震災 (0.25, 0.021); 本田 (0.24, 0.024) \\
\end{minipage}
\\
\hline
\end{tabular}
\caption{
\revisecolor{black}{
Co-occurring words associated with the growth-shape parameter $\alpha_i$ (Japanese version), organized by semantic categories that differ between small and large $\alpha_i$ regimes. Co-occurring words are those appearing within 40 characters of a target new word with a given $\alpha_i$. The left column lists words that tend to co-occur with terms having small $\alpha_i$ (negative correlation), grouped into AKB48 (a Japanese pop idol group)-related terms, place-related terms, game/anime/subculture-related terms, and others. The right column lists words that tend to co-occur with terms having large $\alpha_i$ (positive correlation), grouped into global/public affairs terms, news media/journalism terms, and others. Entries are listed as: Co-occurring word (Kendall's $\tau$, $p$-value). The table shows that small $\alpha$ co-occurrences are dominated by niche, local terms such as AKB48-related terms, place names, and anime/game-related terms, whereas large $\alpha$ co-occurrences are dominated by globally shared, newsworthy terms such as global and political/economic terms. The method for calculating correlations is provided in Appendix Section~\ref{app_sec_cooccurrence}. An English version is provided in the main text (Table~\ref{tab:cooccurrence_alpha_small_vs_large}).
}
}
\label
{tab:cooccurrence_alpha_small_vs_large_japanese}
\end{table*}

\end{CJK*}
\clearpage
\bibliographystyle{apsrev4-2}
\bibliography{model14_pre}

\end{document}